BREAKTHROUGH
INITIATIVES

# Venus Life Finder Mission Study

A suite of mission concepts to explore Venus to study habitability and to potentially find life

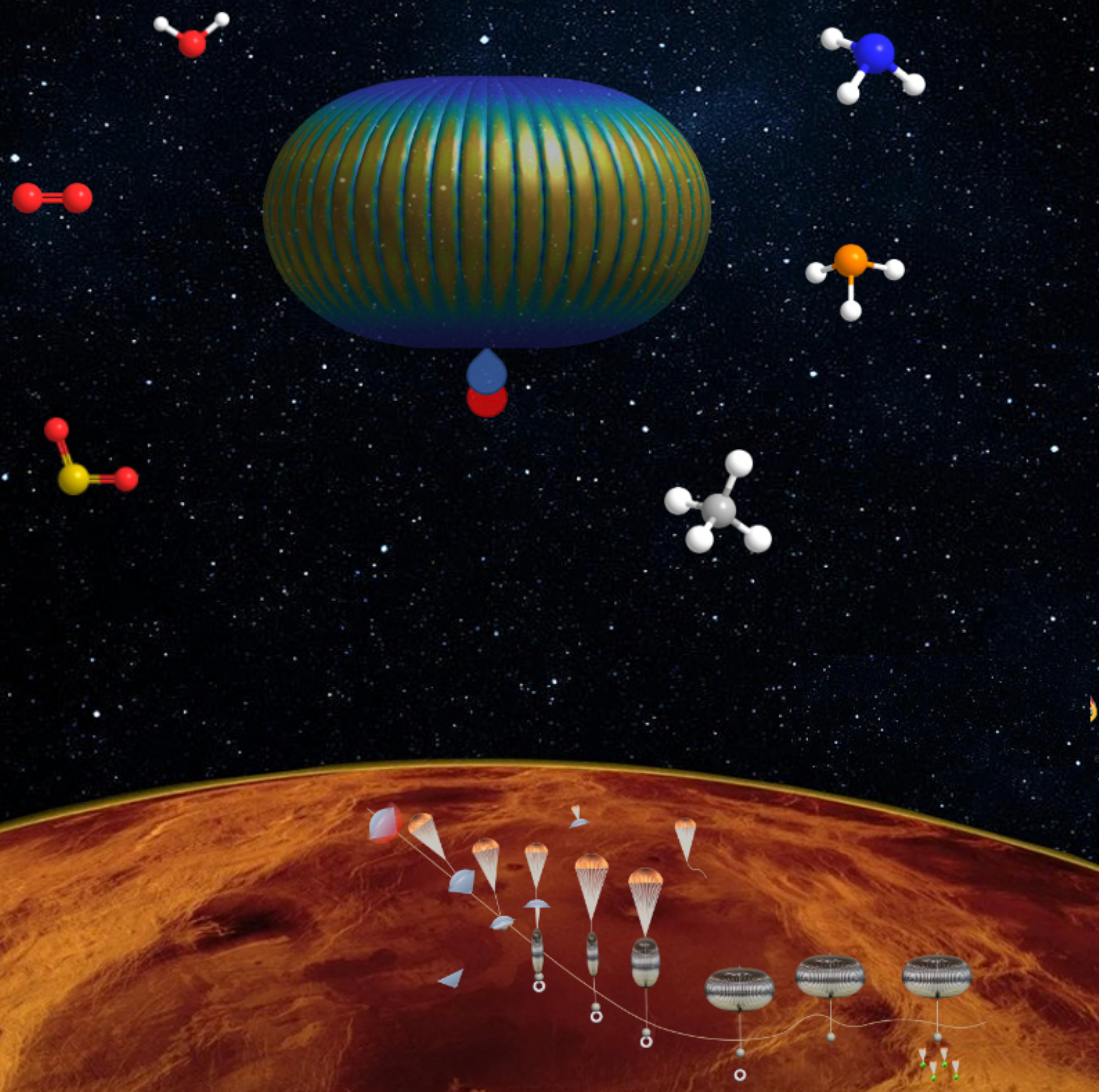

Led by Sara Seager and Janusz Petkowski at the
Massachusetts Institute of Technology

Sponsored by Breakthrough Initiatives

# VENUS MISSION CONCEPT STUDY

Sponsored by the Breakthrough Initiatives

**December 10, 2021**

## TEAM MEMBERS

| Principal Investigator | |
|---|---|
| Sara Seager, Massachusetts Institute of Technology | |
| **Deputy Principal Investigator** | |
| Janusz J. Petkowski, Massachusetts Institute of Technology | |
| **Core Team** | |
| Christopher E. Carr, Georgia Tech | David Grinspoon, Planetary Science Institute |
| Bethany Ehlmann, Caltech | Sarag Saikia, Spacefaring Technologies |
| Rachana Agrawal, Purdue | Weston Buchanan, Purdue |
| Monika Weber, Fluid-Screen | Richard French, Rocket Lab |
| Pete Klupar, Breakthrough Initiatives | Simon P. Worden, Breakthrough Initiatives |

### Acknowledgements

We thank the many individuals who as part of an extended team participated in active discussions that contributed to the report: Ricardo Arevalo (University of Maryland, USA), Archit Arora (Purdue University, USA), Armando Azua-Bustos (Center of Astrobiology, Spain), Laurie Barge (Jet Propulsion Laboratory, USA), William Bains (Massachusetts Institute of Technology, USA and Cardiff University, UK), Jolie Berkow (Massachusetts Institute of Technology, USA), Ewa Borowska (University of Warsaw, Poland), Morgan Cable (Jet Propulsion Laboratory, USA), Henderson "Jim" Cleaves (Tokyo Institute of Technology, Japan), Maxim de Jong (Thin Red Line Aerospace, Canada), Graham Dorrington (Royal Melbourne Institute of Technology, Australia), Daniel Duzdevich (Harvard University, USA), Athul P. Girija (Purdue University, USA), Jane Greaves (Cardiff University, UK), Chris Isaac (SpaceAM, UK), Laila Kaasik (University of Tartu, Estonia), Iaroslav Iakubivskyi (University of Tartu, Estonia), Jordan McKaig (Georgia Tech, USA), Rachel Moore (Georgia Tech, USA), Kenneth Nealson (University of Southern California, USA), Grace Ni (University of Maryland, USA), Mihkel Pajusalu (University of Tartu, Estonia), Ida Rahu (University of Tartu, Estonia), Sukrit Ranjan (Northwestern University, USA), Paul Rimmer (Cambridge University, UK), Andreas Riedo (University of Bern, Switzerland), Robert Weber (Fluid-Screen, USA), Peter Wurz (University of Bern, Switzerland), Rakesh Mogul (Cal Poly Pomona, USA), Niels Ligterink (University of Bern, Switzerland), Jan Spacek (Firebird Biomolecular Sciences, USA)

A special thanks to the following for detailed contributions: Darrel Baumgartner (Droplet Measurement Technologies, USA), Chris Webster and Amy Hofmann (Jet Propulsion Laboratory, USA), Kris Zacny, Isabel King, and Kathryn Bywaters (Honeybee Robotics, USA), Shawn Whitehead (Scoutek, UK), Steven Benner and Gage Owens (Firebird Biomolecular Sciences, USA), Arnan Mitchell (Royal Melbourne Institute of Technology University, Australia), Jan Dziuban (Wroclaw University of Science and Technology, Poland), Piotr Dziuban and Grzegorz Zwolinski (SatRevolution, Poland) Stephen Eisele, John Fuller, and Brett Perry (Virgin Orbit).

Additional thanks to James Longuski (Purdue University, USA), Kevin Baines and Jim Cutts (Jet Propulsion Laboratory, USA), Jamie Drew and Dillon O'Reilly (Breakthrough Initiatives, USA), Reine Johansen, Marcus Murbach, and Periklis Papadopoulos (NASA Ames Research Center, USA), Elizabeth Kimmel, Benjamin Boatright, and Rick Fleeter (Brown University, USA), Jonathan Tobelmann and Elric Saaski (Research International, USA), Peter Gao (Carnegie Institution for Science, USA), Kayla Bauer (Massachusetts Institute of Technology, USA).

We acknowledge our industry contributors: Advanced Space, Droplet Measurement Technologies, Firebird Biomolecular Sciences, Fluid-Screen, Honeybee Robotics, Organix, Scoutek, Spacefaring Technologies, and Thin Red Line Aerospace.

# TABLE OF CONTENTS

# Executive Summary

Finding evidence of extraterrestrial life would be one of the most profound scientific discoveries ever made, catapulting humanity into a new epoch of cosmic awareness.

The Venus Life Finder (VLF) Missions are a series of three direct Venus atmosphere probes designed to assess the habitability of the Venusian clouds and to search for signs of life and life itself. The VLF missions are a focused, optimal set of missions that can be launched quickly and with relatively low cost. The mission concepts come out of an 18-month study by an MIT-led worldwide consortium. The study was partially funded by the Breakthrough Initiatives.

## The New Opportunity

The concept of life in the Venus clouds is not new, having been around for over half a century. What is new is the opportunity to search for life or signs of life directly in the Venus atmosphere with scientific instrumentation that is both significantly more technologically advanced and greatly miniaturized since the last direct in situ probes to Venus' atmosphere in the 1980s.

Venus is a compelling planet to search for signs of life because of the habitable temperatures in the cloud layers and because of many atmospheric chemical anomalies that together are suggestive of unknown chemistry and possibly the presence of life (Figure ES-1).

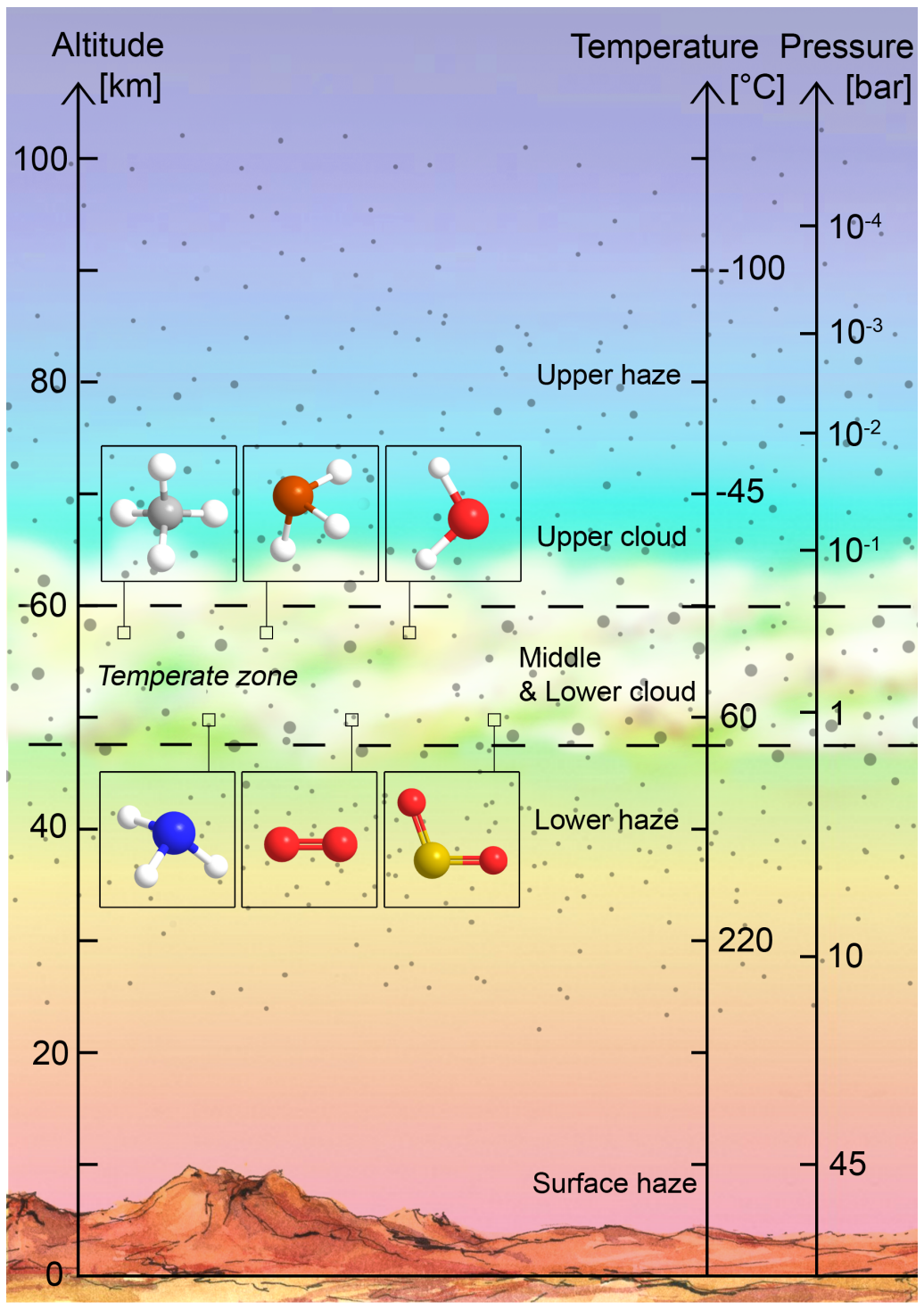

Figure ES-1. Schematic of Venus' atmosphere. The cloud cover on Venus is permanent and continuous and vertically extensive. The middle and lower cloud layers have temperatures suitable for life. Selected molecules relevant to habitability or life and with unexplained presence or vertical abundances are shown. Atom colors are as follows: H = white; C = grey; P = orange; O = red; N = blue; S = yellow. Figure adapted from [14]. Figure credit: J. Petkowska.

The VLF series of missions are directly formulated to assess the habitability of Venusian clouds and to search for signs of life and life itself. In this study we report on a focused set of missions that can be launched quickly and with relatively low cost. While NASA and ESA have recently selected missions to visit Venus at the end of the 2020s (VERITAS, DAVINCI+ and EnVision), these missions are for general studies about the planet's properties and do not address the habitability questions VLF is aimed for.

Remarkably, it has been nearly 40 years since the last Venus in situ measurements. Russian Vega balloons and landers flew in 1985 and the US Pioneer Venus probes flew in 1978. The entire scientific field of Astrobiology has sprung up in the interim. A privately funded series of focused missions to Venus will take advantage of an opportunity for high-risk, high-reward science, which stands to possibly answer one of the greatest scientific mysteries of all, and in the process pioneer a new model of private/public partnership in space exploration.

## New Findings for Life's Survival on Venus

The VLF mission team set out to take a fresh look at the Venus atmosphere's suitability for life, given the very harsh Venus atmosphere conditions, namely the dryness and the extreme acidity of the cloud droplets which are made of sulfuric acid. We created chemistry and biology experiments to guide mission science objectives.

By life we mean single-celled microbial-type organisms.

First, we set out to show that the highly acidic droplets can support organic chemistry and are not necessarily sterile dead zones. The team seeded test tubes of concentrated sulfuric acid with simple organic molecules. The result was an intriguing rich and complex set of organic molecules. What this shows is that sulfuric acid droplets can support a variety of complex chemicals that can in principle be related to life's building blocks [1]. This is very intriguing because all life needs complex organic chemistry. This result is also seen in the petrochemical industry where "red oil" is a waste byproduct in fuel production.

Second, we aimed to find life materials that could survive in concentrated sulfuric acid. Although sulfuric acid is known to be destructive to many biochemicals and engineering materials, there are some that are resistant. We discovered a set of lipids that can not only survive in concentrated sulfuric acid but can self-assemble to form vesicle-like structures (in sulfuric acid concentration of 70% or lower) [2]. More work is needed to determine the membrane permeability and other chemical properties. A protective membrane is critical because most of Earth-life's biochemicals will be instantly destroyed by concentrated sulfuric acid.

Third, we propose that locally- and biologically-produced ammonia ($NH_3$) can neutralize the Venus sulfuric acid cloud droplets such that a subset of the cloud particles may be brought to an acidity level tolerable by acidophiles on Earth [3]. For specificity, concentrated sulfuric acid is billions of times more acidic than the most acidic environment on Earth. Yet with $NH_3$ neutralization, the Venus cloud particles will be brought to about pH = 0 or 1. This theory is motivated by the suggested presence of $NH_3$ by both Venera 8 [4] and the Pioneer Venus probe measurements [5] and leads to a cascade of reactions that help resolve long-standing Venus atmosphere anomalies [3].

Based on the above, our mission priorities include the search for organic compounds in the cloud particles, the measurement of the acidity of the cloud particles, and a sample return to search for cell-like structures. In addition the missions will search for biosignature gases and other indicators of life or habitable conditions.

## A Partnership for a Venus 2023 Launch

We have partnered with Rocket Lab to provide the science payload and science team to go with their 2023 Venus Mission's rocket, cruise spacecraft, and direct probe entry vehicle. The Venus direct entry vehicle aboard Rocket Lab's Photon spacecraft has room for up to 1 kg of science instrumentation for the short-duration (three minute) descent through the cloud layers.

We choose to search for organic molecules in the cloud droplets by an instrument called an autofluorescence nephelometer. This will be the first attempt to search for organic molecules in the Venus atmosphere. The instrument can also constrain the composition of the cloud particles, key to letting us know if the clouds are more habitable than pure concentrated sulfuric acid.

## A Balloon Mission and an Atmosphere Sample Return Mission

Our study found that ultimately a Venus atmosphere sample return is needed to robustly answer the compelling question, "Is there life on Venus?" We envision bringing back about one liter of Venus atmosphere and up to tens of g of Venus cloud particles for exquisitely detailed investigations here on Earth of the kind that cannot be done remotely.

Before pursuing a Venus atmospheric sample return, a cost-effective balloon-born mission to Venus is needed to establish habitability of the Venusian atmosphere and search for signs of life, as well as test technology needed for sample return. We have several international partners interested in joining our balloon mission team.

Collectively, our recommended astrobiology missions offer an efficient and high-impact route to seeking life beyond Earth, possibly enabling a truly historic first discovery.

# 1 Introduction and Motivation

**Abstract:** Long-standing Venus atmosphere anomalies point to unknown chemistry but also leave room for the possibility of life. The anomalies include several gases out of thermodynamic equilibrium, an unknown composition of large, lower cloud particles, and the unknown UV absorber. Here we first review relevant properties of the Venus atmosphere and then describe the atmosphere anomalies and how they motivate the Venus Life Finder mission science.

**New science that emerged from the study:**

- Laboratory demonstration that certain lipids can self-assemble to form vesicle-like structures in sulfuric acid, so far with concentrations explored of 70% or lower [2].
- Review of past Venus measurements that are as yet unexplained.
- A theory on $NH_3$ as a biological neutralizing agent of the Venus sulfuric acid cloud particles, resolving many of the unexplained atmosphere measurements [3].
- Hypothesis for the origin and composition of the unknown UV absorber via a complex organic carbon cycle in the Venus atmosphere [1].

## 1.1 Venus as an Abode for Life

Life in the sulfuric acid clouds of Venus has been a topic of speculation for over half a century. Published papers range widely, from science-fiction-like to invalid conjecture to legitimate hypothesis [6–15]. Today, only Venus' atmospheric cloud layers (a large region spanning from 48 to 60 km altitude) have habitable temperatures—the surface at 735 K is too hot for any plausible solvent and for most organic covalent chemistry (Figure 1-1).

By life we mean single-celled microbial-type organisms. We also assume that microbial life resides inside cloud particles such that it will be protected from a fatal net loss of liquid to the atmosphere, an unavoidable problem for any free-floating microbial life forms.

Life may have originated on the Venus surface, in a manner much like on Earth, and migrated into the clouds. Recent modeling by Way et al. [16,17] suggests the existence of habitable surface and oceans as late as ~700 Mya. Future observations by spacecraft orbiters can search for surface minerals that are indicative of past oceans. It may be, however, that Venus was always too hot for water oceans to condense, but more work is needed due to model sensitivity to cloud and atmospheric circulation feedbacks [18].

In principle life could arise in the clouds independent from the ground [19,20] with material from meteoritic input [21,22] from the asteroid belt (including Ceres, and even from Mars). Life may even have been directly seeded by impacts from Earth ejecta [23–25]. Assuming one of these possible origins, microbial life might yet exist today, and could be ancestrally related or unrelated to microbial life on Earth. Thus, we do not assume that any Venus cloud life, if it exists, is related to life as we know it.

While the Venus clouds are the right temperature for life, the environment is very harsh. The cloud droplets are composed of highly acidic sulfuric acid ($H_2SO_4$), particles that are billions of times more acidic than the most acidic environments where life is found on Earth (such as the Dallol pools in Ethiopia or Iron Mountain in California) [26,27]. No known life of any kind has been observed in droplets of such acidity, although life with protective barriers or shells could theoretically have evolved to cope with such chemistry.

The Venus cloud region is also extremely dry, approximately 50 times drier than the Atacama Desert, one of the driest places on Earth. If Venus cloud life is based on water, life would need robust strategies for water accumulation and/or production and retention, biological adaptations unlike any developed by life on Earth.

The clouds may be scarce in nutrients required for life, in particular metals needed to catalyze reactions. This, however, can be solved by nutrient accumulation and recycling of metals [14].

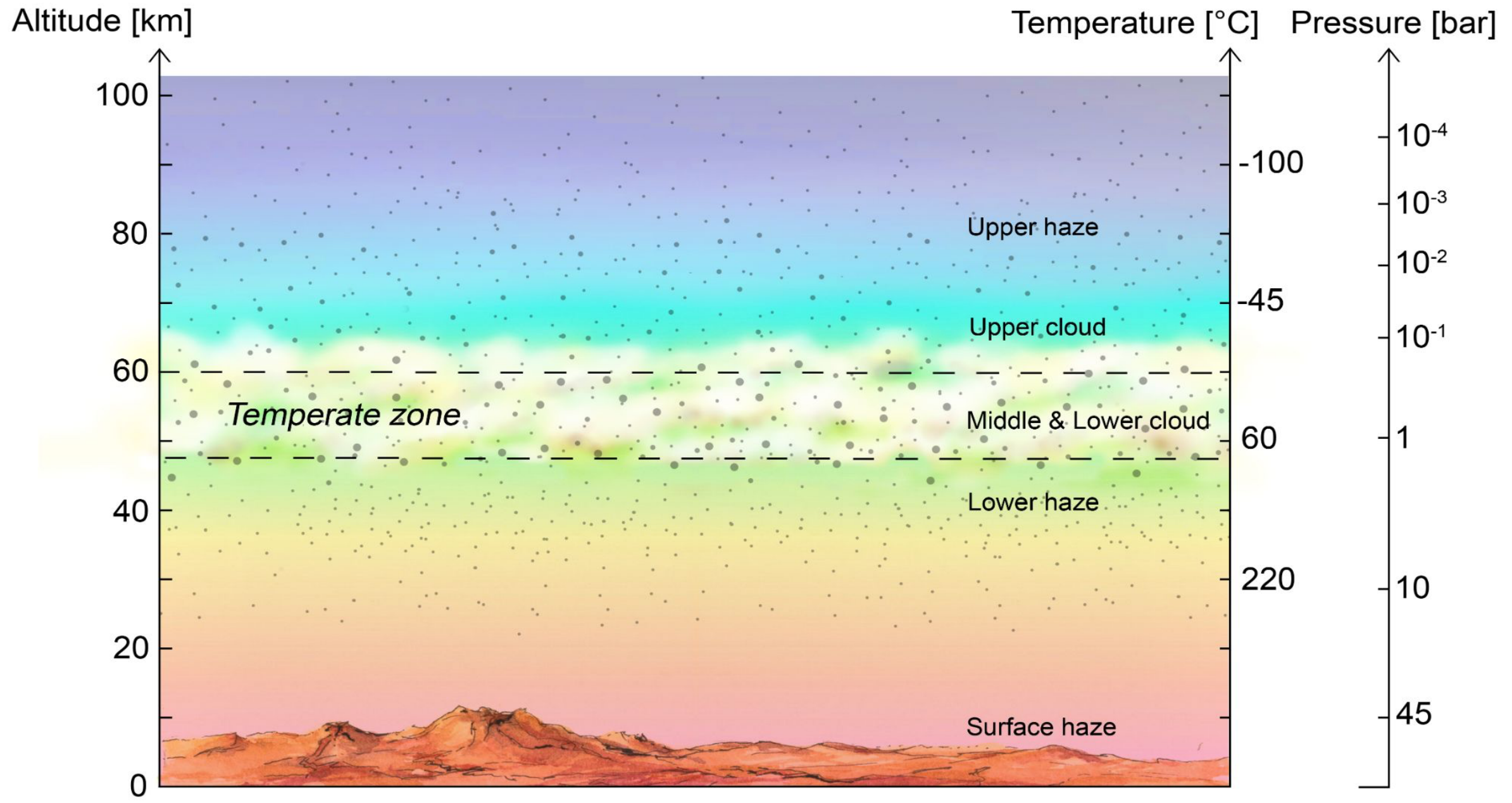

Figure 1-1. Schematic of Venus' atmosphere. The cloud cover on Venus is permanent and continuous, with the middle and lower cloud layers at temperatures that are suitable for life. The clouds extend from altitudes of ~48 km to 70 km ([92] and references therein). Figure adapted from [14]. Figure credit: J. Petkowska.

Nonetheless, in spite of these significant habitability challenges, there are data that motivate the search for life on Venus. To motivate the VLF series of missions, we review the chemistry of the atmosphere and the clouds of Venus, provide a summary of the decades-long unexplained measurements that might have implications for the habitability of the clouds, and discuss speculative, but possible survival strategies for life in the clouds.

## 1.2 Why Venus? Why Now?

There are many scientific reasons to explore the atmosphere, surface, and clouds of Venus, and many possible mission architectures and instrumented platforms to make measurements. Upon detailed study of possible missions focused on astrobiology, and in particular on life detection, it has become clear that only a focused in situ mission and/or a sample return mission carries the likelihood of providing definitive answers to the crucial questions being posed regarding atmospheric chemistry and possible presence of life.

This point is well-illustrated by the fact that the two most recent Venus missions, both orbiters, have merely confirmed and deepened many of the outstanding mysteries of the Venus cloud region described above. Both Venus Express (ESA) and Akatsuki (JAXA) have been successful orbital spacecraft which have returned valuable data on the cloud composition and structure and on atmospheric dynamics and composition. These missions have confirmed the existence of the "unknown ultraviolet absorber" but have not succeeded in identifying the substance(s) responsible for this anomalous absorption. Likewise, Venus Express and JAXA have generally confirmed the overall picture of the Venus clouds and cloud level atmosphere provided by earlier American and Russian entry probes (Pioneer Venus, the Venera and Vega missions) and have filled in many details of cloud structure and dynamics, but they have not resolved the persistent mysteries involving possible trace cloud components, unusual particle shapes, and trace atmospheric gases.

The unresolved cloud-level properties are the very ones which have fueled speculation about possible biological activity, and they have proven particularly impervious to remote investigation. For an illustrative analogy, consider trying to make a definitive determination of the presence of life in a terrestrial location such as the Atacama Desert where microbial life is present but relatively sparse. Satellite remote sensing might hint at some of the right conditions, such as moisture and temperature range, but a definitive positive detection of life would likely require a platform which could directly sample the upper layers of the desert surface, or even bring samples back to specialized laboratories for further study. This illustrates the limitations of remote sensing for biology by orbital missions. What is missing, and the logical next step, is direct sampling of this environment by an entry probe equipped with modern instrumentation.

In this context it is striking to consider that there has never been an in situ investigation of the atmosphere and clouds of Venus employing 21st century scientific instrumentation. The most recent American entry probes were the Pioneer Venus probes which flew in 1978. The Russian Vega balloons flew in 1985. It has been 36 years since any instrument from Earth was flown to directly investigate the atmosphere of Venus. The entire scientific field of Astrobiology has sprung up in the interim. We now know questions to ask which we could not have formulated in the 1980s, but even more important is the progress in scientific instrumentation and miniaturization of electronics during these decades. The time is right for a small, dedicated life-finder mission equipped with modern instrumentation to investigate the clouds and atmosphere of Venus.

At the time of this writing there is a wave of renewed interest in new Venus missions. (DAVINCI, VERITAS, and EnVision). During the VLF study, NASA and ESA announced missions to Venus. These include orbiters which will study the surface geology (the NASA VERITAS mission, to launch in 2028) and/or atmosphere (the ESA EnVision mission, to launch in 2031). Neither DAVINCI nor VERITAS nor Envision will perform in situ analysis of the cloud particle properties. Most importantly, none of these missions have an explicit goal of, or instrumentation chosen for, investigating possible biological activity. There is more interest from Russia, Japan, and India. The Russian Venera-D proposal is an ambitious multi-platform mission which would study the atmosphere, clouds, and surface.

As awareness of and interest in possible life in the Venus clouds has increased in recent years, all of these missions, in their proposal wording, have come to include possible life in the clouds as part of the motivation for their investigations, yet none of them are designed and instrumented in such a way as to explicitly address the question of extant biology. Thus, we do not see ourselves in competition with any of these efforts. As none of these missions are planned to launch before 2028, our more nimble proposed effort may return data before these others launch. In this way, we could lead what looks to be a growing wave of global interest and activity focused on exploring Earth's sister planet. With small, well-targeted efforts focused more narrowly on astrobiology/life-detection goals, we can provide missions which are complementary to the larger, infrequent missions proposed by NASA and other national space agencies.

As we write, the world is poised on the brink of a revolution in space science. Privately funded missions to Mars and Venus, and perhaps beyond, are becoming a reality. We will not supplant other efforts but rather take advantage of an opportunity for high-risk, high-reward science, which stands to possibly answer perhaps the greatest scientific mystery of all, and in the process pioneer a new model of private/public partnership in space exploration.

## 1.3 Venus Atmosphere Anomalies Motivate a Venus Astrobiology Mission

Many intriguing in situ observations of Venus have never been fully explained. Nearly all of these observations might be the result of life's activities, though, alone, life may not be required to explain any of them. These anomalies, both individual and taken together, are significant motivators to return to Venus for in situ observations.

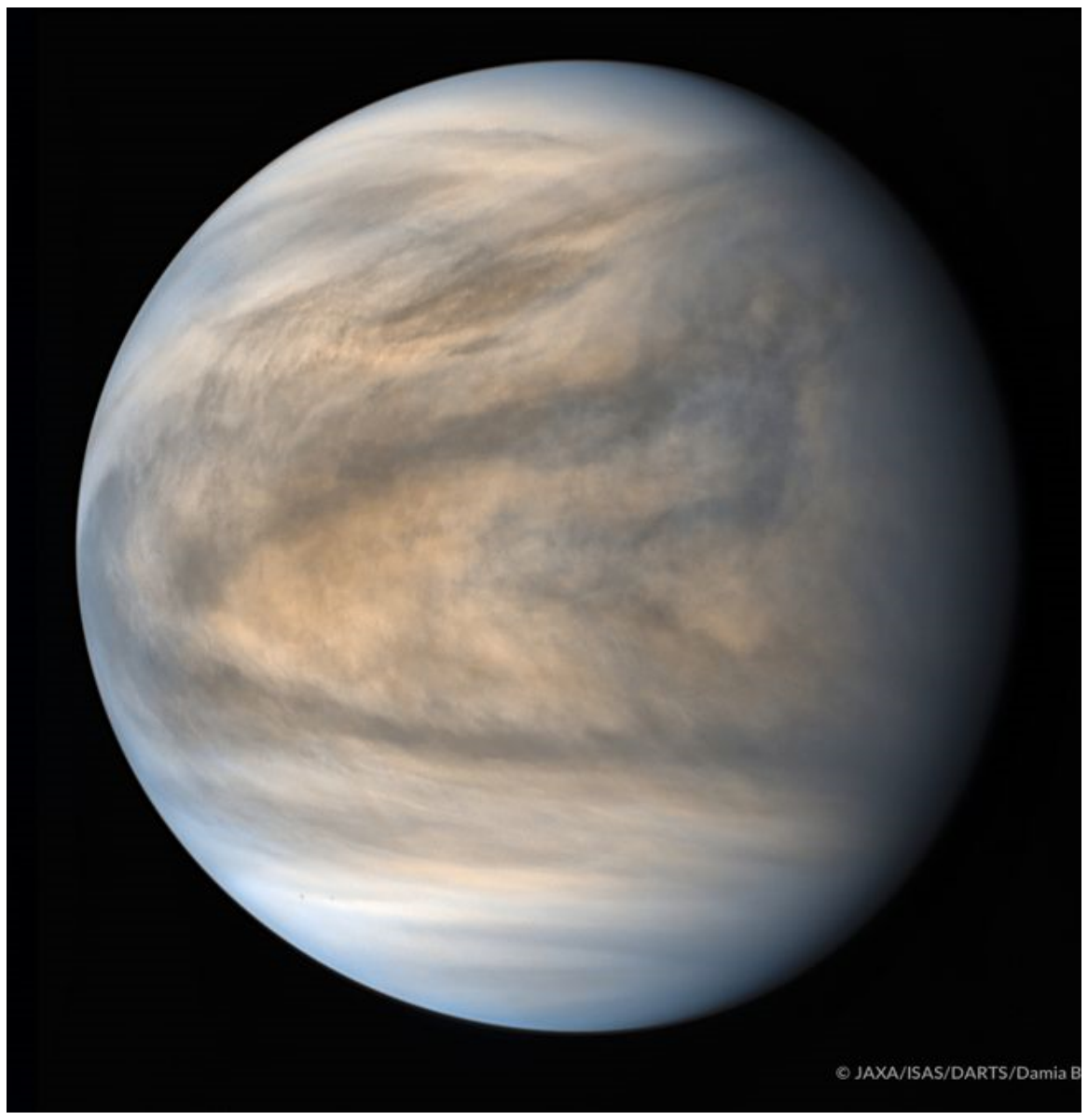

Figure 1-2. Global view of Venus in ultraviolet from Akatsuki, May 2016. A false-color image using the two ultraviolet channels from Akatsuki's UVI camera. Note the distinction between the broad, more turbulent tropical regions and the clear, smoother polar regions. Credit: JAXA / ISAS / DARTS / Damia Bouic.

### 1.3.1 The Anomalous UV Absorber

While Venus appears relatively bland and featureless at visible wavelengths, observers starting in the 1920s noticed unusual high-contrast features in the ultraviolet [28] (Figure 1-2). These features move with the ~4 day super-rotation of Venus' upper cloud deck, yet also display great variability on a wide range of temporal and spatial scales. Much effort has gone into attempting to identify the substance(s) responsible for the absorption between 320-400 nm, but no proposed candidate satisfies all of the observational constraints, leading to the oft-used descriptive term "unknown UV absorber".

After the upper clouds were identified as composed primarily of sulfuric acid droplets [29–31], efforts to identify the UV absorber largely focused on sulfur compounds including $SO_2$, $S_2O_2$, and various allotropes of elemental sulfur. Other proposals have focused on elemental chlorine, which has been identified in the upper atmosphere and shows absorption features in roughly the right spectral range. A good summary of proposed candidates is given in [32].

Despite decades of effort, and observations by two orbiting spacecraft in the 21st century (Venus Express by ESA and Akatsuki by JAXA), none of the proposed candidate molecules have been found to entirely fit the observational data. The candidate molecules either have too low abundance ($S_2O_2$) or do not entirely fit the spectral absorption profile ($FeCl_3$). The mystery of the Venusian UV absorber persists. The unknown absorber is remarkably efficient, capturing more than 50% of the solar energy reaching Venus, with consequent effects on atmospheric structure and dynamics.

Several researchers have suggested that qualities of the unknown UV absorber are possibly consistent with a signature of biological activity in the clouds. The spatial and temporal patterns are somewhat reminiscent of terrestrial algal blooms [12,13]. The great efficiency of absorption, if utilized as a photosynthetic pigment, could provide a large amount of metabolic energy [7]. The UV absorption properties are consistent with a sulfur-based "sunscreen" which both protects the interior of putative organisms from ionizing radiation and captures energy for metabolism [11]. The spectral characteristics of the Venus clouds, including the strong UV absorption, are consistent with the spectrum of certain types of terrestrial bacteria [13].

Most recently [1] proposed organic molecules inside the Venus cloud particles as the UV absorber. The proposal comes from laboratory experiments that started with simple organic molecules, including formaldehyde, in sulfuric acid. A chain of chemical reactions (some initiated by UV radiation), led to a rich variety of organic molecules. The concept is that the simple organic molecules originate from meteoritic delivery, photochemistry, or even possibly life itself [1].

| Region | Altitude (km) | Temperature (K) | Pressure (atm) | Cloud Particle Properties | | |
|---|---|---|---|---|---|---|
| | | | | Average Num. Density (n $cm^{-3}$) | Mean Diameter (μm) | Cloud Particle $H_2SO_4$ Concentration |
| Layers above upper haze | 100-110 | | | | | 100% $H_2SO_4$ |
| Upper haze | 70-90 | 225-190 | 0.04-0.0004 | 500 | 0.4 | 70% $H_2SO_4$ 30% H2O |
| Upper cloud | 56.5-70 | 286-225 | 0.5-0.04 | (1)-1500 (2)-50 | Bimodal 0.4 and 2.0 | liquid 80% $H_2SO_4$ 20% $H_2O$ |
| Middle cloud | 50.5-56.5 | 345-286 | 1.0-0.5 | (1)-300 (2)-50 (3)-10 | Trimodal 0.3, 2.5 and 7.0 | liquid 90% $H_2SO_4$ 10% $H_2O$ |
| Lower cloud | 47.5-50.5 | 367-345 | 1.5-1.0 | (1)-1200 (2)-50 (3)-50 | Trimodal 0.4, 2.0 and 8.0 | liquid 98% $H_2SO_4$ 2% $H_2O$ (or fumic) |
| Lower haze | 31-47.5 | 482-367 | 9.5-1.5 | 2-20 | 0.2 | |
| Pre cloud layers | 46 and 47.5 | 378 and 367 | 1.8-1.5 | 50 and 150 | Bimodal 0.3 and 2.0 | |

Table 1-1. Characteristics of the Venusian cloud particles. Data from [211,212].

Indeed, the petrochemical industry uses concentrated sulfuric acid as a catalyst during octane production from isobutane and butene and finds a rich chemistry in sulfuric acid from the reactivity of hydrocarbon molecules. While the resulting compounds, called "red oil", are an undesirable side product, this chemistry substantiates the idea that the Venus cloud sulfuric acid particles can support diverse organic chemistry.

We will get more information on the possible composition of the UV absorber via NASA's DAVINCI orbiter's high resolution UV instrument.

### 1.3.2 Mode 3 Particle Composition

The composition of a subset of Venus cloud particles, large particles in the lower clouds called "Mode 3" is unknown. (See Table 1-1 for a summary of Venus cloud particle properties vs. altitude.) Adding to the mystery is the fact that the Mode 3 particles as measured by the Large Cloud Particle Size Spectrometer (LCPS) onboard the Pioneer Venus probe appear to be non-spherical [31,33]. Non-spherical means the Mode 3 particles cannot be liquid droplets and are therefore not composed of purely concentrated liquid sulfuric acid. This could indicate unknown chemistry and is intriguing for the presence of life as we know it, which cannot withstand a concentrated sulfuric acid environment.

The nature and composition of the Mode 3 particles is debated with data presently in hand. The key derived parameter is refractive index, which comes from the Pioneer Venus nephelometer which measured backscattered light in a range of angles. The refractive index of the particles in the lower clouds is estimated at ~1.33 assuming spherical droplets [34]. This value is lower than any plausible value for sulfuric acid. The 1.33 index of refraction value can be matched if the particles absorb a small amount of incident light; however [35] find no noticeable absorption in the lower clouds. An alternative explanation is non-spherical particles [31,34], which, again imply non-liquid particles. Data from the Pioneer Venus Optical Array Spectrometer (OAS) [36] also support non-spherical particles. The OAS instrument had three photodiode arrays which measured the shadow of passing particles, which makes the particle size measurement independent of particle composition.

Several studies questioned the existence of the large Mode 3 particles altogether and claimed, e.g., that Mode 3 could in fact be a large "tail" of the liquid Mode 2 particle distribution, once calibration errors were taken into account [37–39]. The unknown composition of the Mode 3 particles leaves room, albeit speculative, for unknown chemistry or life. Microbial cells within the droplets would cause an index of refraction discrepancy. Alternatively, salt formation in a droplet, as a result of acid neutralization (see Section 1.4.2) would alter droplet composition away from pure concentrated $H_2SO_4$ to a more clement chemical environment.

Such decades-long lingering questions on the true nature of the Venus cloud particles motivate each of VLF missions to focus on characterizing the Mode 3 cloud particles.

### 1.3.3 Presence of Non-Volatile Elements in the Cloud Particles

Life as we know it requires metals and other non-volatile species for catalysis, for example transition metals such as iron. (Even for the most ancient enzymes, the protein is largely incidental, serving only to hold catalytic metals in place to facilitate an enzymatic reaction). Detection of metals and other non-volatile species as components of cloud particles would support the potential for habitability of the Venus clouds. In other words, the presence of metals and other non-volatile elements is not a biosignature but is an indicator of habitability.

Both the Vega balloons' and Venera probes' in situ measurement of the elemental composition of the cloud particles suggest that non-volatile elements relevant for habitability are present. Venera 13 and Venera 14 analysis of cloud particles indicates the presence of sulfur, chlorine, and iron [40]. Vega 1 and 2 measurements of the cloud material acquired by the aerosol collector/pyrolyzer suggest significant presence of chlorine, sulfur [41], and phosphorus (P) in the lower cloud [42], but little iron (in contrast to the Venera probe measurements). Other elements suspected to exist are I, Br, Al, Se, Te, Hg, Pb, Al, Sb, and As [43].

In the altitude range of 52 to 47 km the abundance of phosphorus appears to be on the same order as the abundance of sulfur [42,44]. Phosphorus is most plausibly in the form of phosphoric acid droplets ($H_3PO_4$) [45,46]. If the Venera descent probe data is correct and some cloud particles indeed contain > 50% phosphorus species, by mass then by definition the concentration of sulfuric acid in those droplets must be < 50%. Above 52 km, no phosphorus was detected. It is therefore plausible that phosphorus is present as a condensed liquid or solid, phase predominantly in the lower cloud layer [45].

In summary, early measurements by the Vega balloons and the Venera probes suggest the cloud particles are not pure sulfuric acid and the particles likely contain a plethora of other dissolved species (e.g., molecules containing Fe, Cl, P, and others). The exact composition and the concentration of the dissolved species is unknown.

The VLF Habitability and Sample Return missions aim to confirm (and extend the search for ) the presence of non-volatile elements Fe and P in the Venus clouds.

### 1.3.4 Unexpected Atmospheric Gases and Gas Vertical Abundances

A number of unexpected trace gases have been observed to exist in the atmosphere of Venus (Table 1-2). Aside from being relevant to signs of life in their own right, the anomalous gases indicate a chemical disequilibrium when considered together with the main atmosphere constituents. Atmospheric chemical disequilibria have been proposed as a general approach for detecting extraterrestrial biospheres through remote spectroscopy [47–50]. Earth's atmospheric disequilibrium is a result of life's activity, as exemplified by the coexistence of $N_2$, $O_2$ [49]. Beyond the unexpected gases are unexplained vertical abundance profiles of $SO_2$

| Gas | Observation | Altitude | Amount | Comments | Ref. |
|---|---|---|---|---|---|
| $PH_3$ | JCMT | 55-60 km | ~7 ppb | Tentative detection with Earth-based telescopes; possible spatial and temporal variability | [59,70, 71] |
| | ALMA | 55-60 km | ~7 ppb | Tentative detection with Earth-based telescopes; possible spatial and temporal variability | [59,70, 71] |
| | Pioneer Venus | 51 km | N/A | Identification of in the re-analyzed Pioneer Venus LNMS data | [5] |
| $NH_3$ | Venera 8 | 45 km | 0.01 % | Tentative detection by Venera 8 chemical probe at ~2 bar altitude | [4] |
| | Venera 8 | 32 km | 0.1 % | Tentative detection by Venera 8 chemical probe at ~8 bar altitude | [4] |
| | Pioneer Venus | 51 km | N/A | Tentative detection in the re-analyzed Pioneer Venus LNMS data | [5] |
| $O_2$ | Venera 14 | 35-58 km | 18 +/- 4 ppm | Detection by Venera 14 GC | [52] |
| | Pioneer Venus | 52 km | 44 +/- 25 ppm | Detection by Pioneer Venus GC | [51] |
| | Pioneer Venus | 42 km | 16 +/- 7 ppm | Detection by Pioneer Venus GC | [51] |
| $H_2S$ | Venera 14 | 29-37 km | 80 +/- 40 ppm | Detection by Venera 14 GC | [52] |
| | Pioneer Venus | 51 km | N/A | Identification of in the re-analyzed Pioneer LNMS data | [5] |
| HCN | Pioneer Venus | 51 km | N/A | Identification of in the re-analyzed Pioneer LNMS data | [5] |
| $HNO_2$ | Pioneer Venus | 51 km | N/A | Identification of in the re-analyzed Pioneer LNMS data | [5] |
| $HNO_3$ | Pioneer Venus | 51 km | N/A | Identification in the re-analyzed Pioneer LNMS data | [5] |
| $CH_4$ | Pioneer Venus | 51 km | ~1000 ppm | Identified in the original and the re-analyzed Pioneer LNMS data; Possible contaminant. | [5,73] |
| Other hydrocarbons: $C_2H_4$, $C_2H_6$, $C_6H_6$ | Pioneer Venus | 51 km | N/A | Identified in the re-analyzed Pioneer LNMS data; Possible contaminant. | [5] |

Table 1-2. Measured abundances of trace gas species of interest in the Venus clouds and cloud and below-the-cloud atmosphere layers. Taken together, the gases demonstrate chemical disequilibrium in the Venus atmosphere. LNMS is the Pioneer Venus Neutral Gas Mass Spectrometer. GC is the gas chromatographer on either Pioneer Venus or the Venera Probes.

and $H_2O$. The VLF missions aim to confirm and extend the gas measurements.

**Oxygen ($O_2$).** In situ observations of $O_2$ in the Venusian lower clouds and below the clouds have been detected by at least two probes at the 10s of ppm level: Pioneer Venus [51] and Venera 14 [52] (Table 1-2). The $O_2$ detections, however, have been dismissed as artifactual [53,54] because no known physical or chemical process could be found to maintain ppm $O_2$ levels in the hot, reactive lower atmosphere of Venus.

Low altitude $O_2$ is indeed an unexpected gas and is highly reactive, meaning $O_2$ has a very short atmospheric lifetime under the thermodynamic conditions in the Venus atmosphere. In part because of this, $O_2$ can be considered as a potential biosignature gas.

**Phosphine ($PH_3$)**. On modern Earth phosphine ($PH_3$) is a genuine biosignature gas [55], as its production is exclusively associated with microbial life inhabiting strictly anoxic ($O_2$-free) conditions, such as marshlands, swamps, digestive tracts of animals. $PH_3$ is also created by human industrial activity [55–57]. On ancient Earth, tiny amounts phosphine has been proposed to be produced by volcanism and UV-mediated reactions of reduced nickel-iron P-bearing species in meteoritic materials [58].

The recent detection of ppb levels $PH_3$ in the clouds of Venus (at the 55 km altitude) through millimeter-wavelength astronomical observations [59] is surprising as there is no known process capable of producing even a few ppb of $PH_3$ on Venus [45,60]. Volcanically extruded phosphide minerals from the deep mantle have been recently proposed as a potential source of $PH_3$ [61]. However, phosphide-containing minerals, including those from deep mantle plume volcanic eruptions and meteoritic delivery, are an unlikely source of $PH_3$ on Venus [45,62]. For the former, phosphide minerals easily oxidize during their transport to the surface and for the latter too small amounts of phosphides are delivered by meteorites [45].

Since the initial $PH_3$ discovery was announced, several papers have questioned the detection, either on the grounds of data analysis [63–65] or an assignment of mesospheric $SO_2$ rather than cloud-level $PH_3$ [66,67]. In addition, several groups have used IR observations to provide strong upper limits (in the low ppb-sub ppb range) on the abundance of $PH_3$ [68,69]; however these observations are limited to the region above the clouds. The authors of the original discovery stand by their observations and provided a response to the critiques, both on data processing and data interpretation [70,71] and on arguing against $SO_2$ contamination [72].

An independent re-analysis of the Pioneer Venus Neutral Gas Mass Spectrometer (LNMS) data [5] shows evidence of $PH_3$ in the clouds of Venus, via detection of $PH_3$ fragmentation ions. The reanalysis was for the altitude of 51.3 km, and yields a $PH_3$ abundance in the mid-to-high ppb range.

The debate on the presence of $PH_3$ in the clouds of Venus continues and will likely only be resolved by in situ measurements of $PH_3$ gas in the Venus atmosphere.

**Methane ($CH_4$)** is a potential biosignature gas. The low-mass volatile hydrocarbons methane ($CH_4$), ethane ($C_2H_6$), and benzene ($C_6H_6$) were detected in situ in the atmosphere of Venus by the LNMS on the Pioneer Venus Probe [5,73]. $CH_4$ in particular was measured to be present with an unexpectedly high abundance (1000 - 6000 ppm) in the atmosphere below 60 km altitude [73]. In contrast to other gases discussed in this section, the detection of $CH_4$ and other volatile hydrocarbons by the Pioneer Venus probe are likely an artifactual result due to an instrumental contamination and not a genuine atmospheric gas detection [73]. Remote observations with Earth-based telescopes put upper limits for $CH_4$ abundance in the lower atmosphere at $< 0.1$ ppm [74].

**Ammonia ($NH_3$)** is a very interesting gas that is unexpected in an oxidized atmosphere. $NH_3$ has been tentatively detected by two separate probes. In 1972 the Venera 8 descent probe reported the presence of $NH_3$ in the lower atmosphere of Venus, using bromphenol blue as an indicator of a basic atmospheric component [4]. The $NH_3$ measurement has been challenged as erroneous, due to the indicator's potential reactivity with sulfuric acid [75]. The Venera 8 detection of $NH_3$ was also discounted shortly after the measurement; Goettel and Lewis discarded it on the grounds of its unlikelihood in an atmosphere at thermodynamic equilibrium [76]. The argument by Goettel and Lewis is now weakened as a growing list of gases in the atmosphere of Venus indicate thermodynamic disequilibrium [5,77,78].

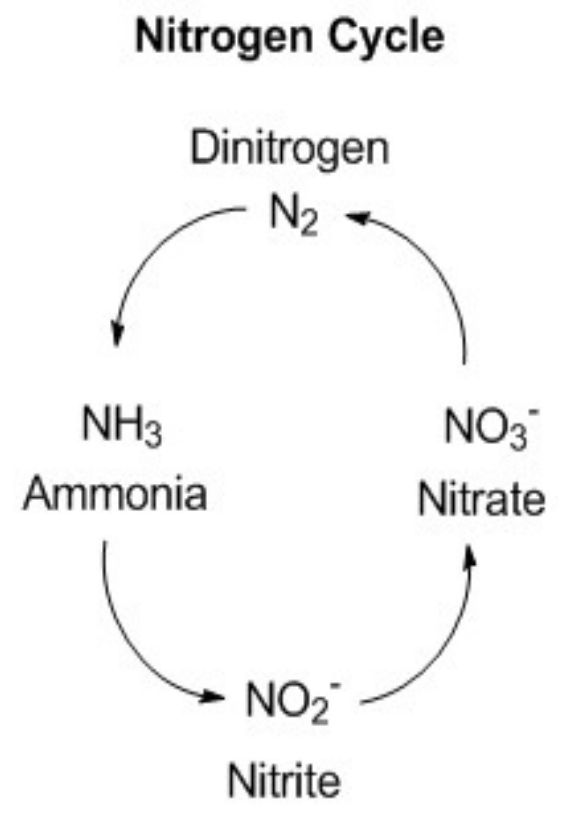

Figure 1-3. The potential nitrogen cycle in the Venusian clouds based on nitrogen species (nitrate, nitrite, ammonia, and $N_2$) tentatively identified in the reanalyzed LNMS data from the Pioneer Venus probe [5]. The identified nitrogen species are also major constituents of the biological nitrogen cycle on Earth.

The recent re-assessment of the Pioneer Venus Large Probe Neutral Mass Spectrometer (LNMS) has also provided suggestive, although not conclusive, evidence for the presence of gas-phase $NH_3$ in Venus cloud layers [5].

While the chemical processes that may generate $NH_3$ in the Venusian clouds is unknown, assuming the tentative detections of the Venera probes and Pioneer Venus are correct, the possibility that $NH_3$ is a biological product remains (see Section 1.4.2). $NH_3$ is a prime target for remeasuring with a new in situ probe.

**General signs of disequilibrium in the clouds.** There are additional gas species identified in the recent reanalysis of the Pioneer Venus LNMS data that indicate chemical disequilibrium in the clouds of Venus. Compounds indicative of unknown chemical processes in the clouds include $HNO_2$, $PH_3$, $H_2S$, $NH_3$, CO, $S_x$ [5].

### 1.3.5 Evidence for a Nitrogen Cycle in the Clouds of Venus

The recent re-analysis of the Pioneer Venus LNMS data shows evidence of nitrogen chemicals at different oxidation states (from -3 to +5): $NH_3$ (-3), HCN (-3), $N_2$ (0), $NO_2^-$ (+3), $NO_3^-$ (+5).

The potential presence of nitrogen chemicals at different oxidation states implies the existence of an active nitrogen cycle in the clouds of Venus (Figure 1-3). Such nitrogen compounds could be key electron donors for anoxygenic photosynthesis (nitrite) or a critical redox pair (nitrate and nitrite) for a postulated hypothetical iron-sulfur cycle in Venus' clouds [13]. Nitrogen species identified in the re-analyzed data from Pioneer Venus are also major constituents of the biological nitrogen cycle on Earth (nitrate, nitrite, ammonia, and $N_2$) [79]. The potential identification of $NH_3$, and other N-species from the terrestrial nitrogen cycle adds another piece of evidence for potential biological activity in the clouds of Venus.

### 1.3.6 Unexplained Gas Vertical Abundance Profiles of $SO_2$ and $H_2O$

The atmosphere vertical abundance profiles of sulfur dioxide ($SO_2$) in the Venus cloud layers and $H_2O$ above the clouds remains unexplained. The presence of $SO_2$ is expected in the atmosphere of Venus and in the clouds. $SO_2$ is a common volcanically-produced gas. The observed abundance profile of $SO_2$ in the Venus cloud layers however is much lower than can be explained by current photochemical models (reviewed by [80]). There is, therefore, missing atmospheric chemistry of some kind.

The observed abundance of water vapor ($H_2O$) above the clouds does not match the $H_2O$ abundance profile predicted by atmospheric photochemistry models [59,81,82]. As in the $SO_2$ case, additional unknown atmospheric chemistry is needed to explain the observations of $H_2O$. Additionally, there appears to be a great variability in observed water vapor abundance values. Repeated measurements by in situ probes vary from 5 ppm to 0.2% (reviewed by [80]), although [83] have dismissed the observations based on contact methods in favor of spectroscopic methods

## 1.4 Revisiting the Venus Cloud Extreme Acidity as a Challenge for Life

The high acidity of the Venus cloud droplets is a major challenge to life of any kind. The concentrated sulfuric acid is billions of times more acidic than the most acidic environments on Earth. The expectation is that Earth-like life simply cannot survive if concentrated sulfuric acid (instead of water) is the solvent for life. Indeed, as part of this study we have shown computationally that the majority of molecules involved in Earth's biochemistry are unstable in concentrated sulfuric acid [84].

For life to be present in the Venus clouds, life must: 1) have a completely different biochemistry than Earth life, relying only on molecules that are stable in concentrated sulfuric acid; or 2) life must reside inside a protective shell made of sulfuric-acid resistant material, such as certain lipids, elemental sulfur, graphite, or wax; or 3) life itself must neutralize the acid to acceptable levels. We first describe the general consensus on acidity of the Venus cloud particles and then comment on each of the three speculative life survival strategies.

### *1.4.1 General Consensus of Venus Cloud Composition and Acidity*

The Venus clouds' main constituent is particles composed of concentrated sulfuric acid droplets. This paradigm is supported by three findings.

1] Photochemical models of the atmosphere are consistent with $H_2SO_4$ clouds. The models predict $H_2O$, $SO_3$, and $H_2SO_4$ to be present throughout the atmosphere (e.g., [81]), and gaseous $H_2SO_4$ [85,86] as well as gaseous $H_2O$ and $SO_2$ (reviewed by [80]) are measured throughout.

The formation of clouds on Venus is photochemically driven (e.g., [87,88]). The theory is that the sulfuric acid vapor is first made at > 70 km.

$$CO_2 + h\nu \rightarrow CO + O$$

$$SO_2 + O + M \rightarrow SO_3 + M$$

$$SO_3 + 2\,H_2O \rightarrow H_2SO_4 + H_2O$$

The $H_2SO_4$ vapor condenses out, creates the droplets, and as the droplets rain down, sulfuric acid thermally dissociates in the lower atmosphere (below 40 km) [87,89]. A fraction of $H_2SO_4$ also likely reforms from the $H_2O$ and $SO_3$ near the bottom of the clouds [87].

The measured and modeled levels of $H_2O$, which together with $SO_3$ will efficiently form $H_2SO_4$, support the theory that the clouds of Venus contain sulfuric acid [51,87,90].

2] Gaseous $H_2SO_4$ has been detected and measured by microwave spectrometry, supporting the photochemical concept of $H_2SO_4$ cloud formation [85,86].

3] The inferred refractive index of the cloud droplets is consistent with the clouds being made of at least 70% sulfuric acid and less than 30% water [29,91]. The concentration of sulfuric acid in droplets is derived through modeling (e.g., [39]) of light scattering to match in situ data. The concentration of $H_2SO_4$ is lower in the top clouds and increases towards the bottom of the clouds as the temperature increases (summarized in Table 1-1; following information from [92] and their Table 1). It is worth mentioning that a concentrated solution of sulfuric acid ($H_2SO_4$-$H_2O$) was found to be in good agreement with ground-based polarization data [30] before the in situ probes.

There is a growing discussion in part of the Venus community whether or not sulfuric acid cloud particles have uniform concentration [93]. The "average concentration" of sulfuric acid in the cloud droplets is ~85% $H_2SO_4$ [92]. However, the concentration of sulfuric acid across the cloud deck likely varies significantly. The concentration reaches ~70% in the top clouds while in the lower clouds the concentration could reach >100% i.e. 'fuming' sulfuric acid or oleum ($H_2S_2O_7$, or a solution of $SO_3$ in $H_2SO_4$) [92].

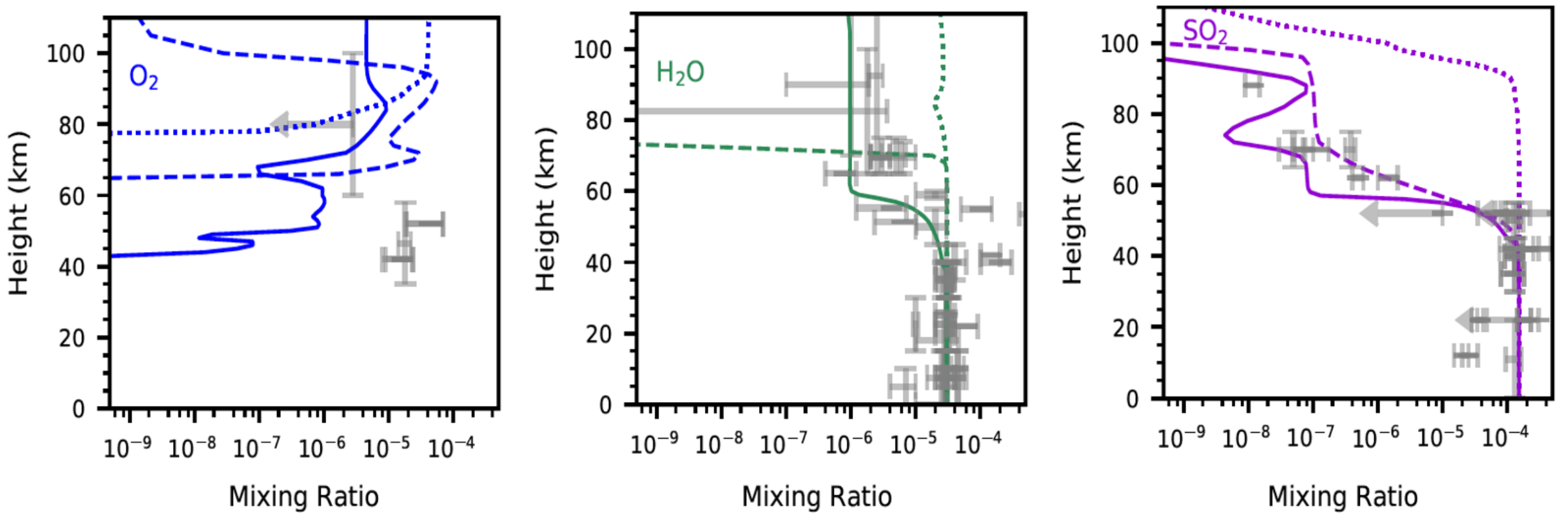

Figure 1-4. Venus atmosphere abundance profiles of three molecular species. The x-axis is the gas fraction by volume, called the mixing ratio. The y-axis is altitude above the surface in km. The lines are gas mixing ratios from our models: with $NH_3$ chemistry (solid lines); without $NH_3$ chemistry (dotted line; model in [80]); without $NH_3$ but with an arbitrary removal rate for $SO_2$ in the cloud layers tuned to fit the data (dashed lines; model in [59] and [80]. Gray points are error bars from observations tabulated in [80]. Left: $O_2$. The model with $NH_3$ chemistry improves upon both the long-standing problem of overabundance of $O_2$ in the upper atmosphere and the presence of $O_2$ in the cloud layers. Middle: $H_2O$. The model with $NH_3$ chemistry supports the presence of water vapor above the cloud layer (> 80 km). Right: $SO_2$. The models with $NH_3$ chemistry (solid line) and without $NH_3$ chemistry but with arbitrary constraints on $SO_2$ (dashed line) both provide a fit to observed values throughout the atmosphere except for the top (> 85 km). A key point is that the model without $NH_3$ and without the $SO_2$ arbitrary removal rate (dotted line) fit the cloud-layer data very poorly whereas the model with $NH_3$ (with no arbitrary constraints; solid line) fits the data much better. The introduction of $NH_3$ to the model explains the persistent Venusian cloud anomalies (the $SO_2$, $H_2O$ profile, the presence of $O_2$ in the cloud layers, the tentative presence of $NH_3$ (not shown), and the potential non-sphericity of the Mode 3 particles (not shown). While the presence of other mineral bases could contribute, none can explain the ppm levels of $O_2$ in the clouds nor the tentative presence of $NH_3$. Figure from [3].

The measured cloud particle refractive index allows for compounds and mixtures other than concentrated $H_2SO_4$. For example, the droplet sulfuric acid concentration could be very variable, between 30% and >100%.

### 1.4.2 *Speculative but Possible Life Survival Strategies*

**New biochemistry.** Earth-like biochemistry cannot survive in concentrated sulfuric acid for longer than a few seconds [84]. Proteins, sugars, certain lipids, and nucleic acids (DNA and RNA) and the great majority of Earth's life small molecule biochemicals are rapidly converted to highly conjugated and crosslinked polymers if concentrated sulfuric acid (instead of water) is used as a solvent for life [1].

We find, however, that concentrated sulfuric acid can, in principle, support different forms of chemical complexity that could hypothetically lead to a completely different biochemistry than on Earth [84]. Both the data repository itself and the demonstration of the predictive power of the models built with it were recently published in open access journals [84,94], (see Appendix A.4).

**Acid-resistant membrane.** As part of the Venus Life Finder Mission study we demonstrated in a laboratory setting that certain lipids can self-assemble to form vesicle-like structures in concentrated sulfuric acid. This major finding demonstrates that life particles with a lipid "cell wall" type structure could potentially survive in the Venus cloud droplets if the concentration of sulfuric acid is around 70% or less (higher concentrations will be investigated in the future) (Appendix A.3).

**Life itself could neutralize the acid to acceptable levels.** In new work supported by this study, we show that locally-produced $NH_3$ can neutralize the acid such that a subset of the cloud particles may be brought to an acidity level tolerable by acidophiles on Earth (pH = 1) [3]. Our highly speculative assumption is that life in the clouds produces acid-neutralizing $NH_3$; in fact, some life on Earth secretes $NH_3$ to neutralize a droplet-sized acidic environment. Pathogens such as *Mycobacterium tuberculosis* and *Candida albicans* can neutralize the interior of phagosomes (acid-containing vesicles inside cells used for digestion of captured organic material), by secreting $NH_3$, thus evading destruction.

The production of $NH_3$ in the Venus atmosphere sets off a chain of reactions that results in gas vertical profiles of $SO_2$, $O_2$, and $H_2O$ that are largely consistent with the observational data (Section 1.3) without resorting to unphysical constraints used by all other existing Venus photochemistry models ([3]; Figure 1-4). In addition, the production of $NH_3$ could also explain the claimed asphericity of the Mode 3 particles, as the neutralization of acidic cloud droplets leads to the formation of ammonium salt slurries (see also [93]) and hence non-spherical particles. More work is needed to investigate any potential abiotic sources.

The VLF Missions aim to search for organic compounds inside the cloud droplets and the VLF Habitability Mission aims to measure the Venus cloud particle acidity.

## 1.5 The Extreme Dryness of the Venus Atmosphere

We would be remiss not to emphasize the extreme aridity of the Venus cloud environment as a significant challenge to life as we know it. In extremely dry environments, terrestrial life can survive as spores or other inactive forms but would not be actively growing and therefore unable to support a sustainable biosphere (which requires at least some cells or organisms to be actively growing).

The Venus atmosphere extreme dryness has been considered a well-known fact for decades (e.g., [95]), having been often described [14,45,84] and most recently reviewed by [96].

Even under the assumption that life resides inside the cloud particle, the water activity is extremely low, because any water molecules inside the particle will be tightly bound to sulfuric acid.

The range of Venus in-cloud water vapor abundance mixing ratios reported in the literature is very large, ranging from 5 ppm to 0.2% or higher (summarized in [80]). The highest values, if confirmed, indicate the presence of local 'habitable' regions with humidity high enough at least for life on Earth. While all global Venus atmosphere models may therefore represent an average of extremely arid "desert" regions, there may exist some localized more humid regions, motivating us to remeasure $H_2O$.

## 1.6 Summary

There is a new opportunity to directly probe the Venusian clouds. Unexplained anomalies, including possible presence of $NH_3$, tens of ppm $O_2$, the $SO_2$ and $H_2O$ vertical abundance profiles, and the unknown composition of Mode 3 particles, have lingered for decades and might be tied to life's activities or be indicative of unknown chemistry itself worth exploring. Indeed, while we may use the presence of life to explain the combined anomalies with some external assumptions, there may as yet be a chemical explanation for each individual anomaly. The acidity of the Venus cloud droplets has not been measured directly and could inform us on the cloud particle habitability. Similarly, no previous mission has directly searched for organic chemistry in the cloud particles. Our proposed series of VLF missions aim to address each of the above to continue where the pioneering missions from nearly four decades ago left off.

# 2 Science Goals and Instrument for the Rocket Lab Mission 2023

**Abstract:** The Rocket Lab mission is a small direct entry probe planned for launch in May 2023 with room for one ~1 kg instrument. The probe mission will spend about 3 minutes in the Venus cloud layers at 48-60 km altitude above the surface. We have chosen a low-mass, low-cost autofluorescing nephelometer to search for organic molecules in the cloud particles and constrain the Mode 3 particle composition. The presence of organic molecules would indicate an environment with complex molecules of the kind needed to support life, though would not uniquely indicate the presence of life. The Mode 3 droplet composition will inform their habitability, and confirm and extend past measurements indicating other than pure concentrated sulfuric acid composition.

**New science objectives for a Venus mission that emerged from the study:**

- The clear objective to focus on astrobiology.
- The new idea to search for organic molecules inside Venus cloud particles by autofluorescence.

**New instrumentation that emerged from the study:**

- A miniaturized nephelometer combined with autofluorescing capability.

## 2.1 Science Goals and Objectives

The overarching science goals are the search for evidence of life or habitability in the Venusian clouds. There are two science objectives.

**Objective 1:** Search for the presence of organic material within cloud-layer particles.

**Objective 2:** Determine the shape and indicies of refraction (a proxy for composition) of the Mode 3 cloud particles.

For **Objective 1**, the motivating concept is that autofluorescence is a way to detect (but not identify) organic compounds inside of the Venus cloud droplets. On Earth, many organic compounds are known to fluoresce when subject to UV radiation (Figures 2-1 and 2-2, and Section A-2). Organic molecules with conjugated double bonds (e.g. aromatic molecules that have delocalized electrons in the rings) when subjected to UV light yield stronger fluorescent signals compared to other molecules in general. While inorganic salts and other materials can also fluoresce, the signal is expected to be much lower than for molecules that contain organic aromatic rings.

Fluorescence behavior will be very different for compounds in water ($H_2O$) vs. sulfuric acid ($H_2SO_4$) or an organic solvent. Ongoing work will eventually inform the optimal UV wavelength, given there is only room for one excitation laser and one detector. Firebird Biomolecular Sciences and Droplet Measurement Technologies are carrying out an assessment of fluorescence of conjugated systems and autofluorescing behavior of organic molecules in concentrated sulfuric acid in order to select the optimal wavelengths (Figure 2-2).

For **Objective 2**, a nephelometer can infer the particle sizes, shapes, and indicies of refraction by measuring scattered light and polarization as a function of angle. We aim to confirm the non-spherical nature of the Mode 3 cloud particles and constrain the particle composition to confirm and build on the hypothesis that the cloud particles are more diverse than concentrated sulfuric acid.

## 2.2 The Instrument: Autofluorescing Nephelometer (AFN)

The chosen instrument is the Autofluorescing Nephelometer (AFN) by Cloud Measurement Solutions and Droplet Measurement Technologies. In addition to the science instrument, the probe will carry temperature, pressure, and altitude sensors.

The AFN is derived from Droplet Measurement Technologies (DMT)'s Backscatter

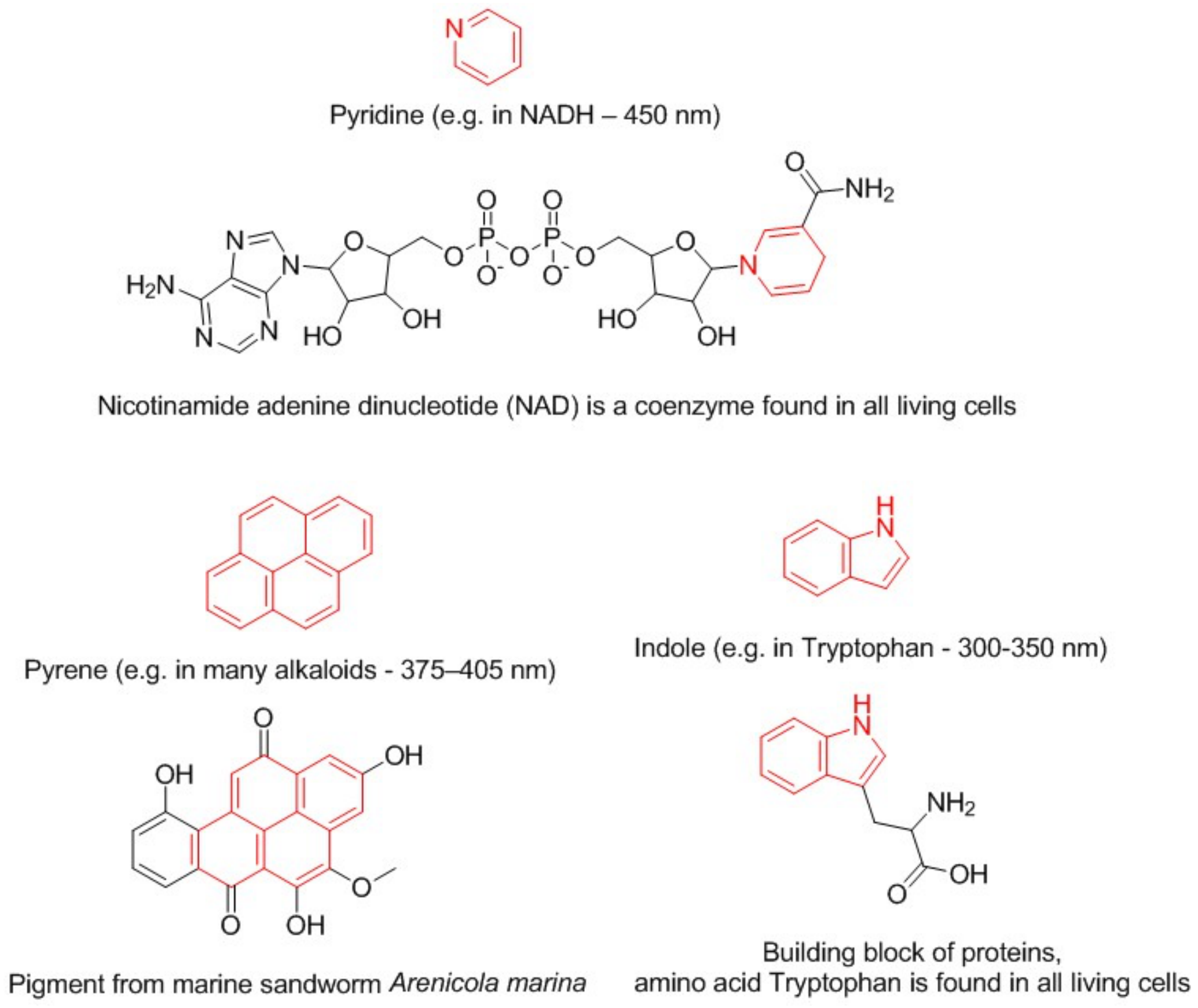

Figure 2-1. Examples of Earth's biochemicals (in water) with autofluorescence properties. NADH, containing pyridine ring (top), is the reduced form of nicotinamide adenine dinucleotide (NAD), a co-enzyme essential to metabolism found in all life on Earth. Tryptophan, containing an indole ring (bottom-right) is an amino acid used by all known life. Other natural organic molecules, like analogs of natural polyaromatic alkaloids (e.g. containing pyrene ring system [213,214]) (bottom-left) also autofluoresce. While we do not expect molecules shown in this figure to exist in the Venus clouds, other organic molecules are stable in concentrated sulfuric acid.

Cloud Probe with Polarization Detection (BCPD) [97] (US Patent US9222873B2) (see Table 2-1 and Figure 2-3). First, we describe the BCPD then the derivative, the AFN, which utilizes an ultraviolet (UV) laser to produce native fluorescence as a measure of the potential presence of organic material.

The BCPD utilizes scattered light and takes advantage of the observation that the dominant component of backscattered light as a function of angle depends upon the size of the particle with respect to the wavelength of the source. The BCPD is based on an earlier Backscatter Cloud Probe (BCP; Figure 2-3), which utilizes a laser in combination with collecting optics to determine the size of individual particles in an atmospheric free stream and to estimate the index of refraction. The BCP has extensive operational history, having operated for more than 60,000 hours on commercial aircraft since 2011 [98]. For the BCPD, two detectors are added capable of discriminating differentially polarized light; non-spherical particles change the polarization of light, allowing sphericity to be measured.

The AFN would use a UV laser and a suitable-wavelength detector for fluorescence detection (Figure 2-2), both highly validated as part of the

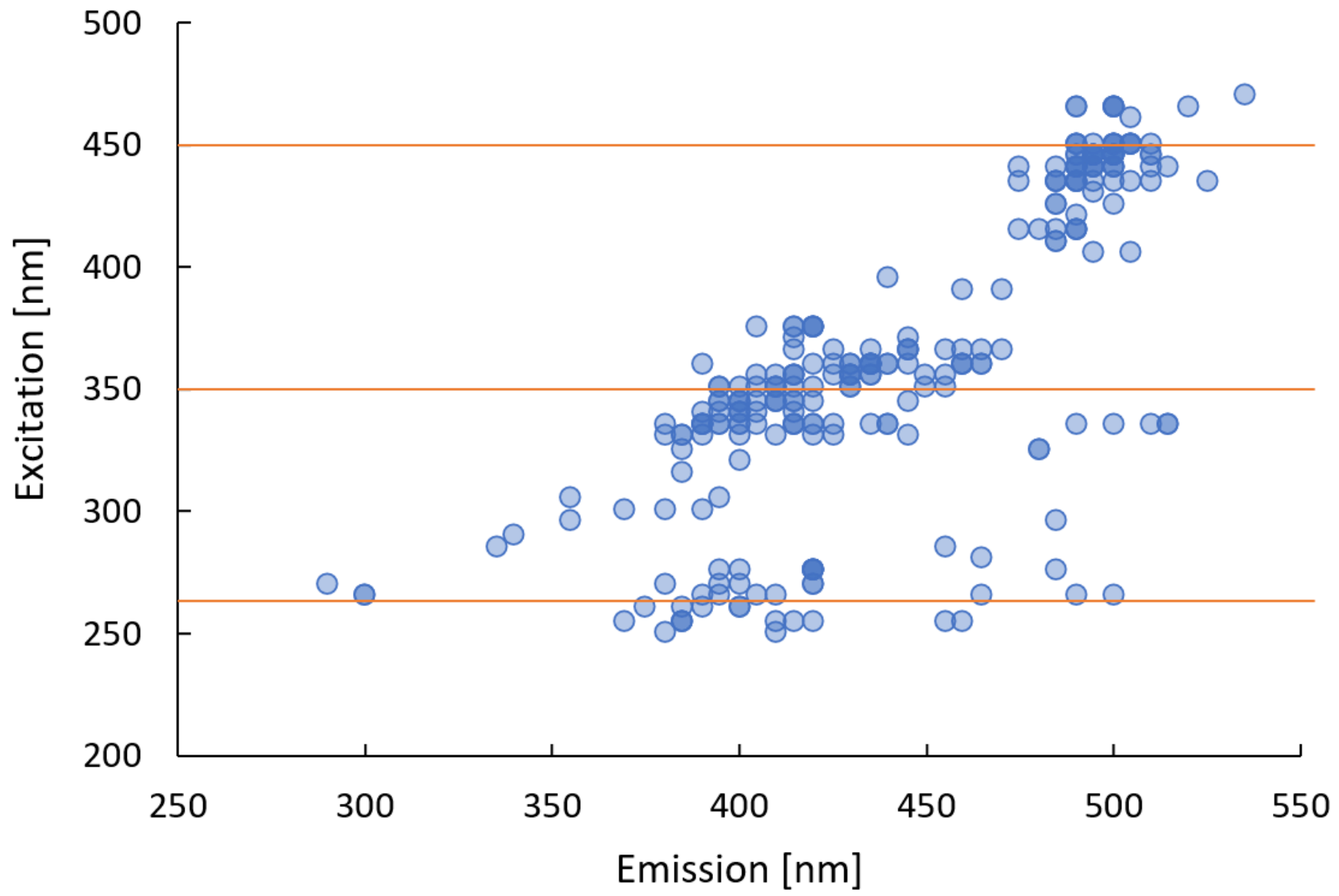

Figure 2-2. Local peak fluorescence emissions vs. excitation wavelengths for organic compounds tested (see Section A-2). The x-axis is emission wavelength (nm) and the y-axis is the excitation wavelength (nm). The blue dots indicate detection of fluorescence. The larger number of instances of recorded fluorescence the darker the shade of the blue dot. The orange lines indicate the suggested excitation laser wavelengths. Data credit: Jan Špaček and Steven A. Benner (unpublished).

| Science Instrument | Limit of Detection | Mass | Volume | Peak Power | Data Rate (per measurement) | Developer |
|---|---|---|---|---|---|---|
| *Autofluorescing Nephelometer | 0 - 1000 particles $cm^{-3}$ | 0.5 kg detector 1.0 kg electronics | 500 $cm^3$ detector 1500 $cm^3$ electronics | 40 W | 200 bytes (2000 bytes per second) | Droplet Meas. Tech. |

Table 2-1. Summary description of the instrument properties. * Denotes a precursor model.

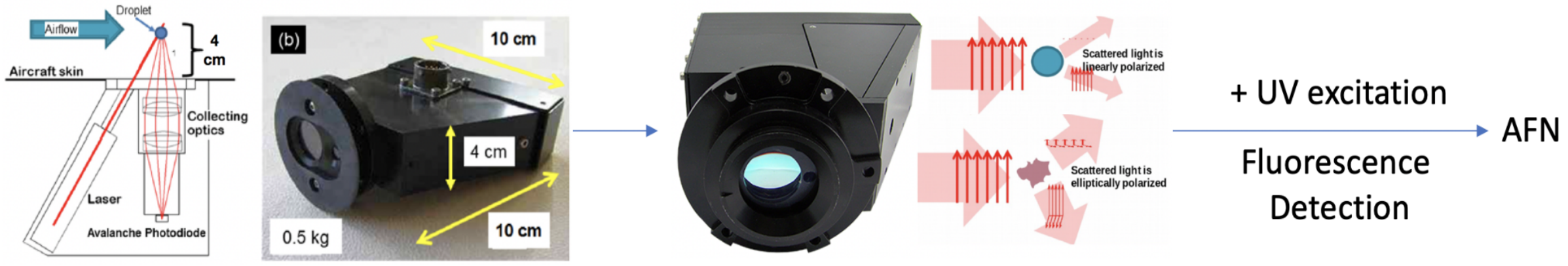

Figure 2-3. Evolution of the Autofluorescing Nephelometer (AFN) from the Backscatter Cloud Probe (BCP) (left of arrow) to the Backscatter Cloud Probe with Polarization Detection (BCPD) (right of arrow). The BCPD is further evolved to the AFN by replacement of the BCPD laser with a UV source and addition of fluorescence-detection compatible optics.

Wideband Integrated Bioaerosol Sensor (WIBS) system, a DMT commercial product. This would produce a highly capable instrument yet one that is much more simple than the WIBS. WIBS has been validated through extensive testing on aerosols and has demonstrated the ability to not only distinguish between biological and non-biological aerosols, but also to discriminate between different types of biological aerosols [99]. For the AFN the goal is simply to identify probable or possible organic material, where the fluorescence is generated by molecular absorption and reemission, for example by tryptophan or NADH (Figure 2-1).

If there are no organics or no organics detected, we will have updated measurements of the particle size distribution and particle shape. A lack of fluorescence would constrain inferred chemical processes within the Venus cloud droplets and atmosphere.

The AFN will operate continuously at Venus cloud-level altitudes. The AFN can sample much faster than 1 Hz, and will set the sampling time based on the fall speed of the probe, as the volumetric sample rate is determined by the velocity of air passing through the sample area. The data rate is on order of 200 bytes per sample, but for data efficiency we may transmit a 50 channel size distribution and information on average refractive index in different size ranges and shape factor in different size ranges. As many as 300 to 15,000 particles per second might be sampled with a data rate on order 2000 bytes $s^{-1}$. High priority data summaries can be used to accommodate lower data rates. The total data amount returned to Earth by the probe is to be determined.

Contamination from the probe heat shield (carbon phenol) will be avoided by probing particles outside of the ablation airflow. However, extensive laboratory testing of many different types of particle compositions related to the expected ablation products should be carried out.

## 2.3 Ongoing Work

As of this writing (December 2021) the team is working on the science requirements, the instrument prototype, and the flight unit.

# 3 ROCKET LAB VENUS MISSION 2023

**Abstract:** Regular, low-cost Decadal-class science missions to planetary destinations will be enabled by high-ΔV small spacecraft, such as the high-energy Photon, to support expanding opportunities for scientists and to increase the rate of science return.

**New opportunity that emerged from the study:**

- A unique opportunity for a small focused mission with a high scientific return.
- A new paradigm for privately-funded space science.

## 3.1 Introduction

Rocket Lab has made the engineering and financial commitment to fly a private mission to Venus in 2023 to help answer the question "Are we alone in the universe?". The specific goals of Rocket Lab's mission are to:

1. Search for signs of life;
2. Demonstrate a high-impact small planetary probe capability;
3. Enable game changing, more regular Decadal-class planetary science using dedicated small launch vehicles and small spacecraft; and
4. Take the first step in a campaign of small missions to better understand Venus.

The mission is planned for launch in May 2023 on Electron from Rocket Lab's Launch Complex 1 (LC-1). The specific launch opportunity will be selected after a more thorough study of the transfer window sensitivities in the detailed design phase. The mission will follow a hyperbolic trajectory with the high-energy Photon performing as the cruise stage and then deploying a small probe into the Venus atmosphere for the science phase of the mission.

## 3.2 The Photon Spacecraft

The high-energy Photon (Figure 3-1), in development by Rocket Lab for the NASA Cislunar Autonomous Positioning System Technology Operations and Navigation Experiment (CAPSTONE) mission launching to the moon in 2021 and NASA Escape and Plasma Acceleration and Dynamics Explorers (ESCAPADE) mission launching to Mars in 2024, is a self-sufficient small spacecraft capable of long-duration interplanetary cruise.

The Photon's power system is conventional, using photovoltaic solar arrays and lithium-polymer secondary batteries. The attitude control system includes star trackers, sun sensors, an inertial measurement unit, reaction wheels, and a cold-gas reaction control system (RCS). S-band or X-band RF ranging transponders support communications with the Deep Space Network (DSN) or with commercial networks and enable traditional deep space radiometric navigation methods. A Global Position System (GPS) receiver is used for navigation near Earth. ΔV greater than 4 km/s is provided by a storable, re-startable bi-propellant propulsion system called Hyper Curie using electric pumps to supply pressurized propellant to a thrust vector-controlled engine. The propellant tanks achieve high propellant mass fraction and can be scaled to meet mission-specific needs.

The high-energy Photon (Figure 3-2) is designed for launch on Electron (Figure 3-3), Rocket Lab's dedicated small launch vehicle. Electron can lift up to 200 kg to a 500 km sun-synchronous orbit from either of two active, state-of-the-art launch sites: LC-1 on the Mahia Peninsula in New Zealand and Launch Complex 2 on Wallops Island, Virginia. Electron is a two-stage launch vehicle with a Kick Stage, standing at 18-meter tall with a diameter of 1.2 m and a lift-off mass of ~13,000 kg. Electron's engine, the 25 kN Rutherford, is fueled by liquid oxygen and kerosene fed by electric pumps. Rutherford is based on an entirely new propulsion cycle that makes use of brushless direct current electric motors and high-performance lithium-polymer batteries to drive impeller pumps. Electron's Stage 1 uses nine Rutherford engines while Stage 2 requires just a single Rutherford vacuum engine. Rutherford is the first oxygen/hydrocarbon engine to use additive manufacturing for all primary components, including the regeneratively cooled thrust chamber, injector pumps, and main propellant valves. All Rutherford engines on Electron are identical, except for a larger expansion ratio nozzle on Stage 2 optimized for performance in near-vacuum conditions. The high-energy Photon replaced the Kick Stage for missions beyond LEO.

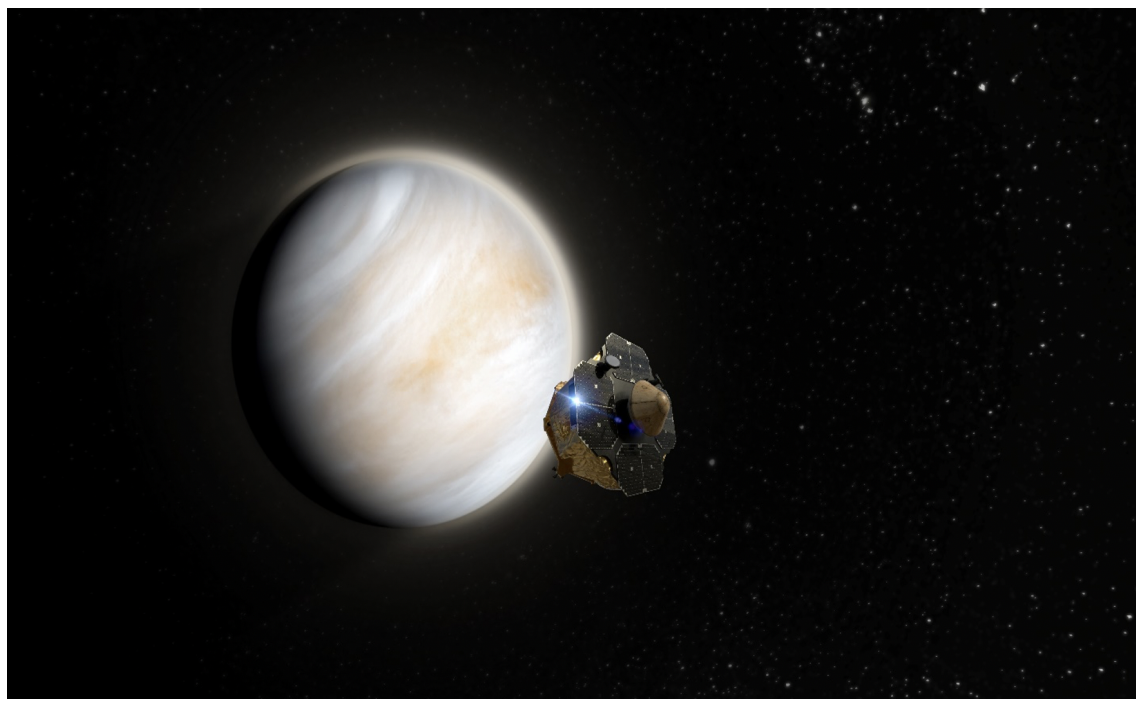

Figure 3-1: Rocket Lab's Electron-launched private mission to Venus will deploy a small probe from a high-energy Photon in 2023.

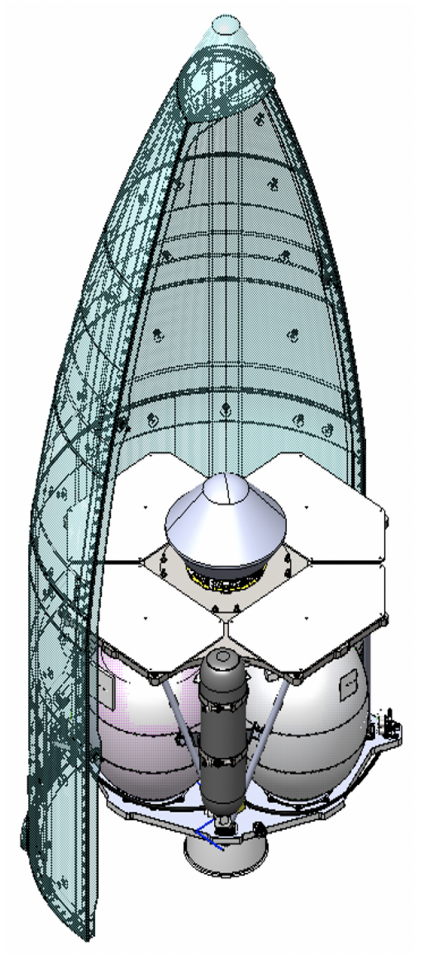

Figure 3-2. High-energy Photon and small Venus entry probe.

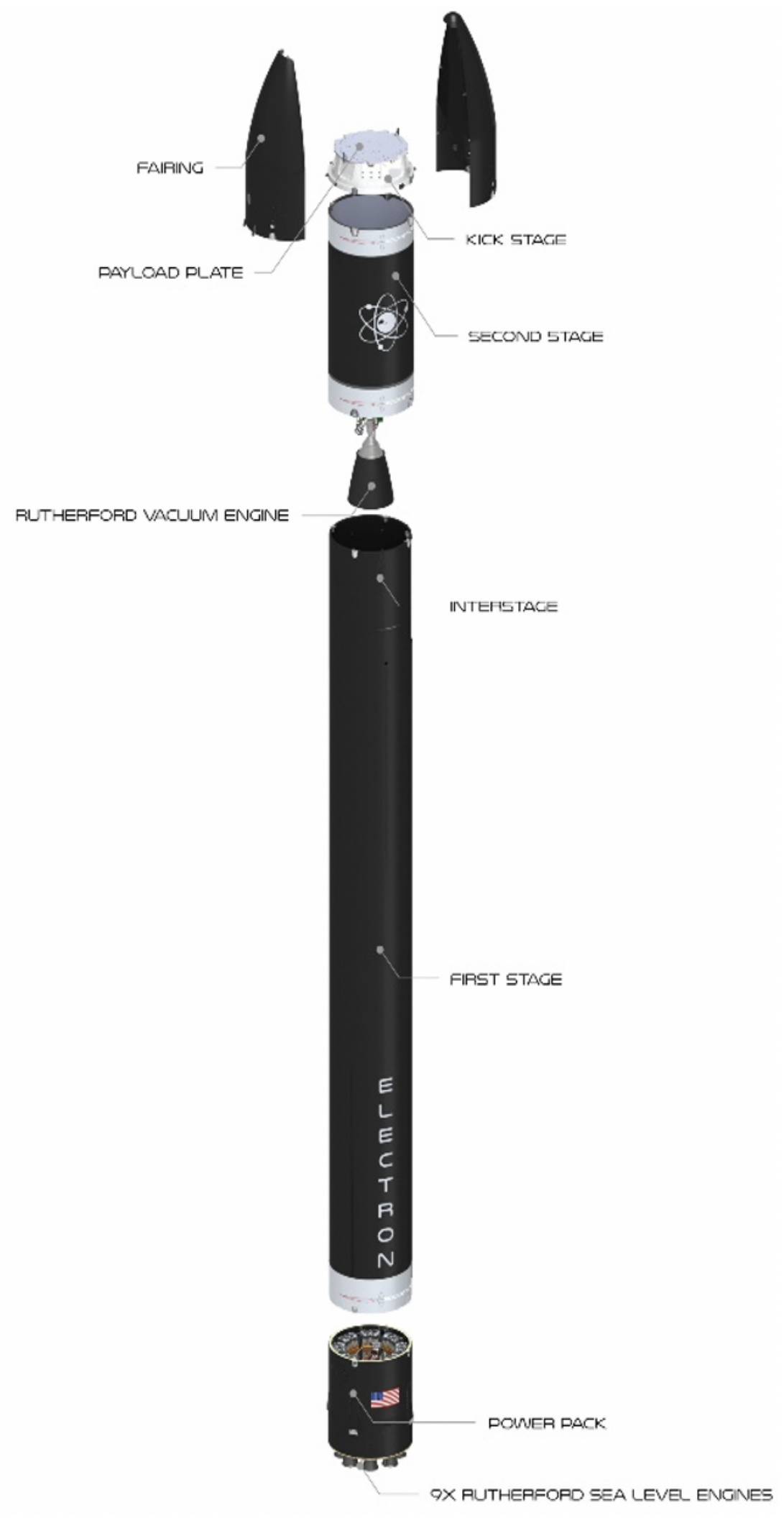

Figure 3-3. Electron small launch vehicle.

### 3.3 Trajectory

Electron first delivers high-energy Photon to a circular parking orbit (Figure 3-4) around Earth. After separating from Electron's Stage 2, high-energy Photon performs preprogrammed burns to establish a preliminary elliptical orbit. High-energy Photon then performs a series of burns through increasingly elliptical orbits, each time raising the apogee altitude while maintaining a nearly constant perigee. Breaking the departure across multiple maneuvers is an efficient approach to Earth escape. By holding burns close to perigee and limiting their duration, propulsive energy is efficiently spent raising apogee while avoiding the burn losses associated with long duration maneuvers. Each phasing maneuver is followed by a planned number of phasing orbits at the new apogee altitude. Phasing orbits provide time for on-orbit navigation, maneuver reconstruction and planning, propulsion system calibration, and conjunction screening. Each planned maneuver includes contingency options to mitigate conjunction events or missed maneuvers. After the nominal apogee raising maneuvers are performed, a final injection burn is executed to place high-energy Photon on the escape trajectory. Trajectory correction maneuvers (TCMs) using the Hyper Curie engine or integrated RCS are used to make fine adjustments to the trajectory and target the appropriate entry interface.

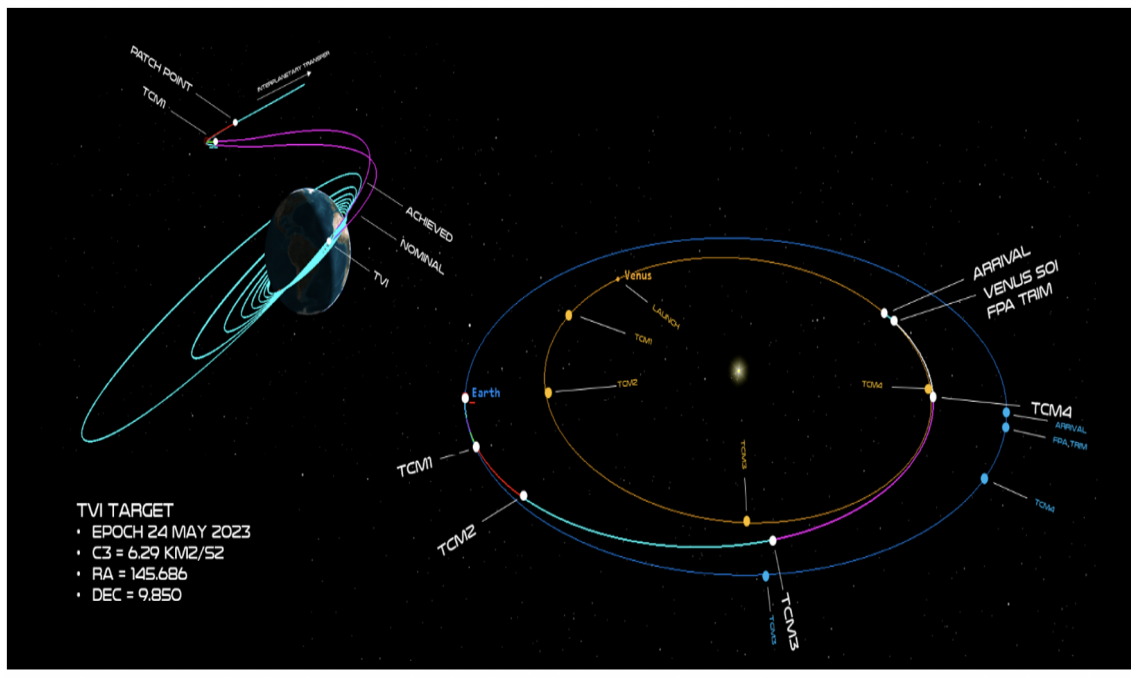

Figure 3-4: Phasing orbits approach to escape trajectory and typical trajectory correction maneuvers are used to target entry interface at Venus.

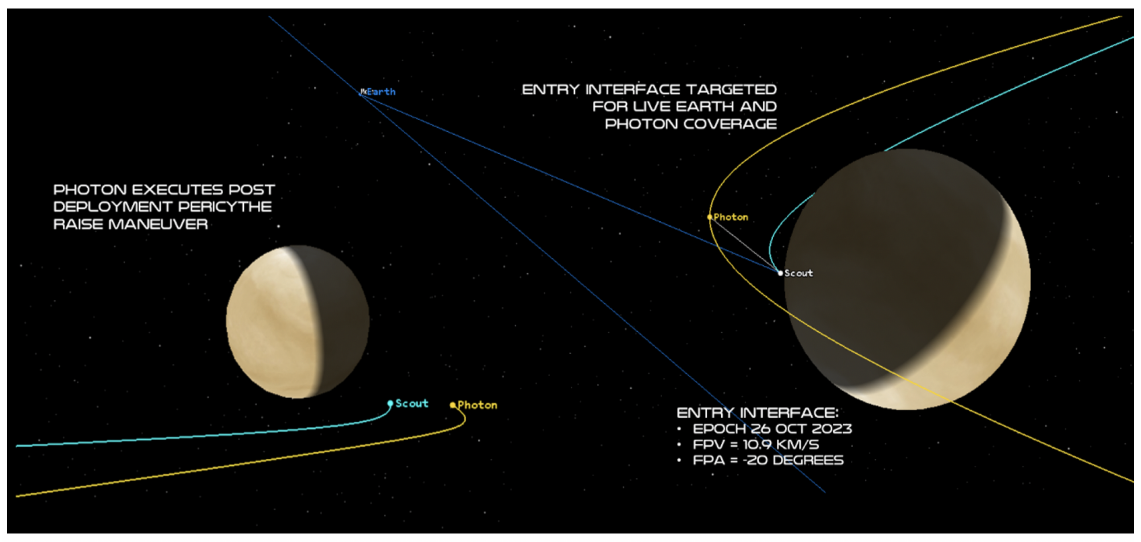

Figure 3-5: High-energy Photon releases the entry probe after targeting the entry interface selected for optimal instrument measurement conditions.

In October 2023, after the cruise phase (Figure 3-5), high energy Photon will target an entry interface to deploy a small (~20 kg) probe directly into the atmosphere with an entry flight path angle (EFPA) between -10 and -30 degrees. The probe communicates direct-to-Earth through an S-band communications link with a hemispherical antenna returning science data captured during the descent and stored on board. A back-up relay communications link via the Photon bus will be traded during the detailed design phase. The entry interface will be selected to satisfy science objectives (day/night, latitude targeting), Earth communication geometry, and other factors. The EFPA will be selected based on an analysis of the entry and descent timeline, the integrated heat load and required thermal protection system (TPS) thickness, probe acceleration (g-loading) limits, navigation precision, and other factors.

## 3.4 The Probe

The small probe (Figure 3-6) will contain up to 1 kg of science payload to explore the habitability of the atmosphere, achieving ~270 sec in the cloud layer between ~45-60 km altitude to perform science operations. The science instrument is an autofluorescing nephelometer described in Chapter 2. The small probe is notionally a ~20 cm diameter, 45-degree half-angle sphere-cone blunt body with a hemi-spherical aft body for static stability in the hypersonic flow regime.

The probe shape will be traded during the detailed design phase based on the stability characteristics in various flow regimes (hypersonics, transonic, subsonic, etc.) and center of gravity location constraints, among other considerations, with the goal of eliminating the need to "spin up" the probe to add gyroscopic stiffness.

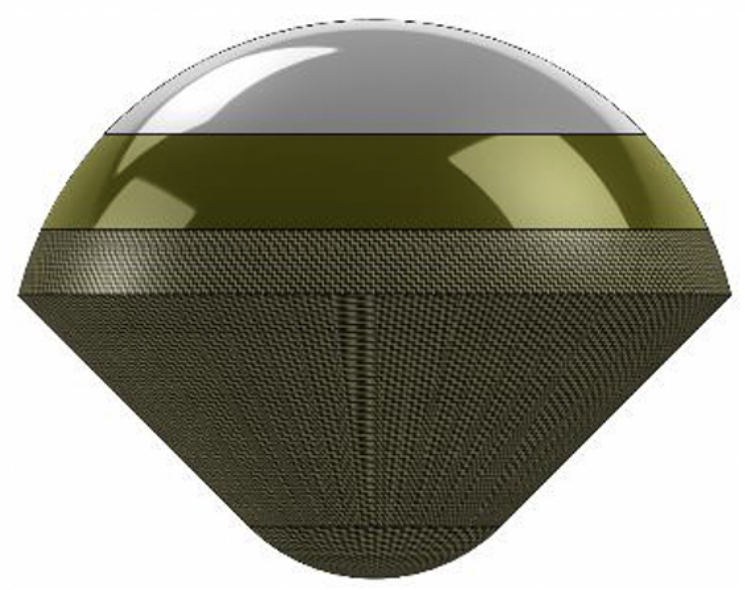

Figure 3-6: The small Venus probe is a 45-degree half-angle sphere cone ~20 cm in diameter (Credit: NASA ARC).

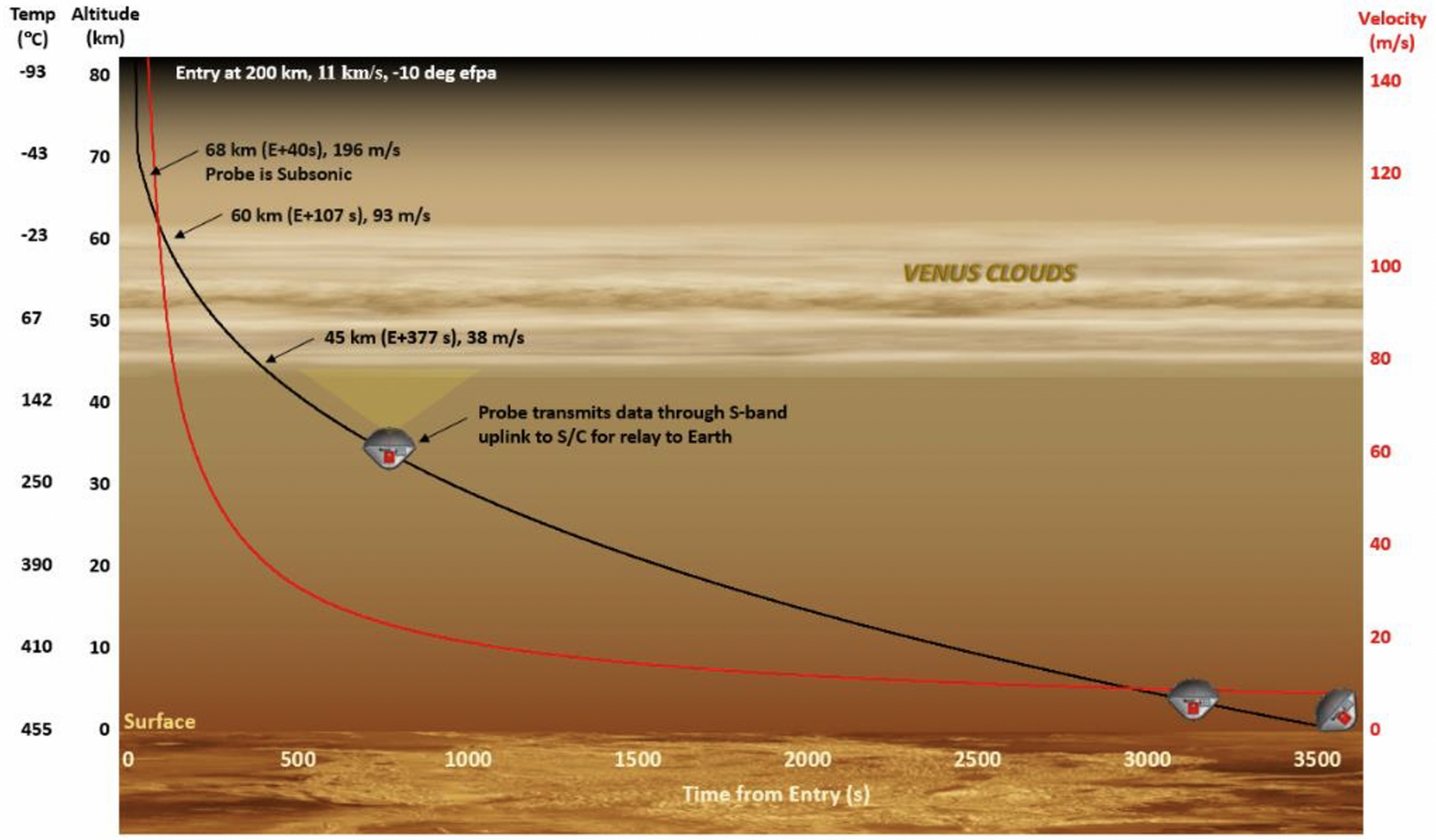

Figure 3-7: The science phase targets the Venus cloud layer between 45 and 60 km altitude, enabling ~270 sec of science observations (Credit: NASA ARC).

The probe diameter will be set based on instrument accommodation and mass growth allowance margins. Two key trades, heat shield separation and the need for a pressure vessel, are still open and will be closed based on mission requirements in the detailed design phase. The probe forebody TPS material is notionally Heat-shield for Extreme Entry Environment (HEEET) with the aftbody materials a radio frequency (RF) transparent TPS such as acid-resistant Polytetrafluoroethylene (PTFE; e.g., Teflon) or silicone impregnated reusable ceramic ablator (SIRCA).

## 3.5 Concept of Operations

The probe will follow the following preliminary Science Phase sequence of events (Figure 3-7), with absolute timing dependent upon the selected EFPA:

- Probe release (potential spin up) after final entry interface targeting
- Coast phase (~5 hours, low energy state)
- Pre-entry (initialization of key systems, TBD timing)
- Relay communications begins and continues throughout science phase
- Entry interface reached
- Heating pulse, RF blackout, peak G's (20 – 80 sec after entry interface)
- No heatshield separation (TBR) (30-90 sec after entry interface)
- Reconfiguration of probe for science data collection (if required)
- Enter clouds (100 – 200 s after entry interface)

- Primary science data collection (275-300 s data collection)
- Leave clouds (375 – 500 s after entry interface)
- Continued data transmission/re-transmission of science data (~20 minutes duration)
- Surface contact (~3500-4000 s after entry interface)
- Communication ends/vehicle passivated

Through the cloud layer and below, the science data will be transmitted direct to Earth at optimized data rates. Finally, a camera is being traded based on the available data budget, mass budget, and overall complexity. The camera could be used to provide context, such as below the cloud layer or to provide additional science value in other ways. However, objectives below the cloud layer, such as the potential to continue science observations with the primary instrument or to return an image of the surface will be performed on a best effort basis only.

# 4 Venus Habitability Mission Science and Instruments

**Abstract:** The most efficient way to investigate if the Venus clouds are habitable or not is a direct in situ probe. A balloon mission with a duration of a week or two can address the most important habitability challenges: the cloud particle droplet acidity and cloud-layer water vapor content and their variability. The balloon mission can also search for and confirm the presence of non-volatile elements needed for life's metabolism, as well as biosignature gases as signs of life.

We present an astrobiology-focused mission concept for a fixed-altitude balloon mission operating at 52 km altitude, with four mini probes to be deployed from the balloon to measure habitable conditions in the lower cloud region. The mission doubles as a preparation for sample return by determining if a subset of the cloud particles are non-liquid and how homogenous the cloud particles are, thereby informing sample collection methods and storage for the return journey to Earth.

**New science applications that emerged from the study:**

- The clear objective to focus on astrobiology, specifically the habitability of the cloud layers and the search for signs of life.
- The objective to measure acidity of Venus cloud particles.
- The objective to search for metals in cloud particles as one of the requirements for habitability.

**New instrumentation that emerged from the study:**

- Two independent single-droplet pH sensors that that measure acidity values not reachable by available technology are now under development.
- Mini probes of about 0.5 kg each, to make measurements of the lower cloud layers.
- The low-mass autofluorescence nephelometer first described in Chapter 2.

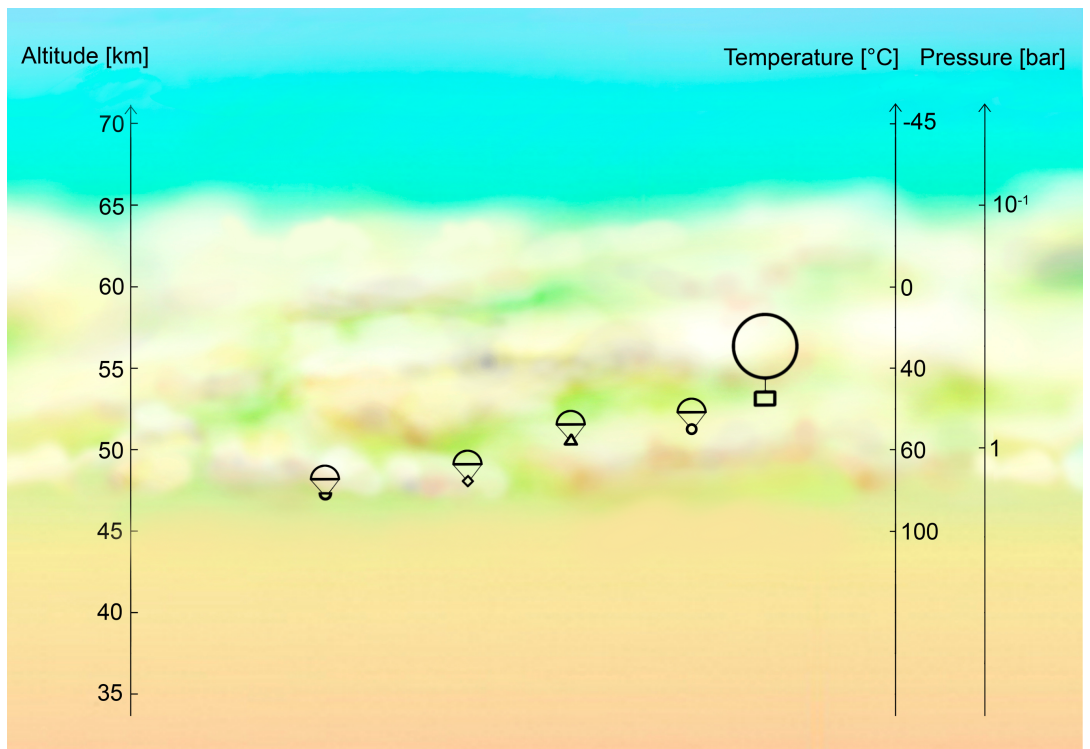

Figure 4-1. Illustration of the Venus Habitability Mission concept. Not to scale. A fixed-altitude balloon with a gondola contains the bulk of scientific instrumentation and releases four deployable mini-probes to sample lower altitudes. Current Figure Credit: J. Petkowska.

## 4.1 Introduction and Motivation

The main motivation of the Habitability Mission (Figure 4-1) is to investigate the cloud droplets with a variety of sensors to explore their ability to support life. The motivation comes from the point that if microbial-type life exists it must be within the temperate cloud layers.

The key challenges to life and hence properties to measure are dryness and acidity. If life exists, it most likely resides inside the cloud particles, protected from a free atmosphere far drier than the limits of life as we know it [14,96]. Yet concentrated sulfuric acid ($H_2SO_4$), the composition of the Venus cloud particles is fatal to Earth life's building blocks. A VLF main goal therefore, is to determine the particles' acidity. The unknown composition and suspected asphericity of the Mode 3 cloud particles [31] leaves room for the Mode 3 particles to be less acidic than pure concentrated sulfuric acid (a liquid, hence generating spherical particles). New work by [80] proposes that a subset of cloud particles may be far less acidic than previously thought, closer to pH levels where Earth extremophiles are known to live at a low pH of 0 or 1. See also [3,93]. A gas sensor can measure atmospheric water vapor in the cloud layers to confirm or refute the presence of humid pockets.

In addition to acidity and humidity, the Habitability mission aims to search for signs of life by way of biosignature gases and cloud particle composition.

| | Goals | Science Objectives | Gondola Instruments | | | | | Mini-Probe Instruments | | | | |
|---|---|---|---|---|---|---|---|---|---|---|---|---|
| | | | mTLS | AFN | pH Sensor | IGOM-A | Weather Instruments Suite (WIS) | IGOM-G | IGOM-A | pH Sensor | MOOSA | Weather Instruments Suite (WIS) |
| Habitability | 1. Measure Habitability Indicators | 1.1 Determine the amount of water vapor in the cloud layers | Green | Red | Red | Red | Red | Red | Red | Red | Red | Red |
| | | 1.2 Determine the pH of single cloud particles | Red | Yellow | Green | Red | Red | Red | Red | Green | Green | Red |
| | | 1.3 Detect and identify metals and other non-Volatile elements in the cloud particles | Red | Red | Red | Green | Red | Red | Green | Red | Red | Red |
| | | 1.4 Measure the temperature, pressure, and windspeed, as a function of altitude | Red | Red | Red | Red | Green | Red | Red | Red | Red | Green |
| Biosignatures | 2. Search for Evidence of Life in the Venusian Clouds | 2.1 Search for signs of life via gas detection | Green | Red | Red | Red | Red | Green | Red | Red | Red | Red |
| | | 2.2 Detect organic material within the cloud particles | Red | Green | Red | Yellow | Red | Red | Yellow | Red | Red | Red |
| | | 2.3 Identify organic material within the cloud particles | Red | Red | Red | Yellow | Red | Red | Yellow | Red | Red | Red |
| Sample Return | 3. Characterize Cloud Particles in Preparation for Sample Return | 3.1 Determine if the cloud particles are liquid or solid | Red | Green | Red | Red | Red | Red | Red | Red | Red | Red |
| | | 3.2 Determine if the cloud particles are homogeneous | Red | Green | Green | Green | Red | Red | Green | Green | Green | Red |

Table 4-1 Venus Habitability Mission science goals and instruments. Color provides a subjective judgement of the ability of an instrument to meet the science objectives. Red = Not applicable, yellow = applicable, green = highly applicable.

## 4.2 Science Goals and Objectives

The overall mission science goals are: to determine habitability of the Venus cloud droplets; to search for signs of life; and to assess the cloud particle properties in order to inform capture, storage and transfer technologies for an atmosphere sample return mission (Table 4-1). In this section we describe the science objectives and relevant instruments.

### *4.2.1 Goal 1: Measure Habitability Indicators*

**Objective 1.1: Determine the amount of water vapor in the cloud layers.** Water, $H_2O$, is of interest because the prevailing view is that Venus clouds are uniformly an incredibly dry environment. Water vapor in the clouds, however, cannot be globally homogeneous based on measured variations in vertical $H_2O$ abundance [80]. If water vapor can be as high as some measurements suggest, Venus may have pockets of high enough humidity for life to thrive (reviewed in [80]). If not, and if life on Venus is based on water, then the ultra-dry and hyper-acidic cloud conditions are extreme for life as we know it here on Earth [14,96]. To survive life would require evolutionary adaptations never developed by life on Earth. The instrument for this objective is the TLS.

In addition to the water vapor we would like to determine the water content of the cloud particles. For this purpose an instrument is still TBD.

**Objective 1.2: Determine the pH of single cloud particles.** The Venus cloud particles are believed to be predominantly concentrated sulfuric acid droplets. The droplets are extremely acidic, billions of times more acidic than the most acidic environments where life is found on Earth. No life as we know it can survive in droplets of such high acidity, although life with protective shells (e.g., certain lipids, sulfur, or graphite) might.

Measuring the single droplet pH for a large number of cloud particles is a key indicator of habitable conditions. There is strong motivation to confirm the high acidity picture and to investigate if the pH of some cloud droplets is much more favorable for life as we know it.

| Gas | Motivation | Instrument |
|---|---|---|
| $O_2$ | Potential sign of life; prior detection | IGOM-G or Tartu $O_2$ sensor, TLS |
| $SO_x$ | Variable profile indicative of cloud particle chemistry | IGOM-G |
| $H_2O$ | Variable profile indicative of cloud particle chemistry, including some anomalously high values | TLS |
| $NH_3$ | Potential sign of life; potential neutralizing agent for cloud droplets; prior tentative detection | TLS or IGOM-G |
| $PH_3$ | Potential sign of life; prior tentative detection | TLS |
| $CH_4$ | Potential sign of life; prior anomalous detection | TLS or IGOM-G |

Table 4-2. Gases of interest. TLS is the Tunable Laser Spectrometer, MEMS IGOM-G is the miniature gas analyzer (PL Patent PL228498B1). The Tartu $O_2$ sensor is described in [199] and in Appendix E. For a list of and commentary on detected gases in the Venus atmosphere see Chapter 1.

New work proposes updrafted mineral dust from the surface or locally-produced $NH_3$ could dissolve in sulfuric acid droplets, buffering them to habitable levels (see Section 1.4 and [3,80,93]). If the droplets are pH of 0 or 1, their acidity is consistent with the most acidic environments where life is found on Earth.

If droplets are less acidic than previously thought, we need to know what fraction. The question of homogeneity is important for the sample return mission: are all particles the same or are a tiny fraction different with properties that make them potentially habitable? The answer will inform sample capture and storage technologies and volume, and will help establish the priority altitude for sample capture.

The instruments relevant to this objective are the single-particle pH meter and the AFN.

**Objective 1.3: Detect and identify metals and other non-volatile elements in the cloud particles.** Life as we know it requires metals and other non-volatile species (e.g., phosphate). Detection of such species as components of cloud particles raises the potential for habitability of the clouds. Possible Vega and Venera detections of Fe and P tentatively suggest a heterogeneous composition of dissolved ions and chemicals in cloud particles. (While our proposed mission will not sample the ~60 km altitude where the "mysterious UV absorber" lies, it is worth noting that the UV absorber might include metals such as Fe in $FeCl_3$).

Measuring the elemental composition of the aerosols as a function of the altitude and searching for organic material within the particles will inform the priority altitudes for sample capture.

The instrument relevant for this objective is the MEMS aerosol analyzer (IGOM-A).

**Objective 1.4: Measure the temperature, pressure, and windspeed as a function of altitude.** The measurements of temperature-pressure profiles and wind speed in the clouds of Venus are not directly astrobiological in nature but nevertheless are worth measuring in their own right. Transient planet gravity waves are encoded in the temperature pressure profiles (e.g., [100]) and measuring them helps substantiate the concept of moving material, including hypothetical spores, up from lower atmosphere layers [14].

The combined temperature and pressure sensor from Emerson's Rosemount Inc. is rated to operate up to 344 bars and 316 °C [101] and is thus suited to cover temperature measurements from 80 km to 20 km and pressure measurements from 70 km to 20 km therefore encompassing the entire cloud deck and stagnant haze layer below the clouds. The instruments for this objective are part of the Weather Instruments Suite (WIS).

| Instrument | Mass (kg) | Volume ($cm^3$) | Average Power (W) | Data Vol. per Meas. (kB) | *TRL |
|---|---|---|---|---|---|
| Mini Tunable Laser Spectrometer (TLS) | 4.60 | 240 | 14.0 | 1000 | 6 |
| MEMS Ion-Gas Micro-Spectrometer for Aerosols (IGOM-A) | 0.34 | 400 | 1.0 | 27 | 4 |
| Autofluorescing Nephelometer (AFN) | 0.80 | 100 | 40.0 | 120 | 3 |
| TOPS pH sensor (pHS) | 0.35 | 844 | 2.0 | 1 | 2 |
| Imaging Unit (IU) | 0.15 | 250 | 0.5 | 100 | 5 |
| Weather Instruments Suite (WIS) | 0.10 | 98 | 1.0 | 0.05 | 5 |
| **Total Gondola Subsystem Mass** | **6.34** | **1932** | **58.5** | **1248** | |

Table 4-3. Scientific instruments considered for the gondola. *TRL does not take into account the Venus environment.

| Instrument | Mass (kg) | Volume ($cm^3$) | Average Power (W) | Data Vol. per Meas. (kB) | *TRL |
|---|---|---|---|---|---|
| MEMS Ion-Gas Micro-Spectrometer for Gas (IGOM-G) | 0.34 | 400 | 0.8 | 27 | 4 |
| MEMS Ion-Gas Micro-Spectrometer for Aerosols (IGOM-A) | 0.34 | 400 | 1.0 | 27 | 4 |
| TOPS pH sensor (pHS) | 0.35 | 844 | 2.0 | 1 | 2 |
| MOOSA pH sensor (MOOSA) | 0.20 | 10 | 2.0 | 1 | 2 |
| Weather Instruments Suite (WIS) - one in each mini-probe | 0.10 | 98 | 1.0 | 0.05 | 5 |
| **Total Mini Probe Instrument Mass** | **1.63** | **2,046** | **9.80** | **56.20** | |

Table 4-4. Scientific instruments considered as payloads for the mini probes. *TRL does not take into account the Venus environment.

### *4.2.2 Goal 2: Search for Evidence of Life in the Venusian Clouds*

**Objective 2.1: Search for signs of life via gas detection.** A number of gases are of key interest for habitability or even as signs of life. The habitability mission aims to detect reduced and anomalous gas molecules as a sign of disequilibrium chemistry and as biosignatures (Table 4-2). The instrument for this objective is the Tunable Laser Spectrometer (TLS) and IGOM-G. The mini TLS will have four channels, meaning only four gases can be measured. See Chapter 1 for motivation for the gases listed in Table 4-2. Here we list the four top choices.

**Objective 2.2 and 2.3: Detect and identify or constrain organic material within the cloud particles.** The presence of organic chemicals is crucial for any life to exist, therefore any detection and identification or constraint of organic compounds in the atmosphere of Venus would bolster support of habitability of the clouds. The search for organic molecules in the Venus cloud particles has not yet been attempted.

Autofluorescence is a way to detect (and potentially identify) organic compounds. On Earth, many compounds are known to fluoresce when subject to UV radiation (see Figures 2-1 and A-2). Ongoing work being carried out by Firebird Biomolecular Sciences and by Droplet Measurement Technologies will eventually inform us of the optimal UV wavelength(s) for the excitation laser and detector(s).

The instrument for Objectives 2.2 and 2.3 is the autofluorescing nephelometer (AFN). To go beyond detection of organic compounds towards identification, we could consider adding a Raman spectroscopy component to the AFN. (A mass spectrometer is beyond the scope of the mission, but see Appendix C.)

#### 4.2.3 Goal 3: Characterize Cloud Particles in Preparation for Sample Return

**Objective 3.1: Determine if the cloud particles are liquid or solid.** The Venus Mode 3 cloud particles have a component of unknown composition and are different from spherical [31]. A non-spherical particle cannot be liquid. The possible non-liquid nature of Mode 3 particles is of significant interest, because of the Rimmer et al. 2021 theory [80] that some Venus cloud particles are solid slurries with a pH of up to +1. A confirmation of the shape of the Mode 3 particles along with measurements to constrain the composition will help determine if the Mode 3 particles might have habitable properties.

With respect to a sample return mission, we need to understand if the cloud particles are solid or liquid as capture methods and capture substrates will differ.

The instrument for this objective is the AFN, with supporting data provided by the pH sensors.

**Objective 3.2: Determine if the Venus cloud particles are homogeneous.** The current thinking is that the clouds on Venus are homogenous, but separated into three particle Modes whose relative fraction differs with altitude. Any further degree of inhomogeneity will inform sample return, as in what volume of sample must we collect? In other words, will a small sample be sufficient, or should we aim to collect many of the "one in a million" particles that might host life?

The instruments for this objective are the AFN, the single-particle pH meter, and the Micro-Electro-Mechanical Systems (MEMS) Aerosol Analyzer (IGOM-A).

### 4.3 Instruments

The balloon gondola will host instruments (Table 4-3) for a one-week duration mission and sample the atmosphere horizontally at a fixed altitude. In addition, mini probes will be deployed downwards from the balloon to sample the atmosphere vertically, at two different locations, to assess the habitability of the clouds across the cloud deck and below. The mini-probes (Table 4-4) will measure the vertical profiles of selected gases, single droplet acidity, and search for the presence of metals, complementing the measurements done by the gondola instruments at the fixed altitude of 52 km. Each mini probe will further include a UHF radio transmitter, batteries, and a parachute.

#### 4.3.1 Single Particle pH Meter

A single-particle pH meter must be custom-developed for the Venus atmosphere, because so far, no pH meter can measure pH in negative numbers accurately nor across the acidity range anticipated on Venus. We aim to measure a pH of -1 or higher. Our main goal is to distinguish between droplets with acidity of << 0 and those with pH > 0. A lingering question is: can the pH of a semi-solid particle be measured equally well as the pH of a liquid particle?

We have engaged two separate groups to pursue two different pH meter concepts.

**The Tartu Observatory pH Sensor** (TOPS) will measure the acidity of single Venusian cloud droplets by the established method of fluorescence spectroscopy, which is widely used for pH measurements (see Figure 4-2).

The general mechanism for the single particle acidity sensor is using a dye-sensitized sensor plate and illuminating it with various wavelengths of light. After cloud particles hit the sensor plate, different spots will fluoresce with different intensities at a given excitation wavelength, allowing for measurement of single particle pH.

Fluorescein is a candidate fluorophore for the TOPS sensor due its high fluorescence intensity [102] and stability over very wide range of sulfuric acid concentrations, from diluted to highly concentrated solutions (e.g. over 10 M sulfuric acid) [103,104]. Under strongly acidic conditions, fluorescein is in its cationic form. Since emission responses of fluorescein forms differ drastically from each other, fluorescein is a good candidate as a fluorescence dye for the Venusian single-particle pH meter [102,104]. The development of the TOPS pH sensor will be carried out by Prof. Mihkel Pajusalu at the Tartu University. For more details about the TOPS single particle pH sensor see Appendix E.

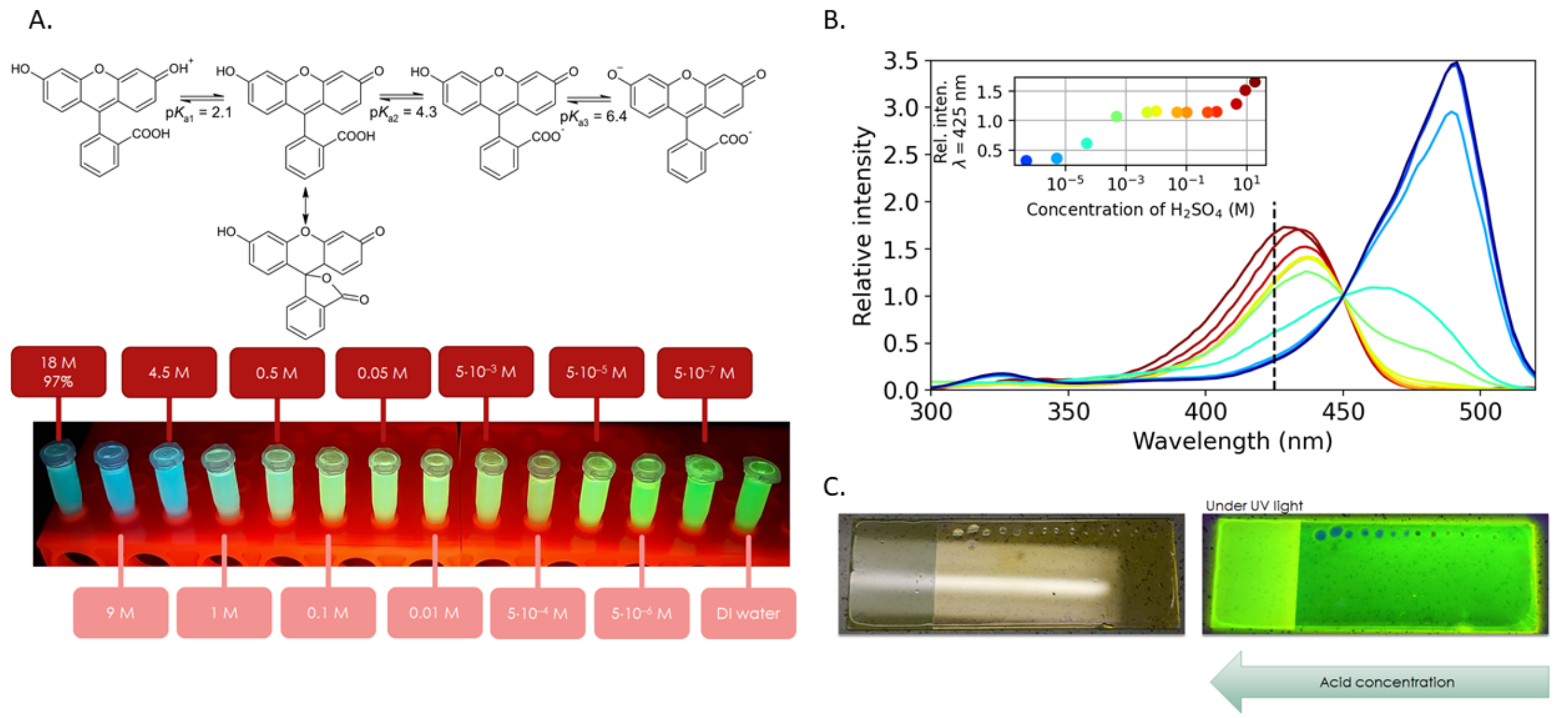

Figure 4-2. Figure E-1. A) Fluorescein in $H_2SO_4$ solutions of varied concentrations under UV light. B) Emission spectra of fluorescein excited in different acid concentrations. Spectra are normalized to the wavelength of 450 nm. Inset: Relative intensities at 425 nm show a clear dependence on $H_2SO_4$ concentration. Note that while this initial experiment is done for fluorescein, further experiments will find the best fluorescence dye (or the mixture of dyes) for this specific application. One potential group of alternatives are rhodamine dyes [215]. C) Glass plate (1×4 cm) covered with polyvinyl alcohol doped with fluorescein, covered droplets of sulfuric acid (18 M – 0 M).

**The Molybdenum Oxide Sensor Array (MoOSA) pH Sensor** can categorize individual cloud particle pH acidity values using an optical method which is mechanically and chemically robust and can measure minute sample volumes (see Figure 4-3).

MoOSA will categorize individual droplets that are deposited on its surface into one of three acidity categories: pH < 0; pH 0-1; pH > 1. Deposition will be brought about by passively exposing one surface of MoOSA to the Venus atmosphere in the altitude range 48-65 km. MoOSA will return counts in each pH category, corresponding to individual depositions on an array of plasmonic $MoO_3$ coated micro-ring resonators (MRRs). The development of the MoOSA pH sensor will be carried out by Prof. Arnan Mitchell at the Micro-Nano Research Facility at the Royal Melbourne Institute of Technology. For more details behind the operation of the MoOSA pH sensor see Appendix E.

### *4.3.2 Autofluorescing Nephelometer*

The autofluorescing nephelometer (AFN) will determine particle shape and indices of refraction (Figure 4-4). The autofluorescence will indicate the presence of organic materials in the cloud droplets. Work is ongoing to determine the optimal wavelength for organic material detection, namely the effect of concentrated sulfuric acid on the autofluorescence behavior of organic molecules. Our preliminary results on fluorescence behavior of organic molecules in concentrated $H_2SO_4$ allow for a recommendation of a specific wavelength for the AFN excitation

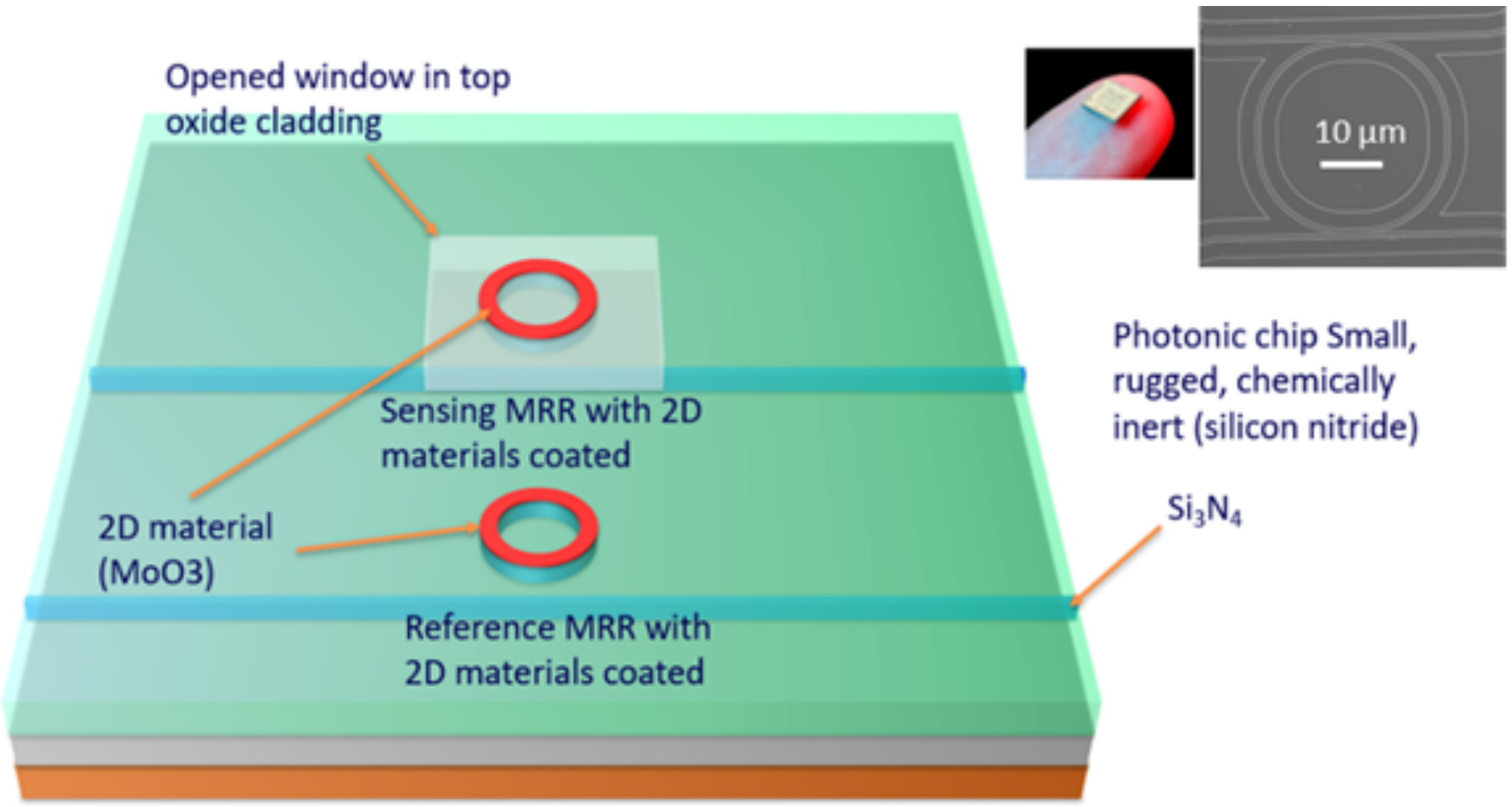

Figure 4-3. Schematic design of the Molybdenum Oxide Sensor Array (MoOSA) pH sensor. Schematic of single top window active MRR above with datum reference MRR below (right) actual manufactured prototype chip with MRR balanced on finger (left). A prototype test unit comprising of a single 20 micron diameter MRR coated with $MoO_3$ has already been tested with nitric acid and showed a reliable linear output response in the range $1 < pH < 6$ [198]. MoOSA is at TRL = 4.

laser and detector (see Figure 2-2 and Appendix A.2).

The AFN would likely be the most power-hungry instrument. Weighing about 800 g with a peak operating power requirement of 40 W. The development of the AFN will be led out by Darrel Baumgartner at Cloud Measurement Solutions and Droplet Measurement Technologies. See Chapter 2 for further description.

### 4.3.3 MEMS Aerosol Elemental Analyzer

The MEMS Ion-Gas Micro-spectrometer with Optical Signal Detection for Aerosols Analysis (IGOM-A) can identify ions dissolved in cloud droplets. The MEMS devices are lightweight (~340 g) and low power with a peak power requirement of 4 W.

The IGOM-A passively collects cloud particles, while the microfluidic channel allows for the flow of the sample to the sample ionizer, which ionizes the sample and induces light emission. The emission from the ionized sample is detected by the MEMS spectrometer which allows for qualitative (based on the wavelength of emission peaks) and quantitative (based on the measured intensity of emission peaks) identification of chemical species in the sample mixture. The result of a single measurement is a VIS/NIR emission spectrum (27 kB per measurement). The IGOM-A must be tailored to specific chemical species. Our main targets are oxidized P species and metals including Fe, Mg, Ca, Mn, Cu, Na, K.

The MEMS device can detect ppm to ppb abundance of specific metals and other ions dissolved in the sample (the limit of detection depends on an ion). One measurement of sample composition requires at least 34 droplets with a mean diameter of 1 µm. Measurements cannot be done for acidic aerosols below circa 100 hPa (0.1 bar). Since the clouds decks span the atmospheric pressure range of approx. 0.5 to 2 bar the MEMS pressure limitations are not a serious issue.

Despite being originally developed for Mars applications, the MEMS IGOM spectrometer is built to withstand acidic conditions of the clouds of Venus. The instrument is designed to work in the corrosive atmosphere of Venus. The ionizing MEMS structure is made of corrosion resistant materials and is assembled to be helium-leak-proof. Like other instruments, IGOM-A must be properly packaged to prevent penetration of the acidic corrosive cloud particles.

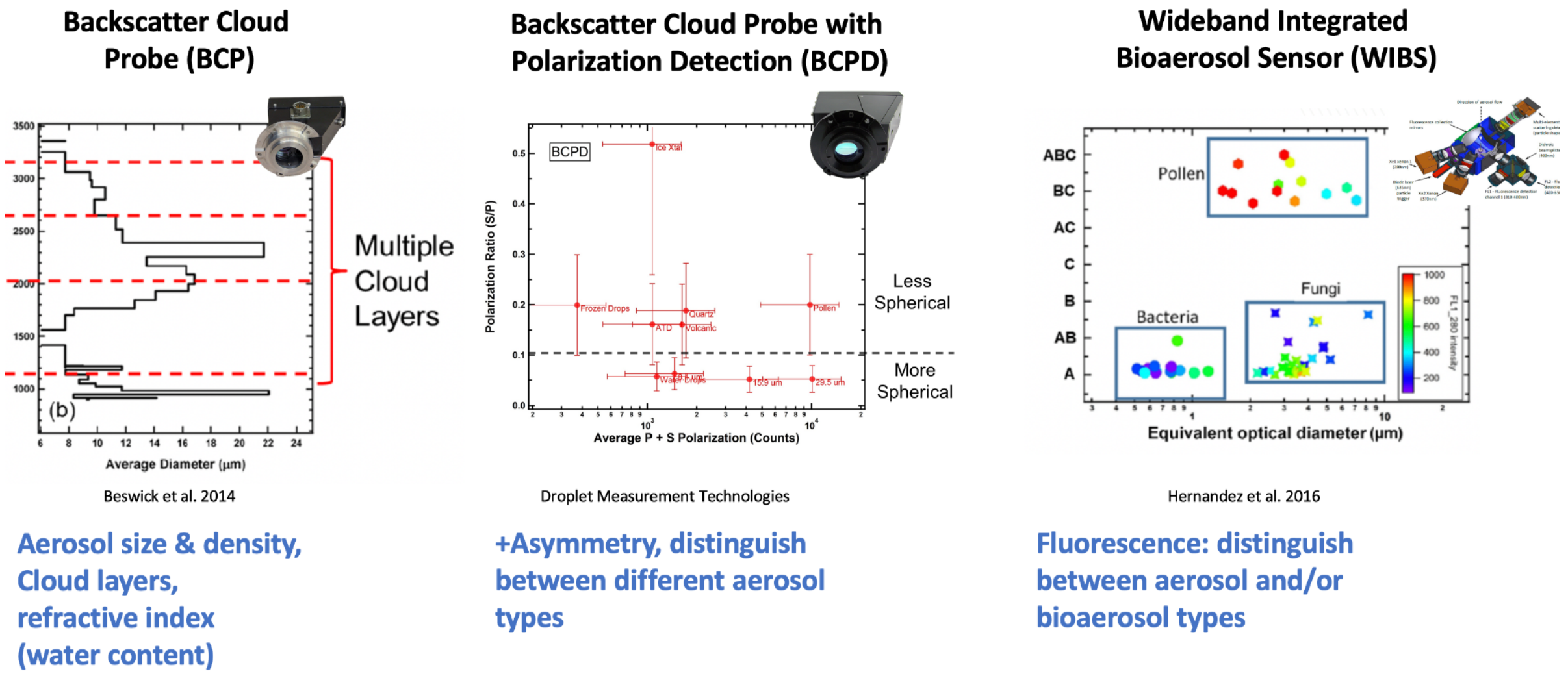

Figure 4-4. Types of measurements enabled by the Autofluorescing Nephelometer (AFN) based on precursor instruments. Left: Like the Backscatter Cloud Probe (BCP), the AFN would be able to map out cloud layers based on particle diameter and concentration. Image from [216]. Center: The AFN also includes polarization detection; prior data from the Backscatter Cloud Probe with Polarization Detection (BCPD) demonstrates the ability to use the polarization channels (S, P) to distinguish between liquid water droplets, frozen droplets, ice crystals, inorganic particles (volcanic, quartz), and large biological particles. Image: Droplet Measurement Technologies. Right: Complex multichannel single particle fluorescence instruments like the WIBS can distinguish between different types of biological particles; in contrast, the AFN would utilize a single fluorescence channel as part of a BCPD-like instrument, enabling fluorescence detection in a simple compact package. Image from [99].

IGOM-A will be built by Prof. Jan Dziuban from Wroclaw University of Science and Technology.

### *4.3.4 Mini Tunable Laser Spectrometer*

The Tunable Laser Spectrometer is a mature in situ gas detection instrument used in a variety of commercial applications including medical sensing, industrial sensing, and Earth science [105,106]. The TLS has seen tremendous success on the Mars Curiosity Rover [107,108].

The TLS infrared laser absorption spectrometer has a small gas cell with highly reflective surfaces so that a laser can bounce back and forth thousands of times to construct an effective path length of up to 10 km. By enabling such a long path length at gas cell pressure below ~100 mbar, the TLS can measure high precision gas abundances (ppb level) at very high spectral resolution ($\lambda/\Delta\lambda$ = 10 million which enables detection of individual rovibrational lines). The unique advantage of TLS is a completely unambiguous detection of a given molecule and the maturity of the instrument.

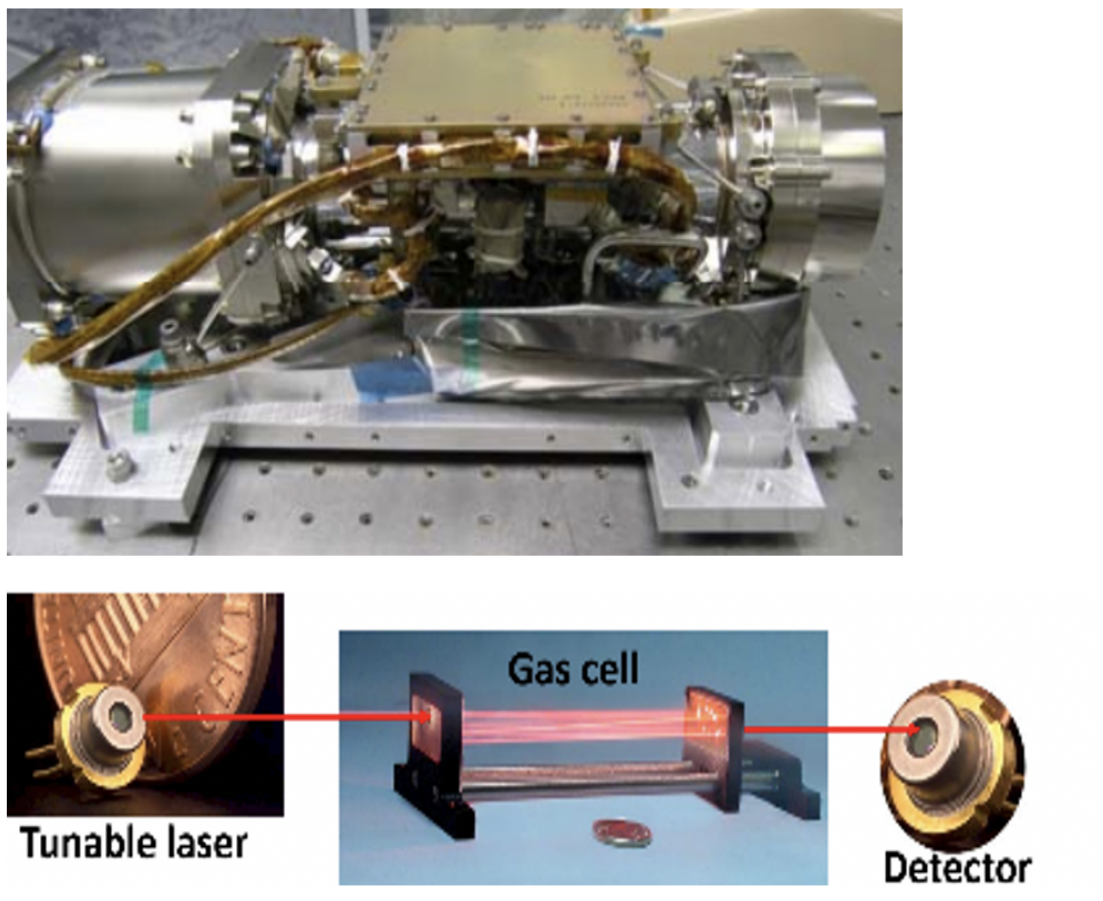

Figure 4-5. The mini TLS instrument and gas cell and detector. Credit: C. Webster.

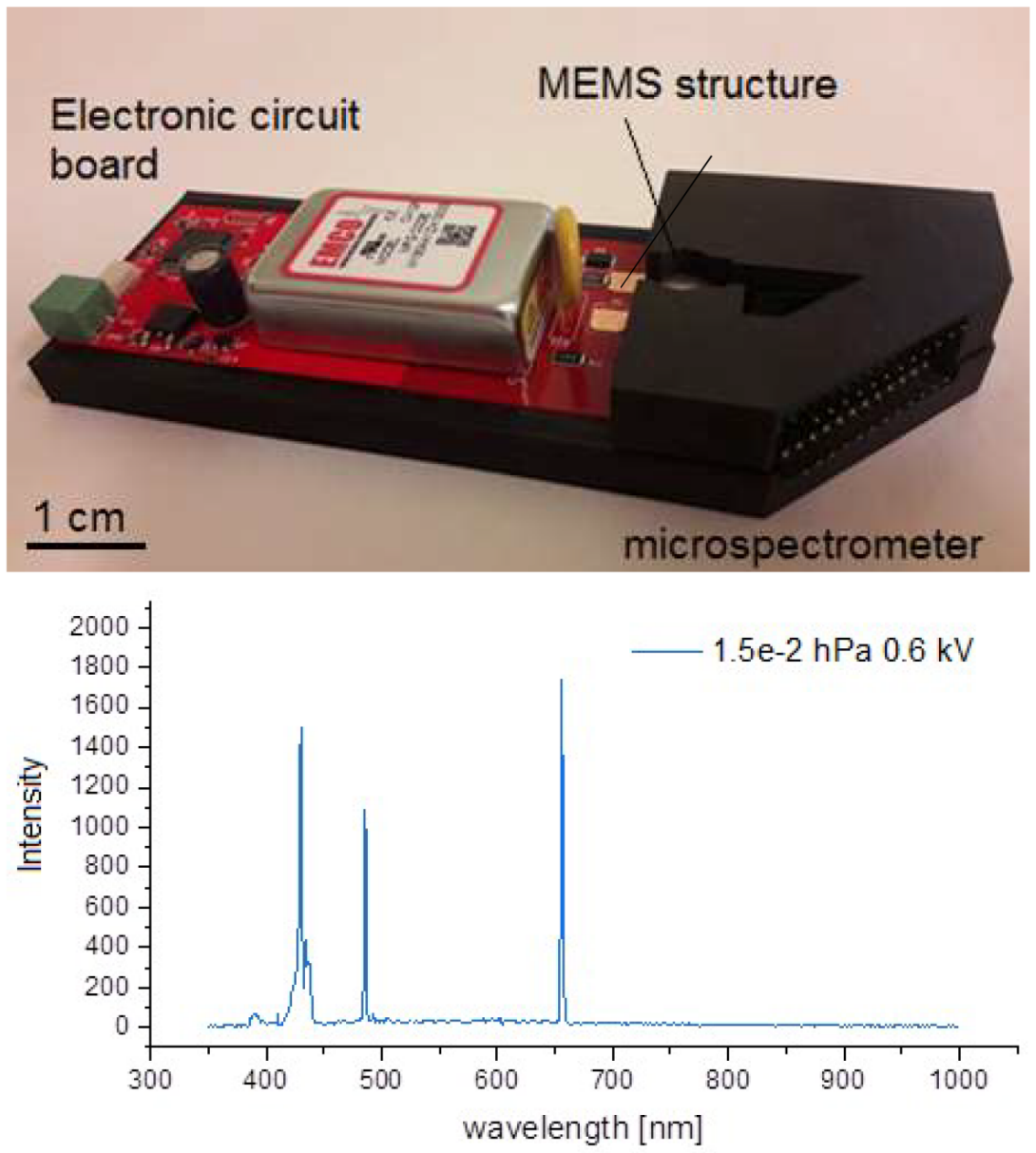

Figure 4-6. Complete MEMS IGOM-G spectrometer (top) and example of the measured spectrum of methane ($CH_4$) gas (bottom).

The mini Tunable Laser Spectrometer is a miniaturized version of the TLS under development at JPL (Chris Webster, PI, Figure 4-5). While the mini TLS has not yet seen space flight, it traces extensive heritage.

The mini TLS considered for the gondola has four channels, each of which can cover a wavelength range dedicated to a gas of interest. Additional molecules also might fall within a specially chosen wavelength range. Because the channels have a narrow wavelength range (on order of 6 wavenumbers) the target gases of interest must be chosen in advance.

The mini TLS data product is a spectrum, recorded as a text file. The mini TLS can operate continuously with a data rate of 500 kilobytes $s^{-1}$, assuming a measurement integration time of 1 s.

#### *4.3.5 MEMS Gas Molecule Analyzer*

A MEMS device similar in design to the one described in Section 4.3.3 can be tailored towards detection of specific gases (Figure 4-6). The MEMS Ion-Gas Micro-spectrometer with Optical Signal Detection for Gas Mixture Analysis (IGOM-G) can detect trace gases in the atmosphere to limit of detection of 0.1 to 0.01 ppm (depending on the expected gas composition and the atmospheric pressure).

The gas sample is passively acquired from the atmosphere, and ionized inside the MEMS electron-impact ionizer to induce light emission. Peaks of emission, detected by a MEMS spectrometer, enables the identification of gases. The IGOM MEMS devices has a sampling time ranging from 1 min for low pressures up to a maximum of 10 mins, including sample acquisition.

IGOM-G can measure $O_2$, to low ppm levels and $NH_3$ and $CH_4$ down to ppb levels. Note that IGOM-G was originally developed to identify the isotopic composition of the Martian $CH_4$.

## 4.4 Concept of Operations Summary

After atmospheric entry the balloon will deploy and inflate, then operate for a nominal lifetime of about one week. The balloon will operate at a fixed altitude and also deploy four mini probes, sets of two identical ones dropped at different Venus latitudes (Figure 4-7). An alternate variable altitude balloon was considered, and operations and vertical excursions are shown in Figure 4-8.

## 4.5 Summary

The Habitability Mission is a streamlined mission concept design to search for signs of life and answer key questions about the habitability of the clouds of Venus. The science centers around the Venus cloud droplets, to determine their acidity and chemical composition, including water content and the presence of dissolved metals. The mission also aims to resolve long standing chemical anomalies of the clouds of Venus, including the presence of ammonia, molecular oxygen and the anomalous shapes of Mode 3 cloud particles, all of which are motivated by the search for signs of habitability and life.

The data on cloud particles' physical and chemical properties returned by the habitability mission will also inform the future development of sample capture, storage and transfer technologies. Together with balloon deployment and operation, the mission will demonstrate critical technologies needed for a successful atmosphere sample return mission.

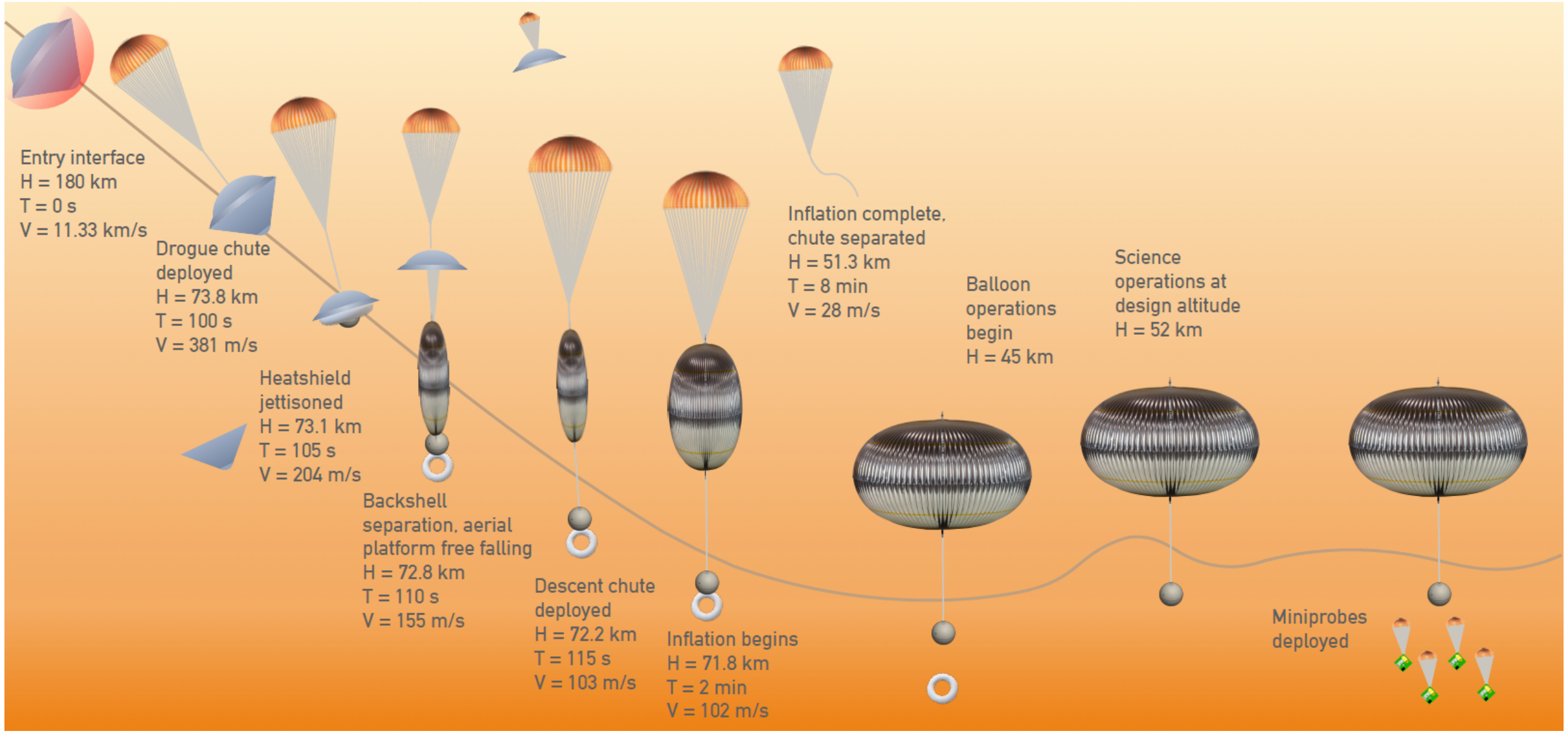

Figure 4-7. Entry, descent, and balloon deployment and operation sequence.

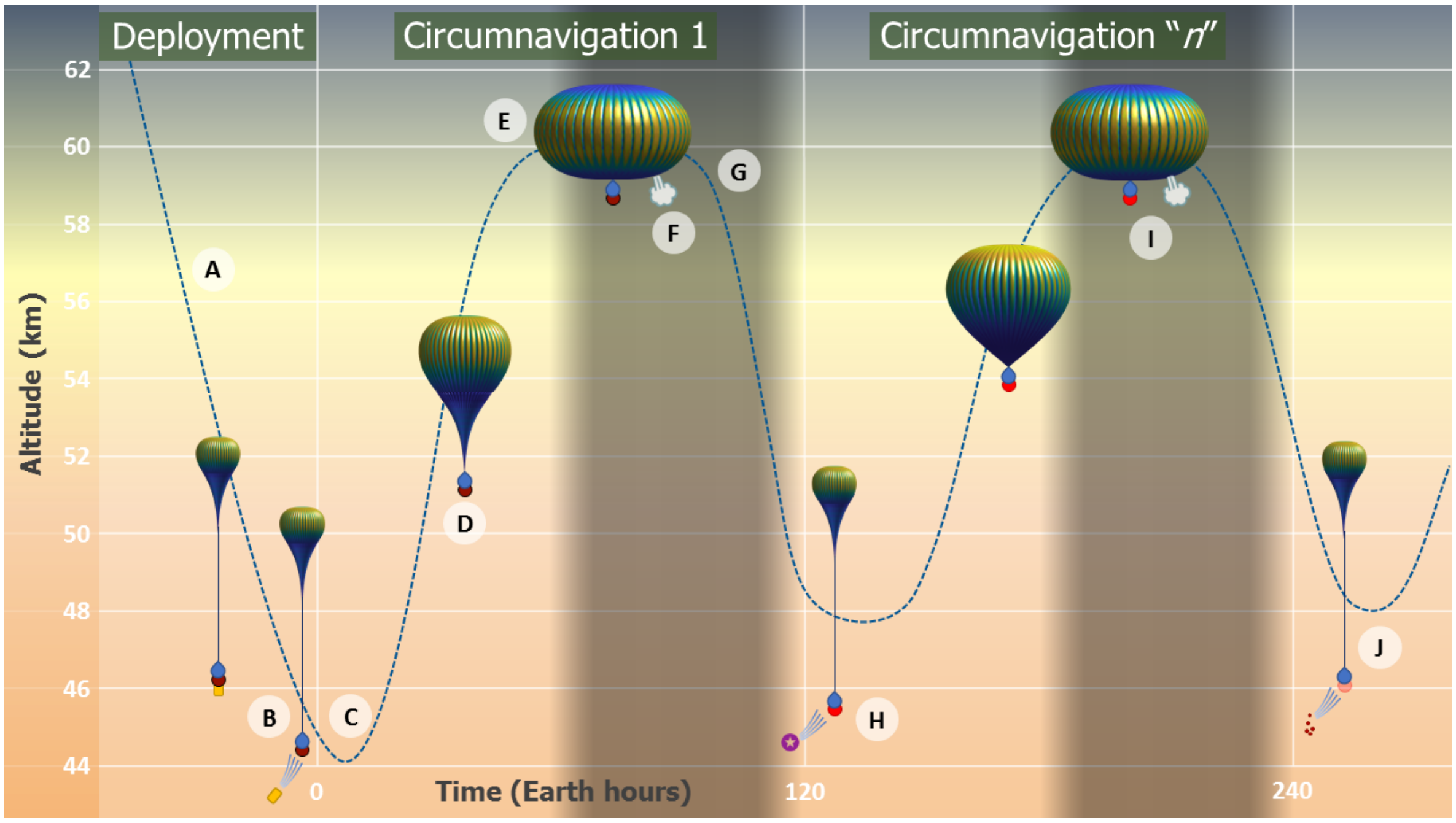

Figure 4-8. Balloon deployment and operation sequence for a simple variable altitude balloon, an alternative to the balloon mission shown in Figure 4-7.

# 5 Venus Habitability Mission Architecture and Design

**Abstract:** The Venus Habitability Mission is designed to search for the evidence of life in the Venusian clouds, measure habitability indicators, and characterize aspects of Venusian cloud droplets and aerosols that might be associated with life. The mission concept consists of an aerial platform that floats at altitude of 52 km to perform science operations. Vertical oscillations of tens of meters are expected and useful. The nominal mission life is one week, with possible extension to two months. The aerial platform will transfer data to Earth via both an orbiting relay spacecraft and a direct-to-Earth transmission.

The complete flight system can be launched on a single launch vehicle and for our design we consider a July 30, 2026 launch date. The launch rocket will deliver the orbiter/cruise vehicle with the entry system to a trans-Venus trajectory. The launch mass of the stacked configuration is 420 kg. The entry system carrying the aerial platform, inflation system, and gondola will enter the Venusian atmosphere 122 days after launch. The aerial platform is a super pressure balloon with a 20 cm-diameter cylindrical gondola attached to the balloon by a 12 m long tether. To reduce the overall cost and risk, heritage systems are used where possible.

**New mission concepts that emerged from the study:**

- A combined fixed altitude balloon platform with the deployment of mini probes to sample a lower altitude.
- A new niche in the balloon platform mission design area with a short duration mission (one week) and miniaturized focused instrumentation.

## 5.1 Background

Since the deployment of Vega balloons in the 1985 [109,110] many studies from a variety of countries and space agencies have suggested the use of balloons for the exploration of Venus [111–126] (for a review see, e.g., [127]). The proposed missions utilizing balloon platforms have diverse science objectives, ranging from specific, focused mission concepts, e.g. balloon-based Venus seismology studies [128] to broad climate characterization [129], to assessment of the habitability of the clouds [130] and the search for signs of life [131,132]. Concepts for balloon exploration of the lower atmosphere of Venus, at ~25 km, and even the near-surface atmosphere (~5 km) have also been considered [133].

Here we describe our habitability mission concept, starting with the instrument payload, followed by the flight systems, mission design, and operations.

## 5.2 Instrument Payload

### 5.2.1 *Aerial Platform Payload*

The Venus Habitability Mission balloon gondola (i.e., aerial platform) will host science instruments (Table 4-3) for a one-week duration mission and sample the atmosphere horizontally at a fixed altitude of 52 km as well as deploy mini probes to sample lower atmosphere layers (Table 4-4). For a science and instrument discussion, see Chapter 4.

We choose 52 km because it is the middle of the lower cloud atmosphere layers—the environment with the longstanding atmosphere anomalies we aim to confirm and investigate whether or not there is any association with habitability or life (see Chapter 1).

### 5.2.2 *Mini Probes: Deployable Instruments*

The Venus Habitability Mission will use mini probes to sample atmosphere layers beneath the fixed altitude balloon. Four mini probes (Table 4-4) will be deployed downwards from the balloon to sample the atmosphere vertically, at two different locations. To assess the habitability of the clouds across the cloud deck and below, the mini probes will measure the vertical profiles of selected gases, single droplet acidity, and search for the presence of metals, complementing the measurements done by the gondola instruments at the fixed altitude of 52 km.

| System | CBE (kg) | Contin-gency (%) | MEV (kg) |
|---|---|---|---|
| Orbiter/cruise vehicle | 300 | 50% | 450 |
| Entry System | 120 | 50% | 180 |
| Aerial Platform System | 43 | 50% | 65 |
| **Total Stacked Mass** | **420** | **50%** | **695** |

Table 5-1 Summary of mass of flight system elements. CBE is current best estimate and MEV is maximum expected

The mini probes include a "bus" consisting of a computer, transceiver, and battery. The above bus subsystems are based on the "ThumbSat" approach by Scoutek. The mini probes include a gliding mechanism to increase residence time. This might be a parachute or components including a parawing, risers, and a chute/drag streamer. The mass of a mini probe will vary from about 0.3-0.6 kg, depending on its science payload (Table 4-4). The subsystem mass will be on order 25 grams, and the gliding mechanism mass will be up to 25% of the total mini probe mass. Ideally the mini probes will be designed to withstand temperatures up to 120 °C and pressures of at least 6 bar (35 km above the surface corresponds to 6 bar and 180 degrees °C), meaning they should not fail until they fall well below the cloud layer.

## 5.3 Flight Systems Description

The key flight system elements are the orbiter, the entry system, and the aerial platform. Table 5-1 summarizes the mass of each element. We build upon past aerial platform concepts, which have been studied for decades (reviewed in Section 5.1).

### 5.3.1 Orbiter

The primary function of the orbiter is as the communication relay between the aerial platform and the Earth ground station. The orbiter is a microsatellite that leverages small spacecraft technologies, including those from the Mars Cube One (MarCO) CubeSats that acted as a relay between the InSight Lander on Mars and the Earth communications ground station [134].

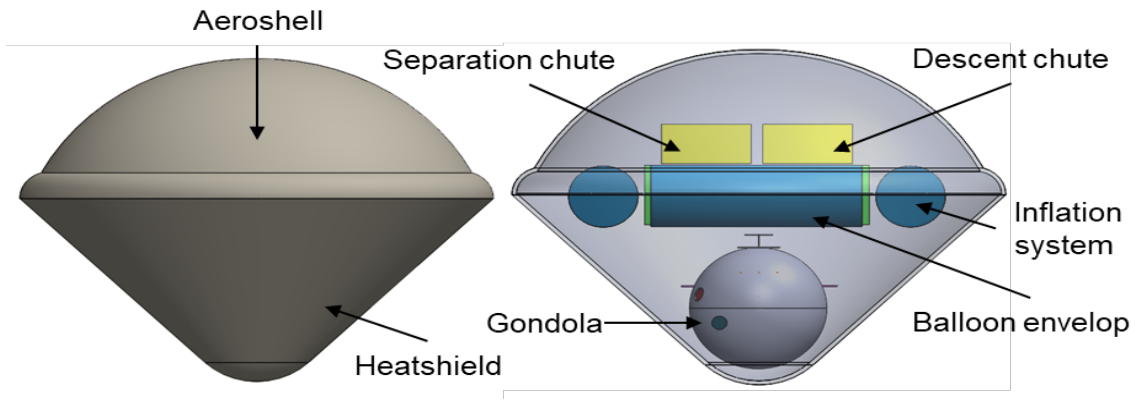

Figure 5-1. Schematic of the entry probe with notional accommodation of the entry probe payload. Structural elements required to hold the different elements together inside the entry vehicle are not shown.

The microsatellite includes a propulsion stage to allow it to enter Venus orbit. During the interplanetary cruise phase, the microsatellite communication system will also operate for the entry system. The orbiter has a medium gain antenna to communicate with the gondola and a high gain antenna to communicate with the Deep Space Network on Earth.

### 5.3.2 Entry Probe

The entry probe geometry is based on the Galileo and Pioneer Venus entry probes [135]. The probe has a maximum diameter of 1.5 m and has a traditional 45-degree sphere-cone geometry. The nose radius is 0.30 m. Major components of the probe are shown in Figure 5-1 which include the heatshield, aeroshell structure, balloon envelope, balloon inflation system, and gondola.

**Probe mechanical design**. The descent probe houses the balloon, four inflation tanks, two parachutes, and a gondola (Figure 5-1).

**Parachute design**. The probe uses two parachutes, both of which are based on the Galileo probe parachutes [136]. The separation chute is a 3-m diameter conical ring sail parachute which will allow the front heatshield and structure to fall off when released. The descent module will then separate from the backshell and the separation chute. After a few seconds of free fall, the 2-meter diameter chute will inflate. The descent probe chute is sized such that it allows at least 6 minutes to inflate the balloon before the system reaches the altitude of 52 km.

| Subsystem | CBE (kg) | +30% MGA | +15% Margin (kg) |
|---|---|---|---|
| Forebody TPS (HEEET) | 25 | 32.5 | 37.4 |
| Forebody Structure | 13 | 16.9 | 19.4 |
| Backshell TPS (PICA) | 5 | 6.5 | 7.5 |
| Backshell Structure | 6 | 7.8 | 9.0 |
| Separation Parachute and Mortar | 6 | 7.8 | 9.0 |
| Separation System | 5 | 6.5 | 7.5 |
| **Aeroshell Total** | **60** | **78** | **90** |
| | | | |
| Inflation System | 10.3 | 13.39 | 15.4 |
| Descent Parachute | 2.6 | 3.38 | 3.9 |
| Aerial Platform Flight System | 43 | 55.9 | 64.3 |
| Engineering Systems | 4.1 | 5.33 | 6.1 |
| **Descent Module Total** | **60** | 78 | **89.7** |
| **Total Entry System Mass** | **120** | **156** | **179** |

Table 5-2. The entry probe mass breakdown.

**Probe mass breakdown**. The mass breakdown for the aeroshell and the descent module is given in Table 5-2. The forebody thermal protection system (TPS) mass fraction is estimated based on the total stagnation-point heat load [137]. The backshell TPS and aeroshell structural mass is scaled based on the Small Next-generation Atmospheric Probe (SNAP) probe design [138]. The parachute mass is based on the Galileo mission and the 2020 Venus Flagship Mission study.

### *5.3.3 Aerial Platform System*

**Balloon.** The balloon is a super-pressure helium-filled balloon 4 m in diameter with a 13 kg mass (excluding the inflation system). The balloon material is built from a complex fabric and polymeric film structure. The balloon is an oblate spheroid, a so-called Ultra High-Performance Vessel (UHPV), a highly unique, proprietary inflatable pressure vessel architecture [139] that has been the focus of numerous NASA contracts ranging in topic from space habitats to cryogenic propellant tanks. It is widely considered to be the most structurally predictable, lowest specific mass containment architecture (collapsible, metallic, composite, or otherwise), furthermore featuring unlimited scalability and structural determinacy.

A variable altitude version of the balloon has been successfully demonstrated in Earth atmosphere [140]. The inflation mechanism is low technology readiness level (TRL) due to the Venus conditions of deployment, which includes a rapid inflation rate, decent at 10s of m/s, sulfuric acid clouds, and autonomous operations. While the balloon construction and operation is proprietary to Thin Red Line, some information can be found in [141].

**Gondola design.** The gondola is a pressure vessel based on the Pioneer Venus Large Probe [142]. The structure is a 20 cm diameter cylinder with hemispherical top ends and with a 2 mm thick shell of titanium. The instruments are placed on beryllium shelves inside the vessel, which are also 2 mm thick.

The pressure vessel is designed to protect the instruments and other subsystems and electronics from the Venus sulfuric acid cloud particles. This reduces the technology development cost of making each of the components resistant to sulfuric acid, at the expense of increased structural mass. With advancements in materials technology, the mass of the pressure vessel is reduced by using light-weight materials such as beryllium or composites. Since the gondola will remain in relatively benign environments (at temperatures less than 150 °C and pressures less than 6 bar), extreme environment technology development is not required. The aerial platform gondola mass breakdown is summarized in Table 5-3.

| Subsystem | Mass (kg) |
|---|---|
| Structure | 2.3 |
| Battery+PDS+Solar Panels | 5.0 |
| Communication | 3.7 |
| Thermal | 0.7 |
| C&DH | 3.1 |
| Science payload | 6.3 |
| **Total gondola subsystem** | **21.14** |

Table 5-3. The aerial platform gondola mass breakdown.

**Thermal control**. The gondola pressure vessel's interior is lined with a 1 mm thick layer of Kapton. The beryllium shelf and instruments deck are separated by a 6 mm thick layer of a phase change material (PCM) to further insulate the instruments. A PCM provides thermal insulation by way of absorbing heat as it transitions from a solid state to liquid or gaseous state. Sodium silicate PCM will be used. The balloon will float at 52 km where the ambient temperature is 60 °C. All instruments will be designed to operate nominally at this temperature. Therefore, for this study, the passive thermal control design is considered sufficient.

**Power**. The aerial platform uses a combination of solar panels and batteries throughout the mission lifetime. A LiCFx battery pack supplies power, which has 450 Wh/kg of specific energy. The total weight of the power subsystem is 5 kg. The energy requirement for the mission is obtained by accounting for a tentative concept of operations (CONOPS) of instruments and communications. The sizing is conservative in terms of the total battery mass.

**Communication**. The antenna is a crossed dipole designed for 2.4 GHz S-band transmission. The antenna is mounted on top of the gondola. The balloon is opaque to radio transmission and thus 75 to 105 degrees elevation is blocked for communication. The dipole antenna has a wide beamwidth, which allows for transmission at elevations above 30 degrees. A synchronous orbit is selected from which balloon is always in line of sight. The data rate is 100 kbps when the orbiter is at periapsis and 100 bps at apoapsis.

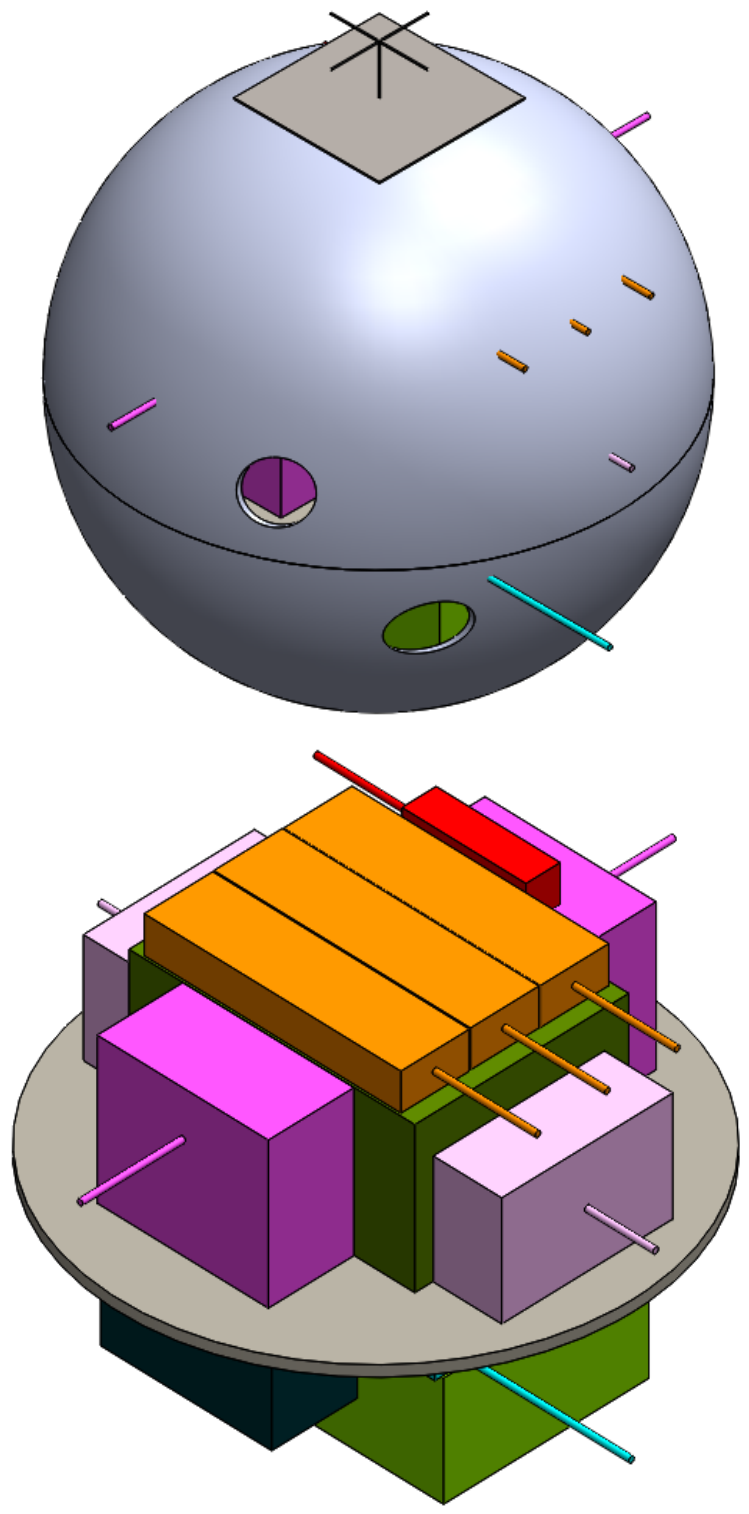

Figure 5-2. Gondola design with instrument accommodation. The gondola is 20 cm in diameter cylinder. Not to scale.

**Gondola instrument accommodation**. The notional accommodation of the gondola instruments is presented in Figure 5-2 and Figure 5-3.

## 5.4 Mission Design and Operations

Mission operations will be split into four phases: 1) launch and cruise to Venus; 2) Venus approach; 3) balloon entry and initial float operations and orbiter capture; and 4) nominal balloon operations.

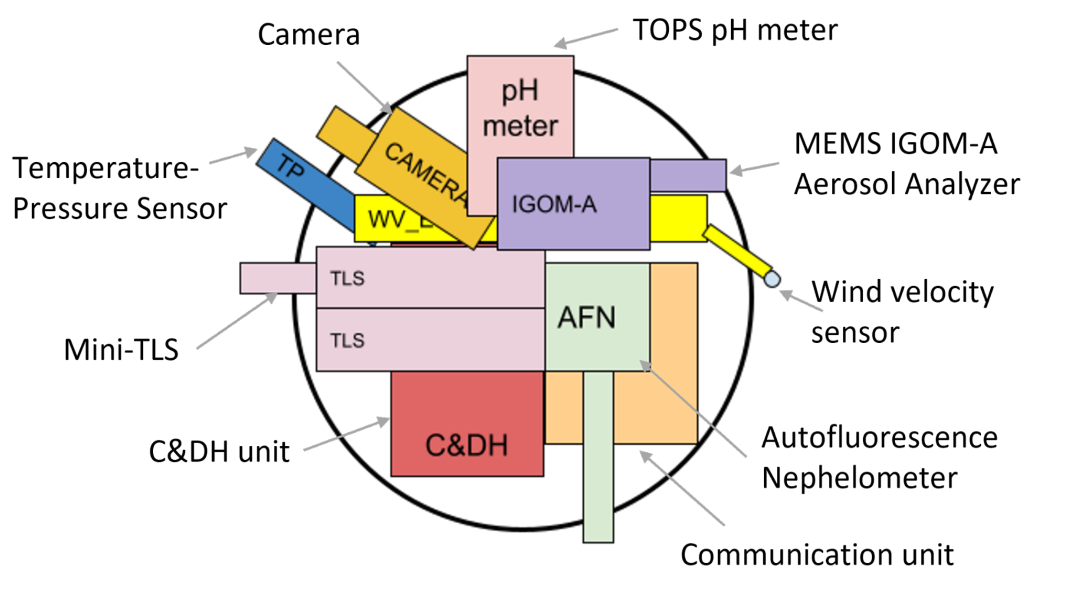

Figure 5-3. Gondola instrument placement schematic.

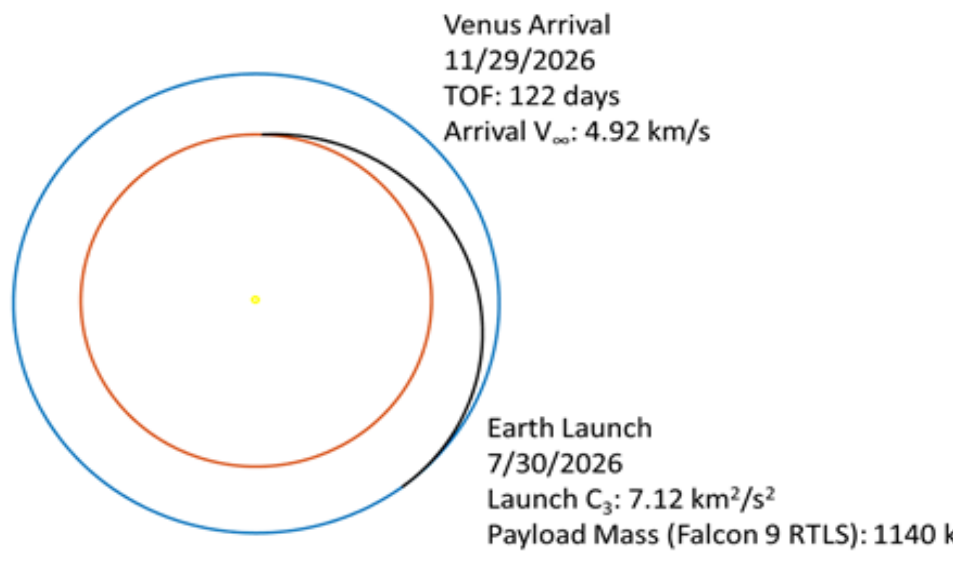

Figure 5-4. Baseline interplanetary trajectory for the Venus Habitability Mission with a 2026 launch.

### 5.4.1 Launch and Cruise to Venus

The habitability mission will notionally launch on July 30, 2026, on, e.g., a Falcon 9 rocket. The launch payload will be an entry system consisting of a constant-altitude balloon attached to an orbiter that will ferry all the vehicles to Venus. The spacecraft will arrive at the vicinity of Venus on November 29, 2026 (Figure 5-4).

### 5.4.2 Venus Approach

Thirty days prior to arrival at Venus, the orbiter and attached entry system will be on a course to intercept the atmosphere at the desired probe entry point on the day side of Venus. The entry point corresponds to a -10-degree entry flight path angle (referenced from 180 km altitude). At this point, the entry system will separate from the orbiter and continue towards entry. This separation imparts a small impulse on the orbiter, and so two days are allotted for orbit determination following entry system separation.

After this orbit determination period, the orbiter performs a propulsive divert maneuver with a ΔV of up to 200 m/s to raise its periapsis out of the atmosphere to an altitude of about 3800 km and adjust the relative geometries of both spacecraft so that the orbiter can pass over the balloon every circumnavigation for data downlink. Due to the large trade space in orbit selection, a specific point design was not completed and so the divert maneuver is only an approximation of the maximum maneuver size that could be expected. The arrival geometry for the Venus Habitability Mission is shown in Figure 5-5.

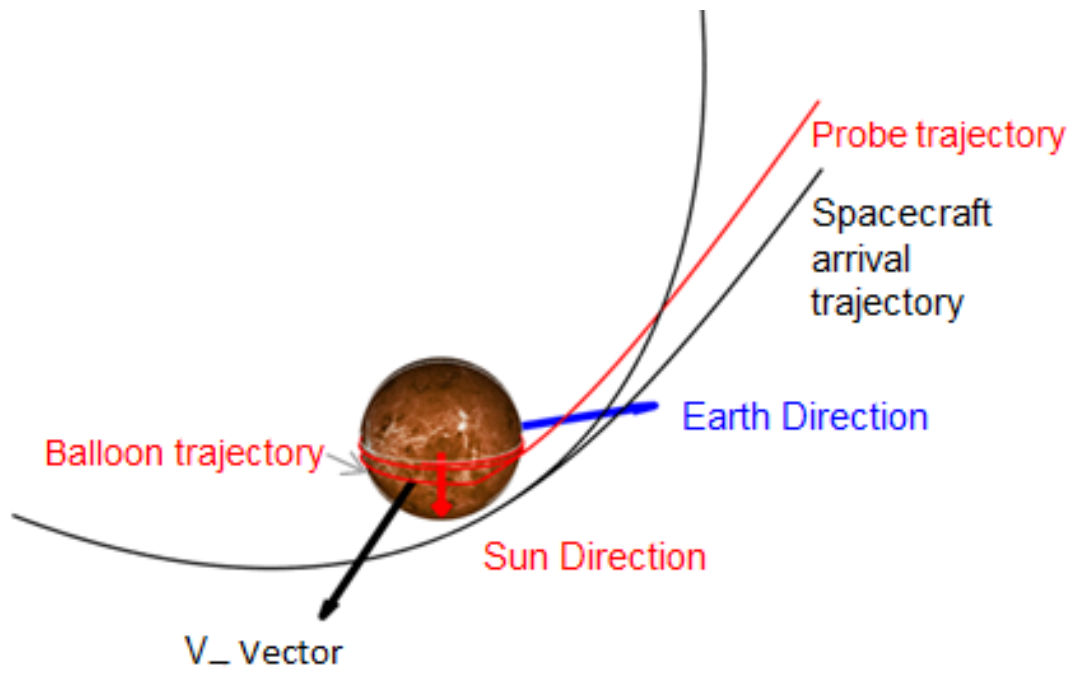

Figure 5-5. Venus Habitability Mission architecture and geometry.

### 5.4.3 Entry and Descent

The probe will enter the atmosphere at a speed of 11.33 km/s at an entry flight-path angle (EFPA) of -10 degrees. The shallow EFPA is chosen to aid data relay to the orbiter and minimize deceleration as well as stagnation-point heat rate. Table 5-4 shows the entry and descent timeline.

Following peak heating and peak deceleration, the separation chute will be deployed at 73.8 km. The total estimated stagnation-point heat load is 21 $kJ/cm^2$. Five seconds later, the heat shield will fall off and another five seconds later the aerial platform system will be released from the backshell. After five seconds of free fall, the aerial platform descent chute will be deployed, and the inflation system will begin to fill the balloon at an altitude of 71.8 km. At E+8 minutes, the balloon inflation

| Time (s) | Altitude (km) | Speed (km/s) | Event |
|---|---|---|---|
| E + 0 | 180 | 11.33 | Entry interface |
| E + 56 | 88.5 | 9.59 | Peak heating, heat rate = 1,502 W/cm$^2$ |
| E + 60 | 84.7 | 7.15 | Peak deceleration, 69 g |
| E + 100 | 73.8 | 0.381 | Separation chute deployed |
| E + 105 | 73.1 | 0.204 | Front heatshield separation |
| E + 110 | 72.8 | 0.155 | Backshell separation, aerial platform free fall |
| E + 115 | 72.2 | 0.103 | Descent chute deployed |
| E + 120 | 71.8 | 0.102 | Balloon inflation begins |
| E + 600 | 53.8 | 0.028 | Balloon inflation complete, chute cut-off |

Table 5-4. Key events during entry and descent sequence followed by balloon inflation and deployment.

will be complete at 53.8 km and the descent chute will be released.

The profiles for altitude vs. time, deceleration, and heating for the probe are shown in Figure 5-6. The entire entry, descent, and balloon deployment CONOPS is summarized in Figure 4-6.

### 5.4.4 Orbiter Insertion

The orbiter will perform its insertion maneuver to Venus orbit capture at a similar time to when the probe is entering the Venus atmosphere. The Venus orbit will be an approximately 5-day orbit—which is about the time that it takes for the balloon to circumnavigate the planet. The ΔV magnitude of the insertion maneuver is approximately 1550 m/s.

### 5.4.5 Balloon Operation and Communication

After entry and initial float operations, the balloon will travel within a design altitude range collecting data from its instruments. There will be an opportunity once per circumnavigation—about every five days or so—to downlink data to the orbiter as the orbiter approaches its periapsis. Depending on the specific orbit selection and the telecommunication system design, the total data volume per circumnavigation can vary from 50 Mbits up to close to 500 Mbits. The science data volume can be improved upon with data compression onboard the balloon. If a synchronous orbit is selected with the balloon always in line of sight with the orbiter, the data rate varies from 100 kbps at the periapsis to 100 bps at apoapsis.

In addition to the orbiter, the balloon can also downlink directly to Earth. Figure 5-7 shows the estimated data rates and data volume.

We determine approximate data volumes to both the orbiter and directly to Earth, by assuming a baseline telecommunications system. The baseline S-band telecommunication architecture uses a 40 W transmitter on the probe with a beamwidth of 71 degrees, and a receiver on the orbiter with a 60-degree beamwidth. Balloon blocks comms from 75 to 105 degrees of elevation. Elevation of 0 to 15 absorbed by atmosphere.

The current mission design has the orbiter and entry vehicle launch and cruise to Venus together, to save cost. In this scenario, the orbiter is performing its insertion maneuver at about the same time as the balloon is descending, meaning that the orbiter will not be able to observe the balloon's entry initial float operations. Observations from the orbiter, however, may be possible shortly after the insertion maneuver is complete. Additionally, this means that any communication and monitoring during orbit insertion and balloon entry time will have to be directly from Earth. Furthermore, the first upload/downlink opportunity from the balloon to the orbiter will occur after nearly one full circumnavigation of the balloon. There is no major downlink opportunity during the first several days of balloon operations. To improve on this situation the orbiter has to already have arrived in orbit before the entry system enters the atmosphere, which we will consider in a future trade.

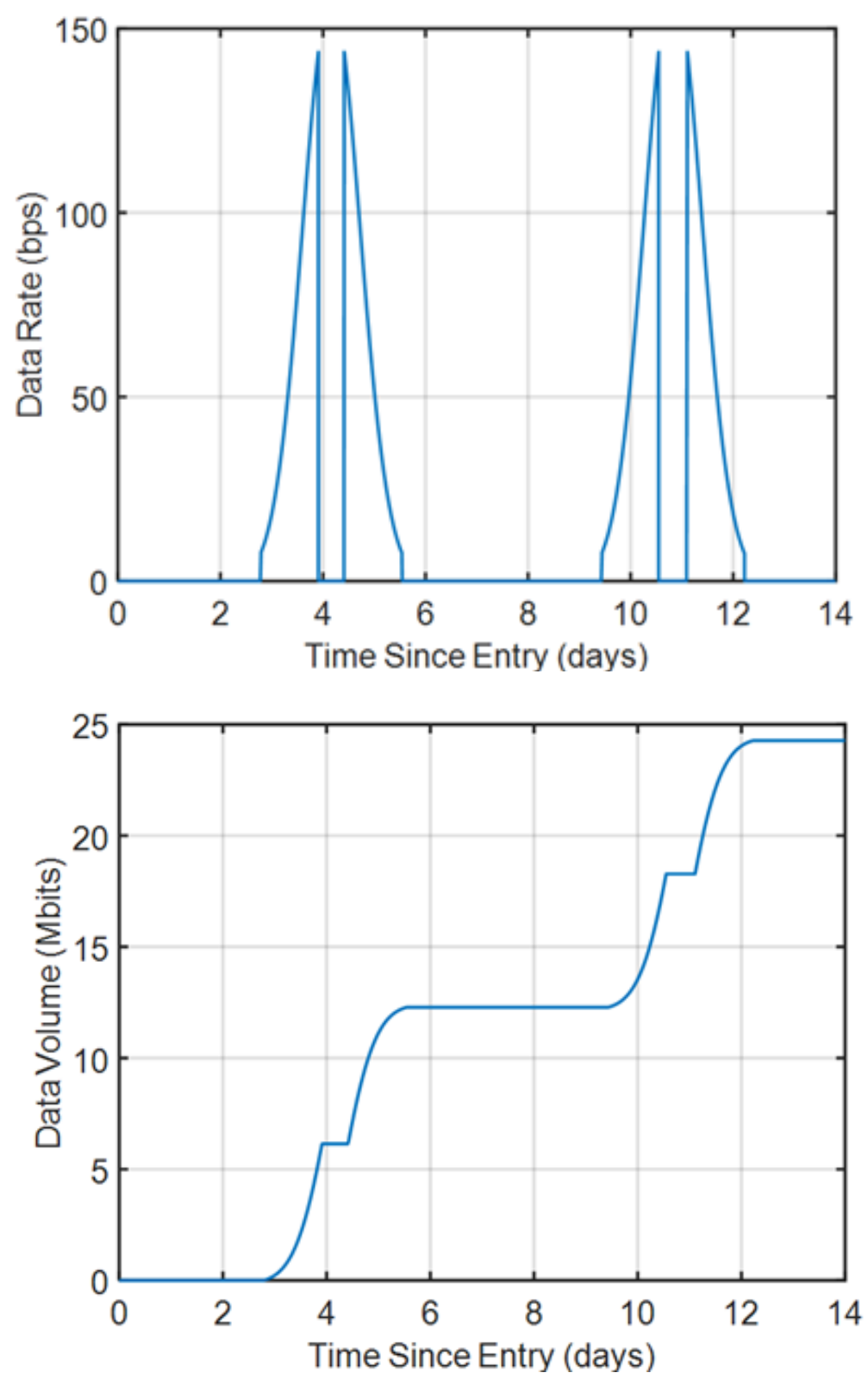

Figure 5.7: Direct to Earth data transfer parameters. Top: data rate. Bottom: total transferred data volume (does not include potential data compression).

After entry and initial float operations, the balloon will travel within a design altitude range. A notional trajectory is shown in Figure 5-8.

## 5.5 Summary

The Habitability Mission Design centers around a fixed-altitude balloon operating for about one week in the middle of the Venus atmosphere lower cloud layer. The concept includes four mini probes to be deployed to sample lower atmosphere layers. Ongoing work will refine the mission concept, further develop the instruments, and improve the communications strategy.

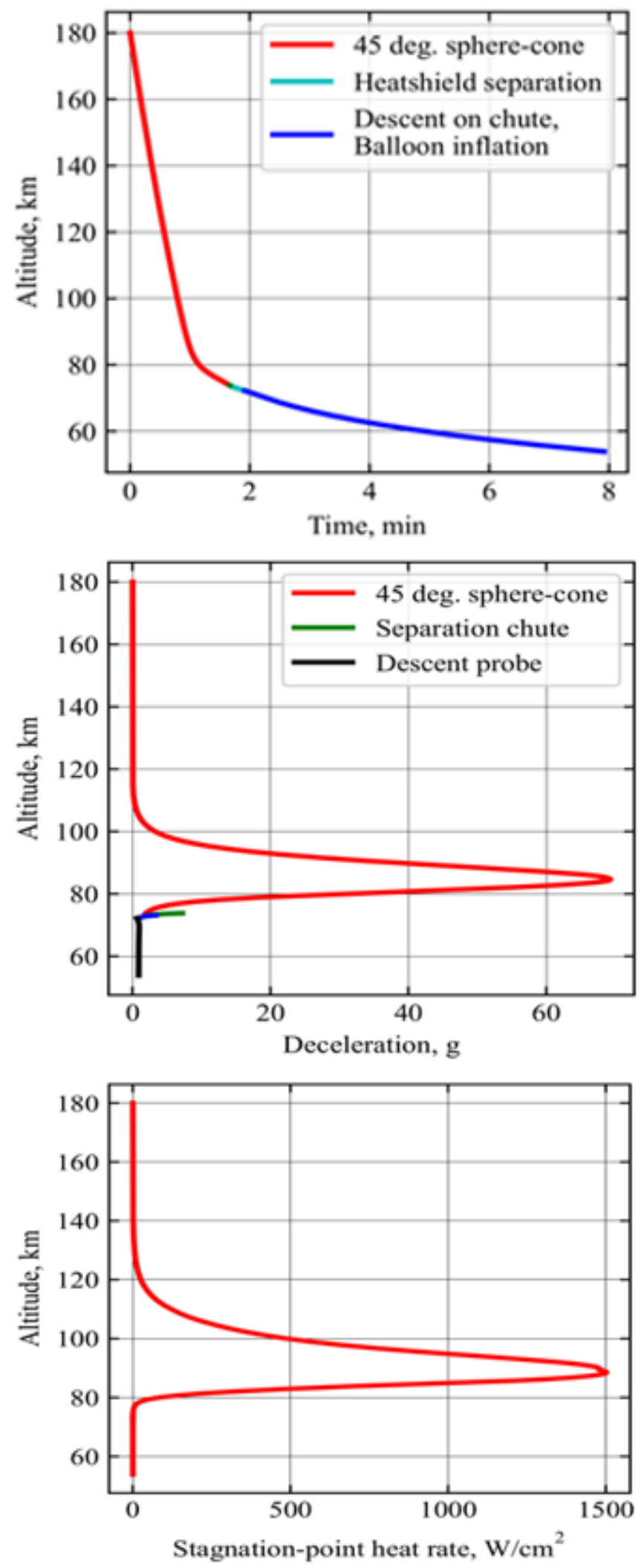

Figure 5-6. Mission entry profiles. Top: altitude vs. time. Middle: deceleration profile. Peak deceleration = 69 g. Bottom: aerodynamic heating profile. The peak heat rate = 1502 $W/cm^2$.

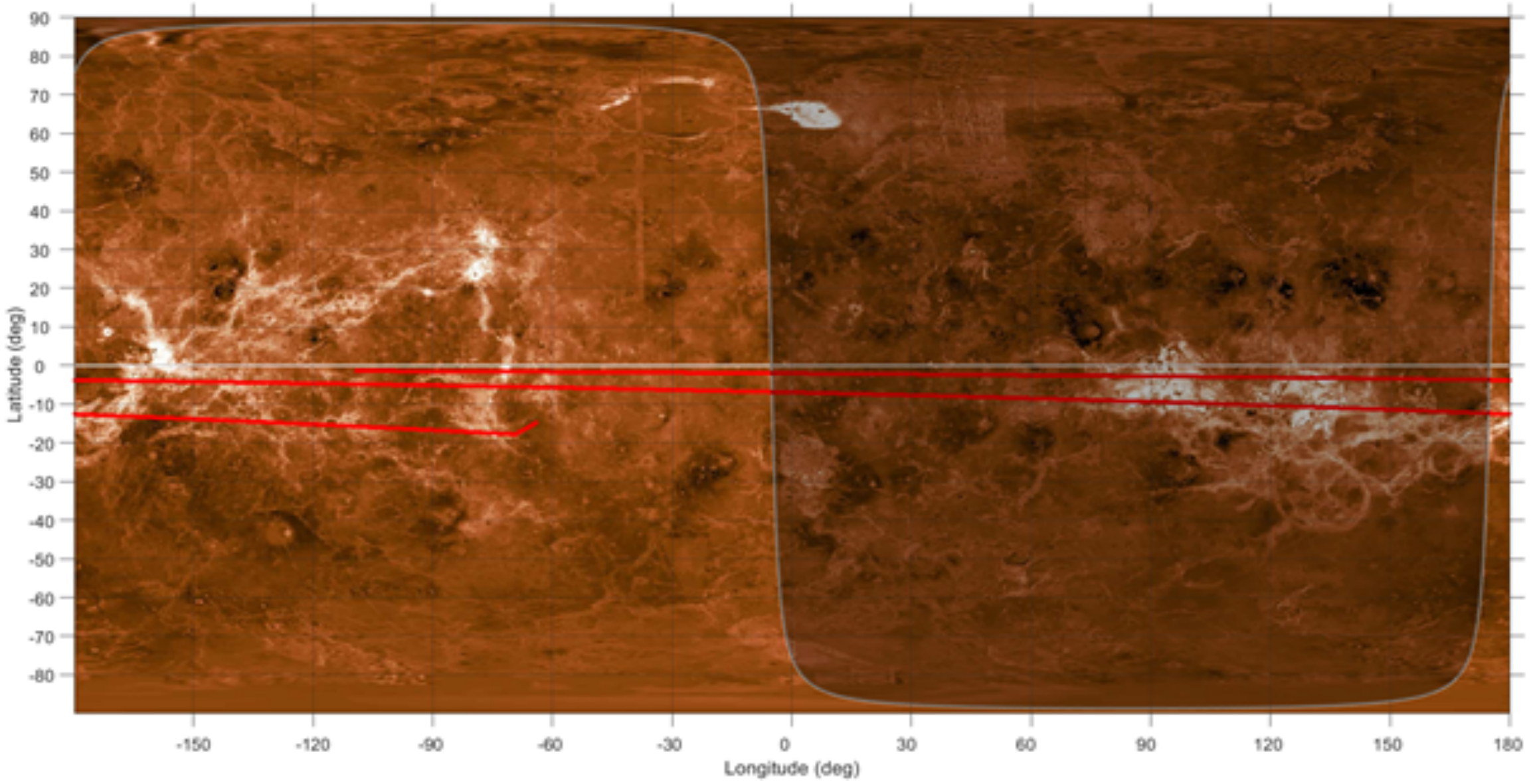

Figure 5-8. Notional balloon trajectory (red line) shown for 28 days after entry using Venus Global Reference Atmospheric Model (GRAM) [217].

# 6 Venus Atmosphere Sample Return Mission

**Abstract:** A sample return of Venus atmosphere and cloud particles is vital for a definitive search for signs of life and life itself. With a sample in hand, we can use the most sophisticated instruments on Earth, instruments with a variety and sensitivity not matched by realistically possible spaceflight instruments. On Earth we can follow flexible and adaptive experimental protocols not possible with a remote spacecraft investigation. We describe a notional sample return mission flight and operations concept as well as sample storage and distribution considerations. We leave the science and Earth laboratory instrument review to Appendix F.

**New mission concept that emerged from the study:**

- Development of a concept of atmospheric and cloud particle sample return, as past designs focused on surface sample return.

## 6.1 Introduction and Motivation

A sample return of cloud particles for detailed study in Earth's laboratories increases the chances of robust detection of the signs of life and even life itself over a remote in situ mission. Analytical techniques in Earth laboratories allow for far more sophisticated and accurate measurements of the kind not available for remote or autonomous in situ measurements in planetary environments. For example, in contrast to automated in situ instruments, Earth laboratories are not limited by mass, power, and data rate capacity. Furthermore, Earth-based laboratories can reach more sensitive limits of detection compared to space-based instruments.

In a laboratory setting on Earth we can easily detect any complex cellular structures and sub-structures, including visualization of an entire cell architecture. On Earth we can detect complex polymeric molecules, those such as genetic polymers that can only be a result of biological activity, more robustly than from in situ. The possibility for false positives or misinterpretation is much lower.

Proper sample storage protocols can be developed so the returned samples can be studied over a significant time period. Any non-destructive analysis techniques allow the same sample to be analyzed multiple times, either by the same or different instrumentation, increasing the science output and reliability of the measurement.

Previous Venus sample return concepts focused on surface sample return, with study reports as early as the mid-1980s [143–147]. A more recent review of past activities is in [148]. The sample return concepts mostly involve a two-component mission, an orbiter that doubles as a return vehicle to Earth orbit as well as an entry and descent vehicle that lands on the Venus surface. The lander would "grab and go", that is spend about an hour on the surface collecting samples by drilling and "vacuuming", as well as possibly capturing surface atmosphere. An ascent vehicle included as part of the lander would inflate a helium balloon system and may also include instruments to capture atmosphere samples during the ascent. Once above about 60 km altitude, the ascent vehicle's solid rocket system would bring the sample payload to orbit where the sample payload would be captured by the orbiter. Return from Venus to Earth is challenging due to the large mass of Venus. Rodgers et al. [149] describe a solar electric propulsion system that takes three years to return from Venus to Earth but is favored over a conventional chemical propulsion system because of a 30% mass savings.

We aim for an atmosphere sample return, both gas and cloud particles, a concept that has not been widely considered [150]. The goal is to capture up to 1 liter of atmosphere and up to 10s of g of cloud particles, with subsets captured from different locales and stored separately.

Our atmosphere sample return concept involves a single super-pressure balloon with sample collecting instrumentation, and an ascent vehicle to send the captured material back to an orbiter which transports the material back to Earth. Our Venus Life Finder (VLF) Sample Return Mission will follow NASA planetary protection protocols.

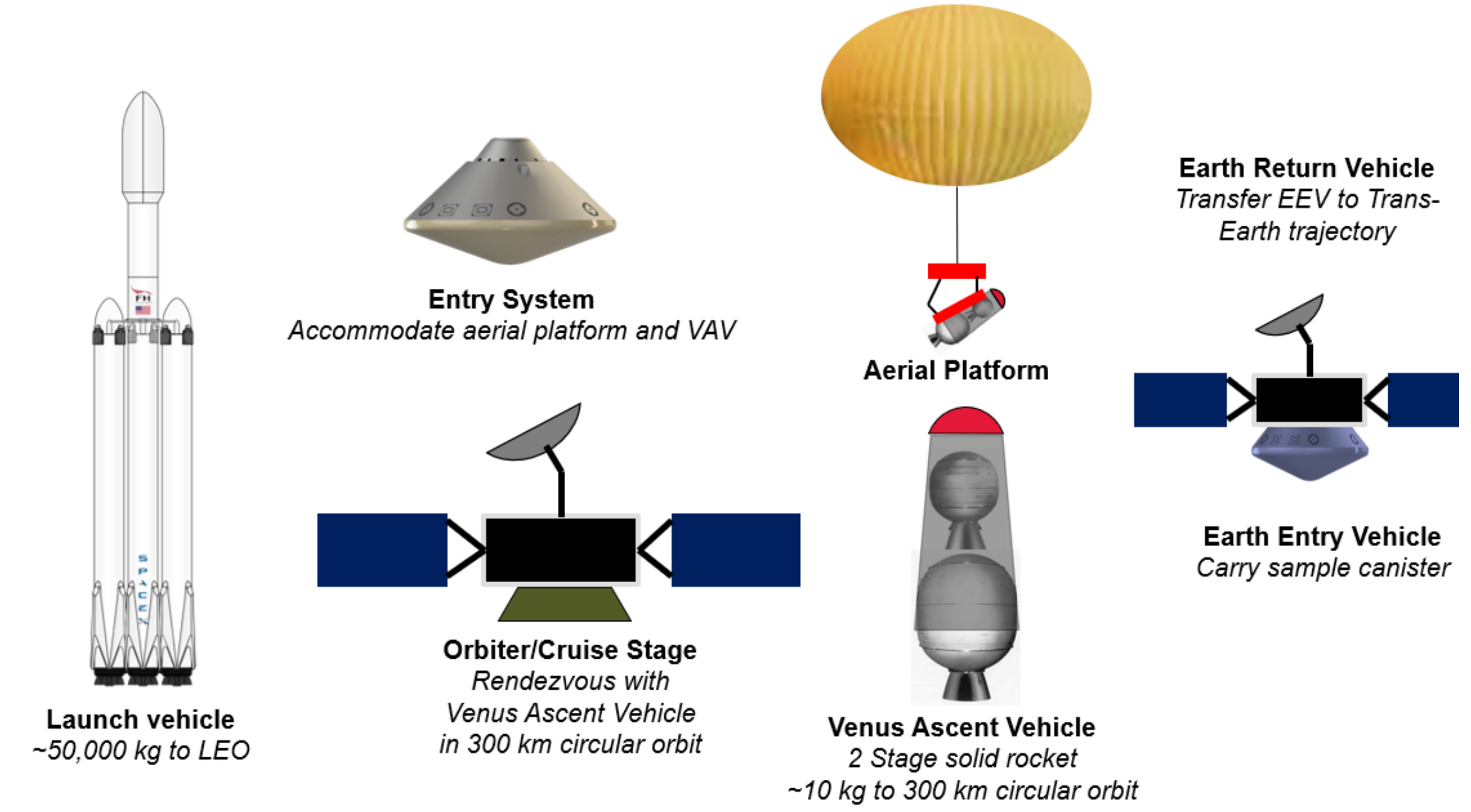

Figure 6-1 Venus atmosphere sample return flight system elements.

## 6.2 Mission Concept Overview

Our mission concept is a balloon aerial platform that floats in the Venus atmosphere cloud layers at the altitude of interest and functions as a launch platform for returning the sample to orbit. This aerial platform collects samples from different altitudes and locations along the balloon's trajectory. The balloon platform also enables the ascent vehicle launch from a higher altitude than where the sample is collected, thus reducing the launch penalty.

The flight system elements include: a heavy lift launch vehicle; an entry system; a combined orbiter and cruise vehicle; an aerial platform; a Venus ascent vehicle (VAV); an Earth return vehicle; and an Earth entry vehicle (Figure 6-1).

The baseline mission architecture is shown in Figure 6-2. All the flight elements stack into one configuration inside the launch vehicle payload fairing and launches on a trans-Venus trajectory. The orbiter/cruise stage carries the entry system with the aerial platform system to Venus. The orbiter performs a deflection maneuver and the entry system deploys, entering and decelerating through the Venus atmosphere. The orbiter performs a Venus orbit insertion maneuver to be captured into a low Venus orbit. After the entry system decelerates and the aerial platform deploys along with the gondola. Once the samples are collected, the VAV launches from the aerial platform into a low Venus orbit. The orbiter will perform orbit transfer and phasing maneuvers to rendezvous with the VAV. The sample container will be transferred to the Earth entry vehicle (EEV).

The Earth return vehicle (ERV) departs from the Venus orbit on an Earth return trajectory. At Earth the EEV deploys and enters Earth's atmosphere. The sample container will be recovered from the EEV and transferred to the sample handling and curation facility. The sample will then be distributed to various laboratories globally for analysis.

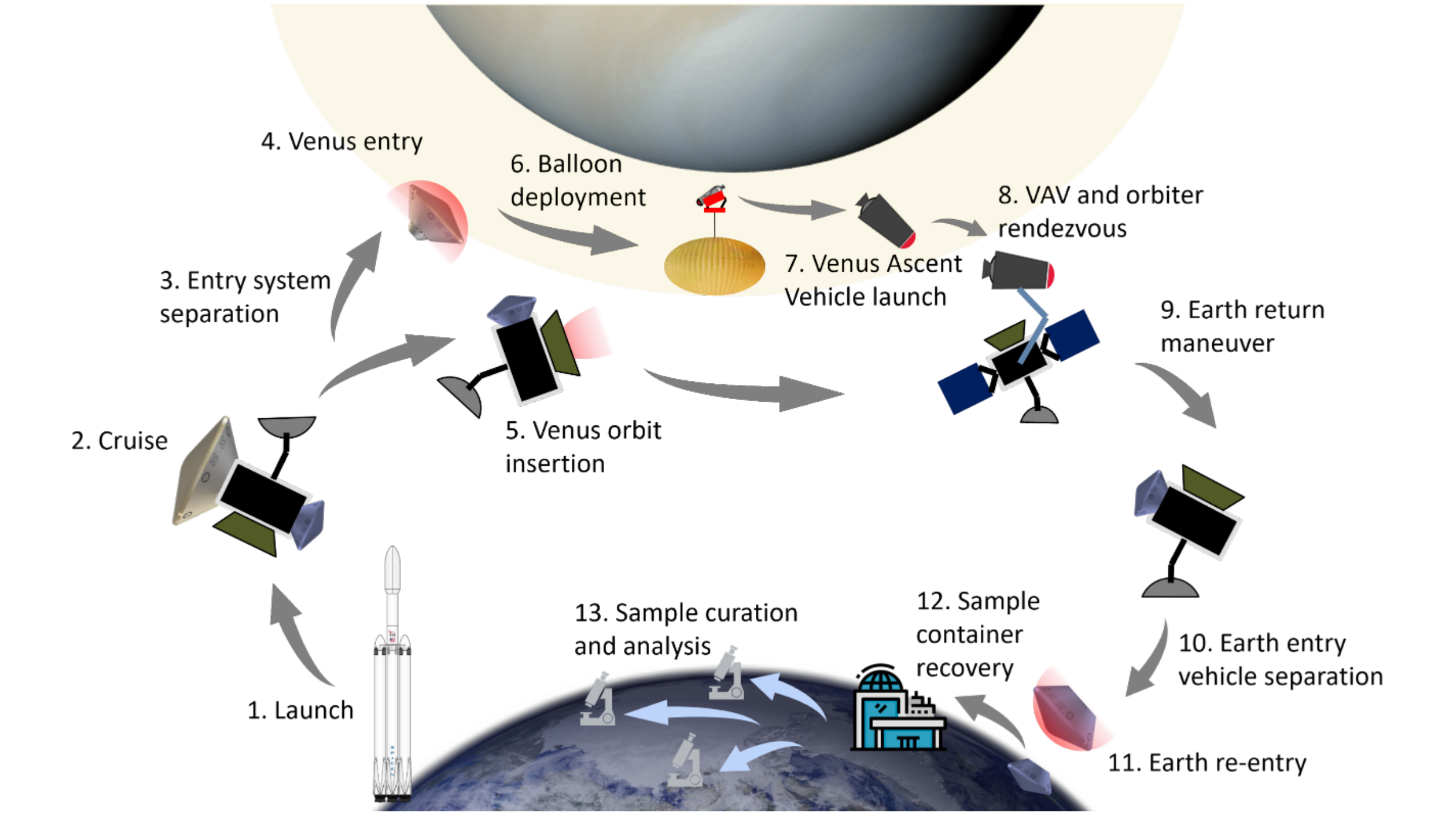

Figure 6-2. The baseline mission architecture for Venus atmospheric sample return.

## 6.3 Primary Instrument Suite

### 6.3.1 Atmosphere and Cloud Particle Capture

Atmosphere and cloud particle capture technology for a Venus sample return mission has not yet been developed. Several options that could be adapted to specific requirements of the atmospheric and cloud particles sample return mission include:

- Use of aerogels (e.g. [151–153])
- Filters
- Electrostatic sticky tape
- Funnels, jars, and bottles (e.g. [154])
- Fog Harp droplet collector (see Appendix C)
- Gas sampling bags

### 6.3.2 Sample Container

A sample container will be used for storage and return sample to Earth. Some of the sample container requirements are: (1) store the different types of samples: gas, aerosol, liquid, and solid; (2) provide thermal protection during the ascent from Venus, cruise phase, and EDL at Earth; (3) provide radiation protection during the cruise phase from Venus to Earth; (4) have electronics for housekeeping and monitoring the samples and relaying the status back to Earth; (5) provide structural support during re-entry on Earth to withstand the high deceleration loads and survive impact.

The design of the sample container depends on several factors. First, the required diversity of samples affects the design of various "sample tubes" to ensure the state of the samples is maintained. Second, the required sample altitudes will affect the ambient temperature and pressure conditions of the sample and the storage conditions. Third, the required volume of sample will affect the mass of the container, which affects the overall flight elements masses. Fourth, whether or not the samples should be frozen sample preservation affects power requirements.

Past sample return missions returned soil, rock, and dust samples. There is no heritage of sample containers that can store gas and liquid samples. There are some options that can be studied further to assess if these can be modified for Venus atmosphere sample return. The sample canister proposed by the CAESAR comet return mission separates solid and gas samples and maintains samples at subzero temperatures [155] (Figure 6-3). The Japanese Aerospace Exploration Agency (JAXA) has developed a low-mass technology that uses passive insulation for sample return from the International Space Station [154]. It is based on steel thermal bottles, which are double-walled vacuum containers. Similar solutions can be studied in the context of the Venus sample return mission (Figure 6-3).

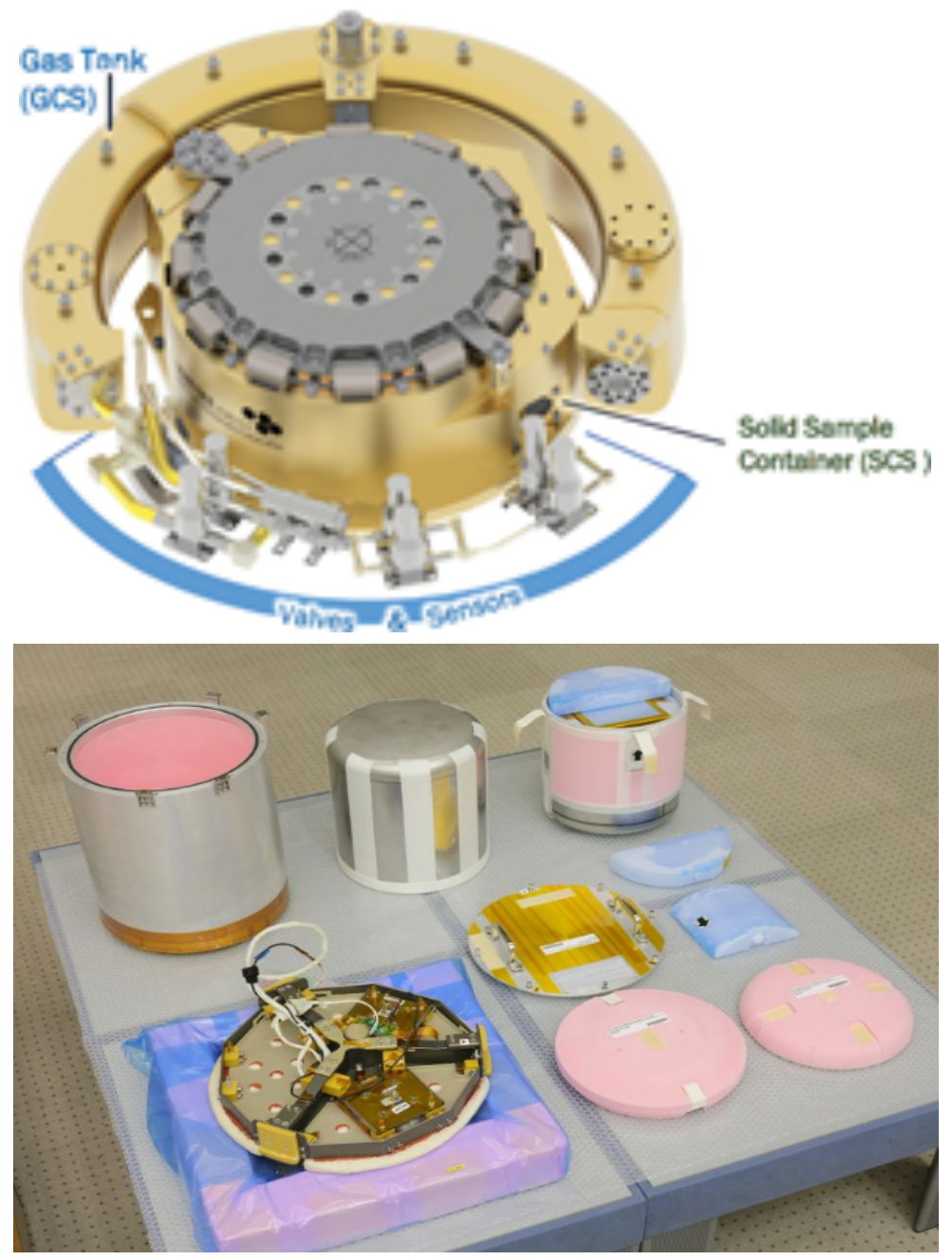

Figure 6-3. (Top) Proposed sample canister for the CAESAR comet return mission [155]. (Bottom) JAXA's sample container based on thermal water bottle technology [154].

For the current mission design, we assume that the sample container (structure plus contents) weighs 10 kg. This is reasonable assumption based on the mass of JAXA's sample container which is 9.7 kg, and which is said to be reduced to 3 kg after further optimization.

## 6.4 Flight Systems Description

This section describes the various flight system elements, summarized in Figure 6-4 with the mass of each of the systems in Table 6-1. The mass is estimated for the baseline trajectory described in section 6.5.1.

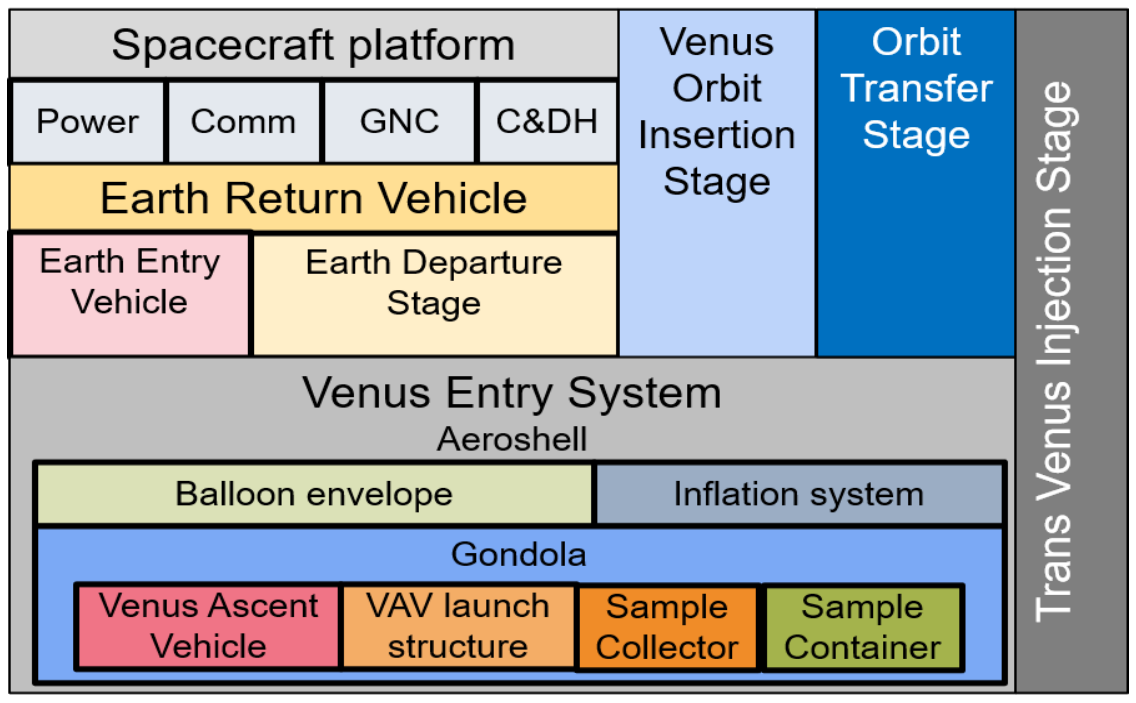

Figure 6-4. Block Diagram of the spacecraft launch configuration.

| System | CBE (kg) | Contingency | MEV (kg) |
|---|---|---|---|
| Earth Entry Vehicle | 60 | 50% | 90 |
| Earth Return Vehicle Wet Mass | 1230 | 50% | 1845 |
| Orbiter at Venus Encounter | 4840 | 50% | 7260 |
| Entry System | 3700 | 50% | 5550 |
| Aerial Platform | 1600 | 50% | 2400 |
| Venus Ascent Vehicle | 880 | 50% | 1320 |
| **Total Mass to TVI** | **8540** | **50%** | **12810** |
| **Total Mass to LEO** | **35500** | **50%** | **53250** |

Table 6-1. Summary of Venus sample return flight system element mass. CBE is current best estimate and MEV is maximum expected value.

### 6.4.1 Cruise Vehicle

The cruise vehicle carries the entry system and orbiter to Venus. The cruise vehicle power, controls, guidance and navigation, communications, and other on-board systems are on a small satellite platform of 200 kg. The orbiter carries the Earth Return Vehicle (ERV) and propulsion stage for Venus orbit insertion. The ERV carries the Earth Entry System (EES) and propulsion stage for Earth return burn from Venus orbit. The small satellite platform is common across the cruise vehicle, orbiter and ERV. The propulsion stages use liquid methane and liquid oxygen as propellants. The dry mass of the propellant stages is assumed to be 10% of the required propellant mass.

### 6.4.2 Orbiter

The primary function of the orbiter is to retrieve the sample container from the VAV. The orbiter is responsible for performing the phasing maneuvers to rendezvous with the VAV. The orbiter hosts actuators to grapple the VAV and transfer the sample container from VAV to the ERV.

### 6.4.3 Entry System

The entry system consists of an aeroshell that decelerates through the atmosphere and a descent module that carries the aerial platform system along with the inflation system. The entry system payload mass is equal to the mass of the aerial platform and inflation system. Table 6-2 shows the mass breakdown for the entry system components.

### 6.4.4 Aerial Platform System

The aerial platform consists of an ultra-high pressurized vessel (UHPV) constant altitude balloon (CAB) that carries the VAV, launch mechanism, and sample capture system. The CAB is sized based on studies by Maxim de Jong (Thin Red Line) [139–141]. The platform also consists of a flight control system that can control the ascent rate of the balloon from the point of complete inflation around 50 km to the altitude from where the VAV will launch, around 66 km.

| System | CBE (kg) | Contingency (%) | MEV (kg) |
|---|---|---|---|
| Aeroshell | 1840 | 50% | 2760 |
| Inflation System | 250 | 50% | 375 |
| Balloon Envelope | 530 | 50% | 795 |
| Gondola Subsystem | 200 | 50% | 300 |
| Venus Ascent Vehicle | 880 | 50% | 1320 |
| **Total Entry System** | **3700** | **50%** | **5550** |

Table 6-2. Entry vehicle component masses. CBE is current best estimate and MEV is maximum expected value.

### 6.4.5 Venus Ascent Vehicle (VAV)

The VAV is a two-stage solid propellant rocket that can launch 10 kg of payload from 66 km altitude into a ~300 km Venus orbit. The first stage is a STAR 30C and second stage is a STAR 24C solid rocket engine. STAR sold motors are a highly mature technology. A similar two-stage solid rocket has been considered for the Mars ascent vehicle for the Mars Sample Return Mission [156]. The total launch mass of the VAV is estimated to be 880 kg. Figure 6-5 shows a notional configuration of the VAV assuming the two STAR engines stacked vertically.

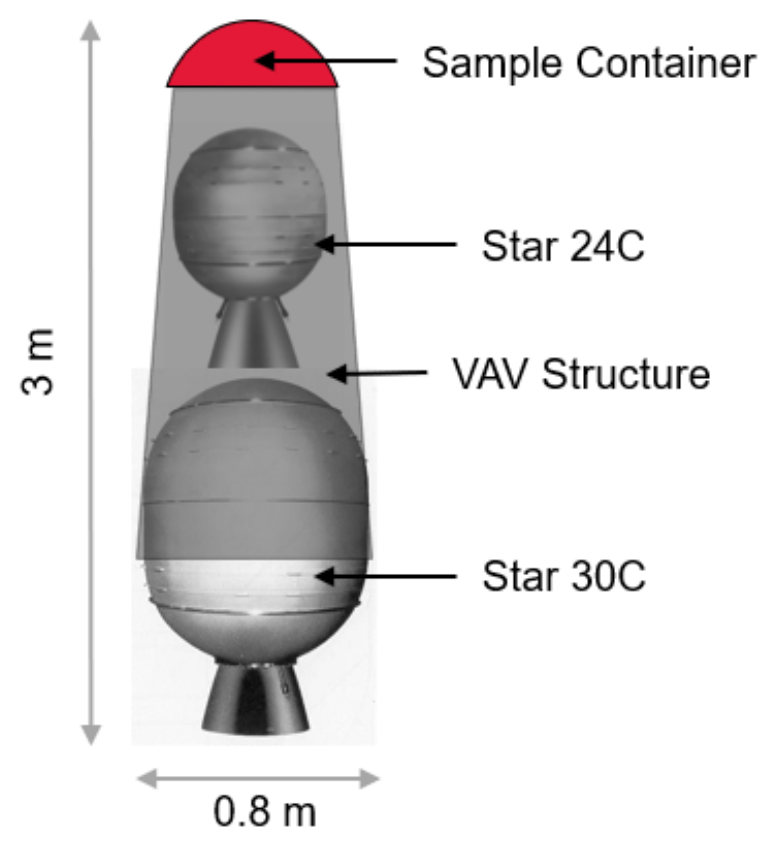

Figure 6-5. Notional configuration of the Venus ascent vehicle.

### 6.4.6 Earth Return Vehicle

The Earth Return Vehicle (ERV) consists of the spacecraft platform and a propulsion stage for a Venus departure burn. The ERV provides power and communications support to the EEV during the cruise phase from Venus to Earth which lasts for 300 days.

### 6.4.7 Earth Entry Vehicle

The Earth entry vehicle (EEV) is a 60-deg half angle sphere cone atmospheric probe based on the Stardust mission's sample return capsule, which weighed 45.6 kg [157]. The EEV is designed to withstand a high g-load of the order of 50 g. It should also survive the impact on land and provide structural support to the sample container. EEV is a mature technology with heritage from the Stardust, Hayabusa, and OSIRIS-REx missions.

## 6.5 Mission Architecture and Design and Operations

### 6.5.1 Interplanetary Trajectory

In this notional design the sample return mission launches on November 4, 2029, on for example, a Falcon Heavy launch vehicle, and cruise for 115 days to Venus. Thirty days before closest approach, the orbiter/cruise stage separates from the entry system and performs a deflection maneuver. The entry system carrying the aerial platform and VAV enters the atmosphere on February 17, 2030. After the entry and deployment events are complete, the orbiter performs an impulsive burn to enter a low Venus orbit with a periapsis of 300 km.

The stay time at Venus is 30 days during which time the aerial platform gondola will collect the samples and transfer the sample container to the VAV. The VAV will ascend from the upper atmosphere into a 300 km circular orbit to rendezvous with the orbiter. The sample container will be transferred from the VAV to the orbiter. The VAV and the Venus orbit insertion stage are left in orbit around Venus, to reduce the propellant requirement for the return journey.

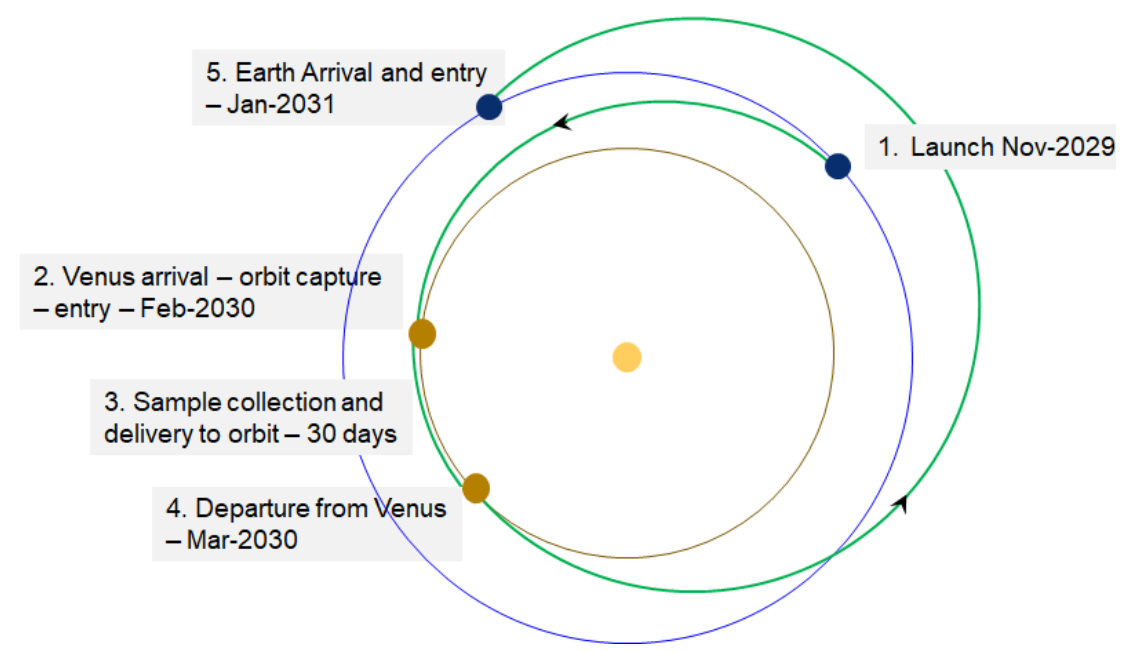

Figure 6-6. Round-trip interplanetary trajectory for the baseline Venus sample return mission.

| Event | $V_\infty$ (km/s) | ΔV (km/s) |
|---|---|---|
| Launch (LEO to TVI) | 3.85 | 3.88 |
| Capture (300 km LVO) | 3.81 | 3.66 |
| Venus Ascent (300 km LVO) | | 8.78 |
| Rendezvous | | 0.10 |
| Venus departure (300 km LVO to TEI) | 4.98 | 4.12 |

Table 6-3. ΔV summary of the round-trip Venus atmosphere sample return mission. CBE is current best estimate and MEV is maximum expected value. LEO is low Earth orbit, TVI is trans-Venus injection, LVO is low Venus orbit, TEI is trans-Earth injection.

The ERV departs from Venus on March 19, 2030, carrying the sample container inside the EEV. The ERV arrives at Earth on January 18, 2031. The total round-trip duration of the mission is 440 days (Figure 6-6).

Table 6-3 shows the ΔV summary for the mission. The launch ΔV is calculated by assuming transfer from a low Earth orbit of 200 km altitude to a trans-Venus injection with C3 of 14.8 $km^2/s^2$. The capture orbit around Venus is a 300 km circular low Venus orbit (LVO). The VAV will ascend to the 300 km LVO. The orbiter will perform maneuvers to rendezvous with the VAV. The ERV will depart from the 300 km orbit on a trans-Earth injection with departure hyperbolic access speed of 4.98 km/s.

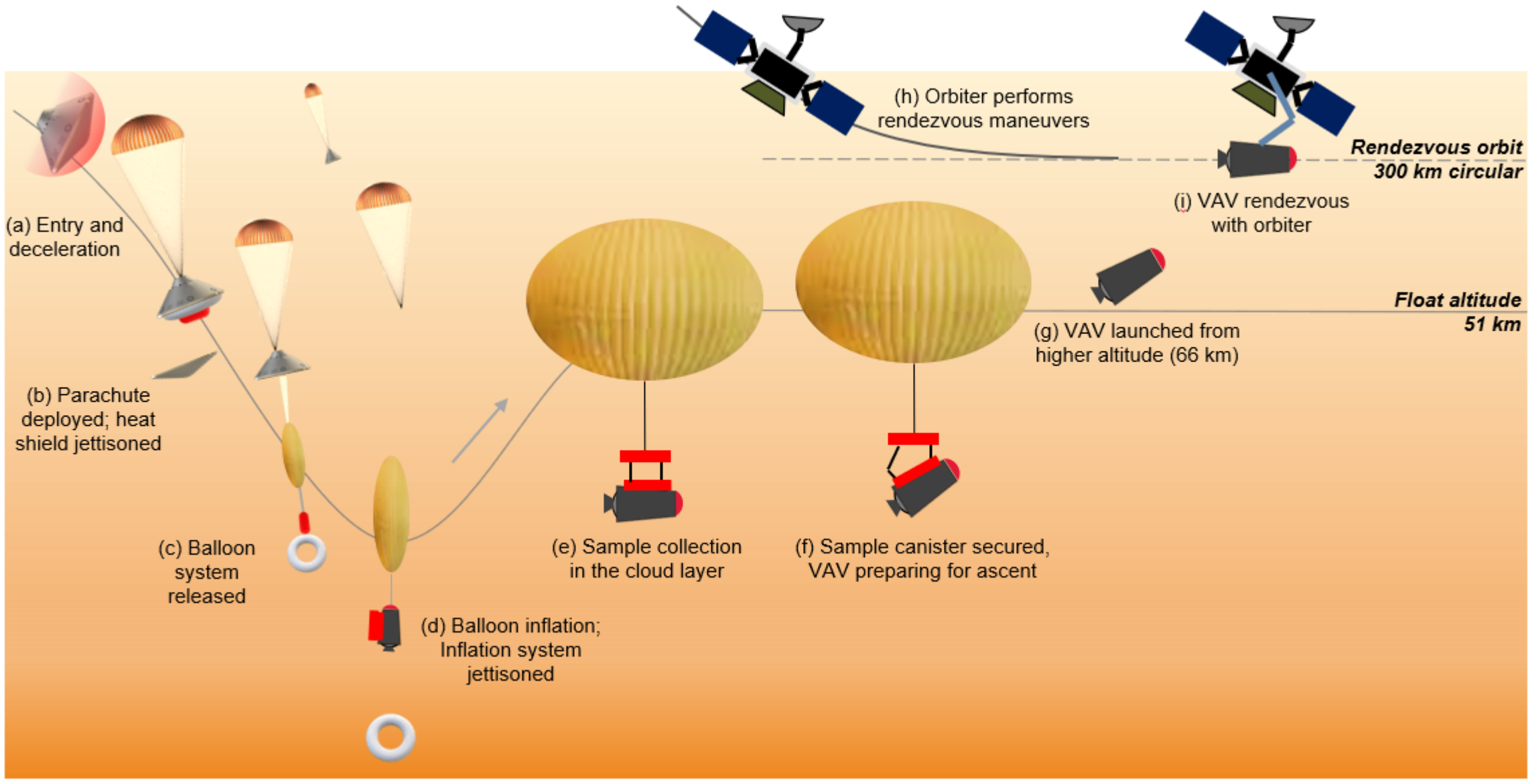

Figure 6-7. Concept of operations for the atmospheric phase of the Venus atmosphere sample return mission.

### 6.5.2 *Atmospheric Operations*

The aeroshell enters the Venus atmosphere at 180 km altitude at entry speed of about 11 km/s. After the system has passed through the peak heat load and peak deceleration phases, the drogue parachute deploys to slow down the descent for payload deployment. The heat shield is jettisoned, followed by deployment of the aerial platform and inflation system. The aft shield is released, and the aerial platform will free fall. The main parachute will be deployed to slow down the descent of the aerial platform.

Next inflation of the balloon will begin. After the balloon inflation is complete, the main parachute is jettisoned, and the inflation system will be dropped. The aerial platform slowly ascends through the atmosphere, collecting samples at various altitudes. After sample collection is complete and all VAV systems are checked out, the aerial platform will rise to 66 km.

The VAV will launched from the platform, with ~45° angle with respect to the horizon. The first stage is fired and VAV reaches 180 km. At 180 km, second stage is fired to reach 300 km altitude. A circularization burn is performed at 300 km to enter circular orbit. The orbiter will rendezvous with the VAV and transfer the sample container to the EEV (Figure 6-7).

## 6.6 Technology Development, Risks and Constraints

### 6.6.1 *Technology Development*

The flight elements consist of both low technology readiness level (TRL) and high TRL systems. We used high TRL technology options where possible even if low TRL technology is more mass efficient. For example, the spacecraft platform along with the various propulsion stages are high TRL technology whereas the alternative of aerocapture for Venus orbit insertion would require lower mass. The EEV is also high TRL, but further development might be needed if Planetary Protection requirements are imposed. The major technologies that need development are listed below.

**Venus ascent vehicle.** We select solid rocket for the baseline mission to reduce the complexity compared to liquid propellant. Although the solid rocket motors are high TRL and low risk, such rockets have never been launched in another planetary atmosphere before, which effectively lower the TRL. Another challenge is launching from a balloon platform. Rockets have been launched from balloons on Earth. But the aerial platform on Venus will experience atmospheric disturbances. We need to analyze the stability of the platform for launch.

**Aerial platform.** The UHPV-based aerial platform will be demonstrated in the earlier VLF Habitability Mission (see Chapters 4 and 5). The UHPV design can be scaled for the balloon for the sample return mission, with the payload mass about 65 times higher than the VLF habitability mission mass. Developing the gondola with the launch platform and sample capture system such that the sample can be transferred to the VAV is a major design challenge.

**Orbital rendezvous**. Although orbital rendezvous is a mature technology in LEO, it has not been achieved around another planetary body before. The requirement of autonomous rendezvous increases the risk. The sample container can be transferred from the VAV to orbiter by using either a robotic arm to grapple the container or docking to the orbiter such that the sample container can be transferred inside the EEV.

### 6.6.2 Risks

The high-risk events and systems are:

1. Venus Ascent Vehicle
2. VAV launch platform
3. Balloon deployment
4. VAV and orbiter rendezvous
5. Maintaining sample integrity during Venus ascent, cruise, and Earth re-entry
6. Sample container survival on impact landing
7. Planetary protection
8. Preventing contamination of the sample during curation and distribution on Earth

## 6.7 Trades

We now turn to describe several key trades.

**Stay time at Venus.** Different round-trip interplanetary trajectories lead to varying stay times at Venus. Trajectories with low C3 and arrival and departure velocities have stay time of 30 days or over one year. We selected 30 days of stay time for the mission design, even though stay times of over one year have the least overall ΔV. Longer stay times require that the sample containers be able to preserve the samples for longer duration, thus potentially increasing the mass of the container and risk of destroying valuable sample material.

**Sample type.** We exclude sample return from the Venus surface in our mission design. Surface sample return increases the technology challenges, requiring critical systems including a lander, balloon, VAV, sample collector, and sample container to be developed to survive and operate in over 450 °C temperature and 92 bar pressure on the surface.

Since we expect life to be within the lower and middle cloud atmosphere layer of Venus, and complex organics would not survive on the surface, a surface sample is not critical for the mission objectives. The significant extra mass and complexity of a surface sample return compared to an atmosphere return appears to outweigh any utility of a surface sample for overall context.

**Sample collection platform.** We chose the aerial platform for sample collection over two other concepts: an orbiter that lowers its orbit into the atmosphere, or an atmospheric probe as it descends through the cloud layers.

An orbiter with low orbit would have the least overall mass and complexity as compared to other collection platforms, because it is a single flight element. But the atmospheric particles will encounter the orbiter sample collector at hypersonic speeds, which could destroy the sample integrity and any biosignatures in the sample.

Like the orbiter with a low orbit, a descent probe would also eliminate need for a balloon system. But because the probe is dropping quickly, it will descend to the lower, denser atmosphere by the time the required samples are captured. This increases the ΔV requirements of the VAV, increasing the mass of the system considerably. It also adds the complexity of launching from a denser atmosphere for the first time and without time to check out the systems before ascent, thus increasing the overall risks.

We therefore selected an aerial platform for the baseline mission because it enables launch from the upper atmosphere where the atmospheric density is very low and thus reduces the risks associated with the ascent and the mass of the VAV. An aerial platform also provides time to check out the VAV systems, collect different types of samples from various altitudes, and sufficient time to use a sample collection method that will preserve the biosignatures in the samples.

**Sample transfer to orbit.** We chose a VAV to transfer the sample from the atmosphere to the orbiter and not the concept of the orbiter lowering into the atmosphere to grab the balloon. The later method, which we call the "scoop and grab" concept is inspired from the Corona spy satellite program of the US, part of which was to drop camera films in a re-entry capsule that the US Air Force caught in mid-air using airplanes [158]. The scoop and grab mechanism involves encounter with the balloon at the relative speeds of at least 7.3 km/s. Such high relative speeds make sample transfer infeasible with current technologies and the scoop and grab mechanism is not considered further. Concepts that can reduce the relative speed between the balloon and "grab" mechanism need to be designed to make this concept feasible for sample retrieval. We therefore discard the scoop and grab mechanism.

**Sample collection altitude.** The sample collection altitude is TBD and will be informed by science results from the small and medium VLF missions. A critical factor is altitude as the temperature affects the thermal design of the balloon and the VAV launch mechanism. The current balloon technology can operate at 45 km altitude (and still operate at lower altitudes, i.e., higher temperatures, but for shorter duration (~15 minutes)).

**Venus ascent vehicle staging.** We select a two-stage VAV for the baseline mission, because a two-stage vehicle is more mass efficient than a single stage vehicle. A two-stage VAV is therefore preferred even though a single stage reduces the complexity and risk associated with the launch. A three stage rocket increases the complexity further.

**Venus ascent vehicle propellant.** We chose a solid propellant rocket for the VAV because it reduces the overall risks and development costs, even though the choice increases the mass. We select Star solid rocket engines for the baseline mission. Given the overall technology development for the VAV, we discarded bipropellants, because even though the Isp is higher for bipropellant compared to solid or monopropellants, bipropellants require cryogenic systems and complex pumping systems.

**Sample container mass.** The sample canister mass affects the VAV design and total mission launch mass. We use a 10 kg sample container for the baseline mission. The actual mass will depend on the types of samples collected, the individual canisters for the different samples, and thermal and structural requirements. We evaluate flight systems' masses for 5 kg and 15 kg sample containers for an estimation of how the total launch mass scales. Table 6-4 shows a comparison of flight system masses for 5, 10 and 15 kg sample container. We see that an increase of 5 kg of container mass leads to about 8% increase in total flight system mass to TVI and LEO. The maximum estimated value for 15 kg sample canister is less than the payload capacity of expendable Falcon Heavy launch vehicle. The final sample container can be chosen according to the launch vehicle capacity.

| Sample Container (kg) | 5 | 10 | 15 |
|---|---|---|---|
| Earth Entry Vehicle | 55 | 60 | 65 |
| Earth Return Vehicle Wet Mass | 1200 | 1230 | 1250 |
| Orbiter At Venus Encounter | 4750 | 4840 | 4950 |
| Entry System | 3200 | 3700 | 4260 |
| Aerial Platform | 1350 | 1600 | 1860 |
| Venus Ascent Vehicle | 680 | 880 | 1100 |
| **Total Mass to TVI** | **7950** | **8540** | **9210** |
| **Total Mass to LEO** | **32900** | **35500** | **38200** |
| **Mass to LEO with 50% margin** | **49350** | **53250** | **57300** |

Table 6-4. Flight system mass estimates for 5, 10 and 15 kg sample container mass.

## 6.8 Sample Storage and Distribution Considerations

In past sample-return missions, such as the 2006 *Stardust* cometary dust sample return [159,160] and the 2020 *Hyabusa2* asteroid sample return [161], "ownership" and distribution of samples lay with the launching space agencies, NASA and JAXA respectively. In our proposed Venus atmosphere sample return ownership might not be so clear. In particular, if the mission is wholly, or even significantly privately funded do the samples belong to the private entity or entities involved?

International law is somewhat complicated. The most relevant document is the Outer Space Treaty of 1967 signed and ratified by most nations. It prohibits expropriation of extraterrestrial bodies, such as Venus by nation-states and contains hortatory language in its preamble: "The exploration and use of outer space, including the moon and other celestial bodies, shall be carried out for the benefit and in the interests of all countries, irrespective of their degree of economic or scientific development, and shall be the province of all mankind." The Treaty does recognize in Article VI that non-governmental entities may be involved in science and exploration – but it is the responsibility of the "appropriate" state party – which generally means the state from which the mission was launched to enforce the Outer Space Treaty's provisions.

The Outer Space Treaty is silent on private ownership of material gathered or extracted from extraterrestrial sources. Not so the so-called "Moon Treaty" which entered into force by UN procedures in 1984 with only five ratifications. It's Article 11 states: "Neither the surface nor the subsurface of the moon, nor any part thereof or natural resources in place, shall become property of any State, international intergovernmental or non-governmental organization, national organization or non-governmental entity or of any natural person." The Treaty specifically states it is applicable to other bodies as well. As of 2021 only 18 states have signed and ratified the Moon Treaty. An additional four have signed, but not ratified the Treaty. Only India, which has signed but not ratified the Moon Treaty, has an independent launch capability. While subject to controversy many, perhaps most believe the Moon Treaty to be a "dead letter" precisely because the extraction, use and economic benefit from such resources is believed to be a driving force behind human expansion into the Solar System.

In the last five years several nations, most notably the United States in 2016 and Luxembourg in 2017 passed laws authorizing private ownership of space resources. In 2020 the US-initiated Artemis accords, now signed by eight nations (Australia, Canada, Italy, Japan, Luxembourg, UAE, UK, and the USA) specifically note the extraction and use of space resources as an objective.

For the proposed Venus atmosphere sample return, where the team may include or indeed be solely undertaken by private concerns the issue of "ownership" will need to be addressed. It will probably be desirable to obtain formal sponsorship through either the USA or Luxembourg – or perhaps both. In addition, mindful of the Outer Space Treaty admonitions that space activities benefit all humankind procedures for making both data, and eventually the returned material available to the world scientific community should be developed.

## 6.9 Planetary Protection and Contamination

The VLF Venus Atmosphere Sample Return Mission will follow NASA planetary protection protocols. We acknowledge that avoiding contamination of the Venusian sample with Earth life during any phase of the sample capture and return mission is paramount and needs to be a major focus for further work.

## 6.10 Signs of Life via Atmospheric Sample Return

The prime objective of sample return is to search for life via cell-like structures, and to search for robust indicators of life such as complex biochemicals including but not limited to genetic polymers. We summarize the biosignatures of interest for a sample return and describe the laboratory instruments and techniques suitable for analysis in Appendix F.

## 6.11 Summary

A sample return of atmosphere and cloud particles from Venus maximizes the chances of a robust detection of life by using the most sophisticated analytical techniques on Earth that are not readily available for in situ planetary studies. In this Chapter we aimed for a notional concept as a starting point. Technology development is most needed for atmospheric sample capture, storage, transport, and sample rendezvous with the orbiter/return vehicle. There is no question that the flight and operations concept outlined here, though ambitious, is fully worth investing in.

# 7 REPORT SUMMARY

**Abstract:** The Venus Life Finder Missions are a series of focused astrobiology missions to search for habitability, signs of life, and life itself in the Venus atmosphere. While people have speculated on life in the Venus clouds for decades, we are now able to act with cost-effective and highly focused missions.

The world is poised on the brink of a revolution in space science. Privately funded space science missions are becoming a reality. Our goal is not to supplant any other efforts but to take advantage of an opportunity for high-risk, high-reward science, which stands to possibly answer one of the greatest scientific mysteries of all, and in the process pioneer a new model of private/public partnership in space exploration.

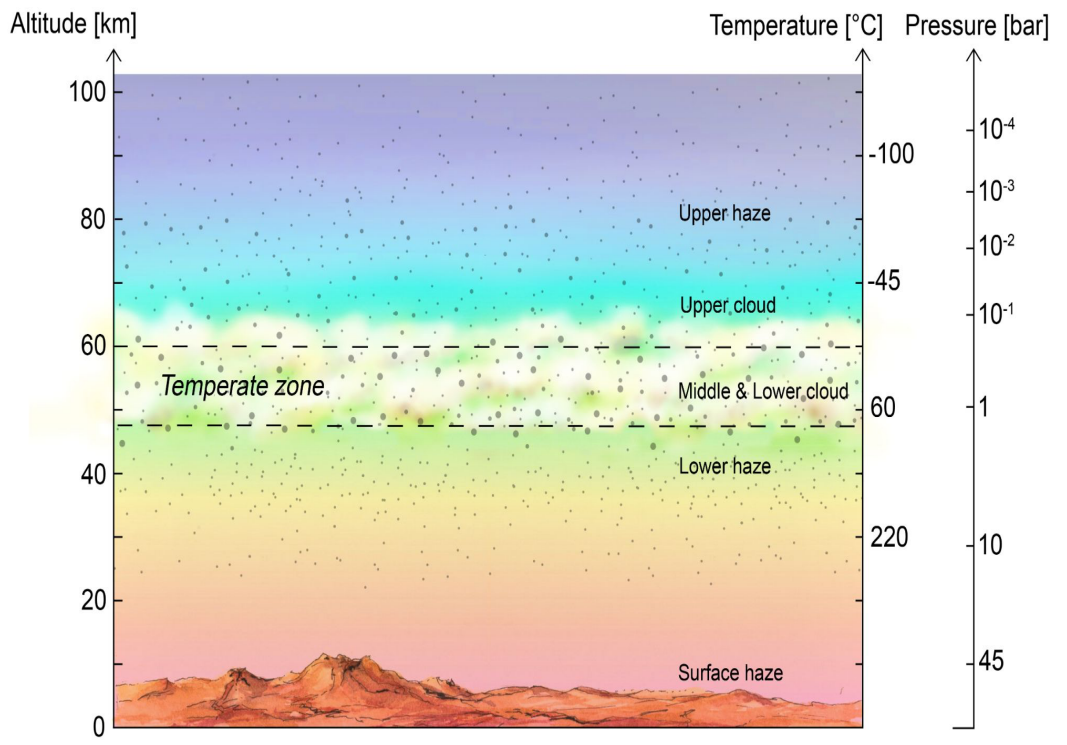

Figure 7-1. Schematic of Venus' atmosphere. The cloud cover on Venus is permanent and continuous, with the middle and lower cloud layers at temperatures that are suitable for life. Figure adapted from [14]. Figure credit: J. Petkowska.

## 7.1 Main Challenges for Life in the Venus Clouds

The Venus cloud environment is very harsh for life of any kind, despite the cloud temperatures being suitable for life (Figure 7-1). The cloud particles are composed of concentrated sulfuric acid ($H_2SO_4$) particles that are orders of magnitude more acidic than the most acidic environments where life is found on Earth. The cloud layers are additionally very dry, approximately 50 to 100 times drier than the Atacama Desert, one of the driest places on Earth. Nonetheless, a subset of the droplets are proposed to be less acidic and past in situ measurements suggest pockets of more humid areas than the current paradigm.

## 7.2 Tentative Signs of Life in the Venus Atmosphere

Fueling the interest in the Venus atmosphere a habitable or inhabited environment are many lingering anomalous in situ measurements that have never been fully explained.

The large particles in the lower cloud layers "Mode 3" are nonspherical and therefore cannot be pure concentrated sulfuric acid which is a liquid. The gas $O_2$ has been found at significant levels (10s of ppm). Observations indicate the possible presence of $NH_3$ and $PH_3$. On Earth these gases and $O_2$ are only associated with life. The combined measurements of $SO_2$, $O_2$, and $H_2O$ indicate a chemical disequilibrium, also a sign of life. See Chapter 1 for a thorough discussion.

The VLF missions focus on confirming and expanding on measurements of the Venus atmosphere anomalies in order to determine the source of the anomalies, and to search for signs of life and life itself.

## 7.3 The Rocket Lab Partnership

The VLF Rocket Lab mission is a partnership between the MIT-led science team and Rocket Lab for the Rocket Lab Mission to Venus in 2023. The VLF mission supports the science instrument and science team with the goal to investigate habitability by way of determining the composition and shape of the Venus Mode 3 particles. The prime focus is the search for organic compounds in the Mode 3 particles. If life is present, it is most likely microbial-type life residing inside the cloud particles. Organic compounds (those with delocalized electrons in ring structures) when subjected to UV-light yield stronger fluorescent signals compared to other molecules in general. A discovery of organic compounds shows that complex molecules suitable for life exists and even leaves room for life's presence.

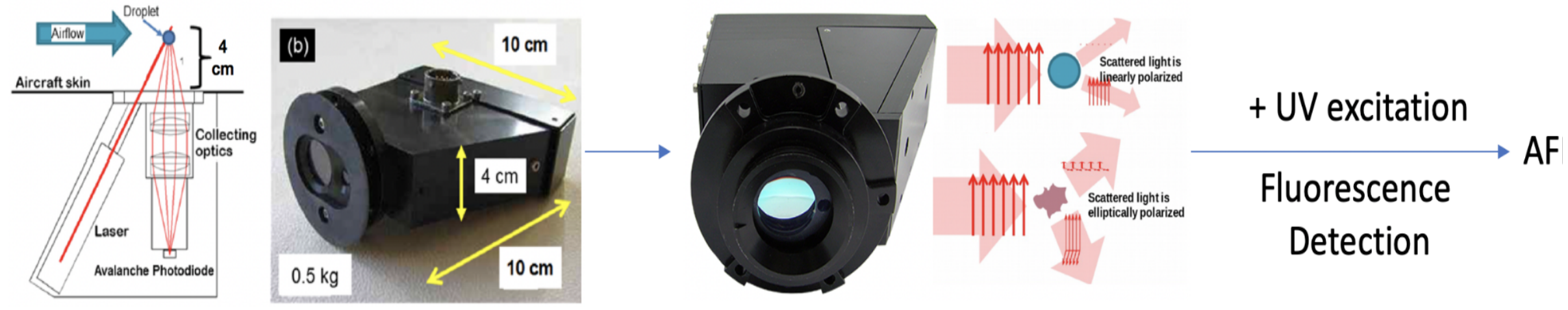

Figure 7-2. Evolution of the Autofluorescence Nephelometer (AFN) from the Backscatter Cloud Probe (BCP) (left of arrow) to the Backscatter Cloud Probe with Polarization Detection (BCPD) (right of arrow). The BCPD is further evolved to the AFN by replacement of the BCPD laser with a UV source and addition of fluorescence-detection compatible optics. Credit: Droplet Technologies Inc. Repeated from Figure 2-3.

The science instrument we have chosen is the Autofluorescence Nephelometer (AFN; Figure 7-2). The AFN will shine a UC laser through a probe window to induce autofluorescence in any material inside cloud particles.

If there is no autofluorescence detected, the AFN will still return useful astrobiology science, as the angular distribution of the laser light intensity and polarization backscattered off of the particles will be used to constrain the composition and shape of the particles. If we can confirm past measurements that indicate some cloud particles are non-spherical, i.e. not liquid, it leaves room for life to be present. The reason is that non-spherical droplets cannot be pure concentrated sulfuric acid and thus their internal composition may be more habitable or may even contain life-like particles.

The Venus direct entry vehicle aboard the Photon spacecraft will spend three minutes in the cloud layers making continuous measurements. The AFN has high heritage via Droplet Measurement Technologies' commercial products that fly on the outside of aircraft. The choice of the AFN delivers new science and is complimentary to the recently selected NASA and ESA missions because none include in situ studies of Venus cloud particles.

## 7.4 A Balloon Mission to Establish Habitability and Search for Signs of Life a

Our proposed VLF Venus Habitability Mission is a 4-m diameter fixed-altitude balloon mission to the Venus cloud layers will use a tailored set of small instruments to search for habitability and signs of life (Figure 7-3). The mission philosophy is to develop a near-term implementable mission.

The mission will: support or refute growing evidence for signs of life in the Venus cloud layers, ascertain the habitability of the Venusian clouds, or lack thereof; and prepare for an atmospheric sample return.

The fixed altitude balloon would operate for one to two weeks at about 52 km altitude. The balloon mission enables spatial distribution of the gases to be measured, due to balloon motion from natural vertical excursions and rapid horizontal winds. To sample lower cloud layers the mission will deploy four mini probes from the balloon to sample lower cloud layers.

The science payload has a mix of mature and novel instruments chosen to reach the science goals.

**A four-channel mini Tunable Laser Spectrometer (TLS)** to measure abundances of key biosignature gases: $O_2$, $NH_3$, and $PH_3$, and the habitability indicator $H_2O$.

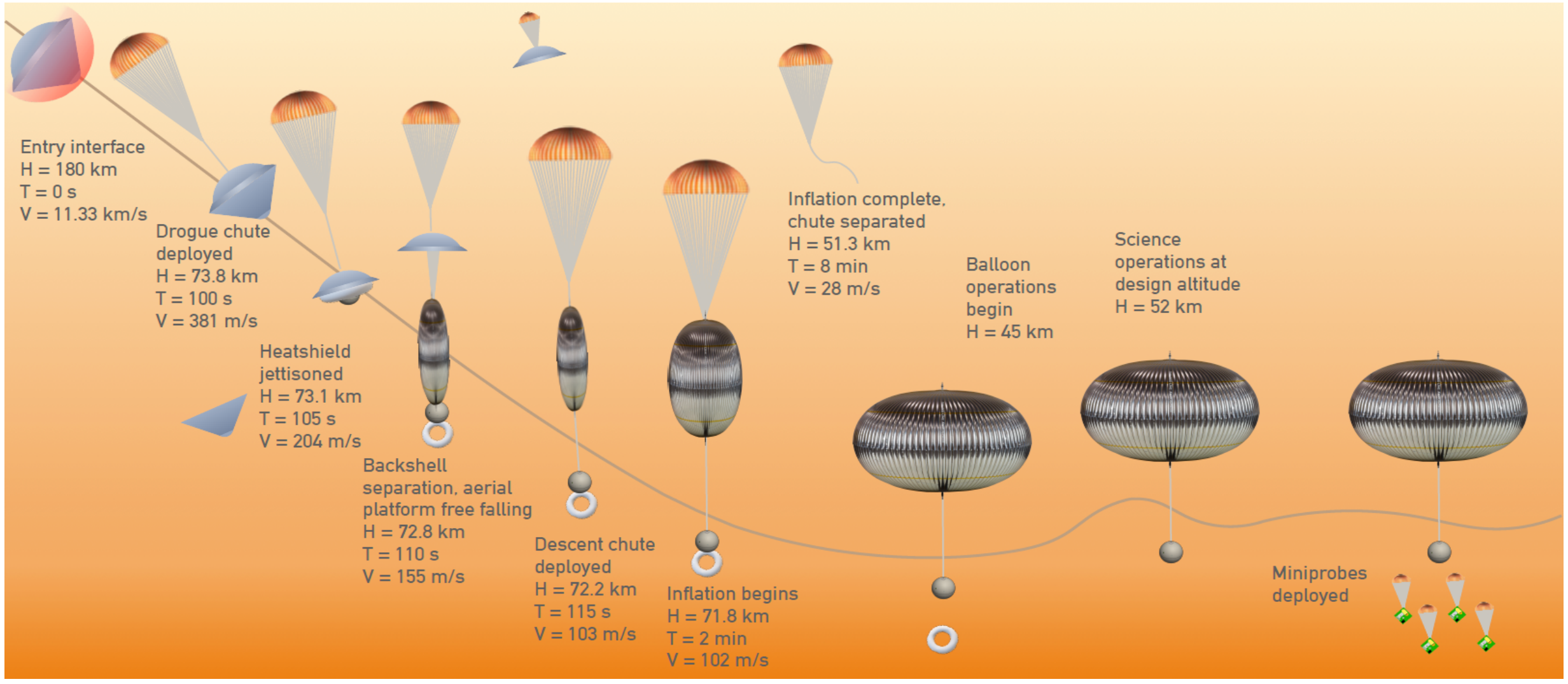

Figure 7-3. Balloon mission entry, descent, and deployment concept of operations. The 52 km fixed-altitude balloon mission is planned to be operational for one week and communicate both through an orbiter as well as direct communication to Earth. Mini probes will be deployed to sample the atmosphere below 52 km. Repeated from Figure 4-7.

**An autofluorescence nephelometer with UV excitation capability.** The AFN will be enhanced in capability compared to the Rocket Lab Mission AFN, by way of additional excitation lasers and a broader wavelength range detector.

**A MEMS device to detect non-volatile elements including metals.** Such elements are needed for all life as we know it to catalyze metabolic reactions. The MEMS device will be tailored to elements of interest.

**A single particle pH sensor** to investigate the hypothesis that cloud particles may have acidities of pH = 1. A measurement of pH = 1 would be a major discovery, because pH = 1 is consistent with the environment for acidophiles on Earth whereas the acidity of concentrated sulfuric acid is orders of magnitude lower and destructive to all life as we know it. As part of this study we motivated development of two independent single particle pH sensors.

**A weather instrument suite** to measure temperature-pressure profiles and wind speed. While not directly astrobiological in nature are worth measuring in their own right. Transient planet gravity waves are encoded in the temperature pressure profiles (e.g., [100]) and measuring them helps substantiate the concept of moving material, including hypothetical spores, up from lower atmosphere layers.

**Four mini probes deployed from the balloon** to sample atmosphere layers below 52 km. Two mini probes would contain a pH sensor and two mini probes would contain a MEMS gas detector. Each would also contain the weather instrument suite.

The total balloon mass would be about 45 kg with 30 kg allocated to the gondola carrying the science payload and mini probes.

During the study we considered a number of trades, including a variety of science instruments and two balloon categories. A fixed altitude balloon is simpler and cheaper than a variable-altitude balloon (by up to a factor of five). Many of the desired instruments do not fit with the small balloon payload mass and are at this stage dauntingly complex to bring to flight readiness. For example, a liquid collector to feed a mass spectrometer and the fluid-screen particle concentrator for a microscope focal plane, are commercially available but require substantial input to become flight ready. Other instruments are relatively high mass, high data volume, and high power (such as the laser desorption mass spectrometer and the fluid-screen plus microscope combination). A balloon-based microscope, while innovative, likely would not reach down to the desired 0.2 μm minimum cell size for life. See Appendix C and D for details on a mission concept that includes these sophisticated instruments.

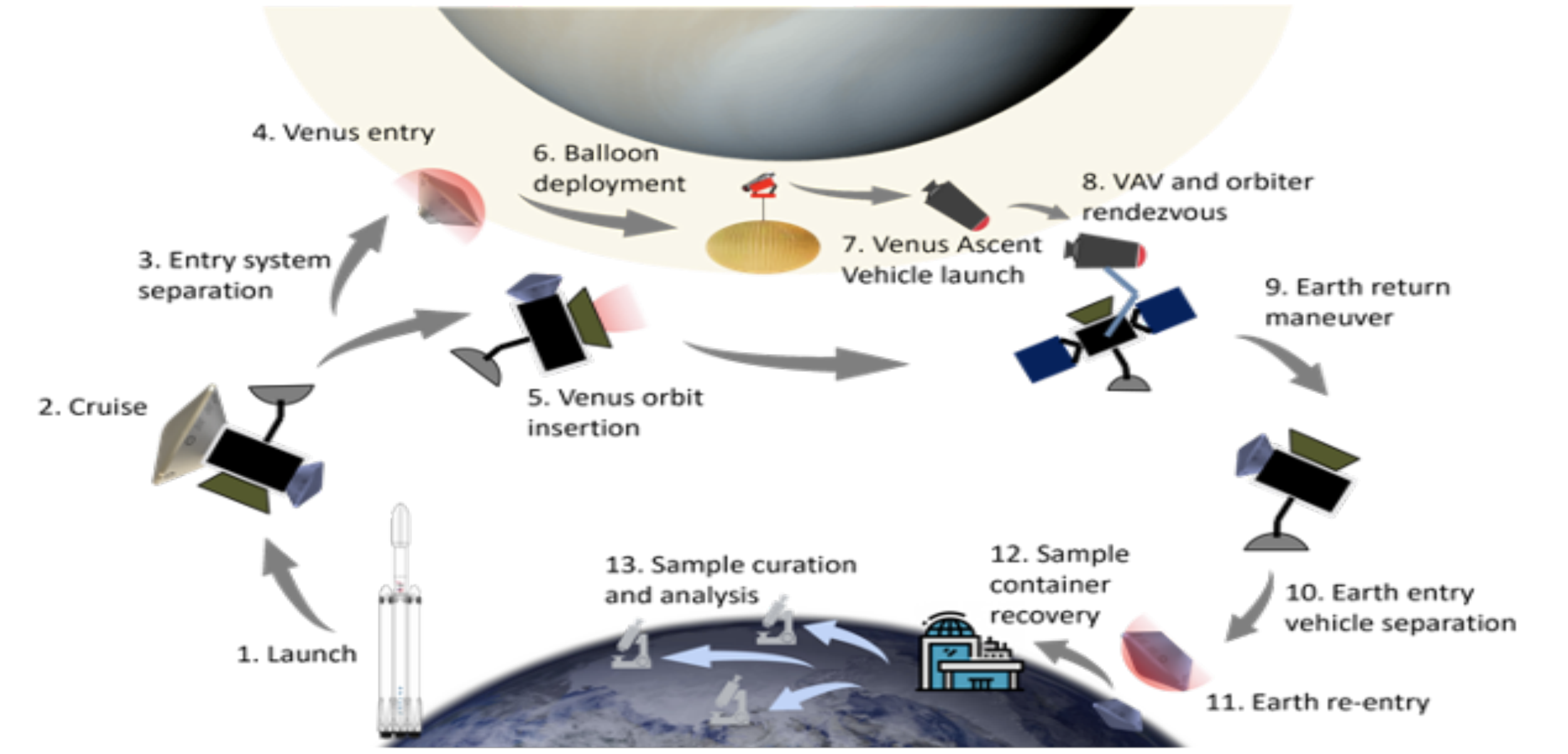

Figure 7-4. Schematic mission architecture for atmospheric sample return from Venus. VAV stands for Venus ascent vehicle. Repeated from Figure 6-2.

## 7.5 Venus Atmospheric Sample Return

The VLF atmosphere sample return Mission aims to return up to one liter of gas and up to tens of g of cloud particles. A sample return without doubt is the most robust way to search for signs of life or life itself in Venus' atmosphere (Figure 7-4). Earth-based laboratories include a wider variety of sophisticated tools with higher sensitivity than space-based instrumentation, and the presence of human investigators allows for a vastly wider range of potential experiments.

Prior to a sample return we need to invest in sample capture and storage technologies. We need to understand whether or not a subset of the Venus cloud particles are solids vs. liquids because this informs the sample capture methods. We also need to investigate the homogeneity of the cloud particles so we understand if we can lump all of the cloud particles together in a storage capsule, or is there the one in a million particle we are looking for that motivates tiny sample capture quantities.

We recommend investment in technology for a sample return, specifically gas and liquid capture and storage technologies.

## 7.6 The Search for Life on Venus Begins on Earth

The VLF mission concept study included biology-related laboratory and computer model experiments.

One of the extraordinary outcomes of this study is that a complex mixture of organic molecules can form from simple compounds (e.g., formaldehyde) reacting with concentrated sulfuric acid. During such reactions simple organic molecules undergo complex chemical transformations, including polymerization, and eventually conversion to mixtures of unsaturated, colorful and acid soluble organic molecules [1]. Such transformations could in principle lead to the formation of autocatalytic cycles analogous to metabolism. Diverse organic chemistry is needed for life.

Also as part of this study we demonstrated in a laboratory setting that certain lipids can self-assemble to form vesicle-like structures in 70% concentrated sulfuric acid. This major finding demonstrates that life particles with a lipid "cell membrane" type structure could potentially survive in the Venus cloud droplets if the conditions are right and the concentration of sulfuric acid is around 70% or less (see Appendix A.3).

Regarding the highly acidic cloud droplets. In new work related to this study we show that locally-produced $NH_3$ can neutralize the acidic cloud particles such that a subset of the cloud particles may be brought to an acidity level tolerable by acidophiles on Earth (pH = 0). The highly speculative assumption is that life in the clouds produces $NH_3$; in fact, some life on Earth secretes $NH_3$ to neutralize a droplet-sized acidic environment. Pathogens such as *Mycobacterium tuberculosis* and *Candida albicans* can neutralize the interior of phagosomes (acid-containing vesicles inside cells used for digestion of captured organic material), by secreting $NH_3$, thus evading destruction.

## 7.7 A Historic Opportunity

Collectively, our recommended astrobiology missions offer a low-cost high-impact route to seeking life beyond Earth, possibly enabling a truly historic first discovery. The near-term Rocket Lab direct entry probe mission offers the potential to uncover a smoking gun with regard to Venus cloud habitability, while enabling instrumentation validation and team-building that would support a more ambitious balloon mission as soon as 2026. Both of these missions would precede and complement planned NASA and ESA missions, while laying the groundwork for a Venus atmospheric sample return.

# Appendix A: Chemistry and Biology Experiments to Inform VLF Mission Science

**Abstract:** The sulfuric acid clouds of Venus are a great chemical unknown. To better understand this challenging environment and to inform VLF mission science we have carried out several experimental studies. 1) We have determined the optimal wavelength for laser excitation of fluorescent organic compounds for the Rocket Lab mission's AFN instrument. 2) We have explored the possibility of the formation of cell-like vesicles in concentrated sulfuric acid. 3) We have planned several other experiments to better understand the chemical reactivity of organic chemistry in concentrated sulfuric acid, with a special emphasis on the false positive and forward contamination assessment. In this appendix we provide a status update for the VLF laboratory experiments, and, where relevant, describe their connection to mission science and design.

## A.1 Introduction and Motivation

The Venus concentrated sulfuric acid ($H_2SO_4$) cloud particle conditions are the great chemical unknown. We aim to explore the chemistry of sulfuric acid more thoroughly and comprehensively than it has been done to date. We aimed for specific laboratory experiments to inform the planned in situ Venus' clouds exploration and the search for life in the clouds (Table A-1).

## A.2 Fluorescence of Organic Compounds in Concentrated $H_2SO_4$

The overall goal of the experiments on fluorescence of organic compounds in concentrated sulfuric acid is to assess if organic compounds fluoresence in concentrated $H_2SO_4$ and if such fluorescence could be used for detection of organic molecules in the Venus' atmosphere.

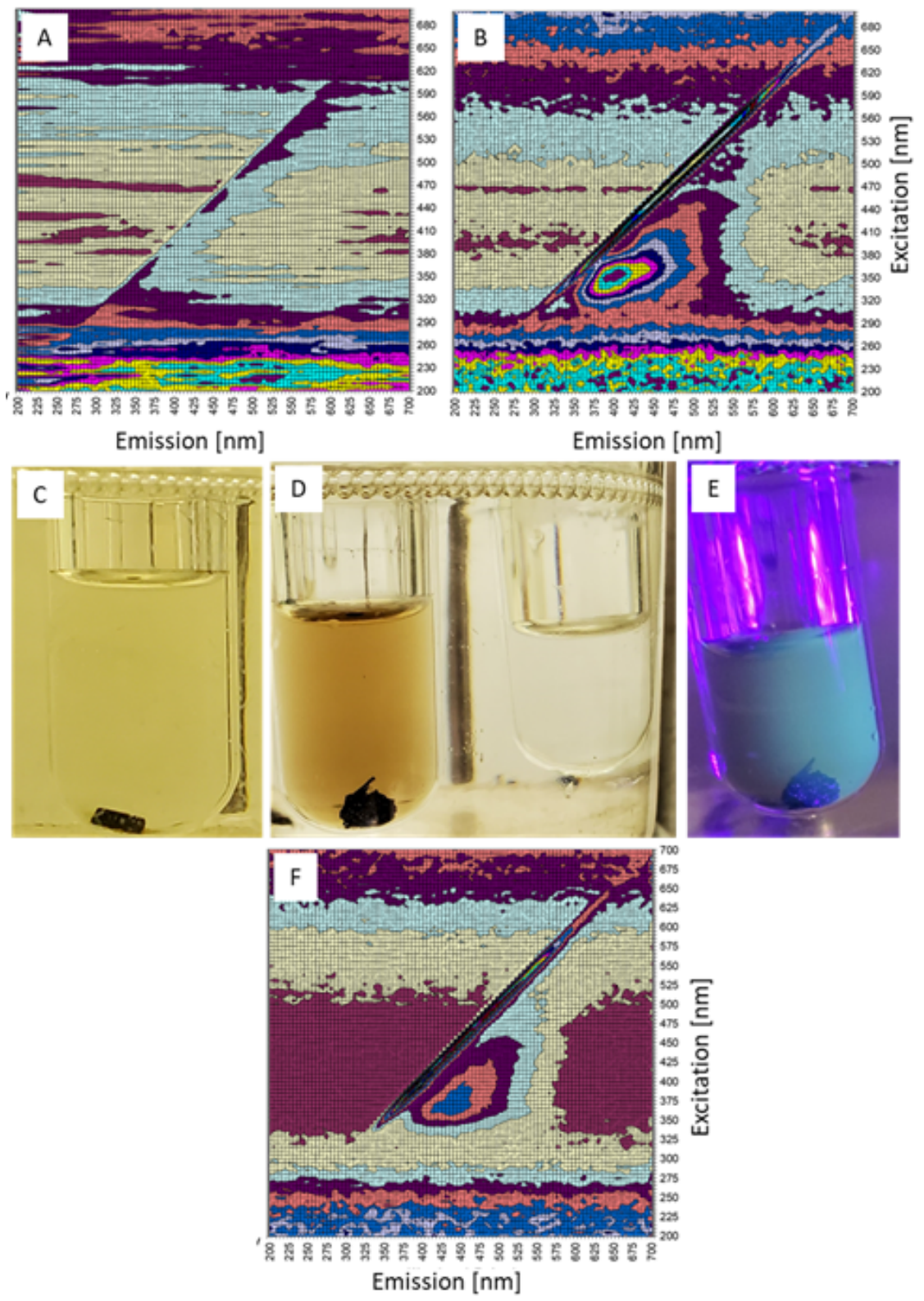

Figure A-1. Fluorescence experiments in 120 °C solution of 90% sulfuric acid and 10% water. Fluorescence of formaldehyde stabilized with methanol at A) t = 0 and B) t=48 h. A piece of Murchison meteorite (0.06 g) at C) t = 15 min and D) t = 16 h with a 90% sulfuric acid only control. The same piece of Murchison meteorite under E) a black light and F) fluorescence of the products dissolved from the Murchison meteorite in 16 hours. Both the formaldehyde and the Murchison meteorite sample react in sulfuric acid and the products exhibit fluorescence.

The specific objective of the study is to determine optimal wavelength for laser excitation of fluorescent species for the AFN instrument selected for the Rocket Lab Mission (Chapters 2 and 3).

For background, the majority of organic compounds react in concentrated sulfuric acid and yield yellow colored fluorescent species, often referred to in the literature as "red oil", conjunct polymers, humic acids, humines, or humic-like acids (e.g., [162–164]). The coloration and fluorescence behavior of organic compounds results from the formation of unsaturated conjugated systems. For further discussion of the "red oil" chemistry see [1].

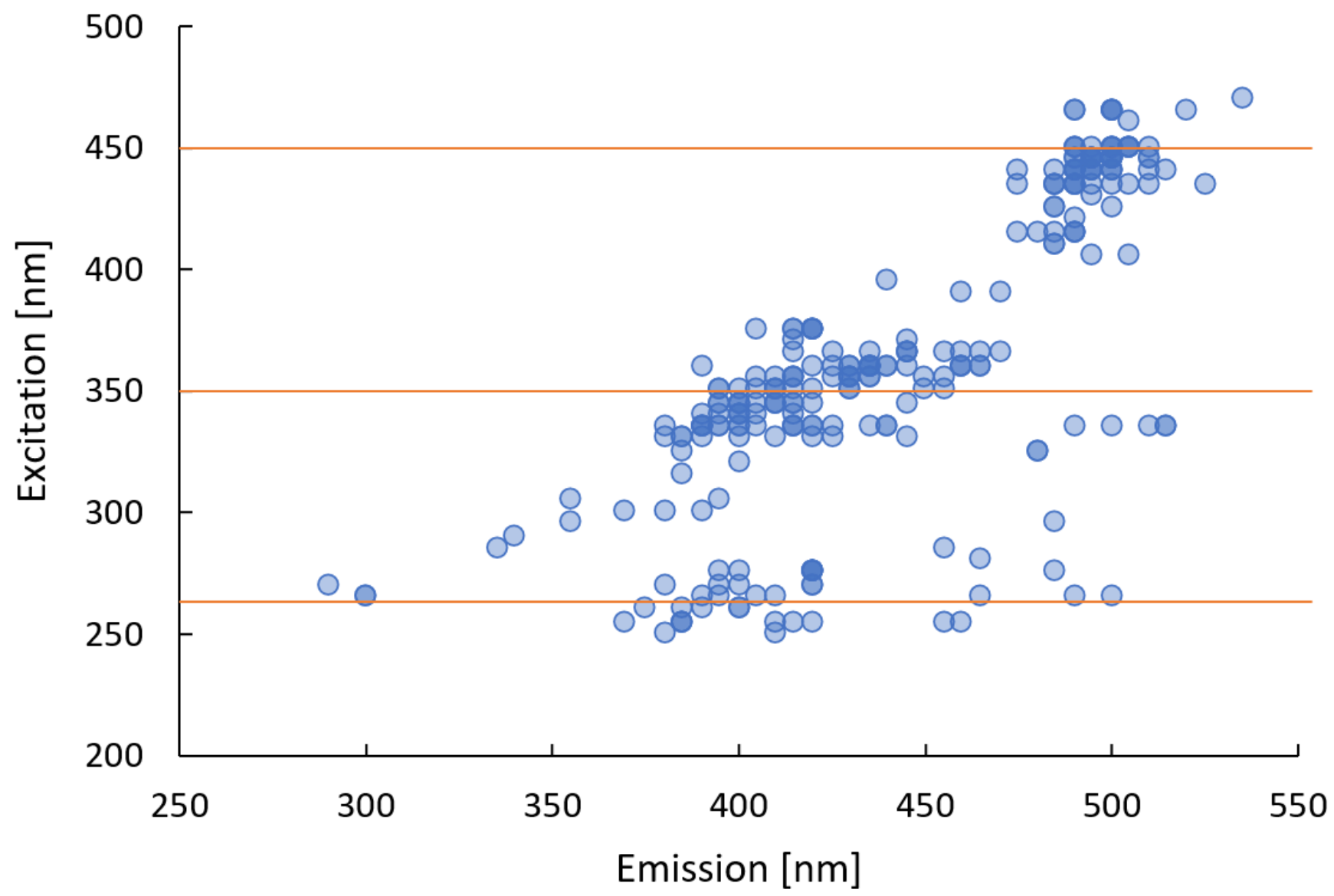

Figure A-2. Local peak fluorescence emissions vs. excitation wavelengths for all organic compounds tested. The x-axis is emission wavelength (nm) and the y-axis is the excitation wavelength (nm). The blue dots indicate detection of fluorescence. The larger number of instances of recorded fluorescence the darker the shade of the blue dot. The orange lines indicate the suggested excitation laser wavelengths. Data credit: Jan Špaček and Steven A. Benner (unpublished). Figure repeated from Figure 2-2.

If there is organic carbon in the Venus atmosphere, it will react with concentrated sulfuric acid in the cloud droplets, resulting in colored, strongly UV absorbing, and fluorescent products that can be detected by the AFN instrument selected for the Rocket Lab mission.

We have exposed several samples containing various organic molecules (e.g., formaldehyde) to 120 °C, 90% sulfuric acid for different lengths of time (Figure A-1).

As a result of the exposure to concentrated sulfuric acid all of the tested organic compounds produced visible coloration, increased absorbance (mainly in the UV range of the spectrum), and resulted in fluorescence (Figure A-1).

The preliminary results on fluorescence behavior of organic molecules in concentrated $H_2SO_4$ allow for a recommendation of a specific wavelength for the excitation laser of the AFN (Figure A-2). The optimal wavelength of the excitation laser is 350 nm, if one laser is used (i.e., for the AFN configuration used in the Rocket Lab mission). For a two-laser excitation configuration the optimal wavelengths are 350 and 440 nm (Figure A-2). In both configurations the selected excitation wavelengths maximize the chances of detection (but not identification) of organic chemicals in the Venusian cloud particles.

Next steps involve determining of the limits of detection of organics by the AFN and the assessment of any potential false positives from the autofluorescence of mineral salts and other inorganic chemicals.

## A.3 Vesicle Formation in Concentrated $H_2SO_4$

The immediate goal for the study of vesicle formation in concentrated $H_2SO_4$ is to identify lipids that can form vesicles in concentrated sulfuric acid. We have chosen 70% $H_2SO_4$ for the initial vesicle formation experiments, to build on the very limited previous studies that suggest this concentration is relevant [165] and to reflect the likely variability of concentration of sulfuric acid in the Venusian clouds, which could be between 30% and >100% (fumic sulfuric acid), depending on the altitude and droplet composition.

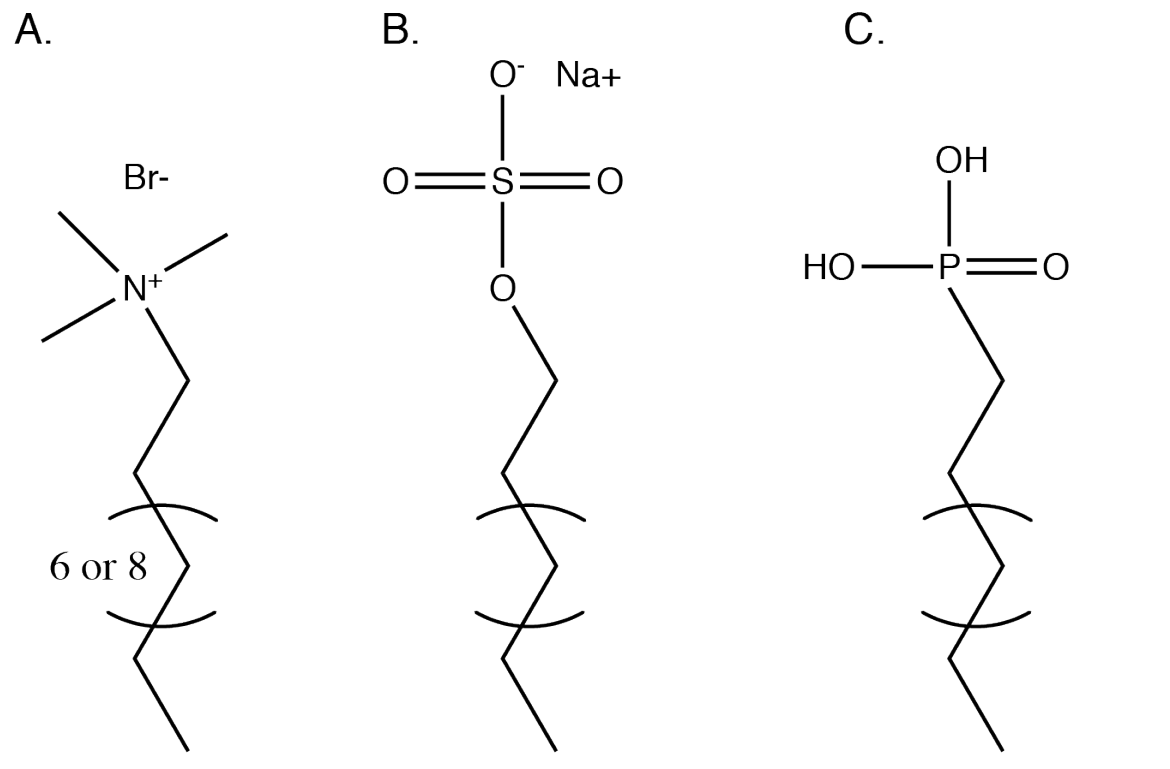

Figure A-3. Lipids selected for testing. (A) Decyltrimethylammonium bromide, (B) sodium n-decyl sulfate, and (C) decylphosphonic acid. These lipids are expected to be stable in concentrated sulfuric below 70% concentration and have chemical properties that may enable vesicle formation.

The longer-term goal is to characterize some of the basic biophysical properties of the vesicles, use them as a model system to study membrane properties in sulfuric acid solvent, and contextualize the results in broader hypotheses about the potential for life on Venus.

For background, Earth life universally employs lipids to form the structure of cell membranes, but unlike for the building blocks of proteins and nucleic acids, the chemical variety among lipids is extensive. Broadly, they harbor hydrophobic tails (usually two chains, and typically with some unsaturation) and a hydrophilic head group (usually with both positive and negative charges, and typically incorporating a phosphate). We cannot experimentally test the possibility of cellular life existing in a Venusian cloud environment using Earth cells because the concentrated sulfuric acid would destroy the Earth's life cell membrane. We seek to address this challenge by identifying a cell-free model system of vesicles. Lipids capable of forming stable membranes can also spontaneously form vesicles in solution.

We have selected a small ensemble of lipid molecules for initial testing using the following criteria:

(1) All chemical moieties are acid resistant. Unsaturated hydrophobic tails and ester linkages are both common in Earth's life lipids but would not be stable in sulfuric acid.

(2) The lipids are commercially available and the lipid tails are single-chained. This is an experimental convenience for our initial study. Single-chained lipids are more soluble than two-chained lipids, and therefore easier to analyze.

(3) The head groups of pairs of lipid species are oppositely charged under acidic conditions. Simple lipids form vesicles when the headgroups can interact electrostatically. If the headgroups carry equivalent charges, then they remain hydrophilic but repel one another and preferentially stay in solution or form micelles (see below). In contrast, headgroup-headgroup interactions mediated by opposite charges can enable vesicle formation.

Figure A-3 shows the initial set of test lipids. We expect trimethyl amine (Figure A-3 (A)) to resist acid hydrolysis and maintain its positive charge, while the oxygens of the sulfate head group (Figure A-3 (B)) interact with any surrounding positive charges. As a third option, we have chosen a phosphonate (Figure A-3 (C)), with a phosphorus-carbon bond instead of the acid labile phosphorus-oxygen bond found in phosphate.

We have confirmed the stability of the test lipids in up to 70% sulfuric acid by proton and phosphorus NMR. In acid concentrations higher than 70% the lipids are likely rapidly converted to highly conjugated and crosslinked polymers. Figure A-4 shows proton NMR data for 50 mM decyltrimethylamine across conditions. The number of peaks, their relative positions, and their area ratios do not change in response to 1% and 70% sulfuric acid indicating that the lipid is stable in that range of sulfuric acid concentration and does not react away. The experiment duration was on order of two hours.

In the next step we have measured whether combinations of test lipids could form vesicles in 70% sulfuric acid. We expect that pairs of lipids with charge-complementary headgroups should form both micelles and vesicles. To test our hypothesis we used dynamic light scattering, which correlates the rate at which a speckle pattern caused by particle light scattering decays to an uncorrelated pattern with the size of the particles diffusing through the dispersing medium. We found that mixtures of test lipids yield size distributions consistent with micelles and vesicles. Figure A-5 shows that a mix of 50 mM decyltrimethylammonium bromide and 50 mM n-decyl sulfate yields a solution with ~10 nm and ~120 nm particles, a distribution that is characteristic of micelles and vesicles in equilibrium. Crucially, the resulting solution is optically clear and uniform, rather than opalescent or opaque, indicating that the aggregates are not oil droplets.

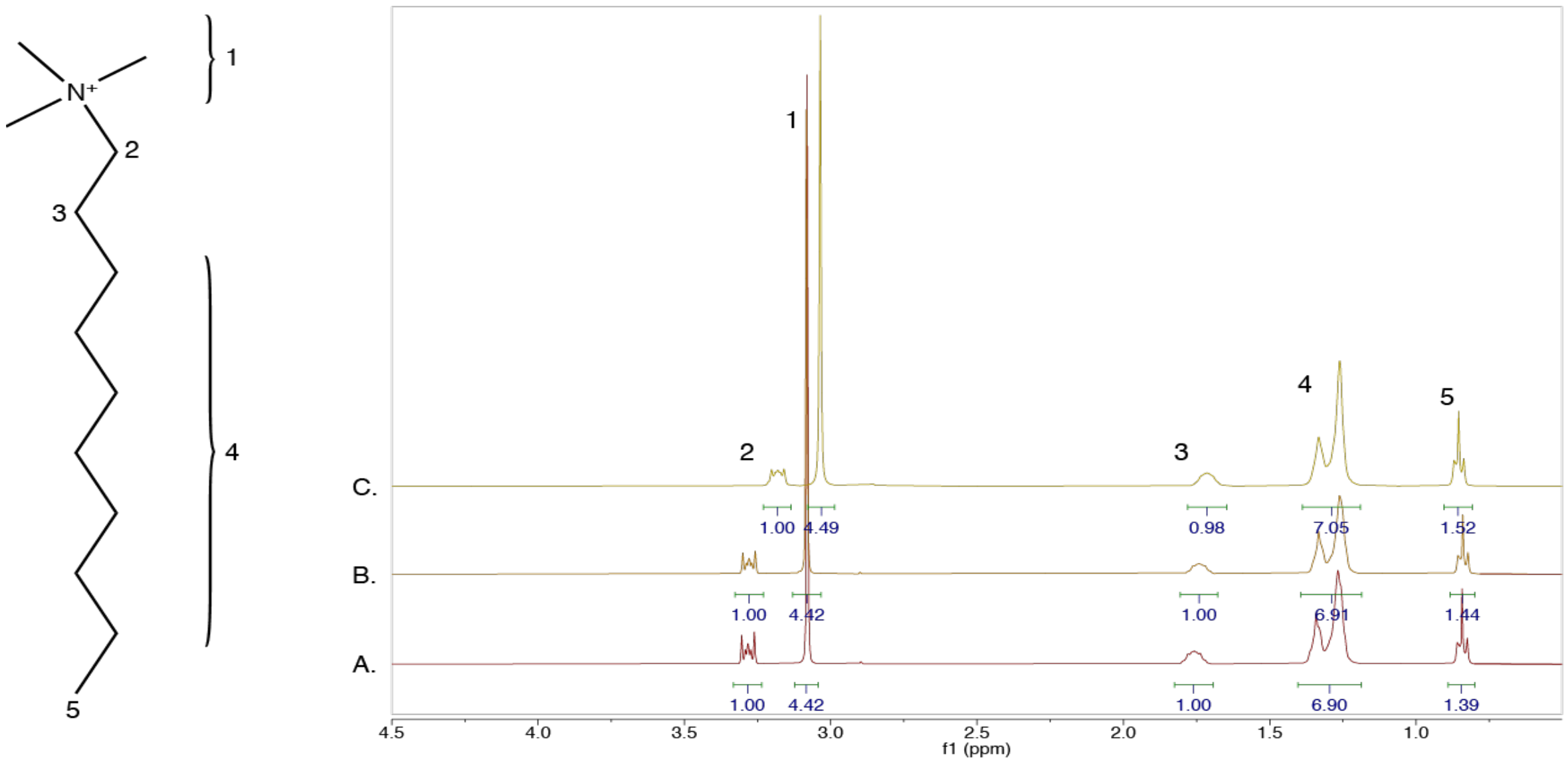

Figure A-4. Decyltrimethylamine is resistant to concentrated sulfuric acid. Proton NMR spectra of 50 mM decyltrimethylamine in (A) $D_2O$, (B), $D_2O$ + 1% $H_2SO_4$, and (C) 30% $D_2O$ + 70% $D_2SO_4$. The measured proton ratios are in excellent agreement with the expected ratios (expected, left to right = 1:4.5:1:7:1.5) and do not change with increasing sulfuric acid concentration. The slight shift in the #2 and #1 peaks is attributable to the significant change in solvent. Spectra were referenced to DSS at 0 ppm. That the peaks do not change as the concentration of sulfuric acid increases is the evidence of Decyltrimethylamine's stabillity in sulfuric acid over the duration of the experiment.

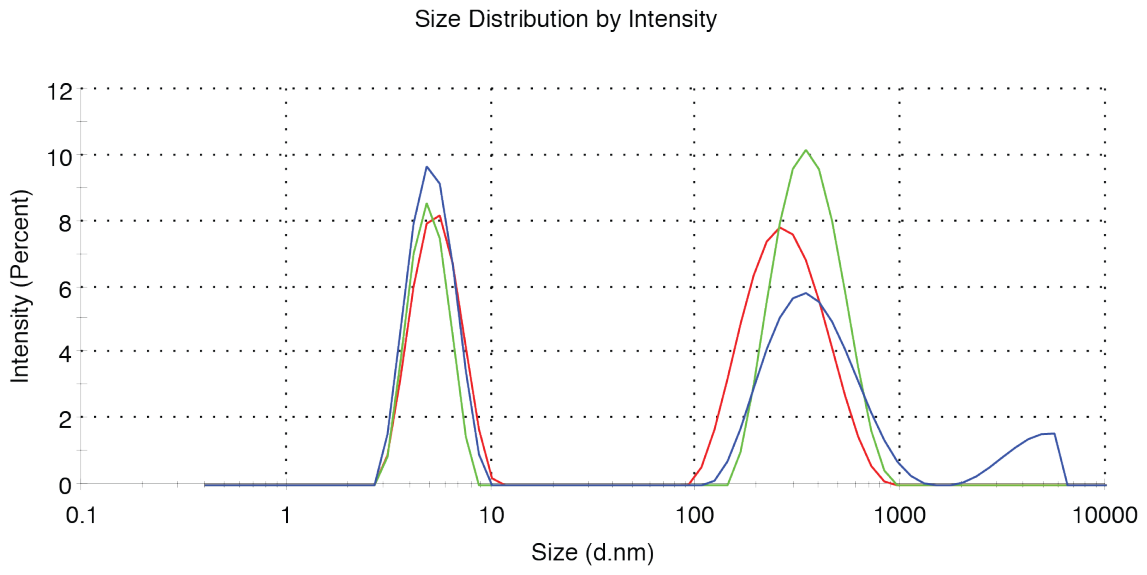

Figure A-5. Dynamic light scattering reveals that micelles and vesicles can form in concentrated sulfuric acid. The x axis is size in nm and the y-axis is intensity in percent. A mixture of 50 mM decyltrimethylammonium bromide and 50 mM sodium n-decyl sulfate was prepared in 70% sulfuric acid (by volume) and characterized by dynamic light scattering. Each trace represents an independent measurement of the same sample. The bimodal distribution is characteristic of micelles (nanometer size range) in equilibrium with vesicles (100s of nanometers range).

Additional dynamic light scattering experiments are ongoing and will determine the acidity range over which such results can be obtained, the critical aggregation concentrations for each relevant preparation of lipids, the lipid combinations that work most efficiently (and the ideal ratios of pairs of lipids), the extent to which any aggregation behavior is entirely concentrated acid-dependent, and, potentially, temperature dependencies. In future work we will assess the chemical and physical factors that affect the membrane formation in sulfuric acid, including mixtures with other organic chemicals, experiments of longer duration, and higher concentrations of sulfuric acid. We will also incorporate microscopy and aim to elucidate how our selected model system can be used to understand the biophysics of lipid bilayer membranes under this extreme solvent condition.

## A.4 Other Work on the Reactivity of Organic Chemistry in Concentrated $H_2SO_4$

We have initiated several other experiments as part of the VLF study.

**Theoretical chemoinformatic assessment of reactivity of organic chemicals in sulfuric acid.** As part of the VLF study we carried out a "big data" analysis of the stability of classes of organic chemicals in conc. sulfuric acid. We have collected the scattered literature data on the reactivity and stability of molecules in concentrated sulfuric acid (spanning more than 120 years of chemical research) into a manually curated data repository (Figure A-6) [94].

The information collected in the repository allowed us to create a model for predicting stability of any organic molecules in a range of sulfuric acid concentrations and temperature conditions. We confirm a commonly held belief that Earth-like biochemistry cannot survive in concentrated sulfuric acid for longer than one second [84]. Proteins, sugars, Earth-like lipids, and nucleic acids (DNA and RNA) and ~75% of Earth's life small molecule biochemicals and natural products are rapidly converted to highly conjugated and crosslinked polymers if concentrated sulfuric acid (instead of water) is used as a solvent for life.

We also, however, find that concentrated sulfuric acid can, in principle, support different forms of chemical complexity that could hypothetically lead to a completely different biochemistry than on Earth [84]. Both the data repository itself and the demonstration of the predictive power of the models built with it were recently published in open access journals [84,94].

**Experimental assessment of chemical stability and reactivity of organic chemistry in concentrated sulfuric acid.** The experiment examines the reactivity of the chemically important molecules for which data in the literature is missing. We aim to test the stability of organic molecules under several different concentrations of $H_2SO_4$ to cover the variable sulfuric acid abundance in the Venusian clouds. The original plan was to test the selected molecules' stability in the following concentrations of sulfuric acid: a) 60% $H_2SO_4$, 40% $H_2O$; b) 75% $H_2SO_4$, 25% $H_2O$; c) 90 % $H_2SO_4$, 10% $H_2O$; d) 100% $H_2SO_4$ ('fuming'

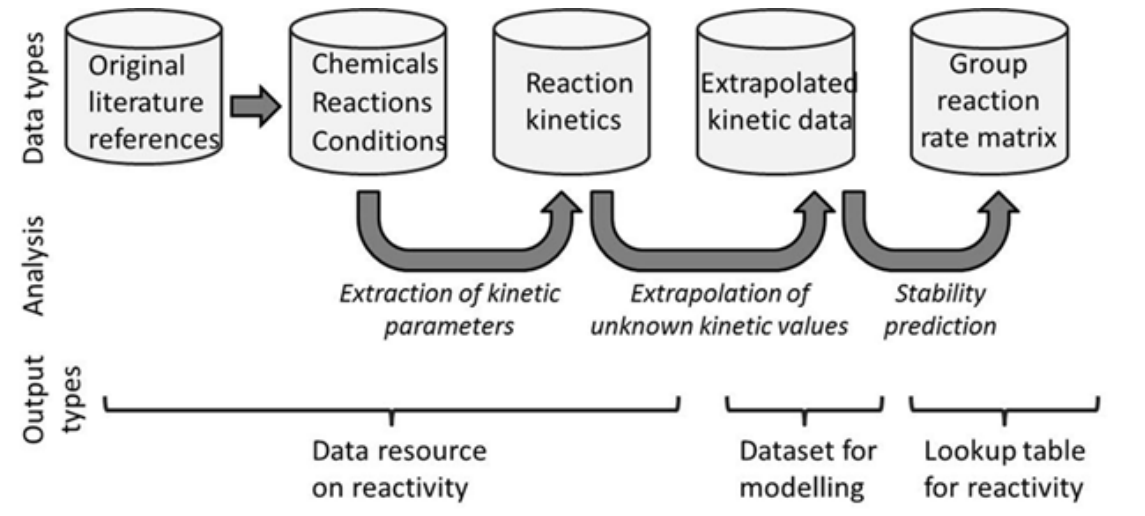

Figure A-6. Development of the database of stability of organic chemicals in concentrated sulfuric acid and the framework for the prediction of reactivity of organic chemicals in concentrated sulfuric acid [94].

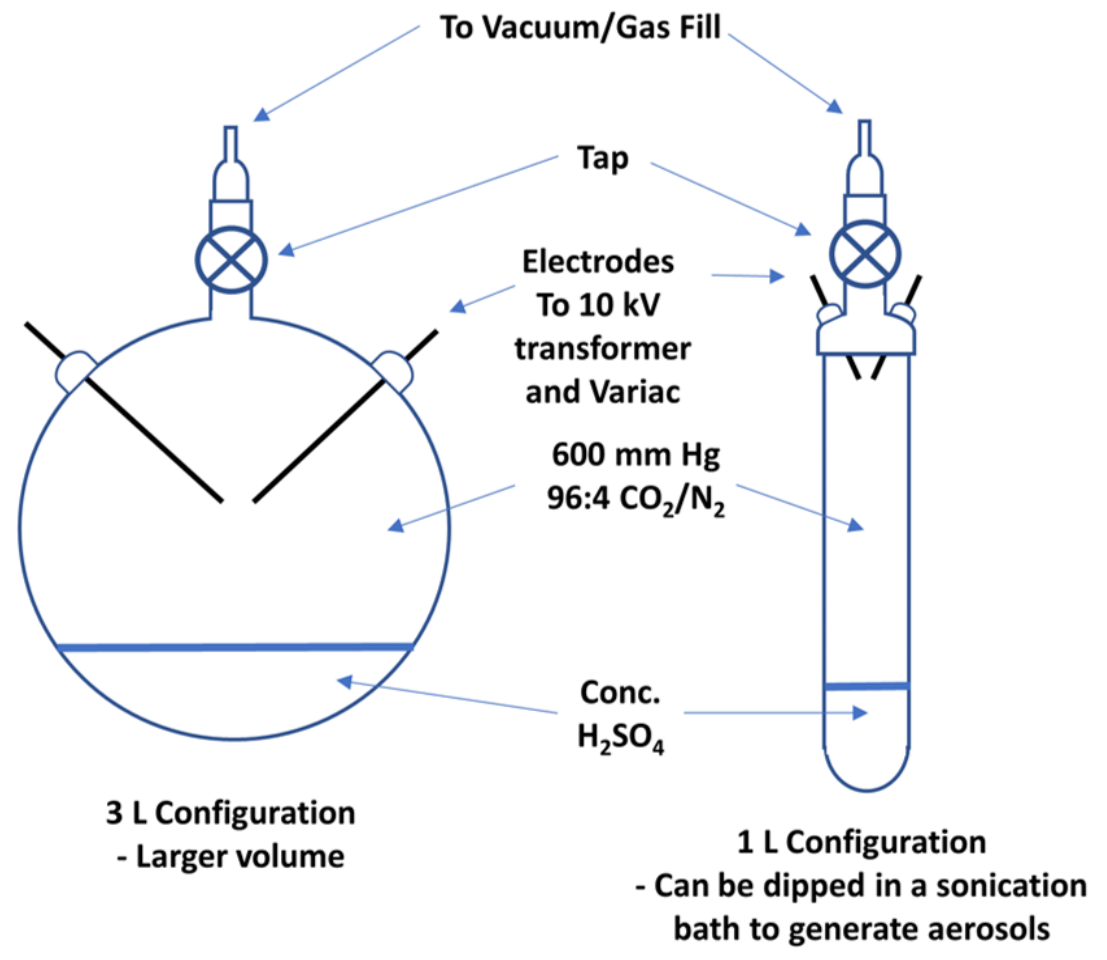

Figure A-7. Miller-Urey experiment in concentrated sulfuric acid. The apparatus for the experiment is built on classic Miller-Urey design [218,219] (in two configurations 3 L and 1 L respectively). There are two sets of initial reaction conditions considered for the experiment: 1) "Water control": $CO_2$, $N_2$ atmosphere + $H_2O$ as liquid (w/o buffer) 2) "Venus environment": $CO_2$, $N_2$ atmosphere + 85% $H_2SO_4$/15% $H_2O$ as liquid (w/o buffer). Analysis will use fluorescent amine analysis as well as ESI-MS, S isotopes should be useful for MS analysis.

| VLF Biology Experiment | Objective | Connection to VLF Mission Science | Reference |
|---|---|---|---|
| **Reactivity of Organic Molecules in Concentrated $H_2SO_4$** | | | |
| Assessment of Properties of Fluorescent Organics in the Venus Atmosphere. | Determine if some organic compounds autofluoresce while in concentrated sulfuric acid. | Inform on the design of the AFN instrument for the Rocket Lab mission by choosing the optimal wavelength(s) for laser excitation of fluorescent organic compounds potentially present in the Venus atmosphere. | Section A.2 J. Spacek and S. Benner in prep. and [1]. |
| Assessment of Chemical Stability and Reactivity of Organic Chemistry in Concentrated $H_2SO_4$. | 1. Assessment of which classes of organic chemicals are reactive and which are stable in conc. sulfuric acid, and to what degree.<br>2. The sulfuric acid reactivity (stability) test focuses on the chemical functional groups that are important chemically and for which the reactivity data in the literature are missing. | Inform instrument range and capability with the information of which classes of organic molecules could be present | Section A.4 [84,94] |
| **Possibility for Life** | | | |
| Vesicle Formation in Concentrated $H_2SO_4$. | 1. Assessment whether "traditional" membranes are stable and can form vesicles in concentrated sulfuric acid.<br>2. Assessment whether sulfuric acid-stable membranes can sequester canonical (Earth-like) biochemistry. | Support for the existence of life in Venus' sulfuric acid cloud particles by demonstrating there are sulfuric acid resistant self-assembling "cell like" membranes. | Section A.3 D. Dzudevich et al., in prep. |
| **False Positives and Forward Contamination** | | | |
| Miller-Urey Experiment in Concentrated $H_2SO_4$. | 1. Assessment if diverse chemistry can be generated in sulfuric acid with input energy.<br>2. Determine which organic molecules could in principle be produced abiotically in liquid sulfuric acid and therefore could be markers of abiotic organic chemistry in liquid sulfuric acid droplets. | False positive assessment: Organic molecules produced during such high energy reactions need not be made by life. | Section A.4 H. Cleaves et al. in prep. |
| Cellular Components' Stability in Concentrated $H_2SO_4$. | 1. Assessment which cellular components of Earth's microbial life, if any, survive in concentrated sulfuric acid and for how long. | Forward contamination assessment. | Section A.4 C. E. Carr., in prep. |

Table A-1. Chemistry and biology experiments to Inform VLF Mission science and instruments.

$H_2SO_4$; oleum). We contracted out the experiments to Organix and analysis is ongoing. Before all of the organic molecules in our list could be tested, we had to stop further experiments due to damage to the analytical equipment from the concentrated sulfuric acid.

**A Miller-Urey experiment in concentrated $H_2SO_4$.** The goal of the study is to assess if diverse chemistry can be generated in concentrated sulfuric acid with input energy. The experiment aims to show conclusively what organic molecules could in principle be produced abiotically in liquid concentrated sulfuric acid and therefore could be markers of abiotic organic chemistry in the Venusian cloud droplets. Organic molecules produced during high energy reactions, similar to Miller-Urey conditions, in the atmosphere of Venus might therefore not be unambiguous signs of life and inform false positives. Nonetheless rich organic chemistry can be related to potential habitability. Figure A-7 shows the schematic of the planned set up.

**Forward contamination assessment.** The goal of the experiment is to assess which cellular components of Earth's microbial life, if any, survive in concentrated sulfuric acid and for how long. The experiment will reveal whether or not forward contamination could pose a risk of false positives.

# Appendix B: Balloon Mission Design Trades

**Abstract:** A balloon-born mission to the cloud layers of Venus is the best way to study the cloud particles for signs of life. Yet, there are many different options with the usual tension between science capability and complexity (i.e., cost.) We summarize the balloon missions we studied and motivate our choices based on various mission design trades. This Appendix relates to Chapters 4 and 5 in the main document.

## B.1 Introduction

In designing an astrobiology-focused balloon mission to the Venus atmosphere cloud layers we faced a large range of options. During the Venus Life Finder (VLF) Mission Concept Study we therefore considered numerous trades, described in this Appendix and summarized in Table B-1. We start with balloon mission options (Table B-2) and then present mission design background and choices for two of the balloon missions, the Habitability Mission (Chapters 4 and 5) and the Venus Airborne Investigation of Habitability and Life Mission (VAIHL; described in Appendix C and D).

## B.2 Balloon Mission Science-Driven Trades

Early on in the VLF study we favored a highly capable mission concept we call the Venus Airborne Investigation of Habitability and Life (VAIHL) Mission. The VAIHL Mission concept includes a non-pyrolyzing mass spectrometer and a microscope with a concentrating mechanism, instrument concepts that have not yet flown on a space mission nor operated on a balloon platform. One element of complexity is how to capture and deliver a concentrated stream of solid and liquid particles to the instruments. In order to sample various atmosphere altitudes, VAIHL is a variable altitude balloon (VAB).

Given our desire to launch a cost-effective balloon mission to Venus in 2026, the VAIHL

| Mission Trade | Options | Attributes |
|---|---|---|
| Spacecraft orbit | Fly-by | Low ΔV |
| | Elliptical synchronous orbit | Continuous coverage |
| | Elliptical Low Venus orbit | High data rate |
| Entry orbit | Direct entry | Low heat load |
| | From capture orbit | Low g-load |
| Balloon | Variable Altitude | Higher science return |
| | Constant Altitude | Low complexity |
| Altitude Range | 52-62 km | Mysterious UV absorbers |
| | 48-52 km | Mode 3 large cloud particles |
| | 42-52 km | Haze and Mode 3 particles |
| | 45-52 km | Haze and Mode 3 particles |
| Comm. Architecture | Orbiter relay | High data rate |
| | Direct-to-Earth | Low cost |
| | Fly-by | Low ΔV |
| Launch Date | 2024 | Less time for instrument development |
| | 2026 | Favorable development schedule |
| Capture orbit inclination | Polar | Maxim two contacts possible between balloon and orbiter over 6 days |
| | Near-equatorial | Frequent and longer coverage |
| Entry Latitude | Low-Latitudes | Balloon trajectory within orbiter range |
| | Mid-Latitudes | Reduced contact time between balloon and orbiter |
| | High-Latitudes | Risk of drifting to polar regions |
| Orbit capture method | Impulsive | Higher propellant requirement, high TRL |
| | Aerocapture | Low TRL |
| | Aero-assist | Lower propellant, longer duration to capture |

Table B-1. Key mission trades. Green highlights are choices for the VAIHL Mission. The Habitability Mission shares the VAIHL choices, but is constant-altitude balloon instead of variable altitude. TRL is technology readiness level.

| Mission | Balloon Type | Alt. Range (km) | Sci. Inst. Mass (kg) | Total Entry Probe Mass (kg) | Mini Probe Inst. Mass (kg) |
|---|---|---|---|---|---|
| VAIHL | VAB[1] | 45 - 52 | 30 | 630 | 0 |
| VAIHL lite | CAB[2] | 51 | 10 | 150 | 0 |
| Hab. | CAB | 52 | 6.4 | 135 | 1.7 |
| Hab. Lite lite | CAB | 60 | 4 | 135 | 4.3 |
| Hab. Heavy | CA-VFB[3] | 48 - 60 | 6.4 | 360 | 1.7 |

[1] Variable Altitude Balloon (VAB): multiple super pressure (SP) envelopes
[2] Constant Altitude Balloon (CAB): single super pressure envelope
[3] Constant Altitude: Variable Float Balloon (CA-VFB): single SP envelope with ascent and descent capabilities via ballast drop and gas release

Table B-2. Comparison of different mission scenarios.

mission is unsuitable due to the complex instrument suite that will take time to mature to flight readiness as well as the high cost of a VAB compared to a fixed-altitude balloon.

Importantly, no space-mission suitable microscope option can sense down to the smallest thought possible size for cellular-type life, 0.2 μm, nor could existing collection options typically reach below one to a few microns. We came to the conclusion that an atmosphere sample return mission is best for a robust search for life, given the huge increase in resolution and sensitivity of Earth laboratory-based instruments as compared to what is possible on a remote balloon platform.

For a balloon astrobiology mission we have therefore chosen the Habitability mission described in Chapters 4 and 5. Our philosophy for the Habitability Mission is to send a suite of small, low-complexity science instruments to operate just long enough to assess the cloud-layer habitability, to search for biosignatures, and to prepare for sample return. A fixed-altitude balloon with mini probes to sample lower altitudes is our cost-effective choice that with an immediate start can be ready for a 2026 launch.

We considered and discarded a few other balloon missions (Table B-2). The Habitability Heavy Mission has similar set of instruments as the Habitability Mission but has instead of a fixed-altitude balloon has a balloon that varies in altitude between 48 and 60 km. The balloon is single envelope, unlike the VAB, and uses ballast drop and air venting to vary altitude for a pre-determined limited number of cycles.

| Launch Date | Arrival Date | Launch C3 ($km^2/s^2$) | TOF (d) | Arr. $V_\infty$ (km/s) |
|---|---|---|---|---|
| 5/20/2023 | 10/24/2023 | 6.09 | 157 | 3.9 |
| 5/28/2023 | 9/23/2023 | 11.47 | 118 | 3.97 |
| 12/26/2024 | 5/8/2025 | 6.96 | 133 | 3.77 |
| 7/30/2026 | 11/29/2026 | 7.12 | 122 | 4.92 |
| 9/17/2026 | 2/27/2027 | 11.89 | 164 | 7.28 |
| 3/11/2028 | 7/1/2028 | 11.06 | 112 | 6.00 |
| 4/1/2028 | 9/18/2028 | 9.04 | 170 | 6.03 |
| 10/22/2029 | 2/4/2030 | 12.84 | 105 | 5.47 |
| 10/27/2029 | 4/4/2030 | 7.78 | 159 | 4.84 |
| 5/16/2031 | 10/22/2031 | 6.11 | 159 | 3.97 |
| 5/26/2031 | 9/19/2031 | 11.79 | 116 | 4.14 |
| 12/24/2032 | 5/7/2033 | 6.83 | 134 | 3.68 |

Table B-3 Direct Earth-to-Venus transfers. TOF is time of flight. Arr. is arrival.

The Habitability Lite mission is similar to the Habitability Mission but the science payload is distributed on about 10 deployed mini probes, with the balloon gondola only carrying a transceiver. The mini probes each have a parachute or a similar mechanism to extend their duration in the cloud layers from a few minutes to a few hours, while the mini probes can still transmit data to the balloon. The balloon then relays data through a flyby space craft or direct-to-Earth transmission. We do not carry this mission forward due to the complexity of simultaneous deployment of a large number of mini-probes and the limited mass carrying capacity of the individual mini-probe that rules out two key instruments and thus severely constrains the science impact of the mission.

| Launch Date | Arrival Date | Lau-nch C3 ($km^2/s^2$) | TOF (days) | Arr. $V_\infty$ (km/s) |
|---|---|---|---|---|
| 5/21/2023 | 10/24/2023 | 2.94 | 155 | 3.87 |
| 12/7/2024 | 5/8/2025 | 5.81 | 152 | 3.58 |
| 1/2/2025 | 5/9/2025 | 4.05 | 127 | 3.81 |
| 7/18/2026 | 10/26/2026 | 5.08 | 131 | 5.00 |
| 8/14/2026 | 10/29/2026 | 6.08 | 107 | 4.95 |
| 8/19/2026 | 2/11/2027 | 10.72 | 176 | 6.36 |
| 9/15/2026 | 2/26/2027 | 9.46 | 165 | 7.15 |
| 3/3/2028 | 6/23/2028 | 11.38 | 112 | 7.00 |
| 3/30/2028 | 9/16/2028 | 6.41 | 171 | 5.90 |
| 10/12/2029 | 4/1/2030 | 6.49 | 170 | 4.84 |
| 10/14/2029 | 1/22/2030 | 15.28 | 100 | 7.67 |
| 11/8/2029 | 4/9/2030 | 5.51 | 153 | 4.87 |
| 4/28/2031 | 10/20/2031 | 5.07 | 175 | 4.11 |
| 5/23/2031 | 10/24/2031 | 3.26 | 154 | 3.86 |
| 5/27/2031 | 8/24/2031 | 19.62 | 88 | 9.31 |
| 12/8/2032 | 5/7/2033 | 5.20 | 150 | 3.56 |
| 1/3/2033 | 5/8/2033 | 4.26 | 125 | 3.71 |

Table B-4 Direct Earth-to-Venus transfers with a 200-km lunar flyby on departure.

The VAIHL Lite Mission is similar to the VAIHL Mission but without the mass spectrometer and microscope, and with a constant-altitude balloon instead of a variable-altitude one. The VAIHL Lite Mission is similar to the Habitability Mission but with no mini Probes, and no solar panels in the power subsystem.

## B.3 Interplanetary Trajectory Selection

The selection of the interplanetary trajectory has a significant impact on the arrival geometry of the spacecraft at Venus, delivered mass, and time of flight. The arrival geometry directly affects which locations on Venus are available for probe entry, if those locations are on the day side or night side, and whether the probe entry and descent will be visible from Earth.

For the baseline architecture, we choose direct transfers from Earth to Venus as the primary option (see list in Table B-3). We favor a 2026 trajectory, which has the lowest C3 until a 2029 opportunity. The 2026 launch date leaves enough time to develop the required flight elements for the Habitability Mission, although the timeline would be tight for the VAIHL Mission.

In addition to a direct Earth-to-Venus transfer, there is also an option to leverage a 200-km lunar flyby on departure from Earth (see list in Table B-4). Such a trajectory has the potential to increase the delivered mass to Venus, which could be enabling for a low-cost, low-mass mission. The 2026 Moon flyby opportunity could easily replace the current 2026 direct transfer with very little modification to the overall mission design, since both interplanetary trajectories have similar approach geometries at Venus.

We note that in addition to or instead of a lunar flyby, the mission design mass can also be increased by varying the probe coast time from separation from carrier to Venus atmosphere entry. Longer probe coast times allow for less propellant on the carrier's divert maneuver but require the probe to survive on its own for longer (requiring larger batteries, better thermal control, etc.).

## B.4 Launch Vehicle Selection

The options for the launch vehicle include the Space-X Falcon 9 Return to Landing Site (RTLS, 1140 kg), the Falcon 9 Autonomous Spaceport Drone Ship (ASDS, 2500 kg), and Rocket Lab's Neutron Rocket (1200-1400 kg). The Habitability mission would fit within a RTLS whereas the VAIL mission would require the Falcon 9 ASDS for the baseline mission.

## B.5 Entry Site Selection

A given interplanetary trajectory will arrive at Venus from a fixed direction and with a fixed velocity. This results in an arrival geometry at Venus that cannot be changed without the use of large propulsive maneuvers or the selection of an altogether different interplanetary trajectory.

This fixed arrival geometry is more consequential for Venus than for other solar system planets due to Venus's long rotational period (243 days). For other planets, the spacecraft's time of flight could be adjusted by several days with minimal impact. This would allow the destination planet to rotate into the correct orientation to target specific arrival longitudes. For the selected 2026 trajectory, the possible entry locations are shown in Figures B-1 and B-2.

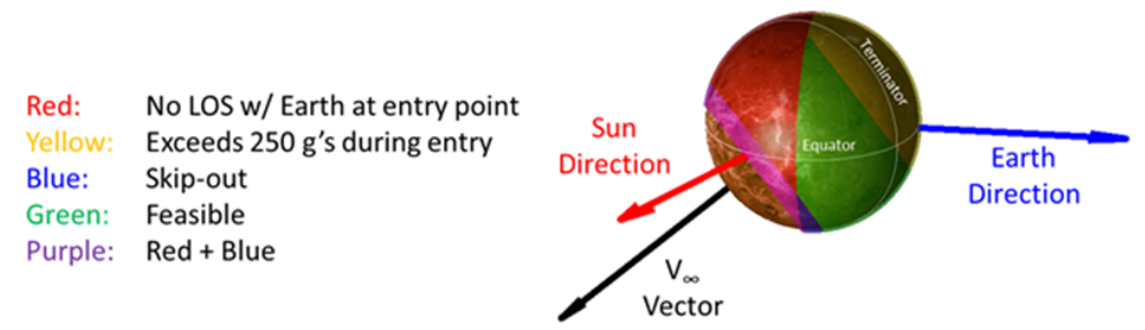

Figure B-1: 2026 trajectory approach geometry.

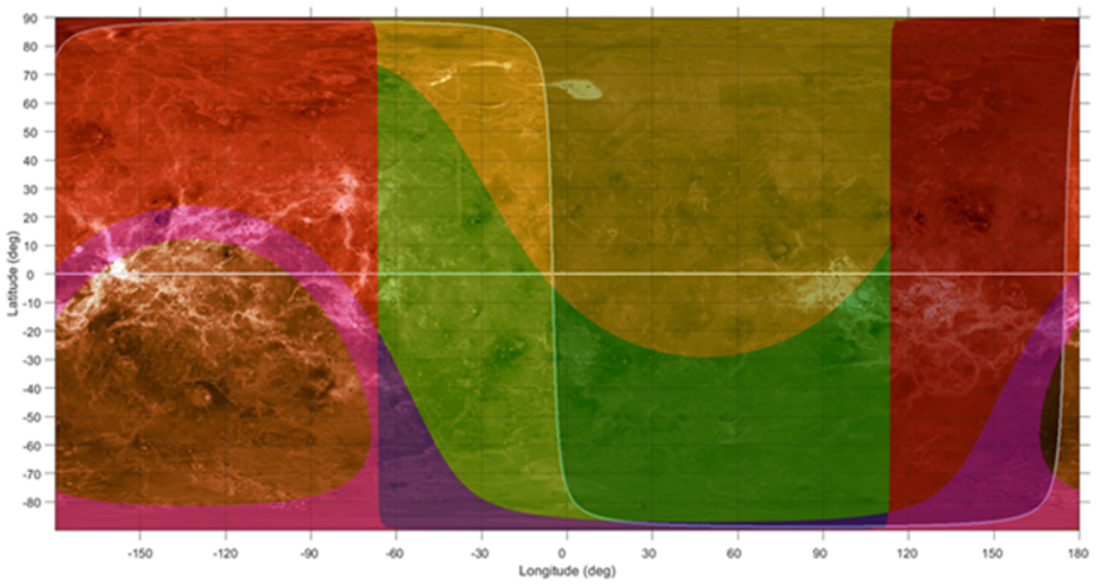

Figure B-2: Surface accessibility map for the 2026 trajectory. Color code same as for Figure B-1.

We select the Venus night side for the probe entry point. Some of the science instruments need to avoid contaminated background light coming from scattered sunlight and this is especially relevant for short-duration missions. We select a shallow entry flight path angle within the mid-latitude region of Venus, though Figure B-2 shows that there are a variety of other feasible entry points.

## B.6 Orbiter vs. Flyby

The orbiter or flyby spacecraft plays a role in data transfer by acting as a communications relay from the balloon to Earth. This avoids the major data communication limitations by a direct-balloon-to-Earth data transfer only

We choose an orbiter for both the Habitability and VAIHL mission concepts to provide the highest possible data rates throughout the balloon's entire lifetime.

While a flyby spacecraft allows for a much smaller design than an orbiter, since it does not need to carry enough propellant or large enough engines for orbit capture, a flyby spacecraft only has its one initial opportunity to communicate with the balloon.

While an orbiter is more expensive than a flyby spacecraft, an orbiter allows for the opportunity to have multiple communication windows with the balloon as the balloon circumnavigates Venus. An orbiter allows for much larger data volume to be transferred back to Earth than a flyby or direct-to-Earth, approximately by a factor of 10.

## B.7 Orbit Selection

The orbiter can enter a wide assortment of different Venus orbits. We choose an elliptical orbit that is as equatorial as possible with a periapsis altitude of about 3800 km and an orbital period approximately matching the circumnavigation time of the balloon (5 days).

The equatorial orbit rationale is to maximize its contact time with the balloon, and therefore increase its data volume. Insertion into an elliptical orbit requires lower propellant than an insertion into a circular orbit.

The relatively high periapsis altitude rationale is to ensure that contact is available for each balloon circumnavigation without needing to constantly adjust the orbit every circumnavigation. While a lower periapsis radius can increase data rates, it also decreases the orbiter's coverage of Venus. The low periapsis can be a problem given the uncertainty of the Venusian atmosphere and where the balloon might drift. A more accurate atmospheric model would allow for designs with increased data rates.

We choose an orbiter orbital period to approximately match the balloon's circumnavigation time to allow for one contact period per orbit, occurring near the periapsis. The orbital period could be decreased to a higher resonance (e.g., 6 hours instead of 5 days); however, this would require more propellant for capture and there would still be only one contact window per circumnavigation. The benefits of a higher resonance are still worth noting. A higher resonance would result in a lower apoapsis radius, which would decrease the distance between the balloon and the orbiter when receiving data, increasing data rates and data volumes per circumnavigation. A higher resonance could also allow for more of a margin for error for balloon contact if the balloon does not travel at the expected longitudinal velocity.

## B.8 Orbit Capture Method

The baseline mission uses a propulsive insertion method for the orbiter to be captured into orbit around Venus. This method requires large quantities of propellant. The alternatives are aerocapture and aerobraking. Drag modulation aerocapture offers an efficient method of inserting multiple small satellites into very-low circular orbits at Venus [166]. However, drag-modulation technology has not yet been demonstrated and lacks flight heritage. Propulsive insertion followed by aerobraking also reduces propellant requirement but takes longer to reduce to the desired orbit size. Since the orbiter needs to act as a communication relay, there cannot be any delay in orbit capture.

## B.9 Entry and Descent Trades

**Entry flight path angle (EFPA).** The shallow EFPA requirement is primarily driven by the orbiter geometry considerations for data relay.

| Balloon System | Float Altitude (km) | Mass |
|---|---|---|
| CAB | 51 | 156.4 |
| CAB | 55 | 145.2 |
| CAB | 60 | 153.4 |
| VAB | 40-51 | 514.0 |
| VAB | 48-52 | 244.3 |
| VAB | 48-58 | 540.4 |
| VAB | 45-52 | 294.3 |

Table B-5. Balloon design options. CAB is constant-altitude balloon and VAB is variable altitude balloon.

The target entry-flight path angle is -9 degrees. With a nominal delivery error of +/- 1 degree, the expected shallow and steep limits are -8 degrees and -10 degrees respectively. The nominal -9 deg. EFPA provides 1 deg. of additional margin against skip-out (-7 deg) over the shallow limit. The peak deceleration and heat rates are shown in Figure 5-6 for the steep limiting case of -10 degrees to provide conservative estimates.

**Parachute sizing**. We size the descent module parachute so as to allow at least six minutes for the balloon to inflate before the vehicle reaches its float altitude of 52 km. Note that the analysis does not include the effects of buoyancy and additional drag created by the balloon as it inflates during the descent. Hence, the reported time of 6 minutes to reach 52 km after the descent chute deployment is conservative. The buoyancy and the additional drag created by the balloon will cause the vehicle to descend more slowly than reported here, allowing sufficient time for the balloon to inflate fully before reaching 52 km. A larger parachute can be used to further increase the margin if desired. Alternatively, if analysis including the effects of balloon buoyancy and drag indicates that the balloon has sufficient time for inflation, a smaller parachute can be used.

**Deployable entry system.** Low-ballistic coefficient deployable entry system such as the Adaptable, Deployable, Entry, and Placement Technology (ADEPT) [167] offers advantages such as lower peak heat rates at Venus compared to rigid aeroshells. While they offer many advantages, we choose the 45-degree sphere-cone rigid aeroshell design as it has extensive flight-heritage at Venus and requires no technology developments.

## B.10 Aerial Platform Trades

**Balloon design**. We considered seven different balloon system configurations: three constant-altitude and four variable-altitude balloons (Table B-5). Driven by mass and science requirements constraints, we choose a CAB operating at 51 or 52 km altitude for the Habitability Mission and a 45-52 km VAB or the VAIHL Mission.

**Gondola structure**. There are two options for the gondola structure: a pressure vessel and a ventilated design. We choose the pressure vessel design for its overall simplicity and heritage, and base our design on the Pioneer Venus Large Probe [142].

A pressure vessel provides the advantage of protecting the instruments and other electronics from the sulfuric acid clouds. A pressure vessel also eliminates the requirement for the instruments to be designed to survive the vacuum of space during the mission cruise phase. The disadvantage of the pressure vessel is structural mass.

An aluminum ventilated box has about a 10 kg lower mass than a pressure vessel (of 66 cm diameter). The instruments in a ventilated box, however, need to be developed to withstand the sulfuric acid and high temperatures of the lower atmosphere. A ventilated box design is used in the Venus Climate Mission Concept Study [129] which includes a ventilation tube with a sulfuric acid filter.

**Gondola power subsystem**. There are three options for the power subsystem for the gondola: (1) solar panels + secondary batteries (rechargeable); (2) solar panels + primary batteries (non-rechargeable); and (3) primary batteries. Secondary batteries have a relatively low specific energy of 150 Wh/kg whereas primary batteries have a higher specific energy of 450 Wh/kg. Table B-6. compares the mass and area for the three power system options for varying mission duration and a 20 W average power requirement.

| Mission Duration (days) | Solar and Secondary Batteries (kg) | Solar and Primary Batteries (kg) | Only Primary Batteries (kg) |
|---|---|---|---|
| 5.0 | 18.7 | 6.7 | 5.3 |
| 10.0 | 18.7 | 9.4 | 10.7 |
| 15.0 | 18.7 | 12.1 | 16.0 |
| 20.0 | 18.7 | 14.7 | 21.3 |
| 25.0 | 18.7 | 17.4 | 26.7 |
| 30.0 | 18.7 | 20.1 | 32.0 |
| 35.0 | 18.7 | 22.7 | 37.3 |
| 40.0 | 18.7 | 25.4 | 42.7 |
| 45.0 | 18.7 | 28.1 | 48.0 |
| 50.0 | 18.7 | 30.7 | 53.3 |
| 55.0 | 18.7 | 33.4 | 58.7 |
| 60.0 | 18.7 | 36.1 | 64.0 |

Table B-6. Power system options considered. Note that the solar panel area with secondary batteries is 3.8 $m^2$ and with primary batteries is 1.8 $m^2$.

We nominally choose a solar panel and rechargeable battery system for the Habitability Mission and a primary battery system for the VAIHL Mission.

Solar panel efficiency and sizing for Venus depends on the altitude of operation as the solar flux decreases with altitude. The solar panel efficiency and lifetime is also affected by the sulfuric acid clouds and needs further consideration for the mission. Because sunlight is diffused in the cloud layer, we can use body-mounted solar panels which are structurally more stable than deployed panels.

## B.11 Flyby Communication Architecture

Although we choose an orbiter over a fly-by spacecraft for communication relay, in this section we present the trades performed while analyzing the fly-by mission design.

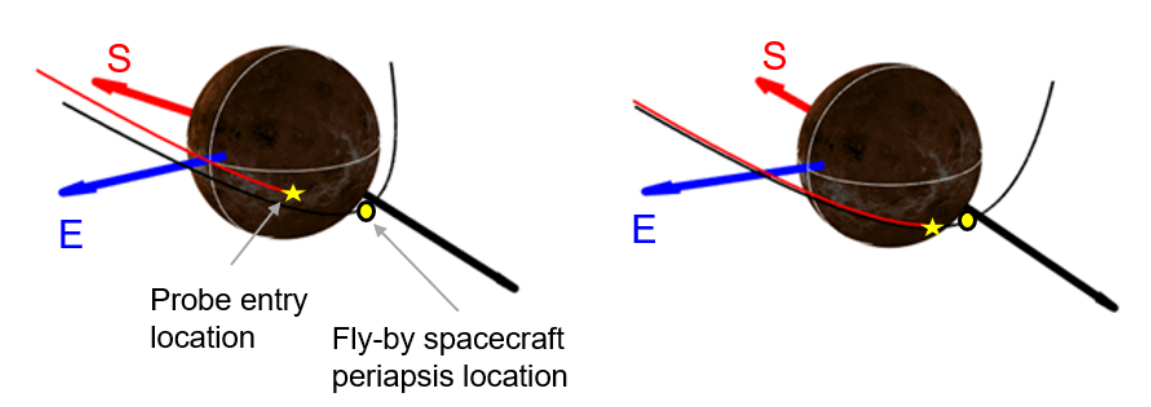

Figure B-3. (Left) -30-degree entry at Venus and (right) -10-degree entry. A steeper entry would allow for better direct Earth communications, but larger distance between probe and carrier during descent greatly decreases data rate performance.

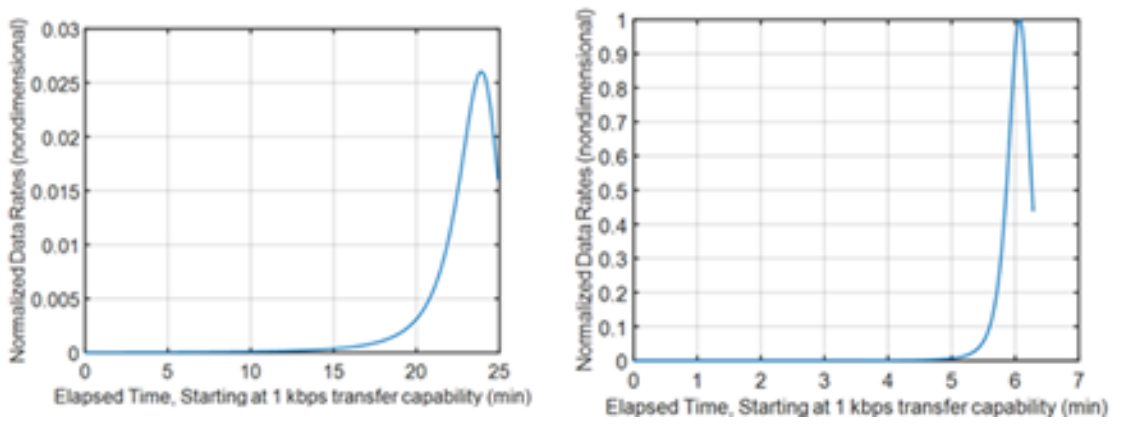

Figure B-4. Data rate for (left) steep entry and (right) shallow entry.

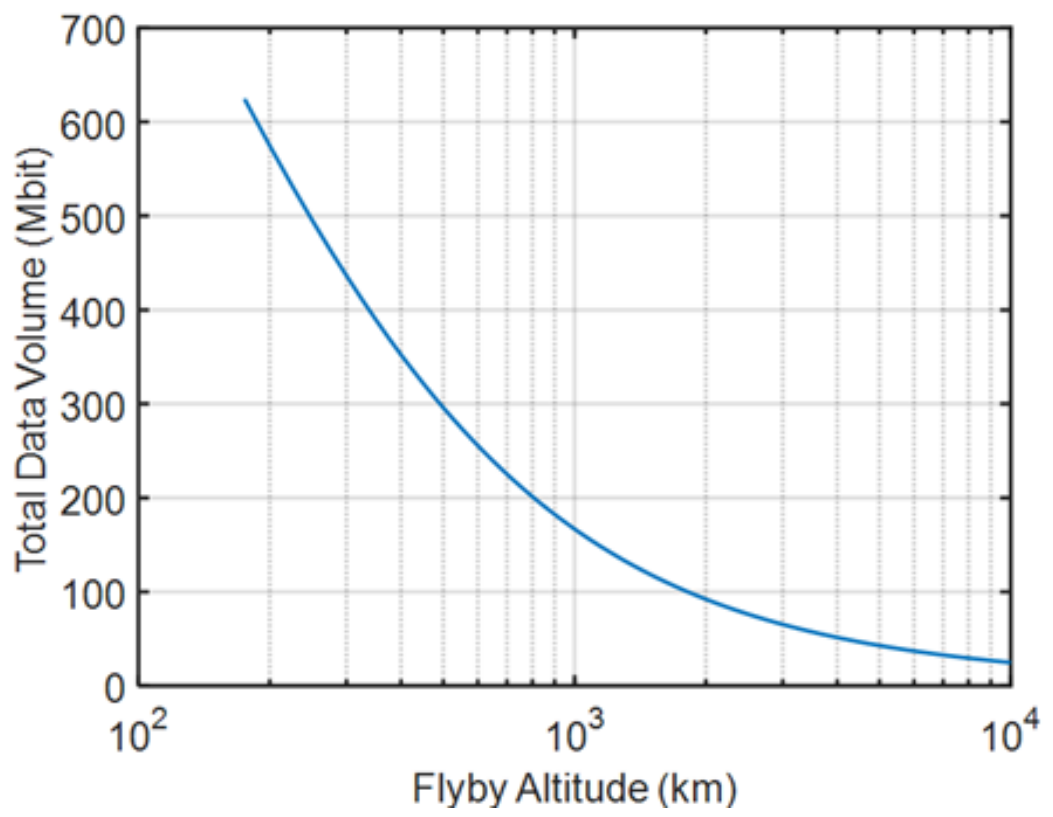

Figure B-5. Larger distances from higher periapsis altitudes result in lower data rates and data volumes.

**Impact of entry site selection on downlink (balloon to Earth) capabilities.** For a given approach trajectory of the spacecraft, the balloon-to-spacecraft geometry is dependent on the entry location of the probe. For the trajectory we have selected, an optimal balloon line-of-sight with Earth requires a steeper entry flight path angle compared to a suboptimal line of sight. Figure B-3 shows the relative locations of the probe during entry and the fly-by spacecraft periapsis. The left image shows the entry location with better line-of-sight with Earth compared to the right image. The probe's horizontal displacement is not significant, and thus balloon inflation takes place closer to the entry location. By the time the flyby spacecraft reaches periapsis, the balloon will not have traveled very far. Therefore, a steeper entry flight path angle means that the probe is farther away from the flyby spacecraft during downlink, resulting in lower data rates. The data rate is lower because it is inversely proportional to the square of the distance between balloon and spacecraft. Figure B-4 shows the data rates for a steep entry (-30 degrees) and a shallow entry (-10 degrees), normalized to the maximum data rate of the shallower entry. The maximum date rate if entering steeply is less than 3% that of the shallow entry.

The entry site also affects direct-to-Earth communication during nominal balloon operations. The balloon is radio-opaque, so communication is impossible when Earth is directly above the balloon or in general at high elevation. We select an entry point farther away from the equator to minimize the duration that Earth spends in this region of high elevation with respect to the balloon. (See Figure 5.8 for direct-to-Earth data rates.)

**Carrier flyby altitude.** The data volume from the balloon to the flyby spacecraft is heavily dependent on the choice of flyby altitude. Higher periapsis altitudes result in reduced data rates and total data volumes. Though higher periapsis altitudes result in longer contact duration, overall data volume is reduced since rates are much slower. Figure B-5 shows the reduction in total data volume transmitted to the flyby spacecraft as the flyby altitude increases. The absolute data volume values on the y-axis depend on the telecommunication system design. But the overall trend remains the same.

For a lower flyby altitude we require better knowledge of the balloon's future location during downlink. This is due to the smaller coverage area and duration for lower altitude. A few degrees of inaccurate prediction will impact the date rate greatly. A higher altitude allows a larger margin in prediction of the balloon's location at the cost of total data volume.

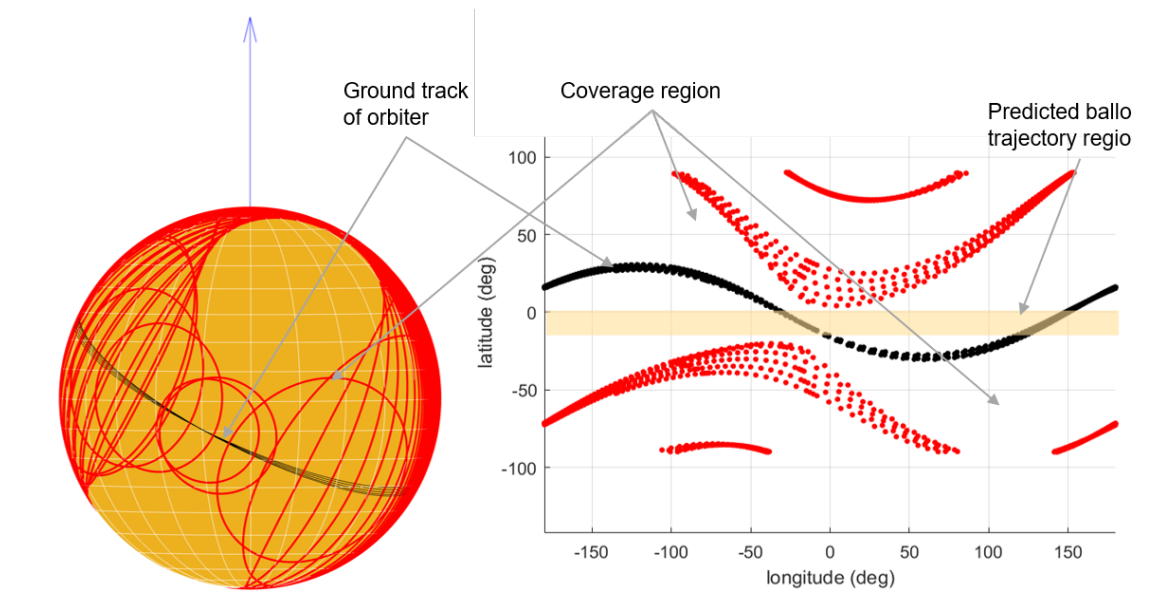

Figure B-6. Orbit ground track and coverage over 30 days for a 6-day orbit with 300 km periapsis altitude and 150 deg inclination.

**Orbit inclination and coverage.** The Venus atmosphere’s rotation is retrograde, which results in the balloon travelling in the retrograde direction. We considered a range of retrograde inclinations for the orbit, from equatorial to polar. The lowest inclination is constrained by the approach hyperbola. The selection criteria are the communication coverage provided by the orbit. The inclination should be such that the periapsis covers the range of latitudes that the balloon travels in. The most favorable inclination would provide coverage along the balloon trajectory throughout the mission duration. This is possible with a low inclination orbit if the balloon travels within the low latitude range. Figure B-6 shows coverage area for a low inclination retrograde 6-day orbit over a period of 30 days. The predicted balloon trajectory is within the coverage region along the track.

# Appendix C: VAIHL Mission Concept Science Objectives and Instrumentation

**Abstract:** The Venus Airborne Investigation of Habitability and Life (VAIHL) mission is a concept to characterize spatial and temporal variability of conditions within the Venus clouds, including whether the clouds may be habitable and possibly inhabited. VAIHL includes novel instrumentation concepts, including a cloud particle capture, concentration, and delivery system to a mass spectrometer and microscope. VAIHL would also inform future missions, including a sample return mission. Here we describe the science objectives and a representative instrument package focused on habitability and the search for life as it would be implemented by an aerial balloon platform.

## C.1 Introduction and Motivation

Major barriers to life in the Venus clouds exist: any life there would not only have to cope with an inferred water activity far below that known to permit replication on Earth ($a_w$~0.6, [168]), extreme acidity from concentrated sulfuric acid droplets ($H_2SO_4$) [92], and a perpetual aerial life cycle with a foundation of extreme heat beneath the cloud decks [13,14]. Yet recent findings point to the possibility of habitable conditions within the clouds of Venus [14,169–172], where pressures and temperatures are similar to those found on the surface of Earth [173,174]. Recent work by Rimmer et al. [80] examined the measured depletion of sulfur dioxide ($SO_2$) above the Venus clouds, and found that removal by hydroxide salts, adequate to explain the observed $SO_2$ depletion, will result in a cloud particle habitable pH of 1 at 50 km altitude. We are motivated to measure water vapor in the cloud layers, and the acidity and water content of the cloud particles.

Venus has indicators off of non-equilibrium gas chemistry that may be suggestive of the presence of life. The debated detection of phosphine [59,67], if confirmed, would represent evidence of unexpected chemistry, and raises the possibility of unknown abiotic or biotic processes in the Venus clouds. In addition, a reanalysis of Pioneer mass spectrometer data from the middle clouds [5] suggests non-equilibrium chemistry, components of a nitrogen cycle, as well as evidence of phosphine as a trace gas. We are motivated to search for the presence of and measure the atmosphere abundances of a number of gases of interest.

In this Appendix we describe the science justification and representative payload for the Venus Airborne Investigation of Habitability and Life (VAIHL) Mission concept. The VAIHL mission is an aerial balloon platform with instrumentation focused on seeking signs of life or life itself in the Venus clouds.

This mission is an expanded and more powerful concept than the VLF Habitability Mission described in Chapters 4 and 5 by including a mass spectrometer and a microscope, together with a cloud particle capture and delivery system.

The VAIHL Mission concept takes advantage of very novel instrumentation not yet considered for space exploration, which nevertheless if developed further could prove to be a game changer in direct life detection in Astrobiology-focused missions.

## C.2 Science Goals and Objectives

There are three science goals for the VAIHL mission. Each of the science goals has several specific science objectives and related instruments (Table C-1).

## C.3 Science Objectives and Representative Instrument Package

A representative set of science instruments (Table C-2) can meet the VAIHL mission science objectives. These instruments would sample gas and aerosols (liquid, solid) and be protected by a pressure vessel to protect them from the Venus environment (Figure C-1). Collection and concentration of cloud particles, as well as their delivery to the mass spectrometer and microscope is a major challenge in an aerial environment, and such a solution has not, to our knowledge, yet been developed for a space mission application. The search for life in the clouds and the detailed analysis of the chemical composition of the cloud particles requires a novel out-of-the-box thinking about the particle capture and concentration before the complicated chemical analysis can commence. We describe our approach below, newly developed as part of this study.

| | Goals | Science Objectives | Instruments: TLS | IGOM-G | AFN | Mass Spec (SSDS + LDMS) | IGOM-A | Microscope (LC + FSC + AFM) | pH Sensor | Weather Instruments Suite (WIS) |
|---|---|---|---|---|---|---|---|---|---|---|
| Biosignatures | 1. Search for Evidence of Life in the Venusian Clouds | **1.1** Search for signs of life via gas detection | Green | Green | Red | Red | Red | Red | Red | Red |
| | | **1.2** Detect organic material within the cloud particles | Red | Red | Green | Green | Yellow | Green | Red | Red |
| | | **1.3** Identify or constrain organic material within the cloud particles | Red | Red | Red | Green | Yellow | Yellow | Red | Red |
| | | **1.4** Detect and characterize morphological indicators of life | Red | Red | Red | Red | Red | Green | Red | Red |
| Habitability | 2. Measure Habitability Indicators | **2.1** Detect and identify metals and other non-Volatile elements in the cloud particles | Red | Red | Red | Yellow | Green | Red | Red | Red |
| | | **2.2** Determine the amount of water vapor in the cloud layers | Green | Red | Red | Yellow | Red | Red | Red | Red |
| | | **2.3** Determine the pH of single cloud particles | Red | Red | Yellow | Red | Red | Red | Green | Red |
| | | **2.4** Measure the temperature, pressure, and windspeed | Red | Red | Red | Red | Red | Red | Red | Green |

Table C-1. VAIHL Mission science goals and selected instrument package. Color provides a subjective judgement of the ability of an instrument to meet the science objectives. Red = not applicable; yellow = applicable; green = highly applicable. For acronyms see Table C-2, below.

### *C.3.1 Science Goal 1: Search for Evidence of Life in the Venusian Clouds*

**Science Goal 1** focuses on searching for signs of life in the clouds of Venus and has four specific science objectives.

To address **Objective 1.1** we consider specific gas targets (Table C-3) and selected low-mass sensitive detectors of key gas targets including a mini Tunable Laser Spectrometer (TLS), with two to four channels minimum, each of which covers a wavelength range dedicated to a gas of interest. The TLS is derived from the TLS that is a component of the Sample Analysis at Mars (SAM) instrument on the Curiosity Rover [107] and few changes are expected to be required to operate in a protected housing in the Venus environment; thus, the TLS is considered to have high maturity for the purposes of the VAIHL Mission. Ten-minute integration times during balloon sampling would allow the mini TLS to interrogate isotope ratios in addition to rapid identification of specific gases. Nominal gas targets include phosphine, methane, ammonia, and water. The TLS has its own dedicated sampling inlet.

To complement the TLS for **Objective 1.1**, we add a MEMS Ion Gas Micro Spectrometer (IGOM-G, Jan Dziuban/Wroclaw University of Science and Technology (WUST)). Like the TLS, gas specific targets must be selected in advance. IGOM-G can focus on a gas with a too-long TLS integration time or a gas with no suitable TLS laser. The IGOM-G has its own dedicated sampling inlet.

| Instrument | Mass (kg) | Vol-ume ($cm^3$) | Avg. Pow-er (W) | Data Vol. Per Meas. (kB) | *TRL |
|---|---|---|---|---|---|
| Mini Tunable Laser Spectrometer (TLS) | 2.30 | 120 | 7.0 | 500 | 6 |
| MEMS Ion-Gas Micro-Spectrometer for Gas (IGOM-G) | 0.34 | 400 | 0.8 | 27 | 4 |
| MEMS Ion-Gas Micro-Spectrometer for Aerosols (IGOM-A) | 0.34 | 400 | 1.0 | 27 | 4 |
| Autofluorescing Nephelometer (AFN) | 0.80 | 100 | $40.0^+$ | 120 | 3 |
| Liquid Collector (LC) | 1.00 | 1000 | 2.0 | N/A | 2 |
| Fluid Screen Concentrator (FSC) | 5.00 | 400 | 5.0 | N/A | 3 |
| Autofluorescence Microscope (AFM) | 5.00 | 3000 | 35.0 | 300000 | 6 |
| Solid Sample Delivery System (SSDS) | 4.10 | 13650 | 60.0 | N/A | 2 |
| Laser Desorption Mass Spectrometer (LDMS) | 10.20 | 14400 | 40.0 | 200 | 4 |
| pH sensor (pHS) | 0.35 | 844 | 2.0 | 1 | 2 |
| Imaging Unit (IU) | 0.15 | 250 | 0.5 | 100 | 5 |
| Weather Instrument Suite (WIS) | 0.10 | 98 | 1.0 | 0.05 | 5 |
| **Total Gondola Subsystem** | **29.68** | **34662** | **194.3** | **300975** | |

Table C-2. Representative instrument package. *TRL does not take into account the Venus environment.

To address **Objective 1.2.** we use the autofluorescing nephelometer (AFN) as a way to detect (but not identify) organics inside of cloud droplets on Venus on a single particle basis, mapping out their temporal and spatial distribution along the path of the balloon through the Venus clouds.

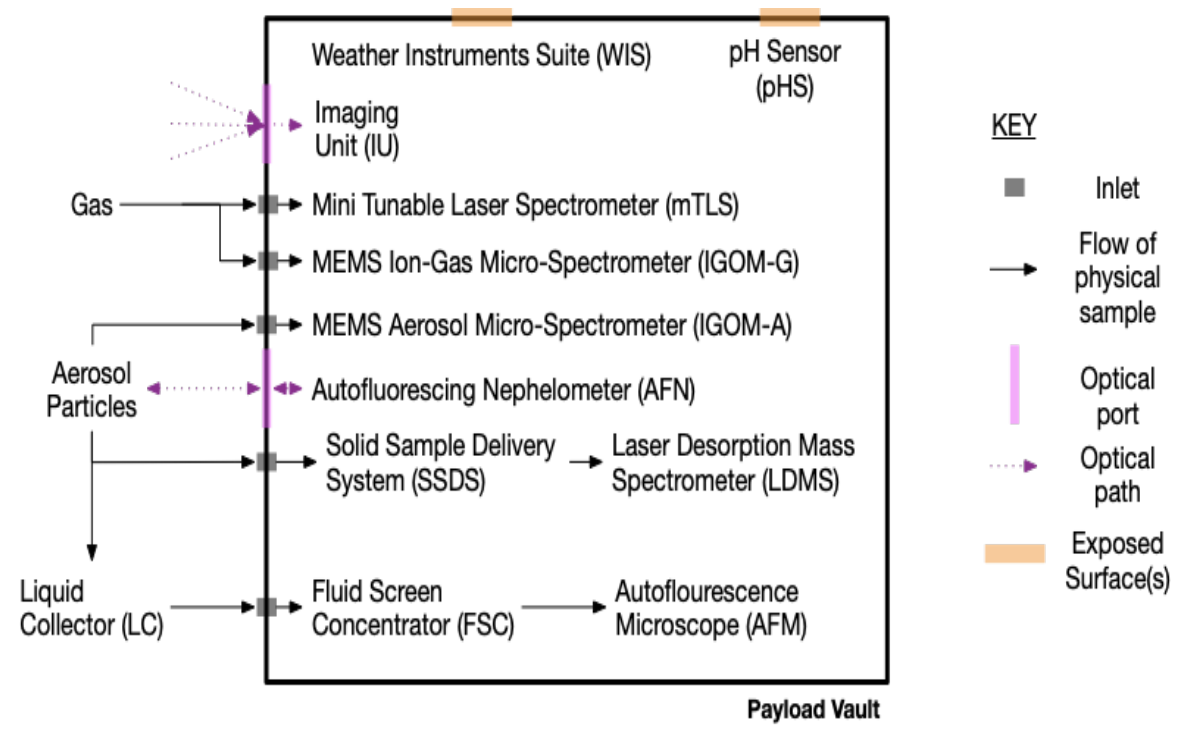

Figure C-1. Gas and aerosol particle (liquid/solid) sample handling. Instruments function as independent units except for certain combinations, where specialized delivery or concentration systems are required, as described in the text. Payloads are protected by a vault to protect them from the Venus environment. All wetted materials (with atmospheric or sample contact) must be selected to accommodate exposure to extreme acidity.

The AFN illuminates particles with a UV laser, and measures backscattered light and any emitted autofluorescence, either through an optical port (e.g., a sapphire window) or, optionally, within a ducted sample channel. The science output of the AFN is single particle size, refractive index, and asphericity, measured through polarization detection, as well as autofluorescence.

A continuous record of single particles and their accumulated statistics would be used to map out cloud layers, differentiate between particle populations on the basis of size, state (liquid, solid, shape), and fluorescence. Fluorescence would be utilized as a proxy for organic moieties known to fluoresce under the UV excitation used for particle detection. To complement measurements by the AFN, we also consider a MEMS Aerosol Micro Spectrometer (IGOM-A, Jan Dziuban/WUST). This device has its own dedicated sample inlet to collect aerosol particles (liquid or solid). This instrument is described further in Chapter 4.

All life requires complex organic chemistry to function therefore **Science Objectives 1.3 and**

| Gas | Pros | Cons |
|---|---|---|
| $PH_3$ | Genuine biosignature produced by anoxic life on Earth<br><br>Not known to be produced abiotically on Venus in significant amounts | Only tentative detection by remote observations from Earth and re-analyzed data from Pioneer Venus<br><br>Possible patchy distribution and/or time variable source<br><br>Low ppb abundance |
| $CH_4$ | Not expected to be present in the Venusian atmosphere; could be an indicator of life. | Only tentative detection by Pioneer Venus, attributed to pyrolysis (ablation product) or might be from spacecraft outgassing<br><br>Low to high ppb abundance in the clouds |
| $NH_3$ | Genuine biosignature on Earth as a product of nitrogen fixation<br><br>Not known to be produced abiotically on Venus in significant amounts<br><br>If present would neutralize Mode 3 "droplets" to pH~1 | Very soluble in sulfuric acid. Mostly present in solution as $NH_4^+$ instead of a gas in the atmosphere<br><br>Likely low ppb abundance in gas phase, if any<br><br>Only a tentative detection by Venera 8 and Pioneer Venus |
| $H_2S$ | Could be considered a weak circumstantial sign of life | Possible photochemical or volcanic product; its detection is not evidence of life |
| $H_2O$ | Quantify the measured anomalous abundance in the clouds/below the clouds<br><br>Useful as an indicator of habitability | Not a biosignature gas, volcanic source on Venus |
| CO | Important gas to constrain photochemical models of Venus' atmosphere | Not a biosignature gas, photochemical product on Venus |
| $O_2$ | Genuine biosignature on Earth<br><br>Anomalous measured abundance of ~20 ppm in the clouds by Venera 13, 14 and Pioneer Venus | |
| $SO_2$ | Anomalous abundance (depletion) in the clouds/below the clouds<br><br>Important member of the sulfur cycle in Venus' atmosphere | Not a biosignature gas, volcanic source on Venus |
| $S_8$ | Important member of the sulfur cycle in Venus' atmosphere | Not a biosignature gas, photochemical product on Venus |
| OCS | Genuine biosignature on Earth<br><br>Important member of the sulfur cycle in Venus' atmosphere and a possible photochemical or volcanic product | Not a biosignature gas, volcanic source on Venus |
| HCl | Important photochemical substrate in the atmosphere | Not a biosignature gas, likely volcanic or surface source on Venus |

Table C-3. Potential gas targets for the mini-TLS detection.

**1.4** aim to detect and identify organic material within the cloud layer particles (1.3) and detect and characterize morphological indicators of life (i.e. cellular structures) (1.4). A mass spectrometer and ultraviolet (UV) microscope are included to detect the presence of organic material within cloud-layer particles, by associating collected samples with AFN-recorded particle characteristics and other context measurements, such as the weather package. While these instruments would contribute to Objective 1.2 (detecting the presence of organics), we describe them below: the mass spectrometer would not only help to detect organic material, but would also identify it **(Objective 1.3);** the UV microscope will not only detect organic material, using fluorescence as a proxy, but would do this in the context of detecting morphological indicators of life (**Objective 1.4**).

To address **Objective 1.3** VAIHL will use a laser desorption mass spectrometer (LDMS). Laser desorption mass spectrometry can analyze almost any kind of solid material, e.g., potential biochemicals (amino acids, lipids, etc.), salts, and minerals or other solid residues present in atmospheric particles without the need of prior pyrolysis of the sample.

A number of space missions have utilized mass spectrometers, all so far with pyrolysis. We choose to avoiding the destructive process of pyrolysis to allow for more reliable identification of parent complex organic molecules.

Two example LDMS instruments being designed for space are *CORALS* [175] and *ORIGIN* [176]. These instruments can detect trace levels (~fmol) of complex organic compounds (>200 Da) from chemically diverse samples without prior knowledge on the sample chemical composition, a feature crucial for chemical characterization of the unknown environment of the clouds of Venus.

For context, the AFN will measure particle sizes and shapes and their possible fluorescence, not for the same stream but for statistical context that will help identify presence of organics. Detection and identification of complex organic chemicals is considered to be indicative of the presence of life (e.g. [177]).

**Objective 1.3** also includes *determining the properties and internal composition of the Mode 3 cloud particles*. The unknown properties allow for the speculation that Mode 3 particles contain metabolically active microbial life. The AFN may detect (but not identify) any potential organic material while the balloon platform can map out the spatial distribution. The detailed chemical characterization of the Mode 3 particles can be achieved with an LDMS (as in Objective 1.3) and possibly with IGOM MEMS devices (as described in Objective 2.1 and in Chapter 4).

**Objective 1.4** aims to directly detect and characterize morphological indicators of life. Any kind of life, no matter its chemical makeup, likely needs barriers (cell walls, membranes etc.) that allow for life to exist in a distinct, separated form from the surrounding environment (i.e. to maintain a cell-like structure). For **Objective 1.4** we aim to collect, capture and concentrate cloud particles for direct visualization with a UV-autofluorescence microscope. On Earth, many natural compounds are known to fluoresce when subjected to UV radiation. Autofluorescence emissions allow for detection and classification of organics within the captured cloud particles and enable direct detection and label-free imaging of microbes or other cell-like structures possibly containing fluorescing organic material.

To achieve **Objective 1.4** we aim to combine the UV-autofluorescence microscope system with a novel microbial capture and separation method called Fluid-Screen [178]. This in turn requires a liquid collector, still under development (Section C.4.2). For context, the AFN will measure particle sizes and shapes (asphericity) and their possible fluorescence, not for the same stream but for statistical context that will help identify presence of organics in association with cloud layers. This system links a liquid sample from the Liquid Collector to the Fluid Screen Concentrator (FSC) which is held at the focal plane of the Autofluorescence Microscope (AFM).

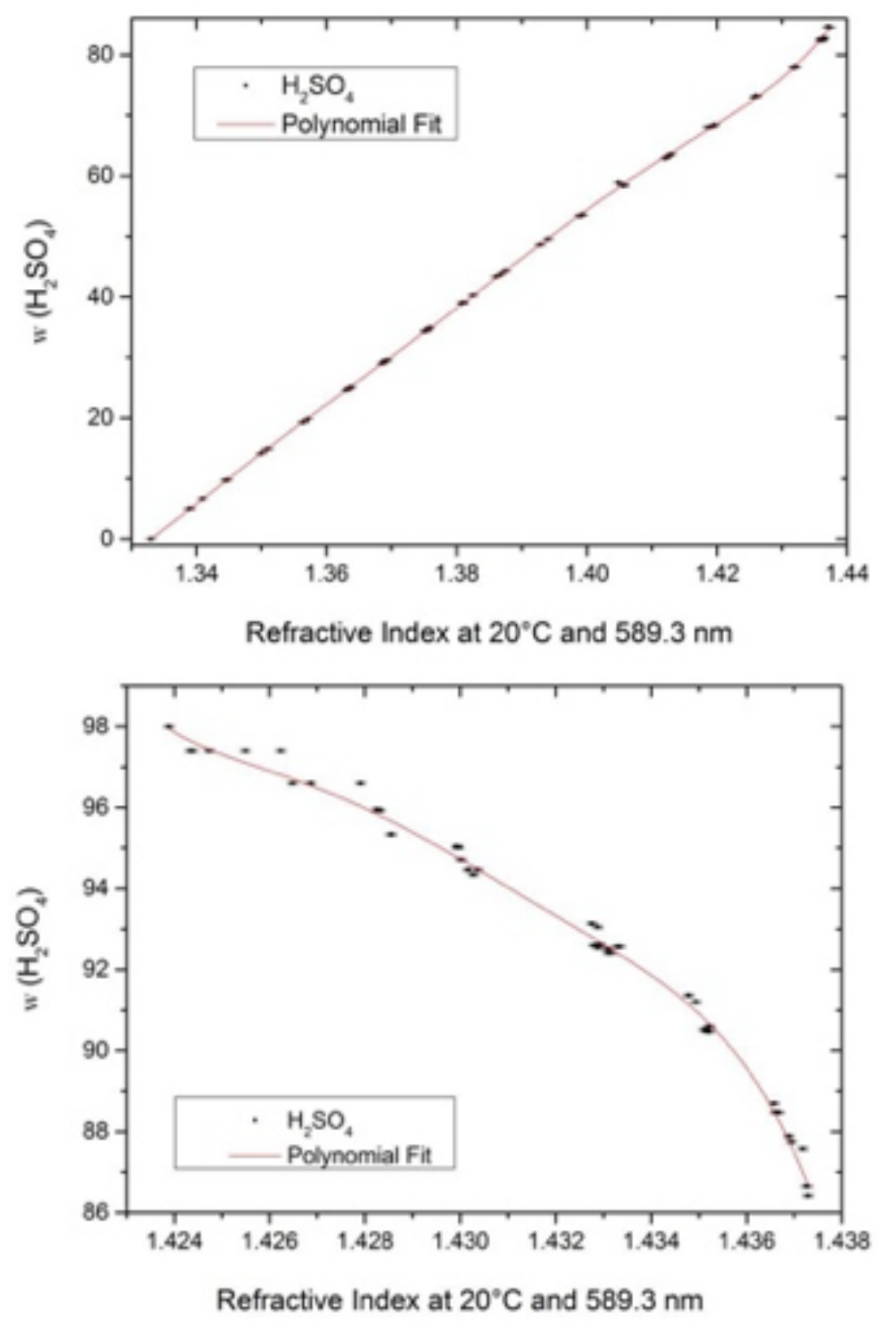

Figure C-2. Refractive index and sulfuric acid concentration in the $H_2SO_4$-$H_2O$ system. Refractive index is linearly related to sulfuric acid weight fraction up to ~80% $H_2SO_4$ (top panel). Above this concentration, refractive index changes reverse direction and the change in refractive index becomes less sensitive per percentage increase in sulfuric acid content (bottom panel). The proposed AFN instrument can estimate the refractive index of individual aerosol droplets, and thus provide one measure of water content in the Venus atmosphere. However, above 80% sulfuric acid (below inferred 20% water content in the sulfuric acid-aqueous dual system), AFN measurements may not well constrain the water content due to the non-linear refractive index response. However, refractive index measurements below 1.42 may indicate unexpectedly high water content (>30%). Image adapted from [220].

### *C.3.2 Science Goal 2: Measure Habitability Indicators*

**Science Goal 2** focuses on assessment of the habitability of the clouds of Venus. Goal 2 has four science objectives.

**Objective 2.1** aims to detect and identify metals and other non-volatile chemical species in the cloud particles if they exist. Life as we know it requires metals and other non-volatile species (e.g. phosphate) for many essential biological functions, including catalysis. Detection and identification of those species as components of cloud particles raises the potential for habitability of the clouds. Cloud particles likely contain non-volatile species dissolved in them (e.g., molecules containing Fe, Cl, P etc.) [92]. The composition and concentration of the dissolved species is not known. **Objective 2.1** also includes detection of metals and non-volatile compounds in the Mode 3 cloud particles.

We propose to use the MEMS (Micro-Electro-Mechanical System) Ion-Gas Micro-spectrometer with Optical Signal Detection for Aerosols Analysis (IGOM-A). IGOM-A allows for identification of ions dissolved in aerosols via an atomic emission spectrum. Our main targets for identification are oxidized P species and metals like Fe, Mg, Ca, Mn, Cu, Na, K. In addition, non-volatile inorganic chemicals can also be detected by the LDMS (see Objective 1.3).

**Objective 2.2** aims to determine the amount of water vapor ($H_2O$) in the clouds of Venus. Water, $H_2O$, is of interest because the Venus clouds are an incredibly dry environment, yet variations in vertical $H_2O$ abundance have been previously reported [179]. The range of water vapor abundances reported in the literature is very large (5 ppm to 0.2%), as summarized by [80] which may represent the presence of more clement local conditions. All global models may therefore represent an average of extremely arid ‘desert’ regions and much more humid ‘habitable’ regions. Confirmation of such more humid regions is critical for the assessment of the habitability of the cloud decks.

**For Objective 2.2** we choose the mini-TLS with a channel dedicated towards detectability of water vapor. The IGOM-G, targeting oxygen, may also help to constrain the amount of water vapor. In addition, AFN, which can measure the refractive index of individual particles, can help constrain the water content of droplets by mapping out changes in refractive index in association with TLS measurements. This comparison would allow extending inference of water vapor abundance to the set of cloud conditions measured by the AFN. However, due to the small change in refractive index for high concentration sulfuric acid (80-100% $H_2SO_4$ in dual $H_2SO_4$-$H_2O$ system, see Figure C-2), water quantity inferences above 80% sulfuric acid may be difficult with AFN data alone. However, any refractive index estimates below ~1.42 may be indicative of droplets with unexpectedly high water-content (>30%). In the future we will consider an additional instrument that can measure the cloud particle water content directly.

**Objective 2.3** aims to determine the acidity of the cloud droplets. The prevailing consensus is that the clouds of Venus are made of concentrated sulfuric acid and are extremely acidic, billions of times more acidic than the most acidic environments where life is found on Earth. No known life of any kind can survive in droplets of such extreme acidity, although life with protective shells might. Recently a new model of cloud chemistry challenged the prevailing consensus [3,80,93]. This model suggests that some fraction of droplets could be much less acidic than previously thought. If the droplets are pH 1, their acidity is consistent with the most acidic environments where life is found on Earth. **Objective 2.3** answers how many particles have pH levels habitable for life as we know it, therefore confirming that the clouds of Venus are much more habitable than previously thought.

To achieve **Objective 2.3** we plan to develop a single droplet pH sensor that would measure the pH of single droplets in situ down to pH = 0 or below. Such a sensor does not exist but we have motivated two separate groups to develop one (see Chapter 4 and Appendix E). The first, from [180] is called a Molybdenum Oxide Sensor Array (MoOSA) and categorizes individual cloud particle pH acidity values into one of three acidity categories: pH < 0; pH 0-1; pH > 1 using an optical method which is mechanically and chemically robust and can measure minute sample volumes. MoOSA passively collects particles as they are deposited on the surface by passive exposure, and are not limited to any specific particle size, although due to the collection process, particles below ~1 micron are not expected to be captured with any significant efficiency. The second single droplet pH sensor, the Tartu Observatory pH Sensor (TOPS), is under development by Prof. Pajusalu at Tartu University. TOPS will measure the acidity of single Venusian cloud droplets utilizing the established method of fluorescence spectroscopy.

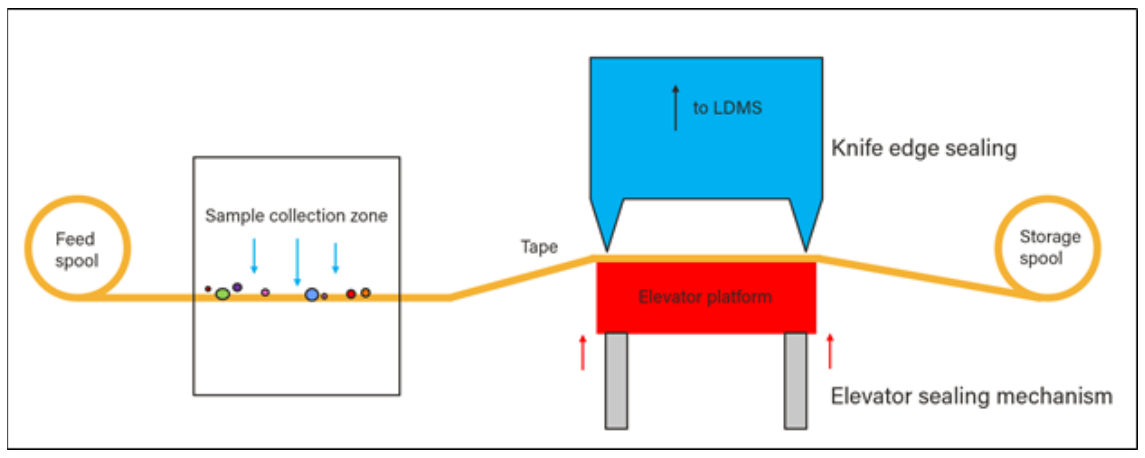

Figure C-3. Schematic of Solid Sample Delivery System (SSDS) designed by Honeybee Robotics. The SSDS is an all-in-one collection, delivery, and sealing mechanism capable of working together with a laser desorption mass spectrometer (LDMS).

**Objective 2.4** focuses on the measurement of temperature-pressure profiles and wind speed in the clouds of Venus. While not directly astrobiological in nature, these auxiliary measurements help establish the basic habitability criteria for the clouds. The combined temperature and pressure sensor from Emerson's Rosemount Inc. is rated to operate up to 344 bars and 316 °C [101] and is thus suited to cover temperature measurements from 80 km to 20 km and pressure measurements from 70 km to 20 km therefore encompassing the entire cloud deck and stagnant haze layer below the clouds.

## C.4 New Concepts Under Development for Venus Atmosphere Sample Collection and Delivery to Instruments

Aerosol collection and distribution on a space mission is a significant challenge within the constraints of a balloon platform. We reviewed existing Earth-based approaches and identified two worth developing for a balloon platform to deliver instrument-relevant sample volumes, both solid and liquid. For the VAIHL mission the goal is delivery of samples to a mass spectrometer, to enable a laser-desorption approach in order to avoid destructive pyrolysis, as well as for delivery of a liquid sample for concentration and microscopy. For these purposes, sedimentation was considered too slow. We focused on inertial methods, which are simple and well understood, although thermal precipitation and electrostatic methods may also be viable, if more complex.

### *C.4.1 Delivery of Material to the Laser Desorption Mass Spectrometer*

As part of this study Honeybee Robotics has developed a custom cloud particle Solid Sample Delivery System (SSDS) for the LDMS. The SSDS (Figure C-3) is an all-in-one collection, delivery and sealing mechanism, and must be directly coupled with the laser desorption mass spectrometer. This sampler uses an inertia-based impact sampler approach to deposit liquid and/or solid particles on a tape. This approach would likely offer adequate capture efficiencies for particles on micron scale or higher. After collection, the sample can be dried, and the tape advanced to position it for analysis with the LDMS. (Future work can address the evaporation efficiency of particles.)

Because the LDMS requires high vacuum, the sample is exposed to a rough vacuum, then a knife edge seal is initiated by a slight upward motion of the “elevator” region, facilitating high vacuum sample exposure to the LDMS.

### *C.4.2 Liquid Collection from an Aerial Platform*

The collection of Venus cloud aerosols from an aerial platform requires a simple and reliable solution to collect and deliver a bulk sample to a microscope. The aerosols may include liquid or solid particles. We evaluated a wide range of aerosol sampling approaches for this application, most based on inertial methods. While we evaluated filter-based approaches, we ultimately decided on non-filter-based approaches to eliminate the complexity of extracting samples from a filter.

Here we describe two different liquid collecting approaches, a Venus Cyclone Sampler (VCS) and a mostly-passive Venus Fog Harp (VFH). We first describe the Venus Cyclone Sampler, which is based on the Research International SASS(R) 2400 Low-Volume Wet Air Sampler (Figure C-4). This system has an extensive field deployment history and utilizes a cyclone-based air sampling approach to achieve a moderately high flow rate (40 L/min) at low power (9W). Air pulled through the system by a blower generates an internal cyclone that promotes inertial deposition of particles against a wetted wall. It provides high concentration factors (40,000/min), and uses a constant-volume fluid loop of 1 ml. The capture efficiency is significant at or above 1 μm, achieving 70% absolute efficiency at 3 μm. Based on Venus atmosphere particle concentrations, we expect to sample ~5M particles/min (upper bound). This instrument is expected to weigh 3.7 kg and occupy ~1L volume (OEM version, Figure C-4, panel D).

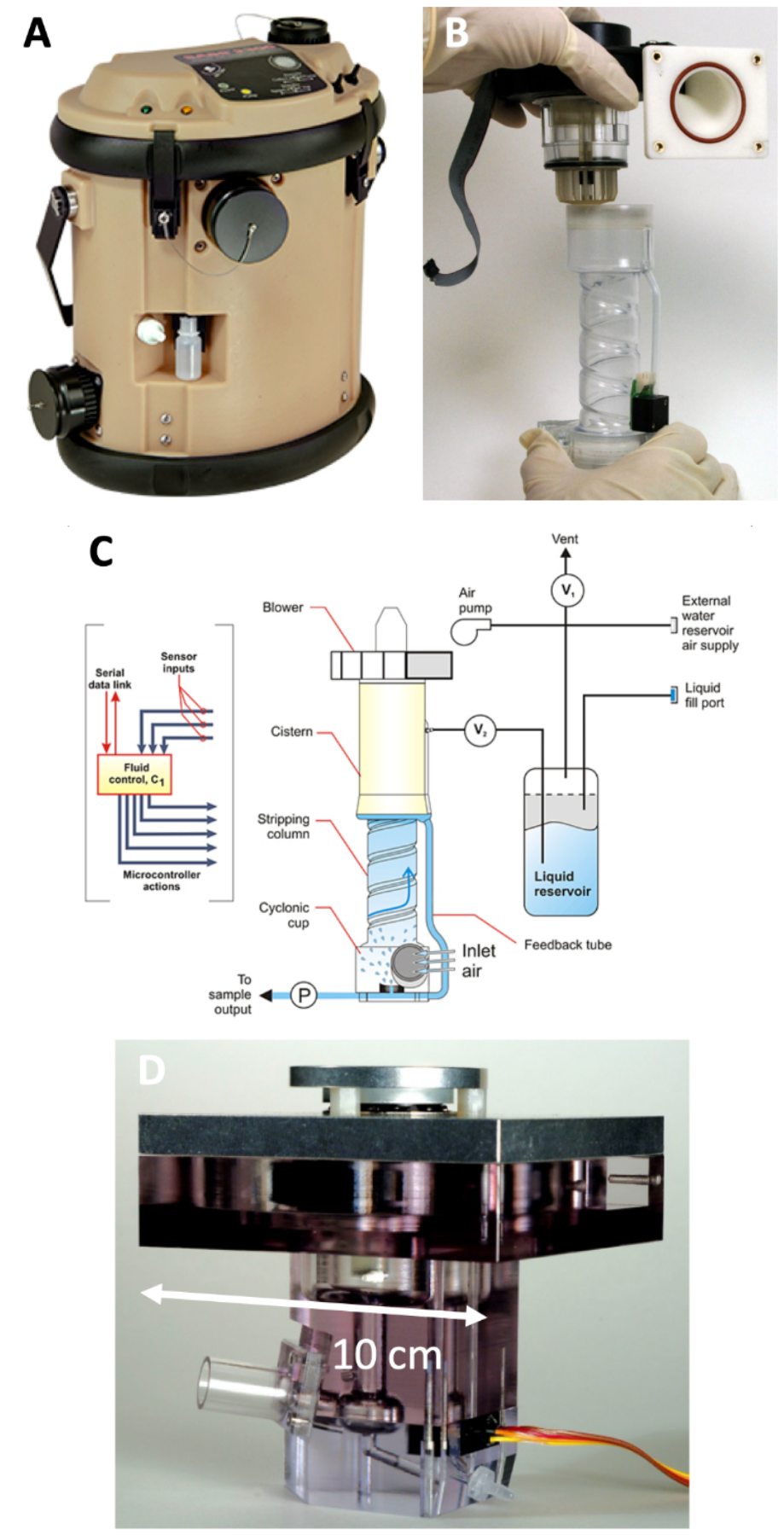

Figure C-4. SASS(R) 2400 Low-Volume Wet Air Sampler. A) Commercial unit. B) Cyclone sub-assembly. C) Functional diagram. D) Module for integration of the sampler into custom equipment. Images: Research International.

The efficiency of this collector is very high, both in terms of particle collection efficiency and in terms of power usage, in comparison to many other aerosol samplers. However, due to extended sampling times to capture possibly rare particles (e.g. life, if present), a lower power solution was also explored, the Venus Fog Harp (VFH).

On Earth, the fog collectors represent a simple and low energy cost method of collecting water, in regions where fresh water is sparse and fog frequently occurs ([181]; Figure C-5). In its standard design a fog collector is a square frame construction placed at some height (~2 meters) above the ground with an outstretched mesh of the active working area 1 x 1 $m^2$ or more, perpendicularly to the wind direction. Wind pushes the fog through the mesh resulting in some fraction of droplets being deposited onto the mesh. When more and more fog droplets deposit, they combine to form larger droplets, run down the mesh material into gutters and eventually into a storage tank. The fog collection efficiency depends on wind speed, mesh structure (shape and material), mesh area and droplet mean volume diameter [182–184]. Relative vertical wind speed in a balloon gondola can reach up to 1 m/s which is enough to gain high collection efficiency [184]. The fog collector's advantage for VLF is its simplicity and lack of moving parts.

Several modifications are needed, however, to adapt a fog collector design for the Venusian environment. Firstly, the mesh material must be resistant to sulfuric acid. Secondly, the mesh structure should be tailored specifically for expected droplet size in Venus clouds to ensure capture of sufficient sample volume. There is a risk that potential life structures and other particles would adhere to the mesh and would not run down to the storage tank. In order to reduce potential particle adherence, a detailed study of the sulfuric acid-resistant mesh is required.

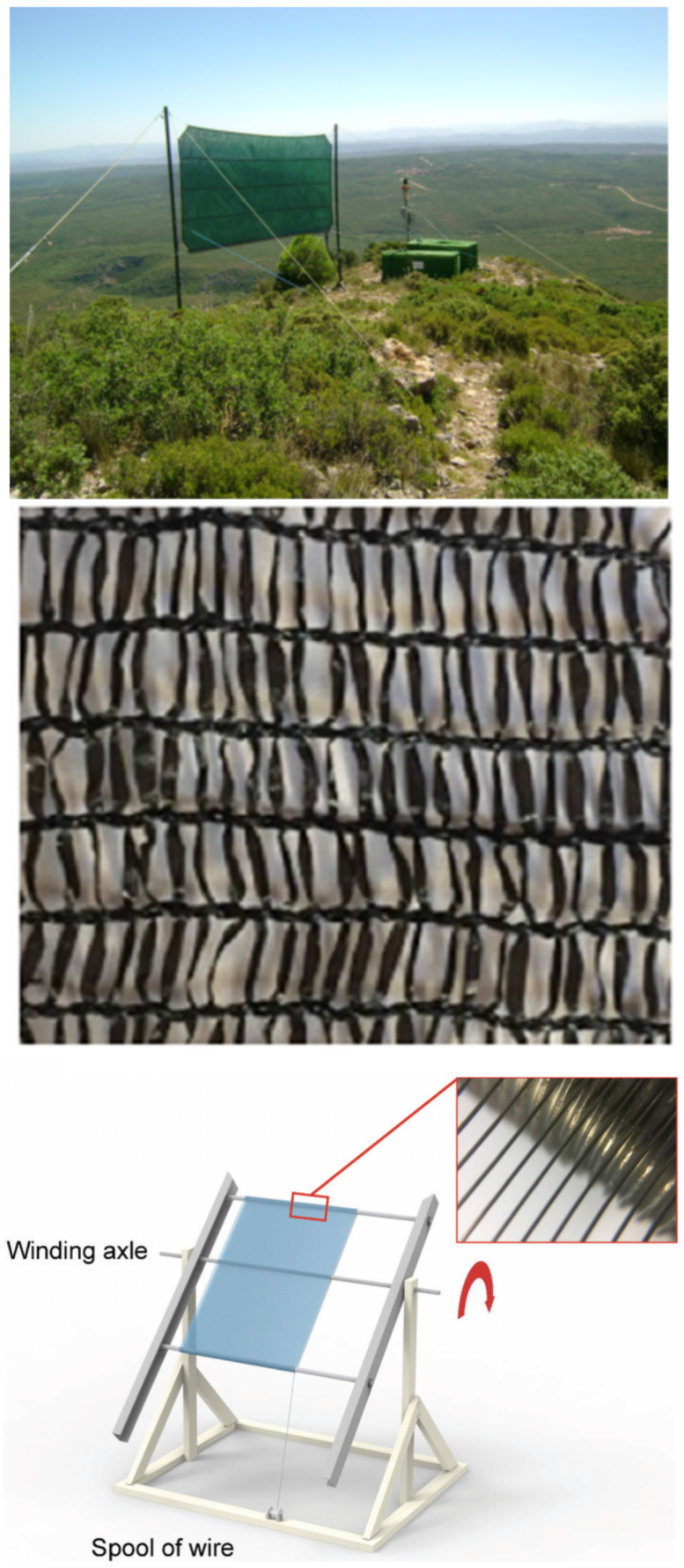

Figure C-5. Inspiration for Venus Fog Harp (VFH). Top: 18 $m^2$ flat-panel fog collector located at Mount Machos (Valencia region, Spain), together with the water tanks [181]. Center: An example of a water droplet collecting mesh (Raschel mesh) [183]. Bottom: A fog harp uses parallel mesh (wires) to enhance collection efficiency [185].

To address these challenges, we propose to utilize a design similar to the fog harp proposed by Shi et al. [185] (Figure C-5, bottom panel), which uses vertical wires to enhance collection efficiency.

The Venus Fog Harp design would utilize a scaled down version of the fog harp, with a nominal size of 20 cm x 20 cm. The use of wires can be adapted to the Venus environment through use of suitable materials, such as stainless steel for collection wires. To further enhance efficiency or reduce the risk of loss of sample if viscous, we can consider possibilities such as vibration or cooling, although this would increase the power requirements.

The potential Venus Fog Harp collection efficiency was assessed as follows: Prior Vega balloon data from Venus [186] demonstrates the potential for winds commonly within 1 m/s, and we assume a more conservative baseline of 0.25 m/s. Next, we determined a nominal collection efficiency using data from the fog harp [185], resulting in an overall collection efficiency of ~3% (aerodynamic efficiency 17%, deposition efficiency 30%, collection from wire to inlet of 60%). Particle size dependence remains to be evaluated although our expectation is that this system will be most effective for larger droplets (>1 μm) because 1) this is the target range for fog collection, and 2) the system is limited by the low efficiency of inertial impaction for small particles.

Using the 3% collection efficiency, assumed flow rate of 0.25 m/s, and area, we can estimate an effective flow rate of 600 L/m, and collection of ~3M particles/min based on estimated Venus conditions. We assume operation will be largely passive but budget 1W average power for vibration, position adjustment (rotation or angular alignment, if incorporated into the design), and periodic operation of an inlet pump to collect sample.

Notably, while the SASS collector would achieve higher particle collection rates than the VFC due to its higher collection efficiency, it would also require substantially more power (perhaps an order of magnitude). Under reasonable assumptions, the VFC would be smaller, less massive, and more power efficient than the SASS collector. For these reasons, we baseline the VFC as our liquid collector (LC) of choice, even while it is low TRL. One limitation is that the sample volume, and thus, density of any identified particles in the atmosphere, would have to be estimated based on integrated wind conditions as measured by the Weather Instruments Suite (WIS). In contrast, the cyclone sampler offers a relatively mature backup option, one that would provide a defined volume sample not dependent upon wind fluctuations.

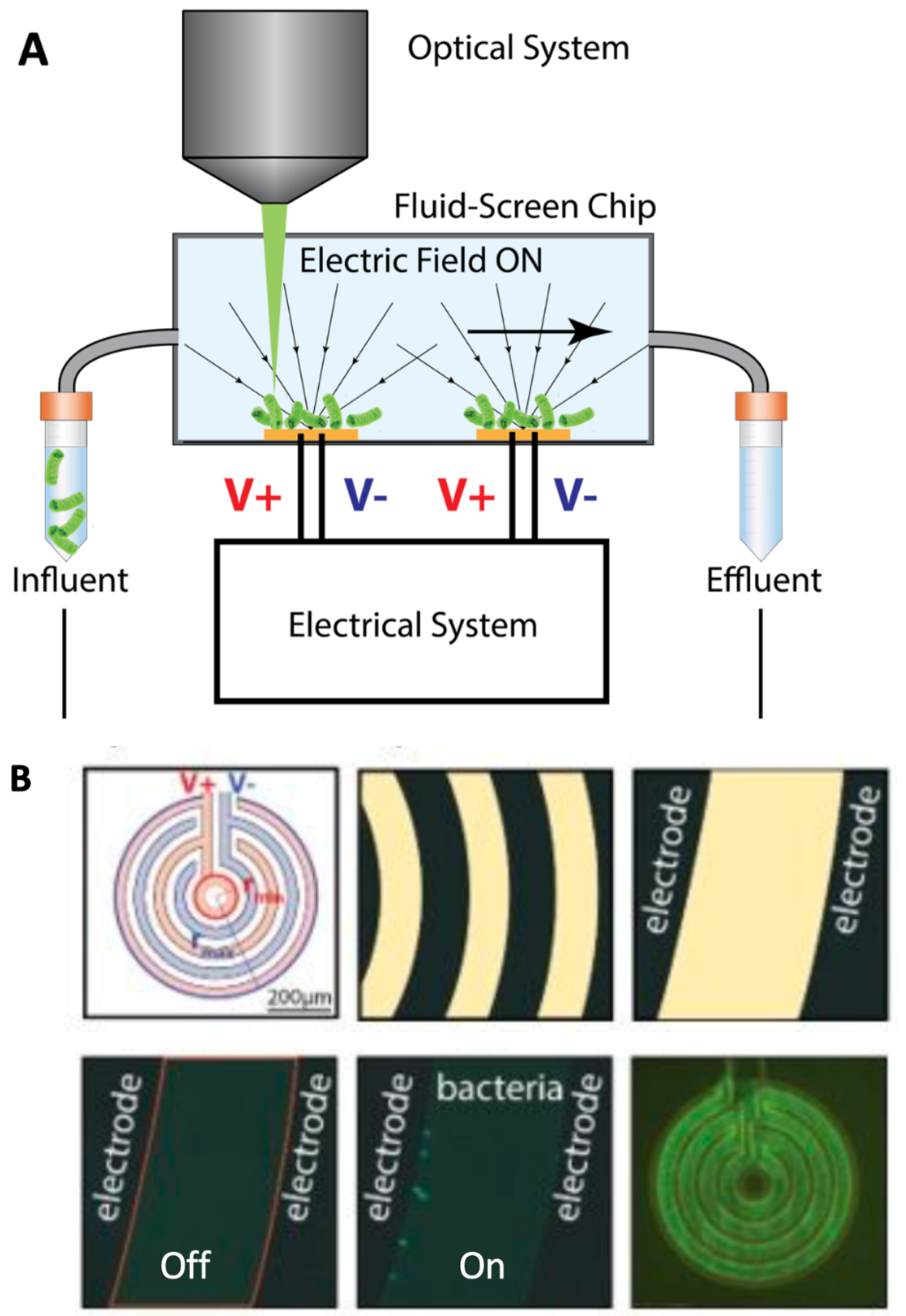

Figure C-6. Fluid Screen Concentrator (FSC). A) Schematic. B) Dielectrophoresis based concentration using electrodes, demonstrating capture of bacteria. Figure modified from [178].

### C.4.3 Liquid Sample Concentration and Fluorescence Microscopy

Once a liquid sample is acquired, it would be passed to the Fluid Screen Concentrator (FSC, Figure C-6) and then imaged using the Autofluorescence Microscope (AFM) to seek morphological indicators of life. The liquid sample from the LC is fed to the FSC through custom microfluidics channels. FSC is a novel dielectrophoresis microbial capture and separation method developed by a company called Fluid-Screen [178].

Dielectrophoresis has shown great promise for decades [187–189]and Fluid-Screen [178] made it work and commercially available. Particles suspended in the liquid sample are captured and concentrated on the basis of surface charge interaction with electrodes. Such charge interaction leads to very fast, efficient, reliable and repeatable capture and separation of microbial cells from other particles (particle size is not a key factor, thus FSC can concentrate across size).

The FSC method is universal and can capture and separate different species of bacteria and fungi to viruses, from various sample matrices. Method verification experiments demonstrated that the FSC captures effectively 100% of bacteria in the test samples, a capture efficiency much higher than previously reported for similar technology [178]. Further testing is required to establish suitable protocols for particle capture in environments mimicking Venusian cloud conditions (specifically concentrated sulfuric acid or solutions with high salt concentrations) that might modify the standard FSC protocols. While in its current form the FSC is not yet space ready, the necessary miniaturization of the system and adaptations are feasible.

FSC will concentrate and hold the particles at the focal plane of the microscope. We baseline an autofluorescing microscope system (AFM) derived from SHERLOC (currently at Mars on the Perserverance Rover and thus has high technology readiness) [190,191]. SHERLOC includes an UV-autofluorescence Raman spectroscopy system that operates at 30 μm resolution in a wide-field scanning capacity. However, in an alternative optical configuration, deep UV fluorescence microscopy can achieve resolutions in the sub-μm resolution, and would detect native UV fluorescence of any organics (< ppb of aromatic organics) within captured cloud particles [190]. The detection of organics with the FSC-AFM system does not require any externally added reagents or fluorescent dyes and solely exploits UV (<250 nm) excitation and 320 nm emission.

# Appendix D: VAIHL Mission Concept Architecture and Design

**Abstract:** The Venus Airborne Investigation of Habitability and Life (VAIHL) Mission explores the lower cloud deck of Venus to search for signs of life or life itself. This mission is complimentary to but more sophisticated and complex than the VLF Habitability Mission described in Chapters 4 and 5. The VAIHL mission will search for the evidence of life in the Venusian clouds, measure habitability indicators, characterize aspects of Venusian cloud droplets and aerosols that might be associated with life, and directly search for life. The mission consists of a variable altitude balloon aerial platform that floats between the altitudes of 45 and 52 km, with the ability to stop at altitudes of interest and perform science operations. The nominal mission life is 30 days. The aerial platform transfers data to Earth via an orbiting relay spacecraft and a direct-to-Earth transmission.

## D.1 Mission Concept Overview

We design a mission to explore the cloud layers of Venus between 45 and 52 km, the altitudes of astrobiological interest. The mission architecture consists of a variable-altitude balloon (VAB) that carries a suite of science instruments to search for signs of life or life itself (see Appendix C on VAIHL Mission science objectives and instruments). The altitude range that the VAB covers extends from 45 km to 52 km, which consists of three regions: the lower haze (below 47.5 km); the lower cloud region (47.5 - 50.5 km); and the lower part of the middle cloud region (50.5 – 56.5 km). The system is designed to function for 30 days, circumnavigating Venus approximately 6 times over the mission duration. This architecture enables spatial, temporal and day-night profiling of samples.

For design purposes, the baseline mission concept launch date is July 30, 2026, on a Falcon 9. The launch mass of the stacked configuration is 1530 kg. Figure D-1 shows the flight system elements which include: the launch vehicle with the stacked configuration inside the payload fairing; the orbiter/cruise vehicle; the aeroshell; and the aerial platform with the flight control system and science gondola. To further mature the concept, a more thorough and high-fidelity analysis is required.

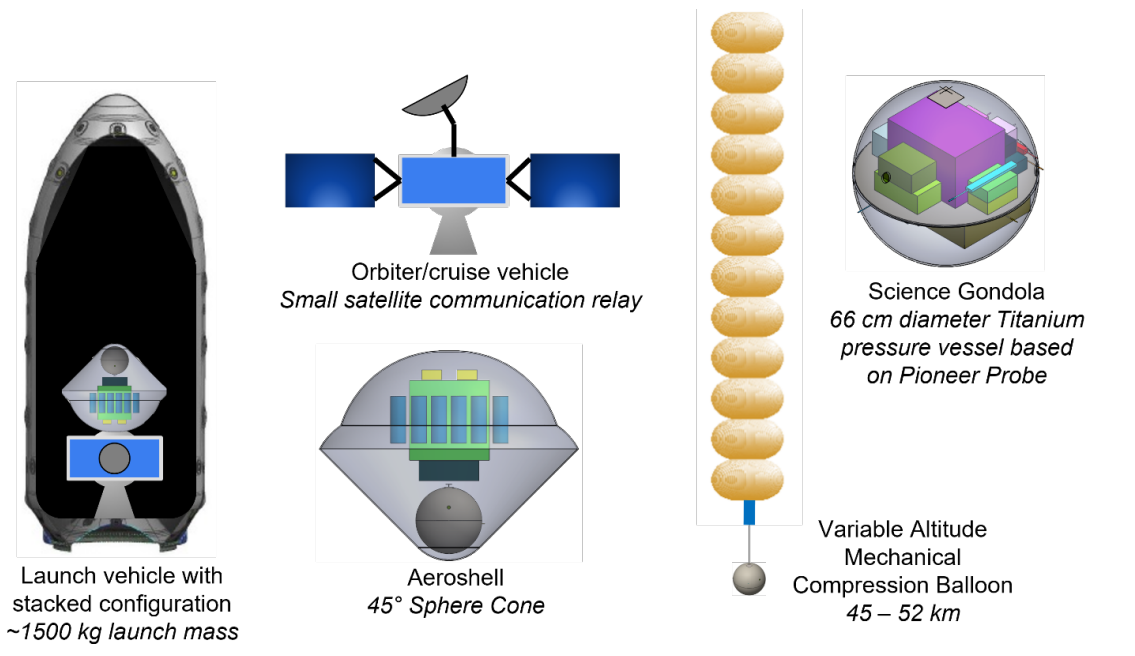

Figure D-1. Flight system elements for the VAIHL Mission.

## D.2 Aerial Platform Instruments

The aerial platform payload consists of sample collection and analysis instruments described in Appendix C.

## D.3 Flight System Description

This section describes the various flight system elements and the mission design. The key flight system elements are orbiter, entry system, and aerial platform. Table D-1 summarizes the mass of each of the flight system elements.

### *D.3.1 Orbiter*

The primary function of the orbiter is of communication relay between the aerial platform and Earth ground station. The orbiter design is similar to that for the Habitability Mission, described in Chapter 5.

| System | CBE (kg) | Contin-gency (%) | MEV (kg) |
|---|---|---|---|
| Orbiter/cruise vehicle | 300 | 50% | 450 |
| Entry System | 420 | 50% | 630 |
| Aerial Platform System | 300 | 50% | 450 |
| **Total Stacked Mass** | **1020** | **50%** | **1530** |

Table D-1. Summary of mass of flight system elements.

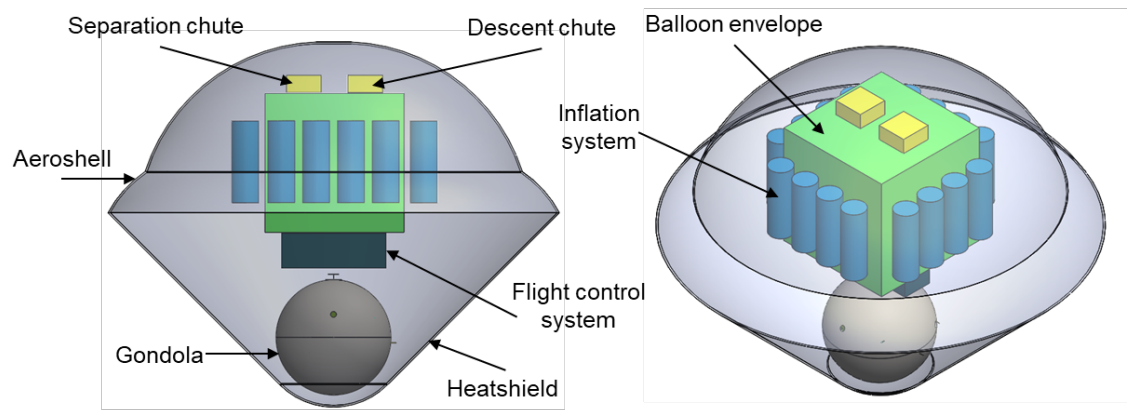

Figure D-2. Entry probe design.

### *D.3.2 Entry System*

**Aeroshell Design.** The VAIHL atmospheric entry probe geometry is based on the Galileo and Pioneer Venus entry probes [135]. The probe has a maximum diameter of 2.60 m and has a traditional 45-degree sphere-cone geometry. Major components of the small probe are shown in Figure D-2 which include the heatshield, aeroshell structure, balloon envelope, flight control system, and gondola which holds the science instruments.

**Probe Mechanical Design.** The descent probe houses the VAB envelope, sixteen inflation tanks, two parachutes, the flight control system, and the gondola. There is no enclosure for the descent module. Some structural elements, however, will be required to hold the different elements together inside the entry vehicle.

**Parachute Design.** The parachute design and inflation CONOPS are similar to that of the Habitability mission (see Chapter 5). The separation chute is a 6-m diameter conical ring sail parachute, and the descent chute is 3.0-meter in diameter. The descent probe chute is sized such that it allows the probe to spend at least 6 minutes before it reaches an altitude of 52 km.

| Subsystem | CBE (kg) | + 30% MGA | + 15% Margin (kg) |
|---|---|---|---|
| Forebody TPS (HEEET) | 150 | 195 | 224 |
| Forebody Structure | 70 | 91 | 105 |
| Backshell TPS (PICA) | 30 | 39 | 45 |
| Backshell Structure | 60 | 78 | 90 |
| Separation Parachute and Mortar | 30 | 39 | 45 |
| Separation System | 20 | 26 | 30 |
| **Aeroshell Total** | 360 | 468 | 538 |
| | | | |
| Aerial Platform Structure | 40 | 52 | 60 |
| Descent Parachute | 10 | 13 | 15 |
| Aerial Platform + Miniprobes | 300 | 390 | 449 |
| Engineering Systems | 10 | 13 | 15 |
| **Descent Module Total** | 360 | 468 | 538 |
| **Total Entry System Mass** | 720 | 936 | 1076 |

Table D-2. Summary of entry system mass breakdown.

**Entry System Mass Breakdown.** Table D-2 shows the mass breakdown for the aeroshell and the descent module. The design methodology is similar to that for the habitability mission probe.

### *D.3.3 Aerial Platform System*

The balloon for the VAIHL Mission is a variable-altitude mechanical compression balloon designed by Thin Red Line Aerospace. The balloon system is comprised of 12 helium super-pressure balloons arranged in a column. A cable passes through the column longitudinally and attaches to a stepper motor at its base. By lengthening and shortening the cable, the flight control system increases and decreases the volume (and hence the buoyancy) of the envelope to vary its altitude at a rate of up to 8 m/s. The diameter of the balloon column is 3.6 m and its height is 25.8 m. The mass of the entire balloon system (flight control system + balloon envelope) is 111 kg. The volume of the envelope is 200 $m^3$ and its maximum expected operating pressure is 9.8 kPa. Zylon (R) PBO fiber pressure restraint tendons line the balloon envelope, which is coated in a proprietary 5-layer laminate including metalized films operable to 135 °C. The envelope is also cloaked in an acid-resistant Polytetrafluoroethylene (PTFE)-class membrane and features planar fabrication to reduce packaging volume as well as the risk of trauma prior to deployment. The system is designed to accommodate a 150 kg payload within the 45-52 km altitude band for about 30 days.

**The mechanical-compression variable-altitude balloon platform** is a robust, continuously super-pressurized, rapid altitude control aerobot, an example of which is shown in Figure D-3. Ascent and descent of the VAB are initiated and maintained by modulation of the balloon's lifting gas density through motorized mechanical compression. As schematically shown in Figure D-3, as the balloon ascends and descends in accordance with its programmed trajectory, the balloon's accordion-like segmented envelope leaves the lifting gas volume free to adapt to the variation in atmospheric density associated with the varying altitude. The number of balloon segments can be adapted to the desired altitude range: a larger number of segments presents a greater ratio of distended to compressed volume, which in turn provides access to a greater range of trajectory altitudes. VAB technology is currently being adapted for lower altitude, elevated temperature Venus application.

**The ultra-high-performance vessel (UHPV).** As seen in Figure D-3, the VAB comprises a stack of oblate spheroidal segments. Each of these segments is a so-called UHPV. While conspicuously presenting the "pumpkin" shape that is well-known in the balloon world, UHPV departs from the conventional pumpkin design by eliminating reliance on gore-based construction. UHPV is a highly unique, proprietary inflatable pressure vessel architecture [139] that has been the focus of numerous NASA contracts ranging in topic from space habitats to cryogenic propellant tanks. It is widely considered to be the most structurally predictable, lowest specific mass containment architecture (collapsible, metallic, composite, or otherwise), furthermore featuring unlimited scalability and structural determinacy. This robust, proven technology forms the building block for a Venus aerobot.

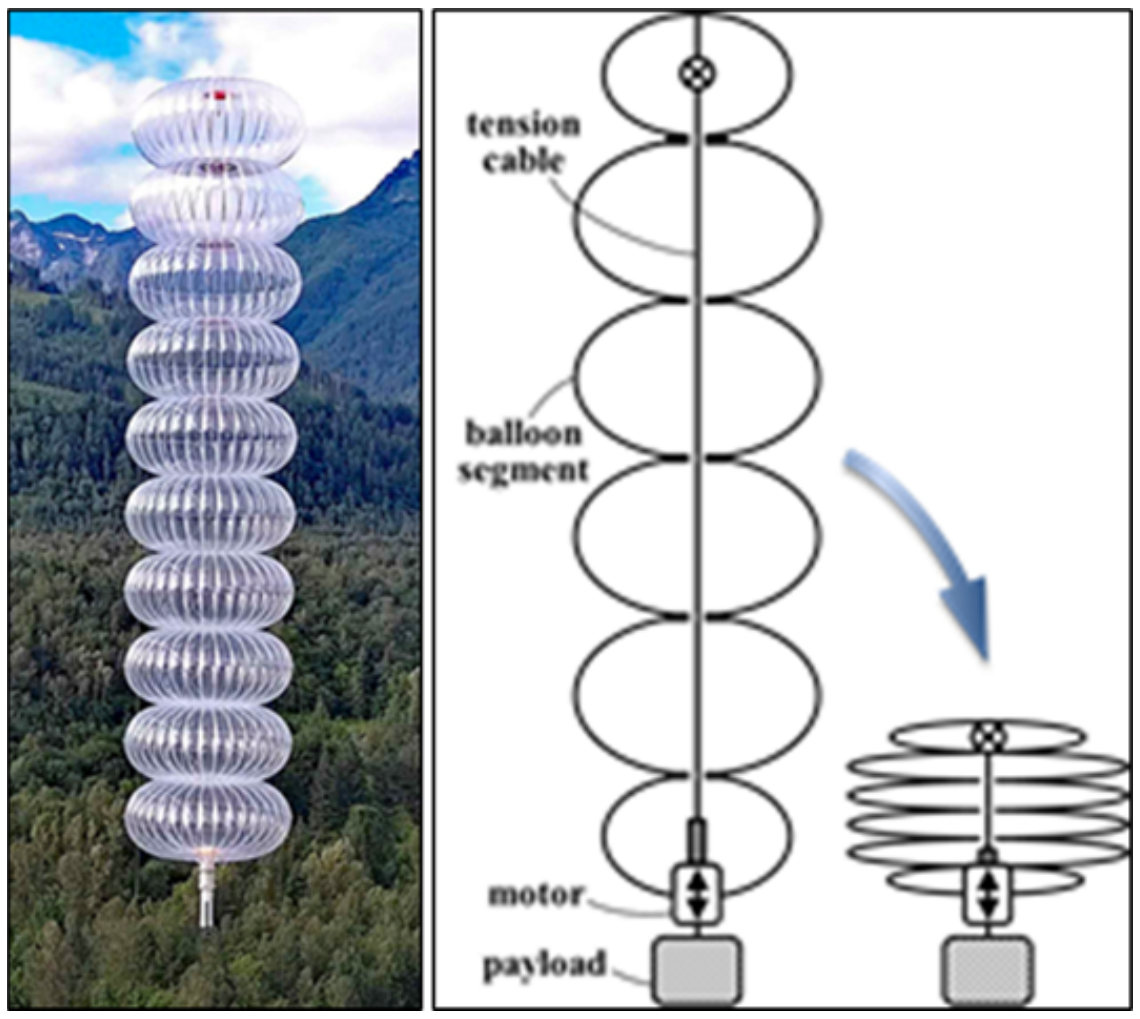

Figure D-3. Mechanical compression variable altitude balloon.

**Gondola design.** The gondola is designed as a spherical pressure vessel based on Pioneer Venus Large Probe heritage [142] (see Figure D-4). The structure is a 65 cm diameter sphere made with a 6 mm thick shell of titanium. The instruments are placed on a beryllium shelf inside the vessel, which is also 6 mm thick. A breakdown of the gondola system mass is provided in Table D-3 [this table is missing].

*Thermal Control.* The gondola pressure vessel is lined with a 1 mm thick layer of Kapton on the inside. The beryllium shelf and instruments deck are separated by a 6 mm thick layer of a phase change material to insulate the instruments further. The insulator is sodium silicate phase change material. The balloon cycles between 45 and 52 km, giving it time to cool down periodically. For this study, the passive thermal control design is considered sufficient for such operations.

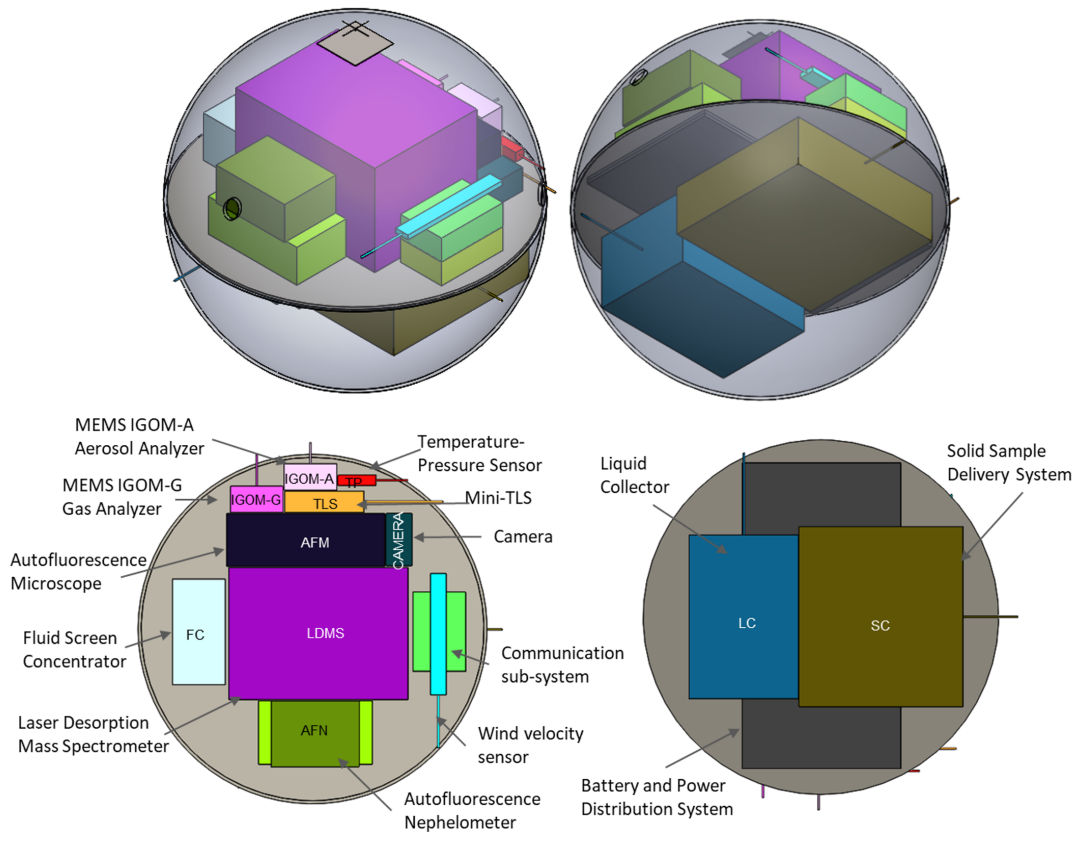

Figure D-4. (Top) Concept for instrument accommodation inside the gondola pressure vessel. (Bottom) Top and bottom deck of instruments and subsystems inside the gondola.

| Subsystem | Mass | | | Avg. Power | | |
|---|---|---|---|---|---|---|
| | CBE (kg) | % Cont. | MEV (kg) | CBE (W) | % Cont. | MEV (W) |
| Structure | 37.4 | 50% | 56.2 | N/A | N/A | N/A |
| Science instruments | 30.0 | 50% | 45.0 | 10 | 30% | 13 |
| Battery+PDS | 30.0 | 50% | 45.0 | 1 | 30% | 1.3 |
| Communication | 3.7 | 50% | 5.6 | 2.5 | 30% | 3.25 |
| Thermal | 7.5 | 50% | 11.3 | 0 | 30% | 0 |
| C&DH | 3.1 | 50% | 4.7 | 5 | 30% | 6.5 |
| **Total Gondola Subsystem** | **111.8** | **50%** | **167.7** | **18.5** | **30%** | **24.05** |

Table D-3. Summary of gondola system mass breakdown.

*Power*. The VAIHL Mission uses batteries throughout the mission lifetime. We use a LiCFx battery pack, which has 450 Wh/kg of specific energy. The total weight of required batteries is 23 kg. Including a power distribution system and contingencies, total power subsystem weight is about 30 kg. The sizing is conservative in terms of the total battery mass. A system with solar panels and rechargeable batteries weighs less than 30 kg and can extend the mission duration beyond 30 days. The addition of solar panels increases complexity due to the deployment mechanism and atmospheric disturbances.

*Communication*. The antenna is a crossed dipole designed for 2.4 GHz S-band transmission. The antenna is mounted on top of the gondola. The balloon is opaque to radio transmission and thus 10 degrees elevation is blocked for communication. The dipole antenna has a wide beamwidth, which allows for transmission at elevations above 30 degrees.

**Gondola instrument accommodation**. The instruments are mounted on both sides of the shelf. Sealed penetrations, windows and inlets are designed on the pressure vessel for the various instruments. There are notionally five inlets that intake the atmosphere samples, each for the: Liquid Collector, Solid Sample Handling System, Mini-TLS, MEMS Gas Analyzer (IGOM-G), and MEMS Aerosol Analyzer (IGOM-A). There are two windows looking out into the atmosphere, each for the Camera and Autofluorescence Nephelometer (AFN). There are two protrusions from the gondola that do not provide visual access or intake the sample. The protrusions are for the Weather Suite, including one for the temperature and pressure sensor and one for the anemometer. Figure D-4 shows the various instruments and subsystems inside the gondola.

## D.4 VAIHL Mission Operations Overview

Mission operations is split into four phases: 1) Launch and Cruise to Venus, 2) Venus Approach, 3) Balloon Entry and Initial Float Operations and Orbiter Capture, and 4) Nominal Balloon Operations.

### *D.4.1 Launch and Cruise to Venus*

The VAIHL baseline mission launches on July 30, 2026, on a Falcon 9. The launch payload is the entry system attached to the orbiter which also functions as a carrier spacecraft for the cruise phase. The spacecraft will arrive at the vicinity of Venus on November 29, 2026 (shown in Figure 5-4). The Venus approach CONOPS are same as that for the Habitability Mission, described in Chapter 5.

| Time (s) | Altitude (km) | Speed, (km/s) | Event |
|---|---|---|---|
| E + 0 | 180 | 11.33 | Entry interface |
| E + 56 | 88.5 | 9.59 | Peak heating, heat rate = 1336 W/cm$^2$ |
| E + 60 | 84.7 | 7.15 | Peak deceleration, 69 g |
| E + 100 | 73.8 | 0.382 | Separation chute deployed |
| E + 105 | 73.2 | 0.177 | Front heatshield separation |
| E + 110 | 72.8 | 0.125 | Backshell separation, aerial platform free fall |
| E + 115 | 72.3 | 0.088 | Descent chute deployed |
| E + 120 | 71.8 | 0.109 | Balloon inflation begins |
| | | | |
| **Time (min)** | **Altitude (km)** | **Speed (km/s)** | **Event** |
| **E + 8** | 51.8 | 0.016 | Balloon inflation complete, chute cut-off |
| **E + 10** | 52 | 0 | Jettison inflation system, balloon moves to desired altitude |

Table D-4. EDL CONOPS. Key events during entry and descent sequence followed by balloon inflation and deployment.

| Event | ΔV (m/s) | Resulting Mass (kg) |
|---|---|---|
| Launch | | 2520 |
| Entry system release | | 1440 |
| Divert maneuver | 200 | 1345 |
| Capture | 1550 | 794 |
| Cleanup (5% of capture) | 78 | 774 |

Table D-5. Summary of propellant requirements, ΔV and mass.

### D.4.2 Entry, Descent, and Inflation

The probe enters the atmosphere at speed of 11.33 km/s at an entry flight-path angle (EFPA) of -10 degrees. Table D-4 shows the entry and descent timeline for the VAIHL mission probe. The entry, descent and inflation CONOPS are similar to that of the Vega Balloon mission [192].

Following peak heating and peak deceleration, the separation chute is deployed at 73.8 km. The total estimated stagnation-point heat load is 18 kJ/cm$^2$. Five seconds later, the front heat shield falls off and another five seconds later the aerial platform payload falls off the back heatshield. After five seconds of free fall, the aerial platform descent chute is inflated, and the inflation system begins filling the aerial platform at an altitude of 71.8 km. At E+8 minutes, the aerial platform inflation is complete at 51.8 km and the descent chute is released. At E+10 minutes, the inflation system is jettisoned, and the balloon's flight control system is engaged, and the balloon adjusts itself to 52 km to begin variable altitude science operations.

Figure D-5 shows the altitude vs. time, deceleration, and heating profiles during the entry and descent phase of the atmospheric probe. The entire entry, descent, and balloon deployment CONOPS is summarized in Figure D-6.

### D.4.3 Orbiter, Propellant Requirements, ΔV and Mass Summary

The propellant requirements, ΔV, and mass summary are shown in Table D-5. If the total payload capacity is at its maximum, the resulting allowable mass for the orbiter is 774 kg, which is more than the allocated mass of 300 kg for the spacecraft bus. This mass, along with allowances in the estimated mass, allows for a wide margin.

### D.4.4 Nominal Balloon Operations

After entry and initial float operations, the balloon will travel through the 45-52 km altitude band collecting data from its instruments. The VAB can ascend and descend at the maximum rate of 8 m/s and can hold any altitude in between. The CONOPS for altitude variations depend on multiple factors: thermal design of balloon and gondola; time required for each instrument to take measurements at each altitude; communication with the orbiter; and desired number of complete cycles between the minimum and maximum altitude.

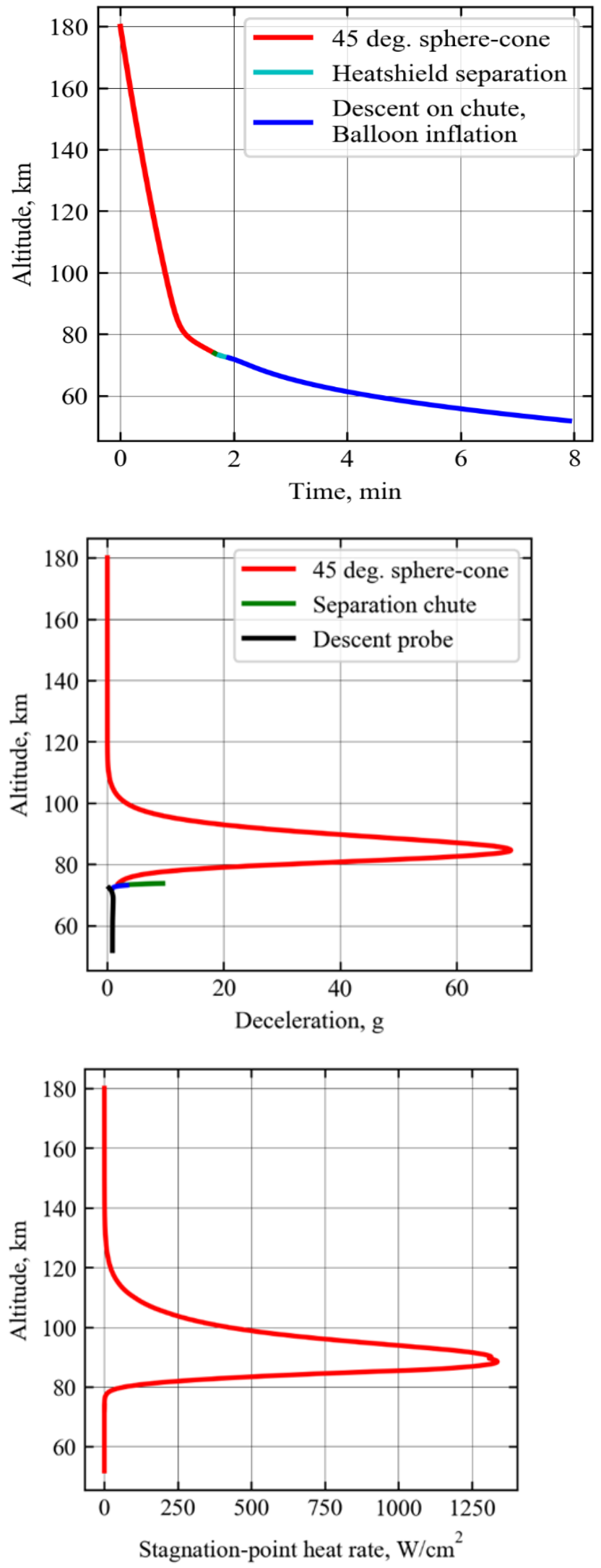

Figure D-5: (Top) Altitude vs time. (Middle) Deceleration profile. Peak deceleration = 69 g. (Bottom) Aerodynamic heating profile. Peak

The communication architecture is similar to that for the Habitability Mission (See Chapter 5). If an orbiter is not used, the data rate will reduce by a factor of at least 10. The balloon can, of course, keep transmitting after the end of the science mission life, thus increasing the total data volume transmitted. The balloon can survive for longer than 30 days.

## D.5. Technology Development

The TRL of the science instruments and aerial platform systems is summarized in Table D-6. The orbiter and aeroshell have mission heritage from ongoing and past NASA missions that includes MarCo and Pioneer Venus probes respectively [134,142]. The VAB has been successfully demonstrated in Earth's atmosphere [140]. Further development is required to make the mission components tolerant to high temperatures at lower altitudes. The inflation mechanism is low TRL due to the conditions of deployment, which include a rapid inflation rate, decent at 10s of m/s, sulfuric acid clouds, and autonomous operations. The aerial platform system also needs to withstand peak deceleration loads of up to 300 g. The gondola pressure vessel is high TRL as it is based on the Pioneer Venus large probe design. Further development is required to make it light weight using latest materials technology

# Appendix E: Venus Habitability Mission Instrument Supplement

**Abstract:** This Appendix to Chapter 4 provides an extended instrument description for the Venus Life Finder Habitability Mission. We focus on the pH sensors, which have not previously been developed for space, and are not common for extremely low acidities. In addition we describe a miniaturized oxygen sensor suitable for a mini probe instrument.

## E.1 Single Particle pH Meters

### *E.1.1 Tartu Observatory pH Sensor (TOPS)*

The Tartu Observatory pH Sensor (TOPS) will measure the acidity of single Venus cloud particles utilizing the established method of fluorescence spectroscopy, which is widely used for pH measurements. The general mechanism for the single particle acidity sensor is using a dye-sensitized sensor plate and illuminating it with various wavelengths of light. After cloud particles hit the sensor plate, different spots will fluoresce with different intensities at a given excitation wavelength, allowing for measurement of single particle pH or acidity. The method is based on the composition of the fluorophore, whose emission response depends on the association or dissociation of protons [193–196]. Fluorescein is a candidate fluorophore for the TOPS sensor due its high fluorescence intensity [102] and stability over very wide range of sulfuric acid ($H_2SO_4$) concentrations, from diluted to highly concentrated solutions (e.g. over 10 M sulfuric acid) [103,104]. Under strongly acidic conditions fluorescein is in its cationic form. Since emission responses of fluorescein forms differ from each other in intensity and wavelength, fluorescein is a good candidate as a fluorescence dye for the Venusian single-particle pH meter [102,104].

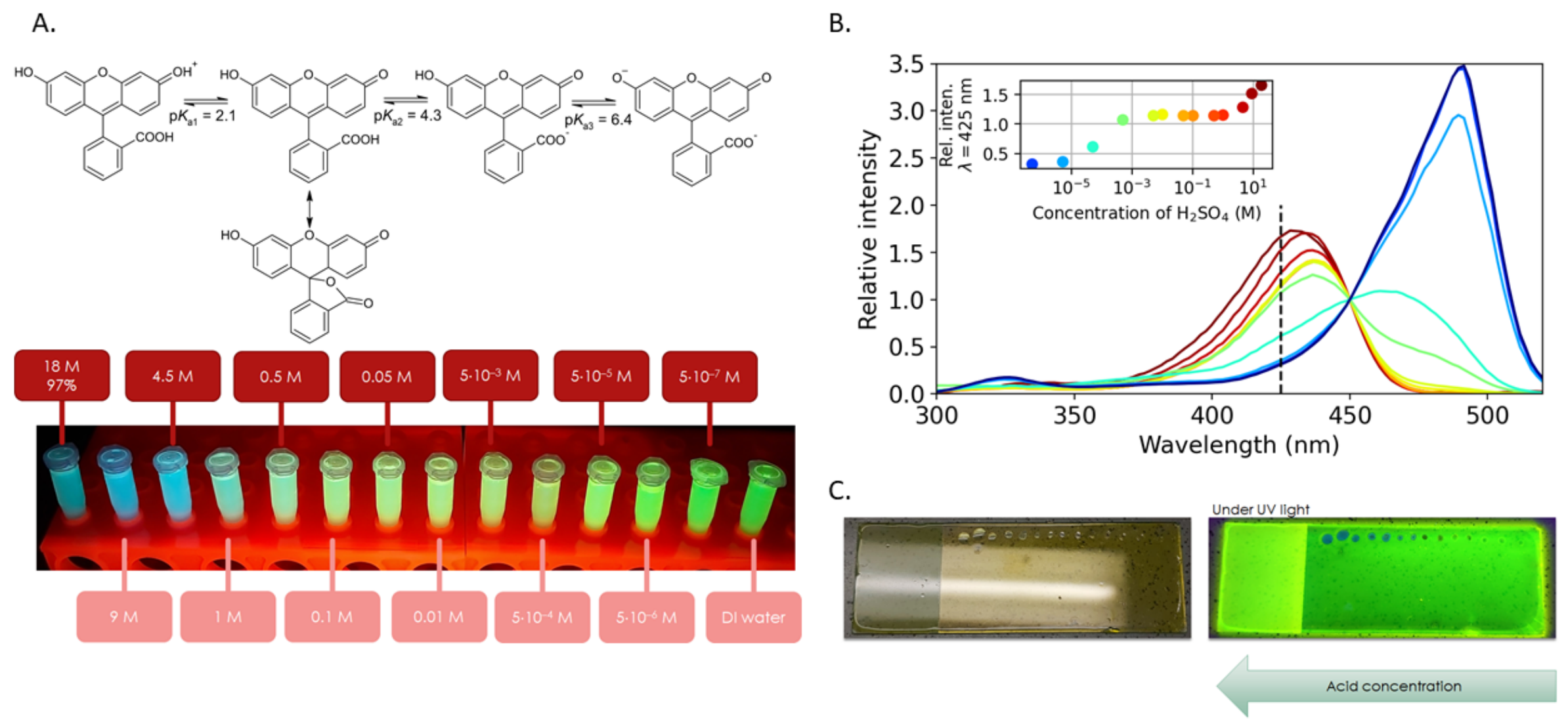

Figure E-1. A) Fluorescein in $H_2SO_4$ solutions of varied concentrations under UV light. B) Emission spectra of fluorescein excited in different acid concentrations. Spectra are normalized to the wavelength of 450 nm. Inset: Relative intensities at 425 nm show a clear dependence on $H_2SO_4$ concentration. Note that while this initial experiment is done for fluorescein, further experiments will find the best fluorescence dye (or the mixture of dyes) for this specific application. One potential group of alternatives are rhodamine dyes [215]. C) Glass plate (1×4 cm) covered with polyvinyl alcohol doped with fluorescein, covered droplets of sulfuric acid (18 M – 0 M). Figure repeated from Figure 4-2.

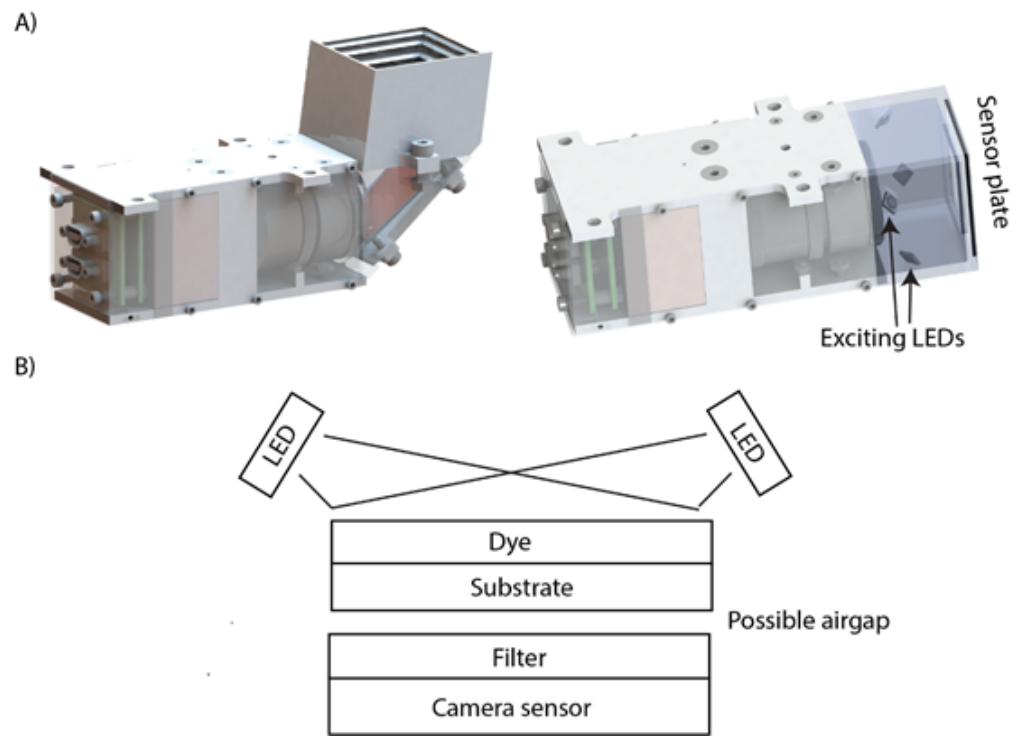

Figure E-2. Two possible paths for development of the TOPS acidity sensor. A) The high-TRL approach. Left: OPIC instrument from the Comet Interceptor mission. Right: OPIC instrument with the periscope replaced by a module with the pH sensor glass mounted and LEDs placed for excitation. B) General concept for a simplified sensor for the low-TRL approach that is expected to reduce the instrument mass and volume by a factor of around 3.

As the sensor would likely be characterizing acidity by measuring the excitation response of the sensor plate when illuminated with light sources with various wavelengths, we measured the excitation spectra of the solutions above (Figure E-1A) and looked for two excitation wavelengths, which could be used for measuring relative fluorescence and to produce a fault-tolerant signal.

The excitation spectrum shows (Figure E-1B) that the acidification of the solution leads to a shift of the peak towards shorter wavelengths. The inset on Figure E-1B shows the relative emission intensity between excitations at 425 and 450 nm, with monotonic behavior based on sulfuric acid concentration.

To develop a reliable pH sensor, the fluorophore must be adhered to a surface, e.g. polyvinyl alcohol (PVA) [197]. A glass plate was covered with PVA doped with fluorescein, and after the surface had dried, small droplets of sulfuric acid with different concentrations (as described previously, Figure E-1A) were put on the plate. The concept was proved to work as the prototype of a pH sensor (Figure E-1C).

We consider two possible paths for development of TOPS acidity sensor. The "High-technology readoness (TRL) design" (Figure E-2A) and the "Low-TRL design" (Figure E-2B).

The high-TRL approach relies on modifying an automatic cometary observation camera OPIC (Optical Periscopic Imager for Comets), currently being developed at Tartu Observatory for the Comet Interceptor ESA F-class mission (Figure E-2). Parts of this system are already TRL 8 and OPIC itself is in I-SRR by ESA. LED and plate module TRL is 3.

The low-TRL approach will enable the instrument mass and volume reduction by up to a factor of around 3. The approach includes one or both of the following: 1) omitting the optics system and placing the sensor plate against the sensor, this approach reduces TRL of the optics solution to TRL 2. And, 2) development of a fully custom camera solution for a TOPS acidity sensor. Similar camera solutions, with high TRL levels, were previously developed at Tartu Observatory for SAMPLR experiment (by Maxar Technologies, part of Commercial Lunar Payload Services part of the Artemis program), European Student Earth Orbiter and ESTCube-1.

### *E.1.2 Molybdenum Oxide Sensor Array pH Sensor (MoOSA)*

The Molybdenum Oxide Sensor Array (MoOSA) can categorize individual cloud particle pH acidity values using an optical method which is mechanically and chemically robust and can measure minute sample volumes.

MoOSA will categorize individual droplets that are deposited on its surface into one of three acidity categories: pH < 0; pH 0-1; pH > 1. Deposition will be brought about by passively exposing one surface of MoOSA to the Venus atmosphere in the altitude range 48-65 km. MoOSA will return counts in each pH category, corresponding to individual depositions on an array of plasmonic $MoO_3$ coated Micro-Ring Resonators (MRRs). Each of these MRRs will be streamed with near-infrared light from a micro-laser (wavelength 1530-1625 nm), through a 'tree' layout printed into the ceramic sensor surface, Figure E-3. Approximately 100-400 MRRs with diameters of between 1-8 microns will be embedded into the surface. Detections of phase shifts in each of the individual MRRs will be made by a corresponding photodiode array. A set of non-exposed MRRs will also provide confirmation of pH reference datum shifts caused by variation of local ambient pressure and temperature, Figure E-4.

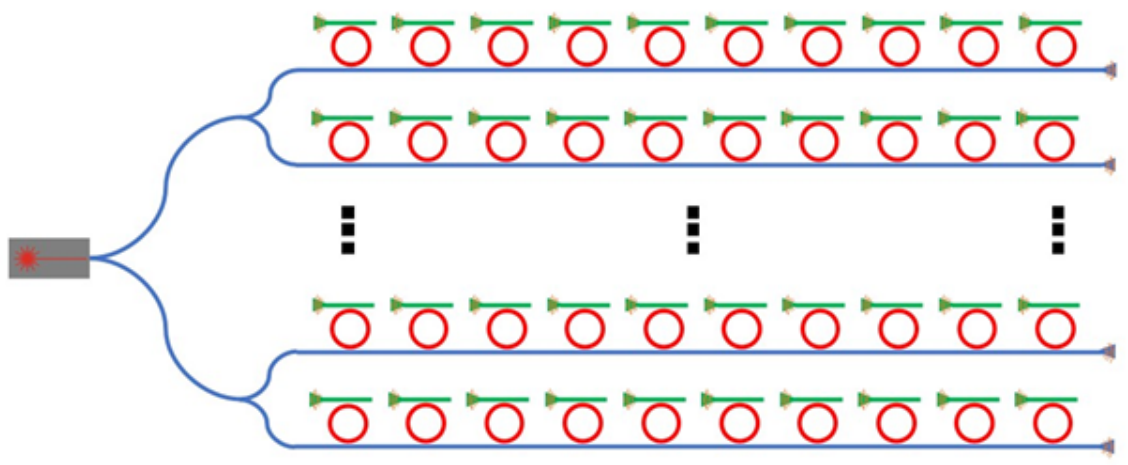

Figure E-3. Schematic of MoOSA showing optical waveguide tree layout (blue) with feeding array of MRRs (red). The orange parts are grating coupler to couple light into the integrated detectors.

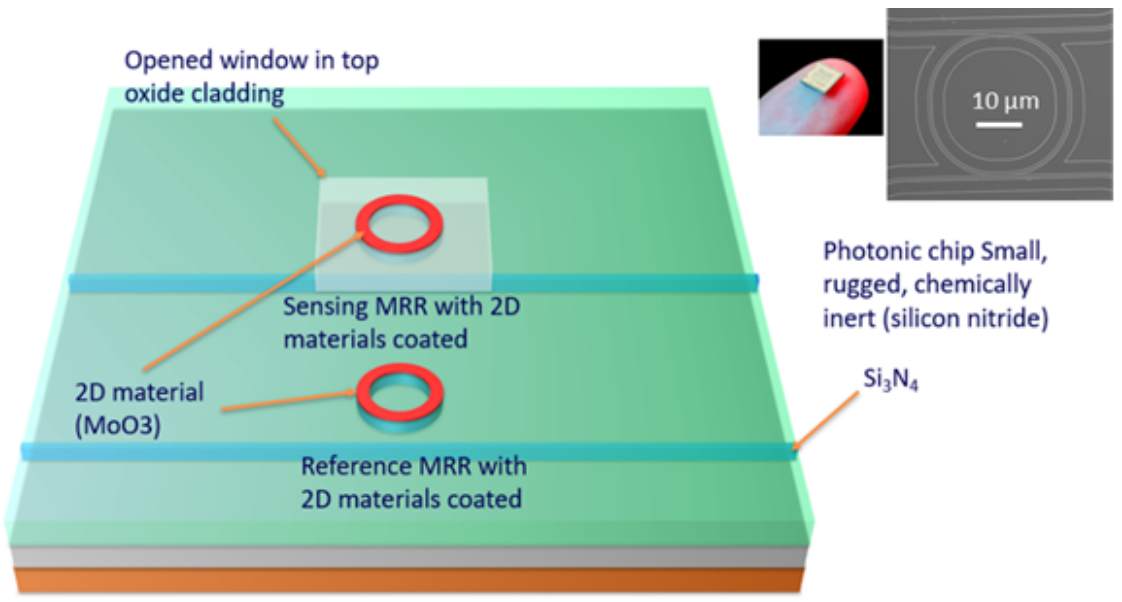

Figure E-4. Schematic of single top window active MRR above with datum reference MRR below (right) actual manufactured prototype chip with MRR balanced on finger (left). Figure repeated from Figure 4-3.

MoOSA is a solid-state integrated unit with no moving parts. The overall dimensions of a flight-ready unit will be less than 40 mm x 25 mm x 10 mm, and it will have a maximum mass of 200 g - including its own battery power source. After the separation of the atmospheric aeroshell of the entry probe, each MoOSA unit will be activated via a short-range blue-tooth link from the atmospheric platform's command unit. Detections in each pH category will be stored within MoOSA and transmitted periodically to the platform's data collection unit. Each MoOSA unit will be vibration tested and built to withstand 200 g deceleration expected at Venus entry and a temperature range from -50 to 100 degrees centigrade.

MoOSA units will be manufactured with the Micro-Nano Research Facility within RMIT University, Melbourne, Australia [180]. The near-infrared micro-laser will be sourced from a local specialist company.

A prototype test unit (see Figure E-4) comprising of a single 20 micron diameter MRR coated with $MoO_3$ has already been tested with nitric acid and showed a reliable linear output response in the range $1 < pH < 6$ [198]. The current NASA-defined readiness status is TRL = 4.

## E.2 Alternative Gas Analyzer: The Tartu $O_2$ Sensor

Molecular oxygen ($O_2$) is a key gas of interest as it is considered a genuine biosignature. A dedicated oxygen sensor will inform us of the abundance of $O_2$ in the clouds of Venus and confirm previous in situ measurements of ppm levels of oxygen by Venera-14 and Pioneer Venus [51,52]. An existing $O_2$ sensor, originally developed at MIT, [199] capable of measuring $O_2$ at ppm levels can be adapted for Venus (Figure E-5). Measurements of $O_2$ in the range of ppm for 1 bar atmosphere (which are the $O_2$ abundance that Venera and Pioneer detected) can be readily achieved with minimal modifications to the existing instrument. (The sensor has also been tested in low vacuum and it can measure equivalent $O_2$ partial pressures).

The design is based on the luminescence lifetime change of complexes of meso-tetra(pentafluorophenyl)porphyrin (Pd-TFPP) and measures it by detecting a phase shift between sinusoidal excitation and emission waveforms. Such means of $O_2$ detection are highly fault tolerant and would work very well in the Venusian atmosphere.

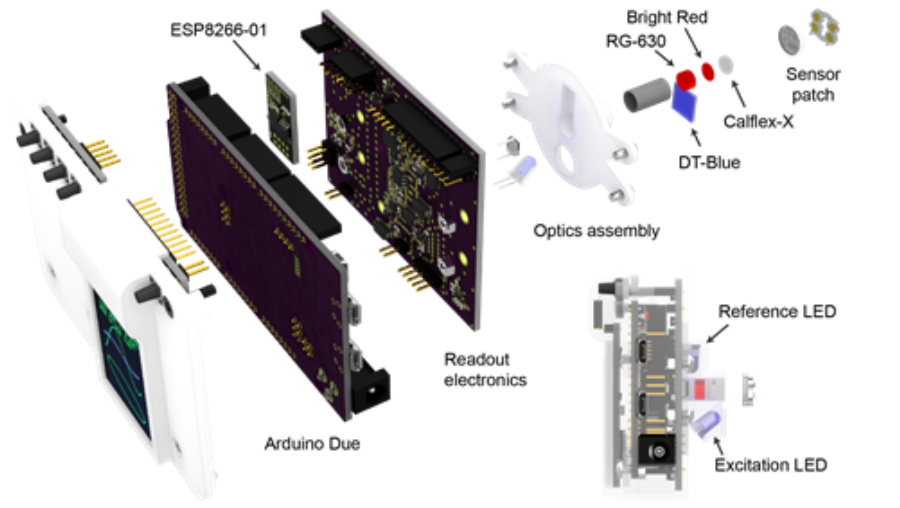

Figure E-5. Current design of the Tartu $O_2$ sensor [199], intended for terrestrial lab use, before adaptation to Venusian cloud conditions.

The original $O_2$ sensor is optimized to work below 100 microbars of oxygen partial pressure with the maximum resolution and theoretical detection limit in tens of nanobars of oxygen (would correspond to 100 ppm and tens of ppb at 1 atmosphere), if well calibrated. The sensor has been demonstrated to work in up to 14 mbar of $O_2$, but the resolution, accuracy and precision starts to significantly decrease past 100 microbars, reducing data quality. The sensitivity can be further limited due to dye aging, temperature fluctuations, electromagnetic interference, and other issues.

For an accurate reading in the Venusian context, a temperature probe has to be added to measure the temperatures of the electronics board and the dye (lab conditions have relatively stable temperatures, so the existing version does not include these features). The existing $O_2$ sensor has been proven to work in the temperature range 20-40 ℃ and has TRL 4 in the context of VLF. Wider temperature ranges have not been validated due to lack of previous need, but it is expected that the sensor can be qualified for a significantly wider temperature range (electronics can be made to industrial or automotive standards to allow -40 to +85 rating, resilience of the dye remains to be determined).

# Appendix F: Venus Atmosphere Sample Return Science and Measurements

**Abstract:** A sample return of Venus atmosphere and cloud particles is vital for a definitive search for signs of life and life itself. With a sample in hand we can use the most sophisticated analytical techniques on Earth that are not readily available for effective cost and risk, or at all, for in situ planetary studies. Any potential detection and structural characterization of complex organic molecules, for example the analogs of genetic polymers, combined with direct visualization of cell-like structures would provide most conclusive evidence for life. Other potential biosignatures, including isotope ratios or biosignature gases, can be used as additional indicators of the presence of life. We review signs of life and analytic techniques to measure them. This Appendix is a supplement to Chapter 6 in the main document.

## F.1. Science on the Return Sample

We review biosignatures from a sample return in an approximate order from the most likely to the least likely to provide conclusive evidence of life. The detection of complex organic molecules, such as analogs of genetic polymers, complemented by the direct visualization of cell-like structures is, in our view, the most conclusive evidence for life. The presence of homochirality is also compelling sign of life. Other potential biosignatures, including isotope ratios or biosignature gases can be used as additional indicators of the presence of life.

**Complex organic molecules** are large organic molecules that have many diverse atoms and bonds. The larger and the more complex an organic molecule is, the less likely the molecule is to be the result of abiological formation. Complex metabolites, especially those that have diverse polymeric structures built from many different monomers, are clear indicators of the presence of extant life. Examples of such complex molecules include genetic polymers like DNA, RNA, and protein polymers built from diverse but specific sets of amino acid monomers that often have complex folded structures.

Any detection of structural regularities in complex chemicals could, in fact, be indicative of biosynthetic pathways and therefore of life's activity. Such clear structural regularities in identified organic molecules (e.g. reminiscent of lipid chain length preferences of life on Earth, or specific preferences for nucleotides in genetic polymers) add additional weight to the biological origin of the analyzed material.

**Distinct compartmentalized structures.** It is likely that any kind of life, no matter its chemical makeup, needs barriers (cell walls, membranes etc.) that allow for life to exist in a distinct form, separated from the surrounding environment (i.e., to maintain a cell-like structure). Detection and characterization of morphological indicators of life is evidence for the existence of life in the examined sample. Detection of such cell-like structures should always be followed up with detailed chemical analysis of such structures.

**Homochirality** is a distinguishing feature of life on Earth. In Earth's biochemistry, 19 of the 20 natural amino acids found in proteins are homochiral, being L-chiral (left-handed), while the D-chiral (right-handed) amino acids are excluded from the protein polymers. Biogenic sugars on the other hand are D-chiral. Such uniformity of chirality is considered to be a biosignature and a sign that the molecules are of biological origin.

**Gases and volatile species associated with life.** All life on Earth generates gaseous products, and basic chemistry suggests the same will be true of any other plausible biochemistry. A good biosignature gas is a gas that is exclusively produced by life and that does not have any significant geological or photochemical sources (or any other formation pathways that could be ascribed to planetary processes). The detection of such gases that "do not belong" in the atmosphere is an indicator of possible biological processes. Simultaneous detection of reduced and oxidized species (e.g. $CH_4$ and $O_2$ on Earth) are taken to be strong evidence of the presence of life [47,48,50]. On Venus such a reduced and oxidized pair could include $O_2$ and $NH_3$ (see Chapter 1) or other reduced gases, including $PH_3$ and $CH_4$.

**Isotope ratios.** Life on Earth generally prefers light isotopes of elements as building blocks of biochemistry. Such biological isotopic fractionation is one of the signposts of life on Earth and in principle a possible indicator for life elsewhere. For example, on Earth the light carbon-12 isotope is preferentially incorporated in biochemicals while the heavier carbon-13 is preferentially excluded. Carbon-12 preference is readily apparent on Earth where one can readily distinguish $CH_4$ produced by bacteria or archaea from $CH_4$ produced by abiotic processes. Distinguishing between biological and non-biological $CH_4$ can be done by analyzing the isotopic composition and studying ratios of carbon-12 to carbon-13 (most $CH_4$ on Earth is biological). Similar isotopic analysis can be done for sulfur. The light sulfur sulfur-32 isotope is used preferentially in biochemistry while heavier isotopes, such as sulfur-33 and sulfur-34 are excluded. Detecting an enrichment of sulfur-32 over the heavier sulfur isotopes could be a sign of life, e.g. an indication of biological sulfate reduction or other sulfur dependent metabolism.

## F.2 Analytical Methods of Identification

We briefly discuss the analytical techniques that can be employed in the search, detection and identification of biosignatures in the returned samples. A major challenge is that most of the analytical methods that are discussed in this chapter are not readily compatible with concentrated sulfuric acid. Separate protocols for sample handling and preparation or instrument adaptation to sulfuric acid is required. The order of discussed analytical techniques follows the subjective order of the importance of biosignatures in life detection discussed above. See Table F-1.

**Ultra-high-performance liquid-chromatography coupled with mass spectrometry (UHPLC-MS)** combines the characteristics of liquid chromatography and mass spectrometry. UHPLC-MS is a modern chromatographic method that achieves very high separation performance, with columns packed with sub-2 μm particles, in conjunction with a system that withstands very high pressures (up to 1500 bar) [200]. The goal of this chromatographic approach is to increase sample throughput and maximum number of resolvable peaks, therefore separate and detect low abundance species (often used in forensic toxicology) [201].

**Gas chromatography and mass spectrometry (GC-MS**) is an analytical method that combines the characteristics of gas-chromatography and mass spectrometry to identify different substances within a test sample [202]. GC-MS is an established and reliable analytical method of detection of unknown or trace species often used in global metabolomic investigations. The high-temperature treatment of the sample, however, might lead to undesirable sample thermal degradation [203].

**ESI-Orbitrap-MS** combines the advantages of soft ionization (electrospray ionization (ESI)) with high-resolution mass spectrometry (orbitrap-MS). The high mass resolution beyond $m/\Delta m > 100,000$ allows resolving isobaric interference and therefore accurate identification and assignment of chemical species within the analyte investigated. ESI-Orbitrap-MS is suitable for, e.g., identification of complex biomolecules.

**Tandem mass spectrometry (MS/MS)** in its most common design combines two mass analyzers (typically quadrupole mass analyzers) that are separated from each other by another reaction compartment that induces e.g., collision or photodissociation reactions. This combination allows the investigation of chemical species that fall within a certain mass range of interest. MS/MS is typically applied for samples that contain mixtures of various molecules.

**Fluorescence spectroscopy** takes the advantage of the induced fluorescence of chemical species, typically by UV light. The light emitted by the analyte is measured with a spectrometer. The technique is especially useful for the non-destructive detection of autofluorescence of organic molecules.

**Fluorescence and UV-VIS-NIR microscopy.** UV-VIS-NIR microscopy allows the investigation of the sample in the several regions of the electromagnetic spectrum, including visible (VIS), near infrared (NIR) and ultraviolet (UV). In fluorescence microscopy the sample is typically illuminated with UV light which induces fluorescence of the chemical species in the sample and allows for detection of organic chemicals, including any potential biological material.

**Nanotomography (Nano-CT)** is a non-destructive technique that uses X-rays to create cross-sections from a 3D-object used to recreate a virtual model without destroying the original sample. The nano-CT technique has very broad applications and has been applied to a wide variety of 3D visualization studies [204], including single cells and tissues, novel materials, and comet samples returned by the Stardust mission [205]).

**Atomic force microscopy (AFM)** allows for a detailed characterization of a surface at micrometer and nanometer level, including surface features and roughness. Surface topology of any nanoscale material/object can be examined. AFM is agnostic towards the identity of the examined material and can cover scales from biomolecule-size to cell-size. AFM is an established method in the laboratory setting, but so far there are no publicly-available results of its successful operation in a planetary environment, possibly due to the extreme mechanical and temperature sensitivity required [206].

**Circular dichroism spectroscopy (CDS)** is an absorption spectroscopy method based on the differential absorption of left and right circularly polarized light. CDS is predominantly used for determination of the secondary structure of optically active species, like proteins, polypeptides, amino acids, and other chiral molecules. Chiral, optically active molecules mainly absorb only one direction (left or right) of circularly polarized light. Such differences in absorption of the circularly polarized light can be measured and quantified. There are several types of CDS. UV CDS is used to, e.g., investigate secondary structure of polypeptide chains. Vibrational CDS or IR CDS is used to investigate the structure of optically active small organic molecules. UV/Vis CDS is used to study metal complexation by organic molecules, including proteins.

**Time-of-flight mass spectrometry (TOF-MS)** is a variant of mass spectrometry that utilizes a pulsed ion source, such as a laser or electron ionization including pulsed extraction. After the ionization process all ions are pushed into the ion optical system of the mass analyzer. The ionized species arrive at the detector system according to their mass-to-charge ratio. Contrary to scanning mass analyzers, such as Quadrupole-MS (QMS) systems, TOF-MS is a non-scanning method, meaning that the chemical fingerprint of the analyte can be measured at once, with every push of the ions into the mass analyzer.

**UV-VIS-NIR, IR spectroscopy (including TLS)** is an analytical technique used to determine the optical properties (absorbance, transmittance, reflectance) of solids, liquids, and gases. UV/VIS/NIR spectroscopy is especially useful for determination of the optical properties of liquids and solids between 175 nm and 3300 nm and for the determination of the concentrations of a wide variety of analytes in solution. Infrared spectroscopy (IR), including the TLS infrared laser absorption spectrometer, excels at trace gas identification.

**High-resolution Raman spectroscopy** is a non-destructive chemical analysis technique which provides detailed information about chemical structure, molecular interactions, crystallinity, phase, and polymorphy of the investigated sample. Typically a Raman spectrum is a distinct chemical fingerprint for a particular molecule or material, and can be used to very quickly identify the investigated substance, including specific chemical bonds, functional groups, lattice modes, etc. [207,208].

| Biosignature | Analytical Method of Identification | Analytical Method Output |
|---|---|---|
| Complex organic molecules (e.g. fatty acids, primary & secondary metabolites, and polymers); Structural regularities indicating common biosynthesis pathways. | UHPLC-MS | Identification of high molecular weight chemicals; separation of isobaric species. |
| | GC-MS | |
| | ESI-Orbitrap-MS | |
| | MS/MS | |
| | Fluorescence Spectroscopy | |
| Distinct compartmentalized structure (e.g. viruses, microbial cells) | UV-VIS-NIR and fluorescence microscopy and spectroscopy | Identification of cells, determination of cellular morphology and density |
| | Nano-CT scan | High resolution 3D structure of particles |
| | Atomic force microscopy (AFM) | Surface topology of any nanoscale material/object |
| Homochirality (e.g. all L or all D building blocks in informational biopolymers); on Earth, simultaneous detection of a group of specific amino acids. | UHPLC-MS | Identification of chiral enrichment; increased abundance of certain amino acids |
| | CDS | |
| Gases and volatile species associated with life (e.g., $CH_4$, $PH_3$, and $NH_3$). | TOF-MS | Detection of chemical disequilibria in the atmosphere |
| | GC-MS | |
| | IR spectroscopy (including TLS) | |
| Isotope ratios (e.g. preferential use of light isotopes of S or C) | Hi-Res Raman spectroscopy | Mapping of functional groups and elemental/isotopic signatures; Isotope ratio of high precision and epsilon accuracy |
| | Laser-ICP-MS | |
| | TIMS | |
| | IRMS | |

Table F-1. Draft list of biomarkers and analytical techniques for a Venus atmosphere sample. We are not searching for Earth-like life and components, so this list a starting point to build on. In addition, experimental equipment and procedures need to be adapted to withstand concentrated sulfuric acid.

**Laser ablation inductively coupled plasma mass spectrometry (LA-ICP-MS).** In LA-ICP-MS a pulsed laser beam removes shot-by-shot a distinct layer of sample material which is subsequently ionized in the ICP system. The generated ions are measured with a MS system, such as TOF. LA-ICP-MS combines the advantages of spatially resolved analysis of the sample at the micron level (lateral, spot size of the focused laser beam) and the high detection sensitivity of the ICP-MS system, allowing for a very favorable limit of detection (ppb level and lower, weight fraction) [209,210].

**Thermal ionization mass spectrometry (TIMS)** is one of the state-of-the-art measurement techniques for accurate measurements of isotope ratios, such as Pb or S. Isotope ratio accuracies at the epsilon level are achieved. The sample is thermally ionized inside the system and the generated ions are subsequently measured with a mass spectrometric system.

**Continuous-flow-isotope-ratio mass spectrometer (IRMS).** Similar to TIMS, IRMS represents a state-of-the-art measurement technique for high accurate element isotope analysis, e.g., for detection of an enrichment of sulfur-32 over the heavier sulfur isotopes. Such enrichment could be a sign of life.